\documentclass[11pt,a4paper]{article}
\pdfoutput=1
\usepackage{jheppub}

\usepackage{anyfontsize}

\usepackage[mode=buildnew]{standalone}
\usepackage{comment}
\usepackage{amsmath,amstext,amsgen,amsbsy,amsopn,amsfonts,amsthm}
\usepackage{tikz}
\usetikzlibrary{shapes,decorations,arrows,automata,backgrounds,petri}
\usetikzlibrary{shapes.misc}
\usetikzlibrary{decorations.pathreplacing}
\usepackage{array}
\usetikzlibrary{patterns}
\usepackage{booktabs}
\usepackage{bm}
\usepackage{graphicx}
\usepackage{caption}
\usepackage{subcaption}
\usepackage{enumerate}
\usepackage{multirow}
\usepackage{hhline}
\usepackage{afterpage}

\tikzset{cross/.style={cross out, draw=black}}
\tikzset{circ/.style={circle,fill=white,draw=black}}
\tikzset{hasse/.style={circle, fill,inner sep=2pt}}
\tikzset{h/.style={circle, fill,inner sep=2pt}}
\tikzset{ns/.style={circle, draw,inner sep=2pt}}

\newtheorem*{constraint}{Constraint}
\newtheorem*{prescription}{Prescription}

\usepackage[latin1]{inputenc}

\newcommand{\M}{\mathcal{M}}
\newcommand{\Mc}{\mathcal{M}_C}
\newcommand{\Mh}{\mathcal{M}_H}
\newcommand{\Or}{\bar{\mathcal{O}}}
\newcommand{\gsl}{\mathfrak{sl}}
\newcommand{\gso}{\mathfrak{so}}
\newcommand{\gsp}{\mathfrak{sp}}
\newcommand{\lam}{\lambda}

\preprint{Imperial/TP/17/AH/06}

\title{Branes and the Kraft-Procesi transition: classical case}
\author[a]{Santiago Cabrera}
\author[a]{and Amihay Hanany}
\affiliation[a]{Theoretical Physics, The Blackett Laboratory, Imperial College London\\
Prince Consort Road, SW7 2AZ United Kingdom}
\emailAdd{santiago.cabrera13@imperial.ac.uk}
\emailAdd{a.hanany@imperial.ac.uk}

 \abstract{Moduli spaces of a large set of $3d$ $\mathcal{N}=4$ effective gauge theories are known to be closures of nilpotent orbits. This set of theories has recently acquired a special status, due to Namikawa's theorem. As a consequence of this theorem, closures of nilpotent orbits are the simplest non-trivial moduli spaces that can be found in three dimensional theories with eight supercharges. In the early 80's mathematicians Hanspeter Kraft and Claudio Procesi characterized an inclusion relation between nilpotent orbit closures of the same classical Lie algebra. We recently \cite{CH16} showed a physical realization of their work  in terms of the motion of D3-branes on the Type IIB superstring embedding of the effective gauge theories. This analysis is restricted to A-type Lie algebras. The present note expands our previous discussion to the remaining classical cases: orthogonal and symplectic algebras. In order to do so we introduce O3-planes in the superstring description. We also find a brane realization for the mathematical map between two partitions of the same integer number known as \emph{collapse}. Another result is that basic Kraft-Procesi transitions turn out to be described by the moduli space of orthosymplectic quivers with varying boundary conditions.}


\begin{document}

\maketitle
\flushbottom

\section{Introduction}

In the previous work \cite{CH16} a new relation was found between brane dynamics in Type IIB superstring theory \cite{HW96} and the geometry of nilpotent orbits in the $\mathfrak{sl}(n)$ algebra over the field $\mathbb{C}$. The aim was to describe the brane realization of the mathematical work by Kraft and Procesi \cite{KP81, KP82}. In \cite{KP81} they developed the theory for the $\mathfrak{sl}(n)$ algebra and in \cite{KP82} they expanded it to the cases $\mathfrak{so}(n)$ and $\mathfrak{sp}(n)$. This paper aims to do the same, to expand the analysis in \cite{CH16} to the other classical cases. In order to do this one utilizes orientifold planes\footnote{The reader is directed to \cite{GK98} for a broad description of the relationship between brane dynamics and effective gauge theories, including the role of orientifold planes.} in the brane configurations, following the construction of \cite{FH00}.

One of the main goals of this paper is to bring attention to the Brieskorn-Slodowy program \cite{B70,Sl80}. This a very interesting way to understand the geometry of a variety and its singularities. As we can read in \cite{Sl80}, whenever a variety $V$ has a regular action of a classical Lie group $G$, if $x\in V$ is a point in the variety, one can find a \emph{transverse slice} $S\subseteq V$ to the G-orbit of $x$. We say that locally the variety \emph{looks like} the direct product $S\times (G\cdot x)$. In this way we can learn a great deal about the geometry of a variety and the nature of its singularities by finding subvarieties which are orbits of its isometry group and computing their transverse slices. In the relevant cases of our study, these slices are always singular varieties.

From the point of view of quantum field theory, the varieties are the different branches of the moduli space (Coulomb branch and Higgs branch) in $3d$ $\mathcal{N}=4$ quiver gauge theories. In the previous paper we showed how slicing the variety in the Brieskorn-Slodowy sense corresponds to performing a Higgs mechanism in the quantum field theory\footnote{The study of mixed branches of $3d \ \mathcal N=4$ (see for example \cite{CartaHayashi16,AC17}) is intimately related to this geometrical notion. The action of going to the mixed branch phase via partial Higgsing can now be understood as a slicing of the moduli space in the Brieskorn-Slodowy sense.}. This is one of the main results that we want to highlight, since it illustrates a new connection between physics and geometry.

From the mathematical point of view, the construction starts with the endeavor of Brieskorn to answer one of the problems raised by Steinberg on 1966: \emph{Study the variety of unipotent elements of a group thoroughly} \cite{St66}. They wanted to understand the geometry of conjugacy classes of Lie groups. Employing Jordan decomposition any element of a group can be decomposed into a product of a \emph{semisimple} element and a \emph{unipotent} one. The conjugacy classes of semisimple elements were already classified, and the remaining task was to analyze the unipotent elements\footnote{Actually Kraft and Procesi \cite{KP82} use the theory of Luna \cite{L73} to show that any attempt on understanding the singularities of closures of conjugacy classes in a classical Lie group can be focussed entirely on the conjugacy classes of unipotent elements.}. As \cite{KP82} points out, for classical Lie groups over the field $\mathbb{C}$, this analysis can be moved to the algebra via the logarithmic map. In the Lie algebra, the problem concerns the variety of \emph{nilpotent} elements.

From this point onwards we can restrict ourselves with the analysis of affine algebraic varieties that are closures of nilpotent orbits of the classical Lie algebras\footnote{Standard texts on nilpotent orbits of Lie algebras are \cite{CM93, Sp82,C85,M02}.}. For a given classical Lie algebra $\mathfrak{g}$ over $\mathbb{C}$ there is a finite number of distinct nilpotent orbits. Furthermore, the closure of a nilpotent orbit of $\mathfrak{g}$ is the union of finitely many nilpotent orbits of $\mathfrak{g}$. Therefore, given any pair of nilpotent orbits $\mathcal{O}$ and $\mathcal{O}'$ such that $\mathcal{O}'\subset \Or$, where $\Or$ is the closure of $\mathcal{O}$, one can always find the slice $S\subseteq \Or$ transverse to $\mathcal{O}'$. This is what Kraft and Procesi did in 1982 for all pairs of nilpotent orbits of any classical Lie algebra\footnote{Fu, Juteau, Levy and Sommers extended this analysis to the exceptional algebras in \cite{FJLS15}.} that are connected by a link in the Hasse diagram of their \emph{partial ordering} structure\footnote{The partial ordering arises with respect to the \emph{inculsion} relation between closures of nilpotent orbits.}. They found that the slices were always singular varieties themselves.

In the present paper we show how these geometrical results have a clear physical interpretation. We describe how the transverse slice can be found in the moduli space by employing the Higgs mechanism. In the case of orbits of the special linear algebra, the analysis becomes particularly straightforward when the $3d$ gauge theory is considered as describing the effective low energy dynamics of a brane configuration in Type IIB superstring theory. This is the system studied in \cite{CH16}. The following pages take the same approach for the remaining classical cases: closures of nilpotent orbits in orthogonal and symplectic algebras. Once again, we introduce a simple formalism that allows us to compute the quiver of the relevant gauge theories, their brane embeddings and the corresponding Kraft-Procesi transitions.

Nilpotent orbits of Lie algebras have seen some relevance in theoretical physics in the past. In particular,  they appear every time there is an embedding of $SU(2)$ into a different group, like for example in the Nahm equations \cite{N79}. An incomplete selection of other examples where they appear could be \cite{Bachas:2000dx,GW09,Gukov:2008sn,Kim:2010bf,GMV11,CDT13,Bourget:2015lua,Heckman:2016ssk}. A recent theorem by Namikawa \cite{N16} establishes that the simplest non-trivial moduli spaces that can be found in $3d$ $\mathcal{N}=4$ effective gauge theories are closures of nilpotent orbits\footnote{See \cite{CH16} for a more detailed explanation of the physical implications of Namikawa's theorem.}.

Section \ref{sec:2} of the present note summarizes the results obtained in \cite{CH16}. Section \ref{sec:3} introduces the brane configurations of \cite{FH00} and the corresponding \emph{orthosymplectic} quivers, employing some conventions developed by  \cite{GW09}. Section \ref{sec:dual} describes the maps between partition sets that are necessary in the classification of brane systems related to nilpotent orbits and their corresponding quiver gauge theories. Section \ref{sec:so4interlude} illustrates our description with examples corresponding to the algebra $\mathfrak{so}(4)$. The new Kraft-Procesi transitions for orthogonal algebras are introduced in this section, discussing the previous examples. In section \ref{sec:sigmarho} we describe the general relation between brane systems and orthosymplectic quivers whose moduli spaces are closures of nilpotent orbits. Section \ref{sec:so5interlude} discusses a second set of examples, in this case corresponding to algebras $\mathfrak{so}(5)$ and $\mathfrak{sp}(2)$. Section \ref{sec:11} gives the general description for Kraft-Procesi transitions of orthogonal and symplectic algebras in term of brane configurations/quiver gauge theory. Section \ref{sec:12} describes the matrix formalism developed to implement efficient computations of the transitions. Section \ref{sec:13} contains all the results obtained from this method and in section \ref{sec:14} we discuss some conclusions. 

%
%
%


\section{Summary of the brane description of Kraft-Procesi transitions for $\mathfrak{sl}$(n)}\label{sec:2}

In \cite{CH16}, we devote \emph{Section 3: Mathematical prelude} to introduce fundamental mathematical tools that are of great importance to our study. These are \emph{nilpotent orbits of Lie algebras} and \emph{hyperk\"ahler singularities}. In particular we review the theory of the polynomial ring of holomorphic functions on hyperk\"ahler singularities. This gives us the necessary apparatus to explain Namikawa's theorem \cite{N16}. The main result of the theorem is that closures of nilpotent orbits are the simplest kind of hypek\"ahler singularities. From the point of view of physics, this means that the simplest moduli spaces that we can find in $3d$ $\mathcal{N}=4$ gauge theories are closures of nilpotent orbits. Furthermore, this theorem implies that other moduli spaces that are not closures of nilpotent orbits can be understood as extensions of them. We do not reproduce this section again here, but we encourage the reader to take a look, since all the derivations are very basic, and yet powerful enough to motivate our research efforts.

\subsection{Brane systems and quivers for unitary theories}

The reader is directed to  \cite{HW96} for a detailed introduction to linear unitary $3d$ $\mathcal{N}=4$ quiver gauge theories and their superstring embedding. In summary, we have a Type IIB superstring configuration with space-time coordinates $x^\mu$, $\mu=0,1,\dots,9$. Let there be D3-branes spanning space directions $\{x^1,x^2,x^6\}$ with position along $\vec{x}:=(x^3,x^4,x^5)$ and $\vec{y}:=(x^7,x^8,x^9)$. Let there also be NS5-branes spanning space directions $\{x^1,x^2,x^3,x^4,x^5\}$  and let us call $(t_i, \vec{w}_i)$ the position of the $i$-th NS5-brane along $(x^6,\vec{y})$. Let there also be D5-branes spanning space directions $\{x^1,x^2,x^7,x^8,x^9\}$ and let $(z_j, \vec{m}_j)$ be the position of the $j$-th D5-brane along $(x^6,\vec{x})$. This configuration allows for 8 supercharges to be preserved \cite{HW96}. If we only consider D3-branes starting and ending in fivebranes along the $x^6$ direction, the low energy dynamics is described by a $3d$ $\mathcal{N}=4$  effective gauge theory living in the worldvolume of the D3-branes.

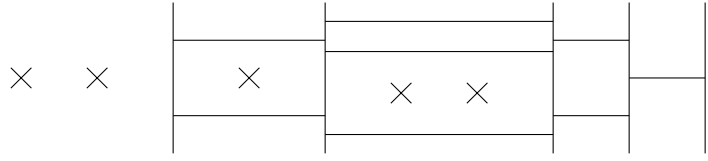
\begin{figure}[t]
	\centering
	\begin{tikzpicture}
		\draw (-1,1) node[cross]{};
		\draw (0,1) node[cross]{};
		\draw (1,0)--(1,2);
		\node [cross] at (2,1){};
		\draw (3,0)--(3,2);
		\node [cross] at (4,.8){};
		\node [cross] at (5,.8){};
		\draw (6,0)--(6,2);
		\draw (7,0)--(7,2);
		\draw (8,0)--(8,2);
		\draw (1,.5)--(3,.5)
			(1,1.5)--(3,1.5);
		\draw (3,.25)--(6,.25)
			(3,1.35)--(6,1.35)
			(3,1.75)--(6,1.75);
		\draw (6,.5)--(7,.5)
			(6,1.5)--(7,1.5);
		\draw (7,1)--(8,1);
	\end{tikzpicture}
	\caption{Generic brane configuration. The horizontal direction corresponds to spatial direction $x^6$. Horizontal lines are D3-branes. The vertical direction corresponds to spatial directions $\vec{x}=(x^3,x^4,x^5)$. Vertical lines are NS5-branes. The direction perpendicular to the paper corresponds to spatial directions $\vec{y}=(x^7,x^8,x^9)$. Crosses are D5-branes.}
	\label{fig:firstExample}
\end{figure}

For the brane configurations that are considered here, a phase transition can always be performed so all D3-branes end only on NS5-branes.  Once the system is in this particular phase, we give different positions to all fivebranes along $x^6$. Let figure \ref{fig:firstExample} be a generic example. The quiver describing the content of the effective gauge theory can be read in the following way:
\begin{itemize}
	\item For each interval $i$ between two neighboring NS5-branes we have a \emph{gauge node} in the quiver with label $n_i$, where $n_i$ is the number of D3-branes that can be found in such interval. The $i$-th gauge node is connected to the $(i+1)$-th one by an edge in the quiver.
	\item For each interval $j$ between two neighboring NS5-branes we have a \emph{flavor node} in the quiver with label $k_j$, where $k_j$ is the number of D5-branes that can be found in such interval. The $j$-th flavor node is connected to the $j$-th gauge node by an edge in the quiver. If $k_j=0$ we can omit the flavor node and the corresponding edge.
\end{itemize}

\begin{figure}[t]
	\centering
	\begin{tikzpicture}
	\tikzstyle{gauge} = [circle, draw];
	\tikzstyle{flavour} = [regular polygon,regular polygon sides=4, draw];
	\node (g1) [gauge, label=below:{$2$}] {};
	\node (g2) [gauge,right of=g1, label=below:{$3$}] {};
	\node (g3) [gauge,right of=g2, label=below:{$2$}] {};
	\node (g4) [gauge,right of=g3, label=below:{$1$}] {};
	\node (f1) [flavour, above of=g1, label=above:{$1$}] {};
	\node (f2) [flavour,above of=g2, label=above:{$2$}] {};
	\draw (g1)--(g2)--(g3)--(g4)
		(f1)--(g1)
		(f2)--(g2);
	\end{tikzpicture}
	\caption{Quiver corresponding to brane configuration depicted in figure \ref{fig:firstExample}. The circles are \emph{gauge nodes} while the squares are \emph{flavor nodes}.}	\label{fig:firstExampleQuiver}
\end{figure}
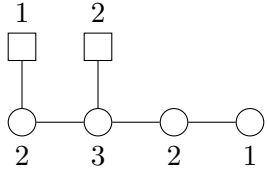

The quiver\footnote{Note that these are precisely the quivers built by Nakajima in \cite{N94}, with $\mathbf{v}$ the array with the ranks of the gauge nodes and $\mathbf{w}$ the array with the ranks of the flavor nodes. The Higgs branch of the quivers is always a hyperk\"ahler variety and it has received the generic name of \emph{Nkajima's quiver variety}.} corresponding to the brane system in figure \ref{fig:firstExample} is depicted in figure \ref{fig:firstExampleQuiver}. The field content of the corresponding $3d$ $\mathcal{N}=4$ effective gauge theory can be read from a generic quiver: 

\begin{itemize}
	\item \textbf{Gauge group:} Each gauge node with label $n_i$ represents a factor $U(n_i)$ of the gauge group $G$:
	\begin{align}
	G=U(n_1)\times\dots\times U(n_{l-1})
	\end{align}
	where $l$ is the number of NS5-branes. 
	\item \textbf{Flavor group:} Each flavor node with label $k_j$ represents a factor $U(k_j)$ of the flavor group $F$:
	\begin{align}
	F=S\left(U(k_1)\times\dots\times U(k_{l-1})\right)
	\end{align}
	The symbol $S(...)$ denotes that a factor of $U(1)$ is removed. This is analogous to the \emph{center of mass} decoupling from the rest of the system. If there where no flavor group, the $U(1)$ factor is removed from the gauge group $G$ instead.
	\item \textbf{Vector multiplets:} For each gauge node with label $n_i$ there are $n_i^2$ vector multiplets transforming under the adjoint representation of $U(n_i)$.
	\item \textbf{Hypermultiplets:} For each edge connecting a gauge node $k_i$ with a flavor node $n_i$ there are $k_i\times n_i$ hypermultiplets transforming in the fundamental representation of $U(k_i)$ and the fundamental representation of $U(n_i)$. For each edge connecting the gauge nodes $n_i$ and $n_{i+1}$ there are $n_i\times n_{i+1}$ hypermultiplets transforming in the fundamental representation of $U(n_i)$ and the fundamental representation of $U(n_{i+1})$.
\end{itemize}

The gauge symmetry group $G$ and the flavor symmetry group $F$ of the example quiver in figure \ref{fig:firstExampleQuiver} are:
\begin{align}
	\begin{aligned}
		F&=S(U(1)\times U(2))\\
		G&=U(2)\times U(3)\times U(2)\times U(1)\\
	\end{aligned}
\end{align}

\subsection{Brane systems and quivers for surface and minimal singularities}

There are two crucial types of hyperk\"ahler singularities that in \cite{KP82} are called \emph{surface singularities} and \emph{minimal singularities}. These are enough to describe the nature of the singularities in closures of nilpotent orbits of $\mathfrak{sl}(n)$. In our case the surface singularities are Kleinian singularities of the form:

\begin{align}
	A_n := \mathbb{C}^2/\mathbb{Z}_{n+1}
\end{align}

The minimal singularities are closures of minimal nilpotent orbits:

\begin{align}
	a_n := \Or_{(2,1^{n-1})}
\end{align}

where $\Or_{(2,1^{n-1})}$ is the closure of the minimal nilpotent orbit of the algebra $\mathfrak{sl}({n+1})$ over the field $\mathbb{C}$ (in physics this variety is known as the \emph{reduced one instanton moduli space}). Here we use the same notation of \cite{CH16} and refer to its \emph{Section 3: Mathematical prelude} for an introduction on closures of nilpotent orbits.

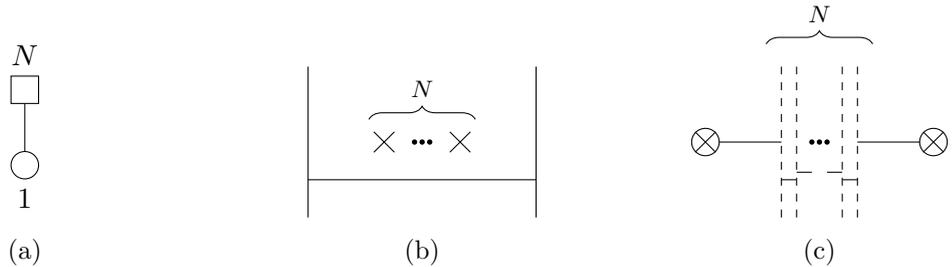
\begin{figure}[t]
	\centering
	\begin{subfigure}[t]{.30\textwidth}
    \centering
	\begin{tikzpicture}
	\tikzstyle{gauge} = [circle,draw];
	\tikzstyle{flavour} = [regular polygon,regular polygon sides=4,draw];
	\node (g1) [gauge,label=below:{$1$}] {};
	\node (f1) [flavour,above of=g1,label=above:{$N$}] {};
	\draw (g1)--(f1)
		;
	\end{tikzpicture}
        \caption{}
    \end{subfigure}
    \hfill
	\begin{subfigure}[t]{.30\textwidth}
    \centering
	\begin{tikzpicture}
		\draw (1,0)--(1,2);
		\draw (4,0)--(4,2);
		\draw (1,.5)--(4,.5);
		\draw (2,1) node[cross] {};
		\draw (3,1) node[cross] {};
		\fill (2.5,1) circle [radius=1pt];
		\fill (2.40,1) circle [radius=1pt];
		\fill (2.60,1) circle [radius=1pt];
		\draw [decorate,decoration={brace,amplitude=5pt}](1.8,1.3) -- (3.2,1.3) node [black,midway,xshift=-0.6cm] { };
		\draw node at (2.5,1.7) {\footnotesize $N$};
	\end{tikzpicture}
        \caption{}
    \end{subfigure}
    \hfill
    \begin{subfigure}[t]{.30\textwidth}
    \centering
	\begin{tikzpicture}
		\draw[dashed] 	(2,0)--(2,2)
				(3,0)--(3,2)
				(2.2,0)--(2.2,2)
				(2.8,0)--(2.8,2);
		\draw (2,.5)--(2.2,.5)
				(2.8,.5)--(3,.5)
				(2.2,.6)--(2.4,.6)
				(2.6,.6)--(2.8,.6)
				(1,1)--(2,1)
				(3,1)--(4,1);
		\fill (2.5,1) circle [radius=1pt];
		\fill (2.40,1) circle [radius=1pt];
		\fill (2.60,1) circle [radius=1pt];
		\draw [decorate,decoration={brace,amplitude=5pt}](1.8,2.3) -- (3.2,2.3) node [black,midway,xshift=-0.6cm] { };
		\draw node at (2.5,2.7) {\footnotesize $N$};
		\draw 	(1,1) node[circ]{}
				(4,1) node[circ]{};
		\draw 	(1,1) node[cross]{}
				(4,1) node[cross]{};
	\end{tikzpicture}
	\caption{}
	\end{subfigure}
	\hfill
 	\caption{Model $3d$ $\mathcal{N}=4$ SQED with $N$ flavors. (a) Quiver. (b) Coulomb branch brane configuration. There are $N$ D5-branes. (c) Higgs branch brane configuration. In (c) a rotation with respect to (b) has been performed: the vertical direction corresponds with spatial directions $\vec{y}$ and the direction perpendicular to the paper corresponds with spatial directions $\vec{x}$. Vertical dashed lines now represent D5-branes (there are $N$ of them) and circled crosses represent NS5-branes.}
	\label{fig:SUNU1}
\end{figure}

From the point of view of physics, the $3d$ $\mathcal{N}=4$ quiver gauge theory known as \emph{SQED with N flavors} has a quiver depicted in figure \ref{fig:SUNU1}(a). The moduli space of this theory has two distinct phases: the Coulomb branch where only the massless vectorplet acquires nonzero vacuum expectation value, and the Higgs branch, where $N-1$ of the hypermultiplets become massless and acquire nonzero VEVs, while the vector multiplet becomes massive. Both branches meet at the singular point where all fields are massless. The Coulomb branch of this theory is in fact the singular variety $A_{N-1}$, this is denoted by: 
\begin{align}
	\begin{aligned}
		\Mc &= A_{N-1}
	\end{aligned}
\end{align}
The Higgs branch is the variety $a_{N-1}$. We write:
\begin{align}
	\begin{aligned}
		\Mh &= a_{N-1}
	\end{aligned}
\end{align}

The brane configuration that corresponds to this quiver is depicted in figure \ref{fig:SUNU1}(b). This corresponds to the Coulomb branch of the moduli space. A phase transition can take us to the brane configuration corresponding to the Higgs branch of the moduli space, depicted in figure \ref{fig:SUNU1}(c).

\begin{figure}[t]
	\centering
	\begin{subfigure}[t]{.30\textwidth}
    \centering
    \raisebox{-.2\height}{
	\begin{tikzpicture}
	\tikzstyle{gauge} = [circle, draw];
	\tikzstyle{flavour} = [regular polygon,regular polygon sides=4, draw];
	\node (g1) [gauge,label=below:{$1$}] {};
	\node (g2) [gauge, right of=g1,label=below:{$1$}] {};
	\node (gd) [right of=g2] {$\dots$};
	\node (g3) [gauge, right of=gd,label=below:{$1$}] {};
	\node (g4) [gauge, right of=g3,label=below:{$1$}] {};
	\node (f1) [flavour,above of=g1,label=above:{$1$}] {};
	\node (f4) [flavour,above of=g4,label=above:{$1$}] {};
	\draw (g1)--(g2)
			(g3)--(g4)
			(g1)--(f1)
			(g4)--(f4)
		;
	\draw [decorate,decoration={brace,mirror,amplitude=5pt}](-.5,-.8) -- (4.5,-.8) node [black,midway,xshift=-0.6cm] { };
		\draw node at (2,-1.3) {\footnotesize $N-1$};
	\end{tikzpicture}}
        \caption{}
    \end{subfigure}
    \hfill
	\begin{subfigure}[t]{.30\textwidth}
    \centering
	\begin{tikzpicture}
		\draw	(1.2,0)--(1.2,2)
				(3.8,0)--(3.8,2)
				(2.2,0)--(2.2,2)
				(2.8,0)--(2.8,2);
		\draw 	(1.7,1) node[cross]{}
				(3.3,1) node[cross]{};
		\draw (1.2,.5)--(2.2,.5)
				(2.8,.5)--(3.8,.5)
				(2.2,.6)--(2.4,.6)
				(2.6,.6)--(2.8,.6);
		\fill (2.5,1) circle [radius=1pt];
		\fill (2.40,1) circle [radius=1pt];
		\fill (2.60,1) circle [radius=1pt];
		\draw [decorate,decoration={brace,amplitude=5pt}](2.1,2.1) -- (2.9,2.1) node [black,midway,xshift=-0.6cm] { };
		\draw node at (2.5,2.5) {\footnotesize $N-2$};
	\end{tikzpicture}
        \caption{}
    \end{subfigure}
    \hfill
    \begin{subfigure}[t]{.30\textwidth}
    \centering
	\begin{tikzpicture}
		\draw[dashed] (1,0)--(1,2);
		\draw[dashed] (4,0)--(4,2);
		\draw (1,.5)--(4,.5);
		\draw (2,1) node[cross] {};
		\draw (3,1) node[cross] {};
		\draw (2,1) node[circle,draw] {};
		\draw (3,1) node[circle,draw] {};
		\fill (2.5,1) circle [radius=1pt];
		\fill (2.40,1) circle [radius=1pt];
		\fill (2.60,1) circle [radius=1pt];
		\draw [decorate,decoration={brace,amplitude=5pt}](1.8,1.3) -- (3.2,1.3) node [black,midway,xshift=-0.6cm] { };
		\draw node at (2.5,1.7) {\footnotesize $N$};
	\end{tikzpicture}
	\caption{}
	\end{subfigure}
	\hfill
 	\caption{Mirror dual of $3d$ $\mathcal{N}=4$ SQED with N flavors. (a) Quiver. There are $N-1$ gauge nodes, with label $n_i=1$. (b) Coulomb branch brane configuration. There is a total of $N$ NS5-branes. (c) Higgs branch brane configuration.}
	\label{fig:SUNU1mirror}
\end{figure}
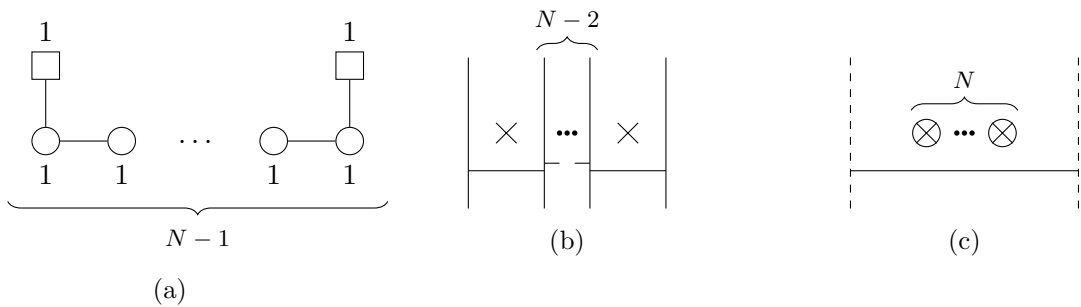 

If one performs $3d$ \emph{mirror symmetry} (this is a duality first found by Intriligator and Seiberg \cite{IS96}; from the point of view of Type IIB superstring theory it consists of S-Duality and a rotation \cite{HW96}) one obtains a dual theory with quiver depicted in figure \ref{fig:SUNU1mirror}(a) and moduli space:

\begin{align}
	\begin{aligned}
		\Mc &= a_{N-1}\\
		\Mh &= A_{N-1}
	\end{aligned}
\end{align}

This gives us a description of both singularities in terms of branes, that is summarized in table \ref{tab:min}. These diagrams are key to identify the nature of the singularities in closures of unitary nilpotent orbits.

The work of Kraft and Procesi \cite{KP82} already relates these two types of varieties: a transverse slice of type $A_n$ is mapped to a transverse slice of type $a_n$ under the action of reversing the order in the partial ordering structure of the closures of nilpotent orbits of the special linear algebra. This shows a really important insight, and the seeds of what would later become $3d$ mirror symmetry can already be recognized in their results.

\begin{table}[t]
	\centering
	\begin{tabular}{| c | c |}
	\hline
	 Singularity & Coulomb branch brane configuration \\ \hline 
	& \\
	$A_n:=\mathbb{C}^2/\mathbb{Z}_{n+1}$ & \raisebox{-.5\height}{\begin{tikzpicture}
		\draw (1,0)--(1,2);
		\draw (4,0)--(4,2);
		\draw (1,.5)--(4,.5);
		\draw (2,1) node[cross] {};
		\draw (3,1) node[cross] {};
		\fill (2.5,1) circle [radius=1pt];
		\fill (2.40,1) circle [radius=1pt];
		\fill (2.60,1) circle [radius=1pt];
		\draw [decorate,decoration={brace,amplitude=5pt}](1.8,1.3) -- (3.2,1.3) node [black,midway,xshift=-0.6cm] { };
		\draw node at (2.5,1.7) {\footnotesize $n+1$};
	\end{tikzpicture} } \\ 
	&  \\ \hline
	&  \\
	$a_n:=\Or_{(2,1^{n-1})}$ &\raisebox{-.4\height}{\begin{tikzpicture}
		\draw	(1.2,0)--(1.2,2)
				(3.8,0)--(3.8,2)
				(2.2,0)--(2.2,2)
				(2.8,0)--(2.8,2);
		\draw 	(1.7,1) node[cross]{}
				(3.3,1) node[cross]{};
		\draw (1.2,.5)--(2.2,.5)
				(2.8,.5)--(3.8,.5)
				(2.2,.6)--(2.4,.6)
				(2.6,.6)--(2.8,.6);
		\fill (2.5,1) circle [radius=1pt];
		\fill (2.40,1) circle [radius=1pt];
		\fill (2.60,1) circle [radius=1pt];
		\draw [decorate,decoration={brace,amplitude=5pt}](2.1,2.1) -- (2.9,2.1) node [black,midway,xshift=-0.6cm] { };
		\draw node at (2.5,2.5) {\footnotesize $n-1$};
	\end{tikzpicture}} \\ 
	& \\ \hline
	\end{tabular}
	\caption{Here we show the surface singularities $A_n$ and the  minimal singularities $a_n$ in terms of branes. To obtain the complementary Higgs branch brane configuration just swap the D5-branes with NS5-branes and vice versa.}
	\label{tab:min}
\end{table}

\subsection{The Kraft-Procesi transition}

\paragraph{Example}

Our first example is the first nontrivial transverse slice that was computed in the case of nilpotent orbits of classical algebras. It is one of the results by Brieskorn and Slodowy \cite{B70,Sl80}: let $\mathcal{O}_{(n)}$ and $\mathcal{O}_{(n-1,1)}$ be the maximal and the subregular nilpotent orbits of $\mathfrak{sl}(n)$ over the field $\mathbb{C}$. Let $\Or_{(n)}$ and $\Or_{(n-1,1)}$ be their respective closures. Then $\Or_{(n-1,1)}  \subset \Or_{(n)}$. Let $S\subseteq \Or_{(n)}$ be the slice transverse to $\mathcal{O}_{(n-1,1)}$, then $S$ is the Kleinian singularity:

\begin{align}
	S&=A_{n-1}
\end{align}

\begin{figure}[t]
	\centering
	\begin{tikzpicture}
	\tikzstyle{gauge} = [circle, draw];
	\tikzstyle{flavour} = [regular polygon,regular polygon sides=4, draw];
	\node (g1) [gauge, label=below:{$n-1$}] {};
	\node (g2) [right of=g1] {$\dots$};
	\node (g3) [gauge,right of=g2, label=below:{$2$}] {};
	\node (g4) [gauge,right of=g3, label=below:{$1$}] {};
	\node (f1) [flavour, above of=g1, label=above:{$n$}] {};
	\draw (g1)--(g2)--(g3)--(g4)
		(f1)--(g1);
	\end{tikzpicture}
	\caption{Self-dual quiver theory. The Higgs and the Coulomb branch of this theory are both $\Or _{(n)}$, i.e. the closure of the maximal nilpotent orbit of $\mathfrak{g}=\mathfrak{sl}(n)$.}
	\label{fig:quiverTSUn}
\end{figure}
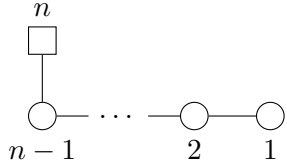

The quiver whose Higgs branch is $\M_H=\Or_{(n)}$ is self dual. This means that its Coulomb branch is also $\M_C=\Or_{(n)}$. It is remarkable to notice that Kraft and Procesi already drew this type of quivers\footnote{More recently, physicists refer to the IR fixed point of this particular quiver \cite{K90} with the name $T(SU(n))$ \cite{GW09}.} and computed their Higgs branches, imposing F-term and D-term conditions back in 1979 \cite{KP79}. The quiver, depicted in figure \ref{fig:quiverTSUn}, has gauge symmetry $G$ and flavor symmetry $F$:

\begin{align}
	\begin{aligned}
	 	F&=SU(n)\\
		G&=U(n-1)\times U(n-2)\times \dots\times U(1)
	\end{aligned}
\end{align}

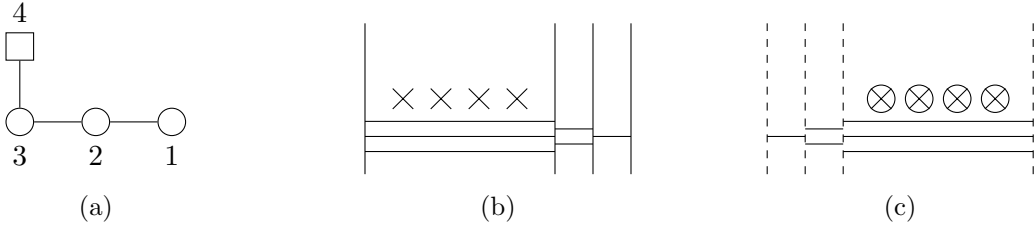
\begin{figure}[t]
	\centering
	\begin{subfigure}[t]{.3\textwidth}
    			\centering
	\begin{tikzpicture}
	\tikzstyle{gauge} = [circle, draw];
	\tikzstyle{flavour} = [regular polygon,regular polygon sides=4, draw];
	\node (g1) [gauge, label=below:{$3$}] {};
	\node (g2) [gauge,right of=g1,  label=below:{$2$}] {};
	\node (g3) [gauge,right of=g2, label=below:{$1$}] {};
	\node (f1) [flavour, above of=g1, label=above:{$4$}] {};
	\draw (g1)--(g2)--(g3)
		(f1)--(g1);
	\end{tikzpicture}
			\caption{}
		\end{subfigure}
		\hfill
		\begin{subfigure}[t]{.3\textwidth}
    			\centering
	\begin{tikzpicture}
		\draw 	(-1,0)--(-1,2)
					(-1.5,0)--(-1.5,2)
					(-2,0)--(-2,2)
					(-4.5,0)--(-4.5,2);
		\draw 	(-1,.5)--(-1.5,.5)
				(-1.5,.4)--(-2,.4)
				(-1.5,.6)--(-2,.6)
				(-2,.3)--(-4.5,.3)
				(-2,.5)--(-4.5,.5)
				(-2,.7)--(-4.5,.7);
		\draw (-2.5,1) node[cross] {};
		\draw (-3,1) node[cross] {};
		\draw (-3.5,1) node[cross] {};
		\draw (-4,1) node[cross] {};
	\end{tikzpicture}
			\caption{}
		\end{subfigure}
		\hfill
		\begin{subfigure}[t]{.3\textwidth}
    			\centering
	\begin{tikzpicture}
		\draw 	[dashed](1,0)--(1,2)
					(1.5,0)--(1.5,2)
					(2,0)--(2,2)						(4.5,0)--(4.5,2);
		\draw 	(1,.5)--(1.5,.5)
				(1.5,.4)--(2,.4)
				(1.5,.6)--(2,.6)
				(2,.3)--(4.5,.3)
				(2,.5)--(4.5,.5)
				(2,.7)--(4.5,.7);
		\draw (2.5,1) node[cross] {};
		\draw (3,1) node[cross] {};
		\draw (3.5,1) node[cross] {};
		\draw (4,1) node[cross] {};
		\draw (2.5,1) node[circle,draw] {};
		\draw (3,1) node[circle,draw] {};
		\draw (3.5,1) node[circle,draw] {};
		\draw (4,1) node[circle,draw] {};
	\end{tikzpicture}
			\caption{}
		\end{subfigure}
	\caption{Model with moduli space: $\M_H=\Or_{(4)}$ and $\M_C=\Or _{(4)}$, where $\Or_{(4)}$ is the closure of the maximal nilpotent orbit of $\mathfrak{sl}(4)$. (a) Quiver. (b) Coulomb branch. (c) Higgs branch.}
	\label{fig:TSU4}
\end{figure}

Let us consider the algebra $\mathfrak{g}=\mathfrak{sl}(4)$, for clarity of the argument. Then, the closure of the maximal nilpotent orbit is $\Or_{(4)}$ and the closure of the subregular nilpotent orbit is $\Or_{(3,1)}$. The slice $S\subseteq \Or_{(4)}$ transverse to $\mathcal{O}_{(3,1)}$ is $S=A_3=\mathbb{C}^2/\mathbb{Z}_4$. The flavor and gauge symmetries are $F=SU(4)$ and $G=U(3)\times U(2)\times U(1)$, the quiver is depicted in figure \ref{fig:TSU4}(a). The brane configurations of the Coulomb and Higgs branches are depicted in figures \ref{fig:TSU4}(b) and \ref{fig:TSU4}(c).

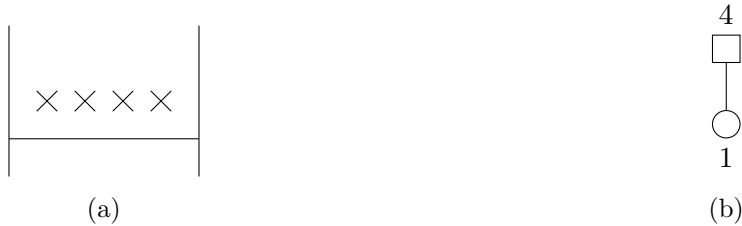
\begin{figure}[t]
	\centering
		\begin{subfigure}[t]{.45\textwidth}
    			\centering
	\begin{tikzpicture}
		\draw 		(-2,0)--(-2,2)
					(-4.5,0)--(-4.5,2);
		\draw 	(-2,.5)--(-4.5,.5);
		\draw (-2.5,1) node[cross] {};
		\draw (-3,1) node[cross] {};
		\draw (-3.5,1) node[cross] {};
		\draw (-4,1) node[cross] {};
	\end{tikzpicture}
			\caption{}
		\end{subfigure}
		\hfill
		\begin{subfigure}[t]{.45\textwidth}
    			\centering
	\begin{tikzpicture}
	\tikzstyle{gauge} = [circle, draw];
	\tikzstyle{flavour} = [regular polygon,regular polygon sides=4, draw];
	\node (g1) [gauge, label=below:{$1$}] {};
	\node (f1) [flavour, above of=g1, label=above:{$4$}] {};
	\draw
		(f1)--(g1);
	\end{tikzpicture}
			\caption{}
		\end{subfigure}
		\hfill
	\caption{SQED with 4 flavors. (a) is the local subsystem around one of the leftmost D3-branes in the Coulomb branch brane configuration of the quiver in figure \ref{fig:TSU4}. The corresponding moduli space is the $A_3$ Kleinian singularity, as indicated in table \ref{tab:min}. (b) Corresponding quiver.}
	\label{fig:SQED4}
\end{figure}

Let us study the Coulomb branch $\M_C$, figure \ref{fig:TSU4}(b).  Let us focus on one of the three leftmost D3-branes. We consider the moduli space generated by the motion of this brane along its $\vec{x}$ position, considering the remaining D3-branes to be spectators. This moduli space is the \emph{slice} $S$ and it is \emph{transverse} to the moduli space generated by the spectators after the D3-brane is \emph{removed}. To see what is the moduli space $S$ we analyze the local subsystem for the D3-brane and the fivebranes that are around it. This is depicted in figure \ref{fig:SQED4}. We see that this brane configuration belongs to the family of brane configurations described in the first row of table \ref{tab:min}. The moduli space is the Coulomb branch of $3d$ $\mathcal{N}=4$ SQED with 4 flavors. This is indeed: 

\begin{align}
	S=A_3
\end{align}

By \emph{removed} we mean that a phase transition is performed to take the D3-brane to the singular point where its $\vec{x}$ position coincides with those of the D5-branes in the same interval:

\begin{align}
	\vec{x}=\vec{m}_1=\vec{m}_2=\vec{m}_3 =\vec{m}_4=(0,0,0)
\end{align}

\begin{figure}[t]
	\centering
	\begin{subfigure}[t]{.49\textwidth}
    \centering
	\begin{tikzpicture}
		\draw 	(-1,0)--(-1,2)
					(-1.5,0)--(-1.5,2)
					(-2,0)--(-2,2)
					(-4.5,0)--(-4.5,2);
		\draw 	(-1,.5)--(-1.5,.5)
				(-1.5,.4)--(-2,.4)
				(-1.5,.6)--(-2,.6)
				(-2,.3)--(-4.5,.3)
				(-2,.5)--(-4.5,.5)
				(-2,.7)--(-4.5,.7);
		\draw (-2.5,1) node[cross] {};
		\draw (-3,1) node[cross] {};
		\draw (-3.5,1) node[cross] {};
		\draw (-4,1) node[cross] {};
	\end{tikzpicture}
	\caption{}
    \end{subfigure}
    \hfill
	\begin{subfigure}[t]{.49\textwidth}
    \centering
	\begin{tikzpicture}
		\draw 	(-1,0)--(-1,2)
					(-1.5,0)--(-1.5,2)
					(-2,0)--(-2,2)
					(-4.5,0)--(-4.5,2);
		\draw 	(-1,.5)--(-1.5,.5)
				(-1.5,.4)--(-2,.4)
				(-1.5,.6)--(-2,.6)
				(-2,.3)--(-4.5,.3)
				(-2,.5)--(-4.5,.5)
				(-2,1)--(-4.5,1);
		\draw (-2.5,1) node[cross] {};
		\draw (-3,1) node[cross] {};
		\draw (-3.5,1) node[cross] {};
		\draw (-4,1) node[cross] {};
	\end{tikzpicture}
	\caption{}
    \end{subfigure}
    \vfill
    \begin{subfigure}[t]{.49\textwidth}
    \centering
	\begin{tikzpicture}
		\draw	(-1,0)--(-1,2)
					(-1.5,0)--(-1.5,2)
					(-2,0)--(-2,2)
					(-4.5,0)--(-4.5,2);
		\draw 	(-1,.5)--(-1.5,.5)
				(-1.5,.4)--(-2,.4)
				(-1.5,.6)--(-2,.6)
				(-2,.3)--(-4.5,.3)
				(-2,.5)--(-4.5,.5)
				(-2,1)--(-2.5,1)
				(-4,1)--(-4.5,1);
		\draw (-2.5,1) node[cross] {};
		\draw (-3,1) node[cross] {};
		\draw (-3.5,1) node[cross] {};
		\draw (-4,1) node[cross] {};
	\end{tikzpicture}
	\caption{}
    \end{subfigure}
\hfill
    \begin{subfigure}[t]{.49\textwidth}
    \centering
	\begin{tikzpicture}
		\draw 	(-1,0)--(-1,2)
					(-1.5,0)--(-1.5,2)
					(-2.5,0)--(-2.5,2)
					(-4,0)--(-4,2);
		\draw 	(-1,.5)--(-1.5,.5)
				(-1.5,.4)--(-2.5,.4)
				(-1.5,.6)--(-2.5,.6)
				(-2.5,.3)--(-4,.3)
				(-2.5,.5)--(-4,.5);
		\draw (-2,1) node[cross] {};
		\draw (-3,1) node[cross] {};
		\draw (-3.5,1) node[cross] {};
		\draw (-4.5,1) node[cross] {};
	\end{tikzpicture}
	\caption{}
    \end{subfigure}
    \caption{Example of a Kraft-Procesi transition. (a) Coulomb branch brane configuration with $\M_C=\Or_{(4)}$. (b) The relevant D3-brane is in the singular point $\vec{x}=\vec{m}_i=(0,0,0)$. (c) The D3-brane is  split into five segments, three of them can acquire nonzero $\vec{y}_i$ position along the directions spanned by the D5-branes. We take the limit when these positions go to infinity, removing these branes from the system. This gives rise to a new brane configuration with a new Coulomb branch $\M_C'\subset\M_C$. (d) A phase transition that annihilates the fixed three branes is performed in order to obtain a Coulomb branch brane configuration where the corresponding new quiver can be read.}
    \label{fig:KPexample}
\end{figure}
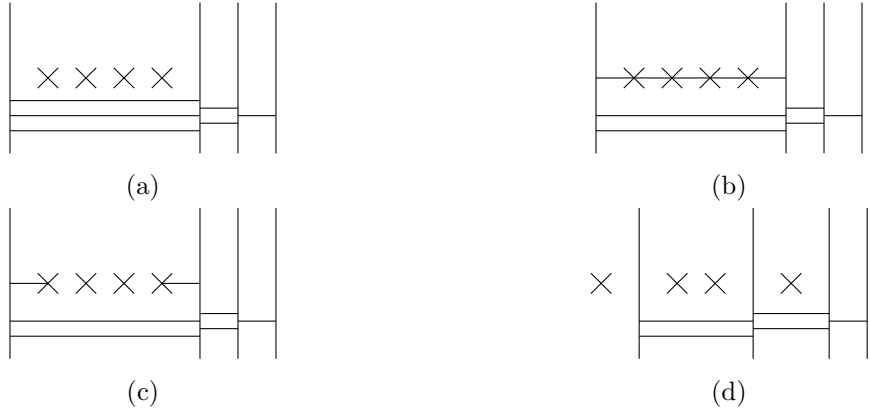

This is depicted in figure \ref{fig:KPexample}(b). The D3-brane can then be split into segments of D3-branes that can move freely in the perpendicular directions $\vec{y}$ spanned by the D5-branes. These threebrane segments can then be taken to infinity in the $\vec{y}$ direction. The resulting system has two fixed D3-branes connected to fivebranes of different nature, the remaining D3-branes are mobile. These are the ones previously considered as spectators, figure \ref{fig:KPexample}(c). After annihilating the fixed threebranes via phase transitions a new quiver can be read from the resulting brane configuration, figure \ref{fig:KPexample}(d). The Coulomb branch of the new quiver $\M_C'$ is a subset of the Coulomb branch of the original quiver $\M_C'\subset \M_C$. $\M_C'$ has the same isometry group as $\M_C$. It is in fact the closure of an orbit of the isometry group, and $S$ is its transverse slice.

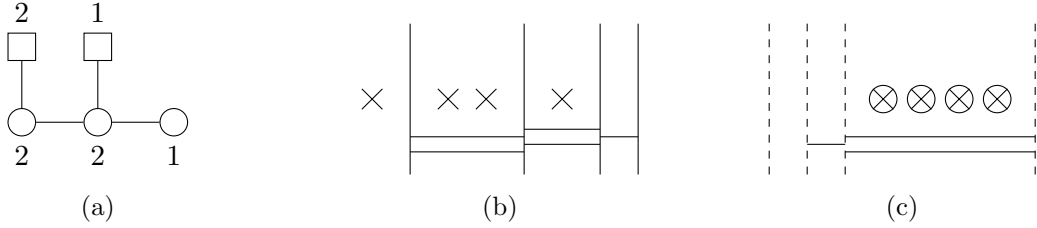
\begin{figure}[t]
	\centering
	\begin{subfigure}[t]{.3\textwidth}
    			\centering
	\begin{tikzpicture}
	\tikzstyle{gauge} = [circle, draw];
	\tikzstyle{flavour} = [regular polygon,regular polygon sides=4, draw];
	\node (g1) [gauge, label=below:{$2$}] {};
	\node (g2) [gauge,right of=g1,  label=below:{$2$}] {};
	\node (g3) [gauge,right of=g2, label=below:{$1$}] {};
	\node (f1) [flavour, above of=g1, label=above:{$2$}] {};
	\node (f2) [flavour, above of=g2, label=above:{$1$}] {};
	\draw (g1)--(g2)--(g3)
		(f1)--(g1)
		(f2)--(g2);
	\end{tikzpicture}
			\caption{}
		\end{subfigure}
		\hfill
		\begin{subfigure}[t]{.3\textwidth}
    			\centering
	\begin{tikzpicture}
		\draw 	(-1,0)--(-1,2)
					(-1.5,0)--(-1.5,2)
					(-2.5,0)--(-2.5,2)
					(-4,0)--(-4,2);
		\draw 	(-1,.5)--(-1.5,.5)
				(-1.5,.4)--(-2.5,.4)
				(-1.5,.6)--(-2.5,.6)
				(-2.5,.3)--(-4,.3)
				(-2.5,.5)--(-4,.5);
		\draw (-2,1) node[cross] {};
		\draw (-3,1) node[cross] {};
		\draw (-3.5,1) node[cross] {};
		\draw (-4.5,1) node[cross] {};
	\end{tikzpicture}
			\caption{}
		\end{subfigure}
		\hfill
		\begin{subfigure}[t]{.3\textwidth}
    			\centering
	\begin{tikzpicture}
		\draw 	[dashed](1,0)--(1,2)
					(1.5,0)--(1.5,2)
					(2,0)--(2,2)						(4.5,0)--(4.5,2);
		\draw 	(1.5,.4)--(2,.4)
				(2,.3)--(4.5,.3)
				(2,.5)--(4.5,.5);
		\draw (2.5,1) node[cross] {};
		\draw (3,1) node[cross] {};
		\draw (3.5,1) node[cross] {};
		\draw (4,1) node[cross] {};
		\draw (2.5,1) node[circle,draw] {};
		\draw (3,1) node[circle,draw] {};
		\draw (3.5,1) node[circle,draw] {};
		\draw (4,1) node[circle,draw] {};
	\end{tikzpicture}
			\caption{}
		\end{subfigure}
	\caption{Model with moduli space: $\M_C=\Or _{(3,1)}$, where $\Or_{(3,1)}$ is the closure of the subregular nilpotent orbit of $\mathfrak{g}=\mathfrak{sl}(4)$. (a) Quiver. (b) Coulomb branch. (c) Higgs branch.}
	\label{fig:T211SU4}
\end{figure}

The resulting quiver corresponding to Coulomb branch brane configuration in figure \ref{fig:KPexample}(d) is depicted in figure \ref{fig:T211SU4}(a). The Coulomb branch of this quiver is also the closure of a nilpotent orbit of $\mathfrak{g}=\mathfrak{sl}(4)$. We have:
\begin{align}
	\begin{aligned}
		\M_C &=\Or_{(4)}\\
		\M_C' &=\Or_{(3,1)}\\
		S &=A_3\\
	\end{aligned}
\end{align}

This is the physical realization of the Brieskorn-Slodowy theory: $S=A_{3}\subseteq \Or_{(4)}$ is the slice transverse to the orbit $\mathcal{O}_{(3,1)}$. It is straightforward to generalize this example to $\mathfrak{g}=\mathfrak{sl}(n)$: one can study similar brane systems to show the physical realization that $S=A_{n-1}\subseteq \Or_{(n)}$ is the transverse slice to the orbit $\mathcal{O}_{(n-1,1)}$.

\begin{figure}[t]
	\centering
	\includegraphics{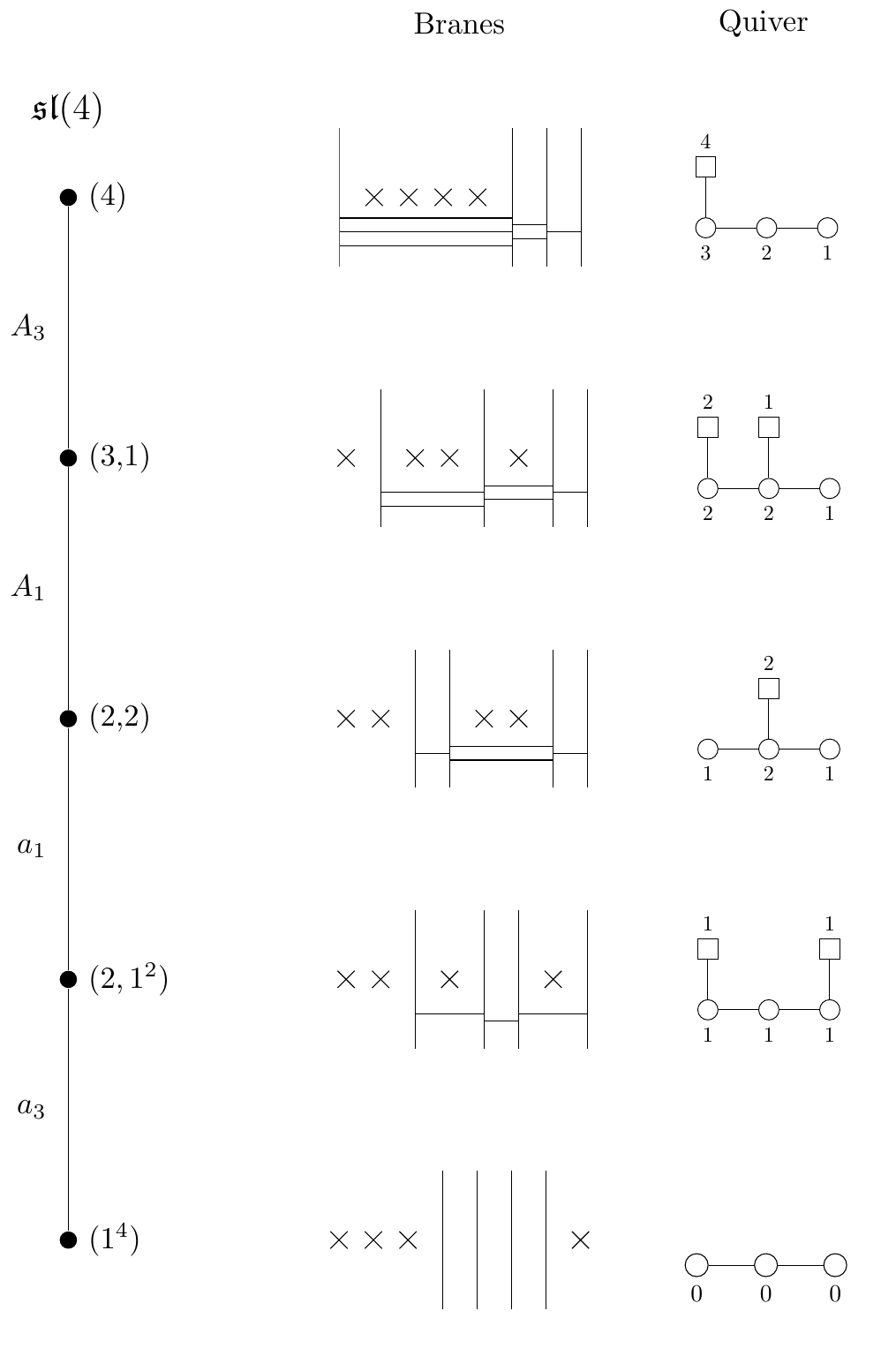}
	\caption{Hasse diagram for $\mathfrak{sl}(4)$. Note that $a_1=A_1$.}
	\label{fig:ExampleHasse}
\end{figure}

This process can be iterated until all quivers are found such that their Coulomb branches are closures of nilpotent orbits of $\mathfrak{sl}(4)$. The transverse slices found in this manner correspond to the transverse singularities established by Kraft and Procesi in \citep{KP81,KP82}. These results can be summarized in a Hasse diagram depicted in figure \ref{fig:ExampleHasse}. The graph at the left of figure \ref{fig:ExampleHasse} represents the partial ordering structure of the different nilpotent orbit closures of $\mathfrak{sl}(4)$, each of them denoted by a node and labelled by a partition. A diagram that represents a partial ordering in a set is given the name of Hasse diagram. The brane configuration whose Coulomb branch is the closure of each orbit is depicted to the right of each node. The corresponding quiver describing the low energy dynamics of each brane system is depicted in the rightmost column. It can be observed that D3-branes that generate either an $A_n$ or an $a_n$ moduli space need to be removed in order to go from one brane system to the next below. These are all Kraft-Procesi transitions. The moduli space generated by the threebranes that have been removed corresponds to the slice $S$ in the upper nilpotent orbit closure that is transverse to the lower one. The type of transverse slice $S$ is used to label each Kraft-Procesi transition. They also label the edges in the Hasse diagram, since each edge corresponds to one transition of this kind.

\paragraph{General definition}

Let $V=\M_C$ (resp. $V=\M_H$) be the Coulomb branch (resp. Higgs branch) of a $3d$ $\mathcal{N}=4$ quiver gauge theory that describes the low energy dynamics of a brane system in Type IIB superstring theory of the type discussed above. Let $S$ denote the moduli space generated by a subset of the D3-branes present in the initial Coulomb branch brane configuration (resp. Higgs branch brane configuration) such that $S\subseteq V$ and $S$ is a hyperk\"ahler singularity of type $A_n$ (i.e. isomorphic to the Coulomb branch of $3d$ $\mathcal{N}=4$ SQED with $n+1$ flavors) or type $a_n$ (i.e. isomorphic to the Higgs branch of $3d$ $\mathcal{N}=4$ SQED with $n+1$ flavors). The \emph{Kraft-Procesi} transition consists on utilizing a Higgs mechanism to remove such subset of D3-branes from the original brane system. The resulting brane system has a new quiver with a new Coulomb branch $V'=\M'_C$ (resp. Higgs branch $V'=\M'_H$) such that $V'\subsetneq V$. $S$ is the \emph{slice} in $V$ \emph{transverse} to $V'$.

Given a Coulomb branch (resp. Higgs branch) brane configuration, it is simple to specify how to find all possible D3-brane subsystems that generate either $A_n$ or $a_n$ singularities. It is enough to identify subsystems of the form of those depicted in table \ref{tab:min}. In other words, letting all fivebranes to have distinct positions along the $x^6$ direction and letting all D3-branes stretch between NS5-branes (resp. D5-branes): a moduli space $S=A_n$ is generated by every D3-brane extending along an interval between two consecutive NS5-branes (resp. D5-branes) such that there are exactly $n+1$ D5-branes (resp. NS5-branes) located within such interval. A moduli space $S=a_n$ is generated by a set of $n$ D3-branes stretching along $n$ consecutive intervals between NS5-branes (resp. D5-branes) such that there is a single D5-brane (resp. NS5-brane) within the leftmost interval and a single D5-brane (resp. NS5-brane) within the rightmost one, and there are no other D5-branes (resp. NS5-branes) in any of the intermediate $n-2$ intervals.

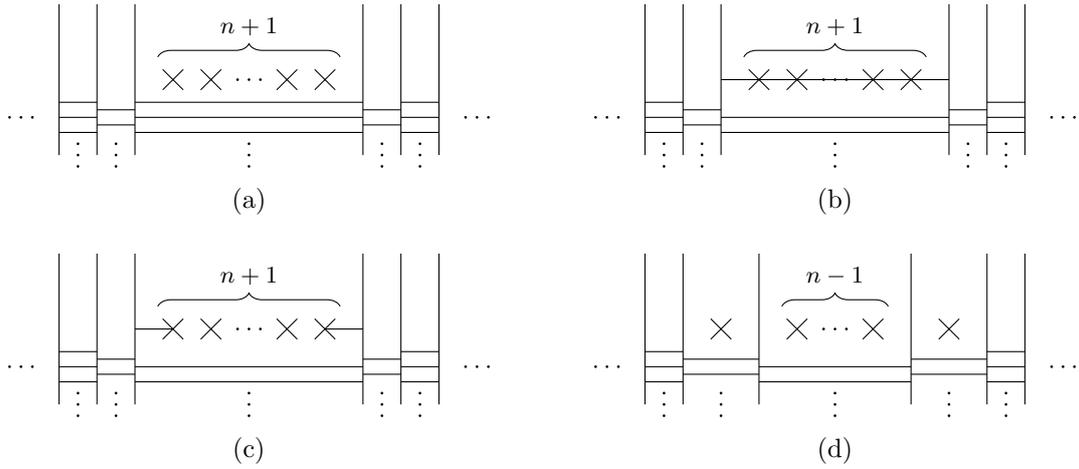
\begin{figure}[t]
	\centering
	\begin{subfigure}[t]{.49\textwidth}
    \centering
	\begin{tikzpicture}
		\draw 	(-.5,0)--(-.5,2)
				(-1,0)--(-1,2)
				(-1.5,0)--(-1.5,2)
				(-4.5,0)--(-4.5,2)
				(-5,0)--(-5,2)
				(-5.5,0)--(-5.5,2);
		\draw 	(-.5,.7)--(-1,.7)
				(-.5,.5)--(-1,.5)
				(-.5,.3)--(-1,.3)
				(-1,.4)--(-1.5,.4)
				(-1,.6)--(-1.5,.6)
				(-5,.7)--(-5.5,.7)
				(-5,.5)--(-5.5,.5)
				(-5,.3)--(-5.5,.3)
				(-4.5,.4)--(-5,.4)
				(-4.5,.6)--(-5,.6)
				(-1.5,.3)--(-4.5,.3)
				(-1.5,.5)--(-4.5,.5)
				(-1.5,.7)--(-4.5,.7);
		\fill (-.75,0) circle [radius=.5pt];
		\fill (-.75,-.15) circle [radius=.5pt];
		\fill (-.75,.15) circle [radius=.5pt];
		\fill (-1.25,0) circle [radius=.5pt];
		\fill (-1.25,-.15) circle [radius=.5pt];
		\fill (-1.25,.15) circle [radius=.5pt];
		\fill (-3,0) circle [radius=.5pt];
		\fill (-3,-.15) circle [radius=.5pt];
		\fill (-3,.15) circle [radius=.5pt];
		\fill (-4.75,0) circle [radius=.5pt];
		\fill (-4.75,-.15) circle [radius=.5pt];
		\fill (-4.75,.15) circle [radius=.5pt];
		\fill (-5.25,0) circle [radius=.5pt];
		\fill (-5.25,-.15) circle [radius=.5pt];
		\fill (-5.25,.15) circle [radius=.5pt];
		\draw (-2,1) node[cross] {};
		\draw (-2.5,1) node[cross] {};
		\fill (-3,1) circle [radius=.5pt];
		\fill (-3.15,1) circle [radius=.5pt];
		\fill (-2.85,1) circle [radius=.5pt];
		\draw (-3.5,1) node[cross] {};
		\draw (-4,1) node[cross] {};
		
		\fill (0,.5) circle [radius=.5pt];
		\fill (0.15,.5) circle [radius=.5pt];
		\fill (-.15,.5) circle [radius=.5pt];
		
		\fill (-6,.5) circle [radius=.5pt];
		\fill (-6.15,.5) circle [radius=.5pt];
		\fill (-5.85,.5) circle [radius=.5pt];
		
		\draw [decorate,decoration={brace,mirror,amplitude=5pt}](-1.8,1.3) -- (-4.2,1.3);
		\draw node at (-3,1.7) {\footnotesize $n+1$};

	\end{tikzpicture}
	\caption{}
    \end{subfigure}
    \hfill
	\begin{subfigure}[t]{.49\textwidth}
    \centering
	\begin{tikzpicture}
		\draw 	(-.5,0)--(-.5,2)
				(-1,0)--(-1,2)
				(-1.5,0)--(-1.5,2)
				(-4.5,0)--(-4.5,2)
				(-5,0)--(-5,2)
				(-5.5,0)--(-5.5,2);
		\draw 	(-.5,.7)--(-1,.7)
				(-.5,.5)--(-1,.5)
				(-.5,.3)--(-1,.3)
				(-1,.4)--(-1.5,.4)
				(-1,.6)--(-1.5,.6)
				(-5,.7)--(-5.5,.7)
				(-5,.5)--(-5.5,.5)
				(-5,.3)--(-5.5,.3)
				(-4.5,.4)--(-5,.4)
				(-4.5,.6)--(-5,.6)
				(-1.5,.3)--(-4.5,.3)
				(-1.5,.5)--(-4.5,.5)
				(-1.5,1)--(-4.5,1);
		\fill (-.75,0) circle [radius=.5pt];
		\fill (-.75,-.15) circle [radius=.5pt];
		\fill (-.75,.15) circle [radius=.5pt];
		\fill (-1.25,0) circle [radius=.5pt];
		\fill (-1.25,-.15) circle [radius=.5pt];
		\fill (-1.25,.15) circle [radius=.5pt];
		\fill (-3,0) circle [radius=.5pt];
		\fill (-3,-.15) circle [radius=.5pt];
		\fill (-3,.15) circle [radius=.5pt];
		\fill (-4.75,0) circle [radius=.5pt];
		\fill (-4.75,-.15) circle [radius=.5pt];
		\fill (-4.75,.15) circle [radius=.5pt];
		\fill (-5.25,0) circle [radius=.5pt];
		\fill (-5.25,-.15) circle [radius=.5pt];
		\fill (-5.25,.15) circle [radius=.5pt];
		\draw (-2,1) node[cross] {};
		\draw (-2.5,1) node[cross] {};
		\fill (-3,1) circle [radius=.5pt];
		\fill (-3.15,1) circle [radius=.5pt];
		\fill (-2.85,1) circle [radius=.5pt];
		\draw (-3.5,1) node[cross] {};
		\draw (-4,1) node[cross] {};
		
		\fill (0,.5) circle [radius=.5pt];
		\fill (0.15,.5) circle [radius=.5pt];
		\fill (-.15,.5) circle [radius=.5pt];
		
		\fill (-6,.5) circle [radius=.5pt];
		\fill (-6.15,.5) circle [radius=.5pt];
		\fill (-5.85,.5) circle [radius=.5pt];
		
		\draw [decorate,decoration={brace,mirror,amplitude=5pt}](-1.8,1.3) -- (-4.2,1.3);
		\draw node at (-3,1.7) {\footnotesize $n+1$};
	\end{tikzpicture}	
	\caption{}
    \end{subfigure}
    \\[12pt]
    \begin{subfigure}[t]{.49\textwidth}
    \centering
	\begin{tikzpicture}
		\draw 	(-.5,0)--(-.5,2)
				(-1,0)--(-1,2)
				(-1.5,0)--(-1.5,2)
				(-4.5,0)--(-4.5,2)
				(-5,0)--(-5,2)
				(-5.5,0)--(-5.5,2);
		\draw 	(-.5,.7)--(-1,.7)
				(-.5,.5)--(-1,.5)
				(-.5,.3)--(-1,.3)
				(-1,.4)--(-1.5,.4)
				(-1,.6)--(-1.5,.6)
				(-5,.7)--(-5.5,.7)
				(-5,.5)--(-5.5,.5)
				(-5,.3)--(-5.5,.3)
				(-4.5,.4)--(-5,.4)
				(-4.5,.6)--(-5,.6)
				(-1.5,.3)--(-4.5,.3)
				(-1.5,.5)--(-4.5,.5)
				(-1.5,1)--(-2,1)
				(-4,1)--(-4.5,1);
		\fill (-.75,0) circle [radius=.5pt];
		\fill (-.75,-.15) circle [radius=.5pt];
		\fill (-.75,.15) circle [radius=.5pt];
		\fill (-1.25,0) circle [radius=.5pt];
		\fill (-1.25,-.15) circle [radius=.5pt];
		\fill (-1.25,.15) circle [radius=.5pt];
		\fill (-3,0) circle [radius=.5pt];
		\fill (-3,-.15) circle [radius=.5pt];
		\fill (-3,.15) circle [radius=.5pt];
		\fill (-4.75,0) circle [radius=.5pt];
		\fill (-4.75,-.15) circle [radius=.5pt];
		\fill (-4.75,.15) circle [radius=.5pt];
		\fill (-5.25,0) circle [radius=.5pt];
		\fill (-5.25,-.15) circle [radius=.5pt];
		\fill (-5.25,.15) circle [radius=.5pt];
		\draw (-2,1) node[cross] {};
		\draw (-2.5,1) node[cross] {};
		\fill (-3,1) circle [radius=.5pt];
		\fill (-3.15,1) circle [radius=.5pt];
		\fill (-2.85,1) circle [radius=.5pt];
		\draw (-3.5,1) node[cross] {};
		\draw (-4,1) node[cross] {};
		
		\fill (0,.5) circle [radius=.5pt];
		\fill (0.15,.5) circle [radius=.5pt];
		\fill (-.15,.5) circle [radius=.5pt];
		
		\fill (-6,.5) circle [radius=.5pt];
		\fill (-6.15,.5) circle [radius=.5pt];
		\fill (-5.85,.5) circle [radius=.5pt];
		
		\draw [decorate,decoration={brace,mirror,amplitude=5pt}](-1.8,1.3) -- (-4.2,1.3);
		\draw node at (-3,1.7) {\footnotesize $n+1$};
	\end{tikzpicture}	
	\caption{}
    \end{subfigure}
\hfill
    \begin{subfigure}[t]{.49\textwidth}
    \centering
	\begin{tikzpicture}
		\draw 	(-.5,0)--(-.5,2)
				(-1,0)--(-1,2)
				(-2,0)--(-2,2)
				(-4,0)--(-4,2)
				(-5,0)--(-5,2)
				(-5.5,0)--(-5.5,2);
		\draw 	(-.5,.7)--(-1,.7)
				(-.5,.5)--(-1,.5)
				(-.5,.3)--(-1,.3)
				(-1,.4)--(-2,.4)
				(-1,.6)--(-2,.6)
				(-5,.7)--(-5.5,.7)
				(-5,.5)--(-5.5,.5)
				(-5,.3)--(-5.5,.3)
				(-4,.4)--(-5,.4)
				(-4,.6)--(-5,.6)
				(-2,.3)--(-4,.3)
				(-2,.5)--(-4,.5);
		\fill (-.75,0) circle [radius=.5pt];
		\fill (-.75,-.15) circle [radius=.5pt];
		\fill (-.75,.15) circle [radius=.5pt];
		\fill (-1.5,0) circle [radius=.5pt];
		\fill (-1.5,-.15) circle [radius=.5pt];
		\fill (-1.5,.15) circle [radius=.5pt];
		\fill (-3,0) circle [radius=.5pt];
		\fill (-3,-.15) circle [radius=.5pt];
		\fill (-3,.15) circle [radius=.5pt];
		\fill (-4.5,0) circle [radius=.5pt];
		\fill (-4.5,-.15) circle [radius=.5pt];
		\fill (-4.5,.15) circle [radius=.5pt];
		\fill (-5.25,0) circle [radius=.5pt];
		\fill (-5.25,-.15) circle [radius=.5pt];
		\fill (-5.25,.15) circle [radius=.5pt];
		\draw (-1.5,1) node[cross] {};
		\draw (-2.5,1) node[cross] {};
		\fill (-3,1) circle [radius=.5pt];
		\fill (-3.15,1) circle [radius=.5pt];
		\fill (-2.85,1) circle [radius=.5pt];
		\draw (-3.5,1) node[cross] {};
		\draw (-4.5,1) node[cross] {};
		
		\fill (0,.5) circle [radius=.5pt];
		\fill (0.15,.5) circle [radius=.5pt];
		\fill (-.15,.5) circle [radius=.5pt];
		
		\fill (-6,.5) circle [radius=.5pt];
		\fill (-6.15,.5) circle [radius=.5pt];
		\fill (-5.85,.5) circle [radius=.5pt];
		
		\draw [decorate,decoration={brace,mirror,amplitude=5pt}](-2.3,1.3) -- (-3.7,1.3);
		\draw node at (-3,1.7) {\footnotesize $n-1$};
	\end{tikzpicture}		
	\caption{}
    \end{subfigure}
    \caption{$A_n$ Kraft-Procesi transition. (a) Generic part  of a Coulomb branch brane configuration with moduli space $\M_C$. The moduli space generated by a D3-brane in the middle interval is the variety $S=A_n$, i.e. it is isomorphic to the Coulomb branch of $3d$ $\mathcal{N}=4$ SQED with $n+1$ flavors. (b) The relevant D3-brane is at the singular point $\vec{x}=\vec{m}_i=(0,0,0)$. (c) The D3-brane is  split into $n+2$ segments, $n$ of them can acquire nonzero $\vec{y}_j$ position along the directions spanned by the D5-branes. Taking the limit when these positions go to infinity, effectively removes these branes from the system. This gives rise to a new brane configuration with a new Coulomb branch $\M_C'\subset\M_C$. (d) The fixed three branes are annihilated via two phase transitions in order to obtain a Coulomb branch brane configuration where the corresponding new quiver can be read.}
    \label{fig:KPdefAn}
\end{figure}

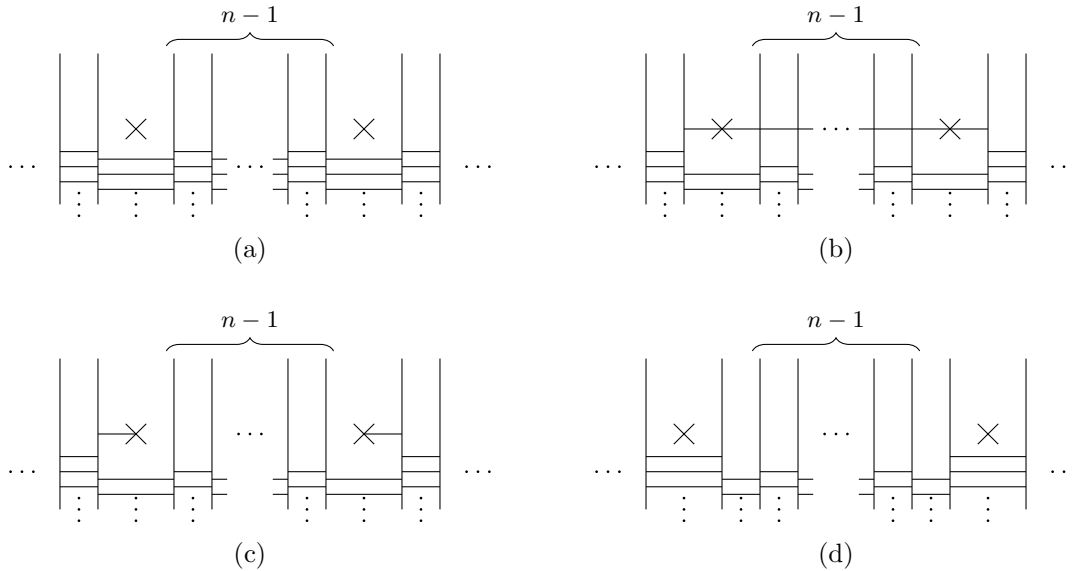
\begin{figure}[t]
	\centering
	\begin{subfigure}[t]{.49\textwidth}
    \centering
	\begin{tikzpicture}
		\draw 	(-.5,0)--(-.5,2)
				(-1,0)--(-1,2)
				(-2,0)--(-2,2)
				(-2.5,0)--(-2.5,2)
				(-3.5,0)--(-3.5,2)
				(-4,0)--(-4,2)
				(-5,0)--(-5,2)
				(-5.5,0)--(-5.5,2);
		\draw 	(-.5,.7)--(-1,.7)
				(-.5,.5)--(-1,.5)
				(-.5,.3)--(-1,.3)
				(-1,.4)--(-2,.4)
				(-1,.6)--(-2,.6)
				(-5,.7)--(-5.5,.7)
				(-5,.5)--(-5.5,.5)
				(-5,.3)--(-5.5,.3)
				(-4,.4)--(-5,.4)
				(-4,.6)--(-5,.6)
				(-3.5,.7)--(-4,.7)
				(-3.5,.5)--(-4,.5)
				(-3.5,.3)--(-4,.3)
				(-2,.3)--(-2.5,.3)
				(-2,.5)--(-2.5,.5)
				(-2,.7)--(-2.5,.7)
				(-2.5,.4)--(-2.7,.4)
				(-2.5,.6)--(-2.7,.6)
				(-3.3,.4)--(-3.5,.4)
				(-3.3,.6)--(-3.5,.6);
		\fill (-.75,0) circle [radius=.5pt];
		\fill (-.75,-.15) circle [radius=.5pt];
		\fill (-.75,.15) circle [radius=.5pt];
		\fill (-1.5,0) circle [radius=.5pt];
		\fill (-1.5,-.15) circle [radius=.5pt];
		\fill (-1.5,.15) circle [radius=.5pt];
		\fill (-2.25,0) circle [radius=.5pt];
		\fill (-2.25,-.15) circle [radius=.5pt];
		\fill (-2.25,.15) circle [radius=.5pt];
		\fill (-3.75,0) circle [radius=.5pt];
		\fill (-3.75,-.15) circle [radius=.5pt];
		\fill (-3.75,.15) circle [radius=.5pt];
		\fill (-4.5,0) circle [radius=.5pt];
		\fill (-4.5,-.15) circle [radius=.5pt];
		\fill (-4.5,.15) circle [radius=.5pt];
		\fill (-5.25,0) circle [radius=.5pt];
		\fill (-5.25,-.15) circle [radius=.5pt];
		\fill (-5.25,.15) circle [radius=.5pt];
		\draw (-1.5,1) node[cross] {};
		\draw (-4.5,1) node[cross] {};
		
		\fill (-3,.5) circle [radius=.5pt];
		\fill (-3.15,.5) circle [radius=.5pt];
		\fill (-2.85,.5) circle [radius=.5pt];
		
		\fill (0,.5) circle [radius=.5pt];
		\fill (0.15,.5) circle [radius=.5pt];
		\fill (-.15,.5) circle [radius=.5pt];
		
		\fill (-6,.5) circle [radius=.5pt];
		\fill (-6.15,.5) circle [radius=.5pt];
		\fill (-5.85,.5) circle [radius=.5pt];
		
		\draw 	(-4,.2)--(-5,.2)
				(-1,.2)--(-2,.2)
				(-2.5,.2)--(-2.7,.2)
				(-3.3,.2)--(-3.5,.2);
				
		\draw [decorate,decoration={brace,mirror,amplitude=5pt}](-1.9,2.1) -- (-4.1,2.1);
		\draw node at (-3,2.5) {\footnotesize $n-1$};

	\end{tikzpicture}
	\caption{}
    \end{subfigure}
    \hfill
	\begin{subfigure}[t]{.49\textwidth}
    \centering
	\begin{tikzpicture}
		\draw 	(-.5,0)--(-.5,2)
				(-1,0)--(-1,2)
				(-2,0)--(-2,2)
				(-2.5,0)--(-2.5,2)
				(-3.5,0)--(-3.5,2)
				(-4,0)--(-4,2)
				(-5,0)--(-5,2)
				(-5.5,0)--(-5.5,2);
		\draw 	(-.5,.7)--(-1,.7)
				(-.5,.5)--(-1,.5)
				(-.5,.3)--(-1,.3)
				(-1,.4)--(-2,.4)
				(-1,1)--(-2,1)
				(-5,.7)--(-5.5,.7)
				(-5,.5)--(-5.5,.5)
				(-5,.3)--(-5.5,.3)
				(-4,.4)--(-5,.4)
				(-4,1)--(-5,1)
				(-3.5,1)--(-4,1)
				(-3.5,.5)--(-4,.5)
				(-3.5,.3)--(-4,.3)
				(-2,.3)--(-2.5,.3)
				(-2,.5)--(-2.5,.5)
				(-2,1)--(-2.5,1)
				(-2.5,.4)--(-2.7,.4)
				(-2.5,1)--(-2.7,1)
				(-3.3,.4)--(-3.5,.4)
				(-3.3,1)--(-3.5,1);
		\fill (-.75,0) circle [radius=.5pt];
		\fill (-.75,-.15) circle [radius=.5pt];
		\fill (-.75,.15) circle [radius=.5pt];
		\fill (-1.5,0) circle [radius=.5pt];
		\fill (-1.5,-.15) circle [radius=.5pt];
		\fill (-1.5,.15) circle [radius=.5pt];
		\fill (-2.25,0) circle [radius=.5pt];
		\fill (-2.25,-.15) circle [radius=.5pt];
		\fill (-2.25,.15) circle [radius=.5pt];
		\fill (-3.75,0) circle [radius=.5pt];
		\fill (-3.75,-.15) circle [radius=.5pt];
		\fill (-3.75,.15) circle [radius=.5pt];
		\fill (-4.5,0) circle [radius=.5pt];
		\fill (-4.5,-.15) circle [radius=.5pt];
		\fill (-4.5,.15) circle [radius=.5pt];
		\fill (-5.25,0) circle [radius=.5pt];
		\fill (-5.25,-.15) circle [radius=.5pt];
		\fill (-5.25,.15) circle [radius=.5pt];
		\draw (-1.5,1) node[cross] {};
		\draw (-4.5,1) node[cross] {};
		
		\fill (-3,1) circle [radius=.5pt];
		\fill (-3.15,1) circle [radius=.5pt];
		\fill (-2.85,1) circle [radius=.5pt];
		
		\fill (0,.5) circle [radius=.5pt];
		\fill (0.15,.5) circle [radius=.5pt];
		\fill (-.15,.5) circle [radius=.5pt];
		
		\fill (-6,.5) circle [radius=.5pt];
		\fill (-6.15,.5) circle [radius=.5pt];
		\fill (-5.85,.5) circle [radius=.5pt];
		
		\draw 	(-4,.2)--(-5,.2)
				(-1,.2)--(-2,.2)
				(-2.5,.2)--(-2.7,.2)
				(-3.3,.2)--(-3.5,.2);
				
		\draw [decorate,decoration={brace,mirror,amplitude=5pt}](-1.9,2.1) -- (-4.1,2.1);
		\draw node at (-3,2.5) {\footnotesize $n-1$};	\end{tikzpicture}	
	\caption{}
    \end{subfigure}
    \\[12pt]
    \begin{subfigure}[t]{.49\textwidth}
    \centering
	\begin{tikzpicture}
		\draw 	(-.5,0)--(-.5,2)
				(-1,0)--(-1,2)
				(-2,0)--(-2,2)
				(-2.5,0)--(-2.5,2)
				(-3.5,0)--(-3.5,2)
				(-4,0)--(-4,2)
				(-5,0)--(-5,2)
				(-5.5,0)--(-5.5,2);
		\draw 	(-.5,.7)--(-1,.7)
				(-.5,.5)--(-1,.5)
				(-.5,.3)--(-1,.3)
				(-1,.4)--(-2,.4)
				(-1,1)--(-1.5,1)
				(-5,.7)--(-5.5,.7)
				(-5,.5)--(-5.5,.5)
				(-5,.3)--(-5.5,.3)
				(-4,.4)--(-5,.4)
				(-4.5,1)--(-5,1)
				(-3.5,.5)--(-4,.5)
				(-3.5,.3)--(-4,.3)
				(-2,.3)--(-2.5,.3)
				(-2,.5)--(-2.5,.5)
				(-2.5,.4)--(-2.7,.4)
				(-3.3,.4)--(-3.5,.4);
		\fill (-.75,0) circle [radius=.5pt];
		\fill (-.75,-.15) circle [radius=.5pt];
		\fill (-.75,.15) circle [radius=.5pt];
		\fill (-1.5,0) circle [radius=.5pt];
		\fill (-1.5,-.15) circle [radius=.5pt];
		\fill (-1.5,.15) circle [radius=.5pt];
		\fill (-2.25,0) circle [radius=.5pt];
		\fill (-2.25,-.15) circle [radius=.5pt];
		\fill (-2.25,.15) circle [radius=.5pt];
		\fill (-3.75,0) circle [radius=.5pt];
		\fill (-3.75,-.15) circle [radius=.5pt];
		\fill (-3.75,.15) circle [radius=.5pt];
		\fill (-4.5,0) circle [radius=.5pt];
		\fill (-4.5,-.15) circle [radius=.5pt];
		\fill (-4.5,.15) circle [radius=.5pt];
		\fill (-5.25,0) circle [radius=.5pt];
		\fill (-5.25,-.15) circle [radius=.5pt];
		\fill (-5.25,.15) circle [radius=.5pt];
		\draw (-1.5,1) node[cross] {};
		\draw (-4.5,1) node[cross] {};
		
		\fill (-3,1) circle [radius=.5pt];
		\fill (-3.15,1) circle [radius=.5pt];
		\fill (-2.85,1) circle [radius=.5pt];
		
		\fill (0,.5) circle [radius=.5pt];
		\fill (0.15,.5) circle [radius=.5pt];
		\fill (-.15,.5) circle [radius=.5pt];
		
		\fill (-6,.5) circle [radius=.5pt];
		\fill (-6.15,.5) circle [radius=.5pt];
		\fill (-5.85,.5) circle [radius=.5pt];
		
		\draw 	(-4,.2)--(-5,.2)
				(-1,.2)--(-2,.2)
				(-2.5,.2)--(-2.7,.2)
				(-3.3,.2)--(-3.5,.2);
				
		\draw [decorate,decoration={brace,mirror,amplitude=5pt}](-1.9,2.1) -- (-4.1,2.1);
		\draw node at (-3,2.5) {\footnotesize $n-1$};	\end{tikzpicture}	
	\caption{}
    \end{subfigure}
\hfill
    \begin{subfigure}[t]{.49\textwidth}
    \centering
	\begin{tikzpicture}
		\draw 	(-.5,0)--(-.5,2)
				(-1.5,0)--(-1.5,2)
				(-2,0)--(-2,2)
				(-2.5,0)--(-2.5,2)
				(-3.5,0)--(-3.5,2)
				(-4,0)--(-4,2)
				(-4.5,0)--(-4.5,2)
				(-5.5,0)--(-5.5,2);
		\draw 	(-.5,.7)--(-1.5,.7)
				(-.5,.5)--(-1.5,.5)
				(-.5,.3)--(-1.5,.3)
				(-1.5,.4)--(-2,.4)
				(-4.5,.7)--(-5.5,.7)
				(-4.5,.5)--(-5.5,.5)
				(-4.5,.3)--(-5.5,.3)
				(-4,.4)--(-4.5,.4)
				(-3.5,.5)--(-4,.5)
				(-3.5,.3)--(-4,.3)
				(-2,.3)--(-2.5,.3)
				(-2,.5)--(-2.5,.5)
				(-2.5,.4)--(-2.7,.4)
				(-3.3,.4)--(-3.5,.4);
		\fill (-1,0) circle [radius=.5pt];
		\fill (-1,-.15) circle [radius=.5pt];
		\fill (-1,.15) circle [radius=.5pt];
		\fill (-1.75,0) circle [radius=.5pt];
		\fill (-1.75,-.15) circle [radius=.5pt];
		\fill (-1.75,.15) circle [radius=.5pt];
		\fill (-2.25,0) circle [radius=.5pt];
		\fill (-2.25,-.15) circle [radius=.5pt];
		\fill (-2.25,.15) circle [radius=.5pt];
		\fill (-3.75,0) circle [radius=.5pt];
		\fill (-3.75,-.15) circle [radius=.5pt];
		\fill (-3.75,.15) circle [radius=.5pt];
		\fill (-4.25,0) circle [radius=.5pt];
		\fill (-4.25,-.15) circle [radius=.5pt];
		\fill (-4.25,.15) circle [radius=.5pt];
		\fill (-5,0) circle [radius=.5pt];
		\fill (-5,-.15) circle [radius=.5pt];
		\fill (-5,.15) circle [radius=.5pt];
		\draw (-1,1) node[cross] {};
		\draw (-5,1) node[cross] {};
		
		\fill (-3,1) circle [radius=.5pt];
		\fill (-3.15,1) circle [radius=.5pt];
		\fill (-2.85,1) circle [radius=.5pt];
		
		\fill (0,.5) circle [radius=.5pt];
		\fill (0.15,.5) circle [radius=.5pt];
		\fill (-.15,.5) circle [radius=.5pt];
		
		\fill (-6,.5) circle [radius=.5pt];
		\fill (-6.15,.5) circle [radius=.5pt];
		\fill (-5.85,.5) circle [radius=.5pt];
		
		\draw 	(-4,.2)--(-4.5,.2)
				(-1.5,.2)--(-2,.2)
				(-2.5,.2)--(-2.7,.2)
				(-3.3,.2)--(-3.5,.2);
				
		\draw [decorate,decoration={brace,mirror,amplitude=5pt}](-1.9,2.1) -- (-4.1,2.1);
		\draw node at (-3,2.5) {\footnotesize $n-1$};	\end{tikzpicture}		
	\caption{}
    \end{subfigure}
    \caption{$a_n$ Kraft-Procesi transition. (a) Generic part  of a Coulomb branch brane configuration with moduli space $\M_C$. The curly brace indicates the presence of $n-1$ NS5-branes in the middle, creating $n-2$ interval with no D5-branes in them. The moduli space generated by one D3-brane in each of the $n$  middle intervals is the variety $S=a_n$, i.e. it is isomorphic to the Higgs branch of $3d$ $\mathcal{N}=4$ SQED with $n+1$ flavors. (b) The $n$ relevant D3-branes are taken to the singular point $\vec{x}_i=\vec{m}_j=(0,0,0)$. (c) The $n$ D3-branes join and then split into three segments, one of them can acquire nonzero $\vec{y}$ position along the directions spanned by the D5-branes. One can take the limit when its $\vec{y}$ position goes to infinity, removing this brane from the system. This gives rise to a new brane configuration with a new Coulomb branch $\M_C'\subset\M_C$. (d) Two phase transitions have been performed to  annihilate the fixed D3-brane segments and obtain a Coulomb branch brane configuration where the corresponding new quiver can be read.}
    \label{fig:KPdefan}
\end{figure}

 Figure \ref{fig:KPdefAn} shows the steps in the $A_n$ Kraft-Procesi transition. Figure \ref{fig:KPdefan} shows the steps in the $a_n$ Kraft-Procesi transition.

The matrix formalism developed in \citep{CH16} \emph{Section 6: The matrix formalism} computes all possible KP (Kraft-Procesi) transitions between nilpotent orbits of $\gsl (n)$, starting from the self-dual quiver in figure \ref{fig:quiverTSUn}. The formalism gives all quivers, brane configurations, dimension of the moduli space and partition $\lambda$ corresponding to the closures of the nilpotent orbits. The purpose of the present note is to develop analogous techniques for orthogonal and symplectic algebras.

\clearpage

\section{Orietifold planes and orthosymplectic quivers}\label{sec:3}

In this section we introduce O3-planes in the brane configurations. Here we use the results presented in \cite{FH00} \emph{Section 2: Some facts concerning O3 planes} (see also \cite{W98} for more details). A single O3-plane that spans directions $\{x^1,x^2,x^6\}$ is added. A new feature of the brane diagrams is that we encounter half D5-branes, half NS5-branes and half D3-branes. Let the O3-plane be infinite in the three directions it spans and let it be located at the origin in the remaining coordinates: $\vec{x}=(0,0,0)$, $\vec{y}=(0,0,0)$. All half fivebranes are also located at the origin, either $\vec{m}_i=(0,0,0)$ for the position of the $i$-th half D5-brane or $\vec{w}_j=(0,0,0)$ for the position of the $j$-th half NS5-brane.

The O3-plane encounters all half fivebranes along the $x^6$ direction. Every time it intersects with one of them it changes its nature \cite{E97,HZ99,HB00}. There are four different kinds of O3-planes: $O3^-$, $\widetilde{O3^-}$, $O3^+$ and $\widetilde{O3^+}$. An $O3^-$ (resp. $O3^+$) changes to $\widetilde{O3^-}$ (resp. $\widetilde{O3^+}$) after passing a half D5-brane and vice versa. An $O3^-$ (resp. $\widetilde{O3^-}$) changes to $O3^+$ (resp. $\widetilde{O3^+}$) after passing a half NS5-brane and vice versa. When reading the quiver from the Coulomb branch brane configuration, the type of O3-plane at the beginning and end of each interval between half NS5-branes determines the nature of the gauge and flavor nodes corresponding to such interval in the way summarized in table \ref{tab:O3}. 

\begin{table}[t]
	\centering
	\begin{tabular}{|c| c |c|}
	\hline
	Orientifold plane & Gauge node & Flavor node \\ \hline
	& & \\
	$O3^-$ & $O_{2n}$ & $C_{k}$ \\ 
	& & \\ \hline
	& & \\ 
	$\widetilde{O3^-}$ & $O_{2n+1}$ & $C_{k}$ \\ 
	& & \\ \hline
	& & \\ 
	$O3^+$ & $C_{n}$ & $O_{2k}$\\ 
	& & \\ \hline
	& & \\ 
	$\widetilde{O3^+}$ & $C_{n}$ & $O_{2k}$ \\
	& & \\
	\hline
	\end{tabular}
	\caption{Correspondence between nodes in the quiver and types of the orientifold plane in the Coulomb branch brane configuration. $n$ is the number of D3-branes in the corresponding interval. $k$ is the number of D5-branes in the interval (In the cases of $O3^+$ and $\widetilde{O3^+}$ $k$ can be a half integer number). $C_r$ denotes the symplectic group $Sp(r)$ of rank $r$. $O_r$ denotes a flavor group $O(r)$. If $k$ is a half integer in an interval with $O3^-$ or $\widetilde{O3^-}$ a phase transition can always be performed that moves half a D5-brane to an adjacent interval, without any brane creation/annihilation. These transitions need to be performed before the quiver can be read. We refer to this transition as the \emph{collapse} of the Coulomb branch brane configuration. (The name \emph{collapse} reflects the effect of this transition on the partition defined by the position of the half D5-branes along half NS5-brane intervals, for a detailed description of the collapse of a partition see section \ref{sec:collapse}).}
	\label{tab:O3}
\end{table}

\begin{figure}[t]
	\centering
        \begin{subfigure}[t]{.49\textwidth}
    \centering
	\begin{tikzpicture}
		\draw 	(1,1) node[cross]{}
				(2,1)node[cross]{}
				(2.5,1)node[cross]{}
				(3,1)node[cross]{}
				(3.5,1)node[cross]{};
		\draw 	(1.5,0)--(1.5,2)
				(4,0)--(4,2)
				(4.5,0)--(4.5,2)
				(5,0)--(5,2)
				(5.5,0)--(5.5,2);
		\draw (2,1)--(2.5,1)
				(3,1)--(3.5,1);
		\draw[dotted](.5,1)--(1.5,1)
				(4,1)--(4.5,1)
				(5,1)--(5.5,1);
		\draw	(1.5,1.5)--(4,1.5)
				(4,1.6)--(4.5,1.6)
				(4.5,1.5)--(5,1.5)
				(1.5,.5)--(4,.5)
				(4,.4)--(4.5,.4)
				(4.5,.5)--(5,.5)
				(1.5,1.8)--(4,1.8)
				(1.5,.2)--(4,.2);
		\end{tikzpicture}
        \caption{}
    \end{subfigure}
    \hfill
	\begin{subfigure}[t]{.49\textwidth}
    \centering
	\begin{tikzpicture}[]
	\tikzstyle{gauge} = [circle, draw];
	\tikzstyle{flavour} = [regular polygon,regular polygon sides=4,draw];
	\node (g1) [gauge,label=below:{$O_4$}] {};
	\node (g2) [gauge,right of=g1,label=below:{$C_1$}] {};
	\node (g3) [gauge,right of=g2,label=below:{$O_2$}] {};
	\node (f1) [flavour,above of=g1,label=above:{$C_2$}] {};
	\draw (f1)--(g1)--(g2)--(g3)
		;
	\end{tikzpicture}

        \caption{}
    \end{subfigure}
 	\caption{Example of a brane configuration and its corresponding quiver. In (a) the horizontal direction corresponds to space direction $x^6$, the vertical direction corresponds to space directions $\vec{x}=(x^3,x^4,x^5)$ and the direction perpendicular to the paper corresponds to space directions $\vec{y}=(x^7,x^8,x^9)$. Vertical solid lines represent half NS5-branes, crosses represent half D5-branes, horizontal lines in the center of the diagram represent O3-planes: a solid line represents an $\widetilde{O3^-}$, the dotted line to the left of the leftmost half D5-brane represents an $\widetilde{O3^+}$, the other dotted lines all represent $O3^+$s. $O3^-$s are located in the spaces between half fivebranes that have been left empty. (b) is the corresponding quiver, where the groups are in agreement with table \ref{tab:O3}; the gauge node $C_0$ has been omited.}
	\label{fig:OSquivers}
\end{figure}
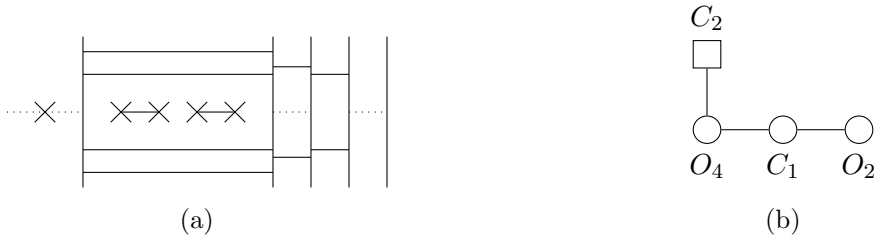

The convention of \cite{GW09} on how to draw the O3-planes is adopted here: a solid horizontal line for $\widetilde{O3^-}$, dotted horizontal line for $O3^+$ and $\widetilde{O3^+}$ and no line for $O3^-$. In figure \ref{fig:OSquivers} we show one example of such quivers. 

\subsection{Nilpotent orbits of $\mathfrak{sp}(n)$ and $\mathfrak{so}(n)$}

As reviewed in \cite{CH16}, \emph{Section 3: Mathematical preulde}, nilpotent orbits of $\mathfrak{g}=\mathfrak{sl}(n)$ are classified employing the set of partitions of $n$: $\mathcal{P}(n)$.
\begin{align}
\mathcal{P}(n)=\{(n),(n-1,1),\dots,(2,1^{n-2}),(1^n)\}
\end{align}
 A similar classification can be employed for nilpotent orbits of orthogonal and symplectic algebras. This description is taken from \cite{CM93}. For $\mathfrak{g}=\mathfrak{so}(n)$ there is a one to one correspondence between the set of nilpotent orbits and the set of partitions of $n$ restricted to partitions where even parts occur with even multiplicity, this set is denoted by $\mathcal{P}_{+1}(n)$. For example:
\begin{align}
 \mathcal{P}_{+1}(5)=\{(5),(3,1^2),(2^2,1),(1^5)\}
\end{align}

For $\mathfrak{g}=\mathfrak{sp}(n)$ there\footnote{By $\mathfrak{sp}(n)$ we denote the algebra of type $C_n$ acting naturally on a vector space of dimension $2n$.} is a one to one correspondence between the set of nilpotent orbits and the set of partitions of $2n$ restricted to partitions where odd parts occur with even multiplicity, this set is denoted by $\mathcal{P}_{-1}(2n)$. For example:
\begin{align}
 \mathcal{P}_{-1}(4)=\{(4),(2^2),(2,1^2),(1^4)\}
\end{align}

There is one exception to the one to one correspondence between partitions and nilpotent orbits for the case of $\mathfrak{so}({2n})$: \emph{very even partitions}. Very even partitions are partitions with only even parts, they correspond to two different nilpotent orbits under the adjoint action of the group $SO({2n})$. They are denoted with superscript $\mathcal{O}^{I}_\lambda$ and $\mathcal{O}^{II}_\lambda$, where $\lambda$ is the very even partition, for example $\lambda=(4^2,2^2)\in \mathcal{P}_{+1}(12)$. Note however that they only correspond to a single nilpotent orbit under the action of the group $O({2n})$, the single orbit is the union $\mathcal{O}^{I}_\lambda\cup\mathcal{O}^{II}_\lambda$; let us denote this union simply by $\mathcal O_{\lambda}$.

\section{Partitions and the dual map}\label{sec:dual}

It is convenient to introduce the \emph{Barbasch-Vogan map} or \emph{dual map} between partitions before giving the general prescription on how to obtain brane systems for nilpotent orbits.  This section reviews the relevant maps between sets of partitions in the context of nilpotent orbits of classical Lie algebras.

Collingwood and McGovern \cite{CM93} refer to the  map $d_{LS}$ introduced by Spaltenstein in \cite{Sp82} as the \emph{Spaltenstein map}; they denote it by $d$. In more recent literature the map is known as the \emph{Lusztig-Spaltenstein map} and is denoted by $d_{LS}$. This is the notation adopted here. The  Lusztig-Spaltenstein map is a composition of the transpose map and the X-collapse, where X is either B,C or D.

An extension to this map is $d_{BV}$, the \emph{Barbasch-Vogan map} \cite{BV85}. This map is described by equation (2.8) of \cite{CDT13} and it appeared before in equation (5) of \cite{A02}. In \cite{CDT13} they adopt the name \emph{Spaltenstein map} for $d_{BV}$, and drop the subscript, denoting it simply by $d$. In the present report we keep the name \emph{Barbasch-Vogan map} and the notation $d_{BV}$.

Before introducing the Lusztig-Spaltenstein and the Barbasch-Vogan maps, let us describe the transpose and the X-collapse maps.

\subsection{Transpose map}

Let $\lambda=(\lambda_1,\lambda_2,\dots,\lambda_k)$ be a partition of length $|\lambda|=k$, i.e. $\lambda_1\geq \lambda_2\geq \dots \geq \lambda_k$ and $\lambda_i\in \mathbb{N}$. The diagram of the partition is a set of squares at positions $(i,j)$ of an $\mathbb{N}^2$ lattice, where $i$ increases downwards, $j$ increases from left to right and $1\geq j \geq \lambda_i$ \cite{M95}. For example, the diagram of $\lambda=(4,2,2,2,1)$ is depicted in figure \ref{fig:partition}.

\begin{figure}
	\centering
	\resizebox{0.2\textwidth}{!}{
		\begin{tikzpicture}
		\draw 	(0,0)--(5,0)
				(0,-1)--(5,-1)
				(0,-2)--(4,-2)
				(0,-3)--(1,-3)
				(0,-4)--(1,-4);
		\draw	(0,0)--(0,-4)
				(1,0)--(1,-4)
				(2,0)--(2,-2)
				(3,0)--(3,-2)
				(4,0)--(4,-2)
				(5,0)--(5,-1);
		\end{tikzpicture}
	}
	\caption{Diagram of the transpose partition $\lambda^t=(4,2,2,2,1)^t=(5,4,1,1)$ after reflecting through the diagonal.}
	\label{fig:transpose}
\end{figure}
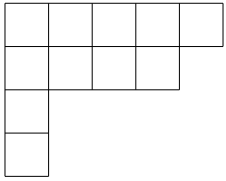

The transpose map reflects along the diagonal of the diagram (i.e. swaps rows and columns), for example $\lambda^t=(4,2,2,2,1)^t=(5,4,1,1)$, see figure \ref{fig:transpose}.

\begin{figure}
	\centering
	\resizebox{0.2\textwidth}{!}{
	\begin{tikzpicture}
		\draw (0,0)--(1,0)--(2,0)--(3,0)--(4,0);
		\draw (0,-1)--(1,-1)--(2,-1)--(3,-1)--(4,-1);
		\draw (0,-2)--(1,-2)--(2,-2);
		\draw (0,-3)--(1,-3)--(2,-3);
		\draw (0,-4)--(1,-4)--(2,-4);
		\draw (0,-5)--(1,-5);
		\draw 	(0,0)--(0,-5)
				(1,0)--(1,-5)
				(2,0)--(2,-4)
				(3,0)--(3,-1)
				(4,0)--(4,-1);
	\end{tikzpicture}
	}
	\caption{Diagram of partition $\lambda=(4,2,2,2,1)$.}
	\label{fig:partition}
\end{figure}
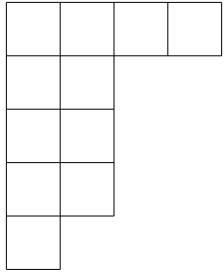

\subsection{X-collapse}\label{sec:collapse}

The X-collapse map is described in  \emph{Section 6.3.3} of \cite{CM93}. Let us review this definition here, starting with $X=B$. The B-collapse takes a partition $\lambda\in \mathcal P(2n+1)$ to the largest partition $\lambda_B$ in $\mathcal P_{+1}(2n+1)$ dominated by $\lambda$. The words \emph{largest} and \emph{dominated} refer to the \emph{natural partial ordering} of the partitions \cite{M95}. Let $\lambda,\lambda'\in \mathcal P (m)$, the \emph{natural partial ordering} is defined as:
\begin{align}
	\lambda>\lambda' \Leftrightarrow \sum_{i=1}^l \lambda_i \geq \sum_{i'=1}^l \lambda_{i'} '\ \ \forall l
\end{align}

One says then that $\lambda$ is larger than $\lambda'$, or that $\lambda$ dominates $\lambda'$. To perform the B-collapse of partition $\lambda\in \mathcal P (2n+1)$ there are two possibilities. If $\lambda \in \mathcal P_{+1}(2n+1)$ then $\lambda_B=\lambda$. If $\lambda \not \in \mathcal P_{+1}(2n+1)$ then at least one of the even parts in $\lambda$ has odd multiplicity. Let the largest of such parts have value $2k$. Substitute the last occurrence of $2k$ with $2k-1$. Then add one to the first part $\lambda_j$ such that $\lambda_j<2k-1$. An extra part with value zero can be added at the end of the partition if needed. Repeat such process until all even parts in the partition have even multiplicity.

The D-collapse is analogous to the B-collapse. It acts on a partition $\lambda \in \mathcal P(2n)$ to obtain $\lambda_D$, the largest partition in $\mathcal P_{+1}(2n)$ dominated by $\lambda$. The way of obtaining $\lambda_D$ is identical to the way in which $\lambda_B$ is obtained in the B-collapse.

In the C-collapse the same steps are taken but in this case to remove odd multiplicities of odd parts. It takes a partition $\lambda\in \mathcal P (2n)$ to $\lambda_C$, the largest partition in $\mathcal P _{-1}(2n)$ dominated by $\lambda$.

Some examples of X-collapse are:
\begin{align}
	\begin{aligned}
		(4,2,2,2,1)_B&=(3,3,2,2,1)\\
		(5,4,4,4,3,2,2,2)_D&=(5,4,4,3,3,3,2,2)\\
		(5,4,4,4,3,2,2,2)_C&=(4,4,4,4,4,2,2,2)\\
	\end{aligned}
\end{align}

Note that for the case of $\mathfrak{sl}(n)$, the A-collapse map is the identity map.

\subsection{Lusztig-Spaltenstein map}\label{sec:LS}

The Lusztig-Spaltenstein map $d_{LS}$ is the composition of the transpose map and the X-collapse:
\begin{align}
	d_{LS}(\lambda)=(\lambda^t)_X, \ \ X\in \{B,C,D\}
\end{align}

For example, let $\lambda=(4,2,2,2,1)\in \mathcal P _{+1}(11)$, then:

\begin{align}
	d_{LS}(4,2,2,2,1)=(4,2,2,2,1)^t_{\ B}=(5,4,1,1)_B=(5,3,1,1,1)
\end{align}

A partition $\lambda$ is called \emph{special} if $(d_{LS})^2(\lambda)=\lambda$. The corresponding nilpotent orbit is called \emph{special nilpotent orbit}. One should think of the Lusztig-Spaltenstein map as a variation on the transpose map that ensures that the resulting partition stays within the set of allowed partitions for a particular classical algebra. When restricted to the set of special partitions of a classical algebra the Lusztig-Spaltenstein map is a bijective map of the set into itself that reverses its natural partial ordering. Note that for the $\mathfrak{sl}(n)$ case the $d_{LS}$ map reduces to the transpose map.

\subsection{Barbasch-Vogan map}\label{sec:BV}

The Barbasch-Vogan map, denoted here by $d_{BV}$, takes a partition of a classical algebra $\mathfrak{g}$ to a special partition of its Langlands dual or GNO dual \cite{GNO76} algebra $\mathfrak{g}^\vee$. Remember that $\mathfrak{sp}(n)^\vee=\mathfrak{so}(2n+1)$ and vice versa; the remaining classical algebras are dual to themselves. The map as defined\footnote{The presentation here follows the clear description given by equation (5) of \cite{A02} combined with two of the identities in Lemma 3.3 of the same source.}  in \cite{A02} is:

A-type:
\begin{align}
	d_{BV}(\lambda) = \lam^t
\end{align}

B-type:
\begin{align}
	d_{BV}(\lambda) = (\lam^t)^-_{\ \ C}
\end{align}

C-type:
\begin{align}
	d_{BV}(\lambda) =  (\lam^t)^+_{\ \ B}
\end{align}

D-type:
\begin{align}\label{eq:D}
	d_{BV}(\lambda) = (\lam^t)_D
\end{align}

Note that in the $\gsl (n)$ and $\gso (2n)$ cases, where the algebras are self-dual, the map reduces to the Lusztig-Spaltenstein map. The $+$ and $-$ superscripts denote augmented and reduced partitions: let $\lambda=(\lam _1, \lam_2, \dots, \lam _k)$ be a partition, then
\begin{align}\label{eq:aug}
	\begin{aligned}
		\lambda^+&:=(\lam _1+1, \lam_2, \dots, \lam _k)\\
		\lambda^-&:=(\lam _1, \lam_2, \dots, \lam _k-1)\\
	\end{aligned}
\end{align} 

One should think about the Barbasch-Vogan map as the map that acts on the set of allowed partitions of a given classical algebra, reverses their natural partial ordering and takes them to the Langlands dual algebra. When restricted to the set of special partitions the  Barbasch-Vogan map is the order-reversing bijection from special partitions of $\mathfrak{g}$ to the set of special partitions of $\mathfrak{g}^\vee$.
This idea is made apparent below, in section \ref{sec:BVexample}, during the discussion of the $\gso (5)$ and $\gsp (2)$ examples.

\paragraph{Comment about notation}

Note that while in papers like \citep{CDT13,GW09} the superscript $\vee$ over a partition $\lam$ is used to denote a generic partition of the algebra $\mathfrak g ^\vee$, in \cite{CHMZ14} $\lambda^\vee$ is used to denote the Barbasch-Vogan\footnote{Note that there is a mistake in \cite{CHMZ14} in the definition of $d_{BV}(\lambda)$ for C-type, the prescription for $(\lambda^+)^t_{\ B}$ is given instead of the prescription for $(\lambda^t)^+_{\ B}$.} image of partition $\lambda$, i.e. $d_{BV}(\lam)$.

\section{$SO(4)$ Interlude}\label{sec:so4interlude}
\subsection{Example: Branes for the maximal orbit of $\mathfrak{so(4)}$}\label{sec:4}

\subsubsection{Partitions}
Let $\mathfrak{g}=\mathfrak{so}(4)$. Let the set $\mathcal{P}_{+1}(4)$ be ordered in a column:
\begin{align}\label{eq:P4}
	\mathcal{P}_{+1}(4)=\left\lbrace\begin{array}{c}
		(3,1)\\
		(2^2)\\
		(1^4)
		\end{array}\right\rbrace
\end{align}
This establishes that there are four nilpotent orbits under the adjoint action of the group $SO(4)$, one corresponding to partition $(3,1)$, two corresponding to the \emph{very even} partition $(2^2)$, and one corresponding to $(1^4)$:
\begin{align}
	\mathcal{O}_{(4)},\ \mathcal{O}_{(2^2)}^{I},\ \mathcal{O}_{(2^2)}^{II},\ \mathcal{O}_{(1^4)}
\end{align}

The closures of these nilpotent orbits are hyperk\"ahler singularities. Higgs branches of orthosymplectic quivers have been found\footnote{See \cite{FH16} for a description of Higgs branches as unions of two cones $\bar{\mathcal{O}}_{(2^{2n})}^{I}\cup\bar{\mathcal{O}}_{(2^{2n})}^{II}$ with $(2^{2n})\in\mathcal{P}_{+1}(4n)$ a very even partition.} to be closures of nilpotent orbits under the action of $O(4)$, this means that we have quiver gauge theories with the following Higgs branches:
\begin{align}
	\bar{\mathcal{O}}_{(4)},\ \bar{\mathcal{O}}_{(2^2)}^{I}\cup\bar{\mathcal{O}}_{(2^2)}^{II},\ \bar{\mathcal{O}}_{(1^4)}
\end{align}

As in the special linear case, the closure of nilpotent orbits is the union of finitely many nilpotent orbits. Hence, there is a partial order in the varieties, determined by the inclusion relation. The variety $\bar{\mathcal{O}}_{(1^4)}$ is trivial, a single point. We have:
\begin{align}
\begin{aligned}
	\Or_{(4)}&=\mathcal{O}_{(4)}\cup\mathcal{O}_{(2^2)}^{I}\cup\mathcal{O}_{(2^2)}^{II}\cup\mathcal{O}_{(1^4)}\\
\Or_{(2^2)}^{II}&=\mathcal{O}_{(2^2)}^{II}\cup\mathcal{O}_{(1^4)}\\
	\Or_{(2^2)}^{I}&=\mathcal{O}_{(2^2)}^{I}\cup\mathcal{O}_{(1^4)}\\
	\Or_{(1^4)}&=\mathcal{O}_{(1^4)}\\
\end{aligned}
\end{align}

To obtain the quiver corresponding to the closure of each nilpotent orbit we need to compute the Barbasch-Vogan dual partitions of (\ref{eq:P4}). According to equation (\ref{eq:D}) the $d_{BV}$ map is the composition of the transpose and the D-collapse. The result of transposition of (\ref{eq:P4}) is:

\begin{align}\label{eq:Pt4}
	\mathcal{P}_{+1}(4)^t=\left\lbrace\begin{array}{c}
		(2,1^2)\\
		(2^2)\\
		(4)
		\end{array}\right\rbrace
\end{align}

Some partitions in $\mathcal{P}_{+1}(4)^t$ do not belong to $\mathcal{P}_{+1}(4)$.
The D-collapse takes care of this issue, producing the set:

\begin{align}\label{eq:dP4}
	d_{BV}\left(\mathcal{P}_{+1}(4)\right)=\mathcal{P}_{+1}(4)^t_{\ D}=\left\lbrace\begin{array}{c}
		(1^4)\\
		(2^2)\\
		(3,1)
		\end{array}\right\rbrace
\end{align}

\subsubsection{Brane configuration}
Brane systems of \cite{CH16} are fully determined by the number of D5-branes, the number of NS5-branes and the linking numbers of each of these branes. After introducing the O3-planes the determining parameters are the number of half D5-branes, denoted $n_d$, the number of half NS5-branes, denoted $n_s$, and the linking number of each of these branes, denoted $\vec{l}_d$ and $\vec{l}_s$ respectively\footnote{Note that in \cite{CH16} $n_s$ and $n_d$ refer to entire fivebranes. We use the convention in \cite{GW09} for the linking numbers of half fivebranes. We consider the $O3^+$, $\widetilde{O3^+}$ and $\widetilde{O3^-}$ to have the same contribution to the linking numbers as half D3-branes. $O3^-$ gives no contribution.}. The difference with \cite{CH16} is that in this case the linking numbers of half fivebranes can change via phase transition such as \emph{splitting} \cite{FH00}. As a result, the linking numbers depend on whether the half fivebranes are away from the O3-planes or on top of them. Therefore, the ambiguity can be removed if one restricts to brane systems in which all half fivebranes are always located at the origin, either $\vec{m}_i=(0,0,0)$ for half D5-branes or $\vec{y}_j=(0,0,0)$ for half NS5-branes, coinciding with the position of the orientifold planes. There is also an extra parameter that one needs to fix in addition to those in \cite{CH16}: the type of each orientifold plane. However, fixing the type of a single orientifold plane unequivocally determines the type of all the others.

In order to get a brane configuration corresponding to a quiver whose Higgs branch is the closure of a nilpotent orbit of $\mathfrak{so}(2n)$ we follow the following prescription. For:
\begin{align}
	\M_H=\Or_\lambda\subset\gso (2n)
\end{align}
the parameters of the model are:
\begin{align}
	n_d=n_s=2n
\end{align}
also:
\begin{align}
	\vec{l}_d=(2n-1,2n-1,\dots,2n-1)
\end{align}
Remember that in the $\gsl (m)$ case the $\vec{l}_s$ parameter is related to $\lambda^t$; in the present case it is related to $d_{BV}(\lambda)$ (which for A-type coincides with $\lambda^t$). 
Let us focus on one example. Let $\lambda=(3,1)$, this corresponds to the maximal nilpotent orbit of $\mathfrak{so}(4)$. We have:
\begin{align}
	\begin{aligned}
	n_d&=4\\
	n_s&=4\\
	\vec{l}_d&=(3,3,3,3)
	\end{aligned}
\end{align}
The \emph{dual} partition is:
\begin{align}
	d_{BV}(\lambda)=(1^4)
\end{align}
The linking numbers of the half NS5-branes are given by $d_{BV}(\lambda)$, we have:
\begin{align}
	\vec{l}_s=(1,1,1,1)
\end{align}
Now one needs to fix the nature of the orientifold planes. Let the rightmost O3-plane be $O3^-$. This ensures that the flavor symmetry of the quiver is $O(4)$ as described below. Note that since $n_s=n_d=2n$, the leftmost plane becomes $O3^-$, this is one of the physical realizations of self-duality of $\mathfrak{so}({2n})$.

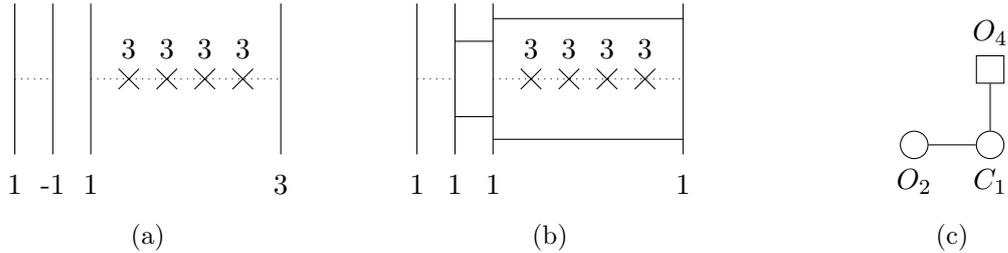
\begin{figure}[t]
	\centering
\begin{subfigure}[t]{.3\textwidth}
    \centering
	\begin{tikzpicture}
		\draw 	(1,0)--(1,2)
				(1.5,0)--(1.5,2)
				(2,0)--(2,2)
				(4.5,0)--(4.5,2);
		\draw 	(1,0) node[label=below:{1}] {} 
				(1.5,0)node[label=below:{-1}] {} 
				(2,0)node[label=below:{1}] {} 
				(4.5,0)node[label=below:{3}] {} ;
		\draw 	(2.5,1) node[cross] {}
				(3,1) node[cross] {}
				(3.5,1) node[cross] {}
				(4,1) node[cross] {};
		\draw 	(2.5,1) node[label=above:{3}] {}
				(3,1) node[label=above:{3}] {}
				(3.5,1) node[label=above:{3}] {}
				(4,1) node[label=above:{3}] {};
		\draw [dotted] (1,1)--(1.5,1)
				(2,1)--(4.5,1);
	\end{tikzpicture}
        \caption{}
    \end{subfigure}
      \hfill
    \begin{subfigure}[t]{.3\textwidth}
    \centering
	\begin{tikzpicture}
		\draw 	(1,0)--(1,2)
				(1.5,0)--(1.5,2)
				(2,0)--(2,2)
				(4.5,0)--(4.5,2);
		\draw 	(1,0) node[label=below:{1}] {} 
				(1.5,0)node[label=below:{1}] {} 
				(2,0)node[label=below:{1}] {} 
				(4.5,0)node[label=below:{1}] {} ;
		\draw 	(2.5,1) node[cross] {}
				(3,1) node[cross] {}
				(3.5,1) node[cross] {}
				(4,1) node[cross] {};
		\draw 	(2.5,1) node[label=above:{3}] {}
				(3,1) node[label=above:{3}] {}
				(3.5,1) node[label=above:{3}] {}
				(4,1) node[label=above:{3}] {};
		\draw [dotted] (1,1)--(1.5,1)
				(2,1)--(4.5,1);
		\draw 	(2,.2)--(4.5,.2)
				(2,1.8)--(4.5,1.8)
				(1.5,.5)--(2,.5)
				(1.5,1.5)--(2,1.5);
	\end{tikzpicture}
        \caption{}
    \end{subfigure}
    \hfill
	\begin{subfigure}[t]{.3\textwidth}
    \centering
	\begin{tikzpicture}[]
	\tikzstyle{gauge} = [circle,draw];
	\tikzstyle{flavour} = [regular polygon,regular polygon sides=4,draw];
	\node (g2) [gauge,label=below:{$O_2$}] {};
	\node (g3) [gauge,right of=g2,label=below:{$C_1$}] {};
	\node (f3) [flavour,above of=g3,label=above:{$O_4$}] {};
	\draw (g2)--(g3)--(f3)
		;
	\end{tikzpicture}
        \caption{}
    \end{subfigure}
 	\caption{Orthosymplectic model with $n_s=n_d=4$, $\vec{l}_d=(3,3,3,3)$, $\vec{l}_s=(1,1,1,1)$ and rightmost orientifold plane of type $O3^-$. (a) is the first step to achieving the right brane configuration, the linking numbers of the half D5-branes are correct, but the half NS5-branes have $\vec{l}_s=(1,-1,1,3)$. Linking numbers of half NS5-branes are displayed below them, linking numbers of half D5-branes are displayed above them. In this case the rightmost O3-plane, the leftmost one and the one between the second and third half NS5-branes are $O3^-$. The dotted lines represent either $O3^+$ or $\widetilde{O3^+}$. In this configuration the $O3^-$ turns into $O3^+$ after crossing through half an NS5-brane. The $\widetilde{O3^+}s$ are found after an $O3^+$ crosses half a D5-brane. In (b) two physical D3-branes have been added and the linking numbers are the desired ones. The quiver can be read from (b), and is depicted in (c). The group $C_0$ has been omitted.}
	\label{fig:SO4maximal}
\end{figure}

The brane system is now fully determined. To draw the Coulomb branch brane configuration that allows us to read the quiver we proceed exactly as in \cite{CH16}. We draw all the half NS5-branes separated along the $x^6$ direction. Then, all half D5-branes are placed in the interval between the two rightmost half NS5-branes, also separated along the $x^6$ direction. After the positions for all the fivebranes have been chosen, the types of orientifold planes are also fixed. This is depicted in figure \ref{fig:SO4maximal}(a).

\paragraph{Linking numbers.} Let us summarize the chosen convention for the linking numbers: the linking number of a half NS5-brane (resp. D5-brane) is the sum of all half D3-branes ending on it from the right minus the sum of all half D3-branes ending on it from the left, plus all half D5-branes (resp. NS5-branes) located to its left. We also have to add $O3^+$, $\widetilde{O3^-}$ and $\widetilde{O3^+}$ to the total of D3-branes on each side. This gives linking numbers $\vec{l}_d=(3,3,3,3)$ and $\vec{l}_s=(1,-1,1,3)$ for the system in figure \ref{fig:SO4maximal}(a).\\

Now we start adding entire D3-branes in the intervals between the half NS5-branes, starting from the left, to achieve the desired values of $\vec{l}_s=(1,1,1,1)$. The first half NS5-brane from the left already has linking number 1. The second from the left has linking number -1, by putting a physical D3-brane (two half D3-branes) between it and its neighbor to the right we set it to $-1+2=1$. The third brane has now linking number -1, so we repeat the process and add a new physical D3-brane between the third and the fourth half NS5-branes. This gives the desired linking numbers 1 and 1. The result is in figure \ref{fig:SO4maximal}(b). The quiver can now be read, figure \ref{fig:SO4maximal}(c), employing the results in table \ref{tab:O3} since we have constructed a Coulomb branch brane configuration were all half D3-branes end on half NS5-branes.

\subsubsection{Higgsing of the example}

There are many interesting dynamical effects that can be studied in the configuration presented in figure \ref{fig:SO4maximal}(b). One is the phase transition to the Higgs branch brane configuration. This was introduced in \cite{HW96} for brane configurations without the O3-planes and in \cite{FH00} for brane configurations with O3-planes. After finding the Higgs branch brane configuration an S-duality can be performed to find a new quiver whose Coulomb branch is $\M_C=\Or_{(3,1)}$. Instead of performing S-duality, one can perform a Kraft-Procesi transition in order to find a new Higgs branch brane configuration, corresponding to a model with Higgs branch $\M_C=\Or_{(2^2)}$. The slice $S\subseteq \Or_{(3,1)}$ transverse to $\mathcal{O}_{(2^2)}$ can also be computed from the branes dynamics of the Kraft-Procesi transition. The Kraft-Procesi transition is the new material in this paper and it is studied in the next section. Here we review the initial phase transition from the Coulomb branch brane configuration in figure \ref{fig:SO4maximal}(b) to its Higgs branch brane configuration.

The main idea behind the transition to the Higgs branch brane configuration is to make all half D3-branes end on half D5-branes. To achieve this, one can pull half D5-branes to the extremes of the configuration so brane creation can happen, then align the half D3-branes so they end on the half D5-branes, then change the perspective and rearrange the half D3-branes. The effects of entire fivebranes splitting into half fivebranes that appear in \cite{FH00} can be avoided by keeping the half fivebranes at the origin (i.e. the position of the orientifold plane) at all times. Then, the only rule that determines whether there is brane creation or brane annihilation during a phase transitions is:

\begin{constraint}
	Linking numbers of half fivebranes are preserved.
\end{constraint}

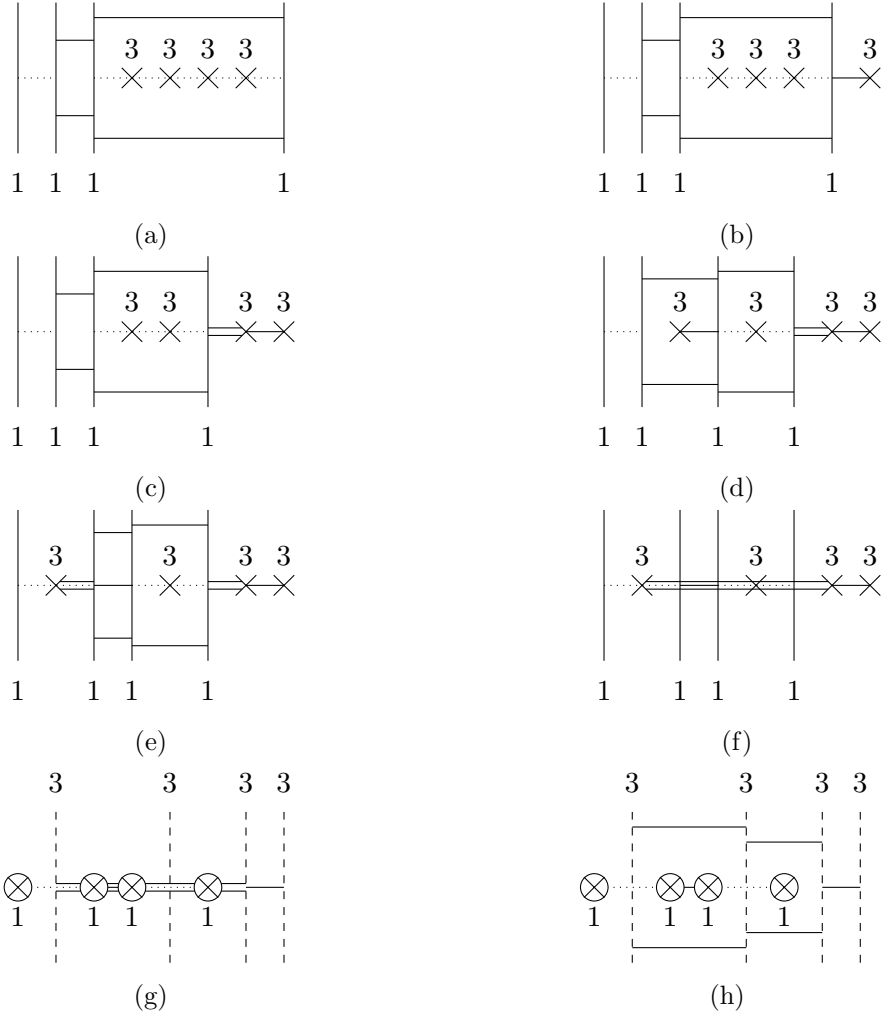
\begin{figure}[t]
	\centering
    	\begin{subfigure}[t]{.49\textwidth}
    	\centering
	\begin{tikzpicture}
		\draw 	(1,0)--(1,2)
				(1.5,0)--(1.5,2)
				(2,0)--(2,2)
				(4.5,0)--(4.5,2);
		\draw 	(1,0) node[label=below:{1}] {} 
				(1.5,0)node[label=below:{1}] {} 
				(2,0)node[label=below:{1}] {} 
				(4.5,0)node[label=below:{1}] {} ;
		\draw 	(2.5,1) node[cross] {}
				(3,1) node[cross] {}
				(3.5,1) node[cross] {}
				(4,1) node[cross] {};
		\draw 	(2.5,1) node[label=above:{3}] {}
				(3,1) node[label=above:{3}] {}
				(3.5,1) node[label=above:{3}] {}
				(4,1) node[label=above:{3}] {};
		\draw [dotted] (1,1)--(1.5,1)
				(2,1)--(4.5,1);
		\draw 	(2,.2)--(4.5,.2)
				(2,1.8)--(4.5,1.8)
				(1.5,.5)--(2,.5)
				(1.5,1.5)--(2,1.5);
	\end{tikzpicture}
        \caption{}
    \end{subfigure}
    \hfill
    	\begin{subfigure}[t]{.49\textwidth}
    	\centering
	\begin{tikzpicture}
		\draw 	(1,0)--(1,2)
				(1.5,0)--(1.5,2)
				(2,0)--(2,2)
				(4,0)--(4,2);
		\draw 	(1,0) node[label=below:{1}] {} 
				(1.5,0)node[label=below:{1}] {} 
				(2,0)node[label=below:{1}] {} 
				(4,0)node[label=below:{1}] {} ;
		\draw 	(2.5,1) node[cross] {}
				(3,1) node[cross] {}
				(3.5,1) node[cross] {}
				(4.5,1) node[cross] {};
		\draw 	(2.5,1) node[label=above:{3}] {}
				(3,1) node[label=above:{3}] {}
				(3.5,1) node[label=above:{3}] {}
				(4.5,1) node[label=above:{3}] {};
		\draw [dotted] (1,1)--(1.5,1)
				(2,1)--(4,1);
		\draw	(4,1)--(4.5,1);
		\draw 	(2,.2)--(4,.2)
				(2,1.8)--(4,1.8)
				(1.5,.5)--(2,.5)
				(1.5,1.5)--(2,1.5);
	\end{tikzpicture}
        \caption{}
    	\end{subfigure}
	\hfill
    	\begin{subfigure}[t]{.49\textwidth}
    	\centering
	\begin{tikzpicture}
		\draw 	(1,0)--(1,2)
				(1.5,0)--(1.5,2)
				(2,0)--(2,2)
				(3.5,0)--(3.5,2);
		\draw 	(1,0) node[label=below:{1}] {} 
				(1.5,0)node[label=below:{1}] {} 
				(2,0)node[label=below:{1}] {} 
				(3.5,0)node[label=below:{1}] {} ;
		\draw 	(2.5,1) node[cross] {}
				(3,1) node[cross] {}
				(4,1) node[cross] {}
				(4.5,1) node[cross] {};
		\draw 	(2.5,1) node[label=above:{3}] {}
				(3,1) node[label=above:{3}] {}
				(4,1) node[label=above:{3}] {}
				(4.5,1) node[label=above:{3}] {};
		\draw [dotted] (1,1)--(1.5,1)
				(2,1)--(3.5,1);
		\draw	(4,1)--(4.5,1);
		\draw	(3.5,1.05)--(3.95,1.05)
				(3.5,.95)--(3.95,.95);
		\draw 	(2,.2)--(3.5,.2)
				(2,1.8)--(3.5,1.8)
				(1.5,.5)--(2,.5)
				(1.5,1.5)--(2,1.5);
	\end{tikzpicture}
        \caption{}
    	\end{subfigure}
    	\hfill
    	\begin{subfigure}[t]{.49\textwidth}
    	\centering
	\begin{tikzpicture}
		\draw 	(1,0)--(1,2)
				(1.5,0)--(1.5,2)
				(2.5,0)--(2.5,2)
				(3.5,0)--(3.5,2);
		\draw 	(1,0) node[label=below:{1}] {} 
				(1.5,0)node[label=below:{1}] {} 
				(2.5,0)node[label=below:{1}] {} 
				(3.5,0)node[label=below:{1}] {} ;
		\draw 	(2,1) node[cross] {}
				(3,1) node[cross] {}
				(4,1) node[cross] {}
				(4.5,1) node[cross] {};
		\draw 	(2,1) node[label=above:{3}] {}
				(3,1) node[label=above:{3}] {}
				(4,1) node[label=above:{3}] {}
				(4.5,1) node[label=above:{3}] {};
		\draw [dotted] (1,1)--(1.5,1)
				(2.5,1)--(3.5,1);
		\draw	(4,1)--(4.5,1)
				(2,1)--(2.5,1);
		\draw	(3.5,1.05)--(3.95,1.05)
				(3.5,.95)--(3.95,.95);
		\draw 	(2.5,.2)--(3.5,.2)
				(2.5,1.8)--(3.5,1.8)
				(1.5,.3)--(2.5,.3)
				(1.5,1.7)--(2.5,1.7);
	\end{tikzpicture}
        \caption{}
    	\end{subfigure}
    	\hfill
    	\begin{subfigure}[t]{.49\textwidth}
    	\centering
	\begin{tikzpicture}
		\draw 	(1,0)--(1,2)
				(2,0)--(2,2)
				(2.5,0)--(2.5,2)
				(3.5,0)--(3.5,2);
		\draw 	(1,0) node[label=below:{1}] {} 
				(2,0)node[label=below:{1}] {} 
				(2.5,0)node[label=below:{1}] {} 
				(3.5,0)node[label=below:{1}] {} ;
		\draw 	(1.5,1) node[cross] {}
				(3,1) node[cross] {}
				(4,1) node[cross] {}
				(4.5,1) node[cross] {};
		\draw 	(1.5,1) node[label=above:{3}] {}
				(3,1) node[label=above:{3}] {}
				(4,1) node[label=above:{3}] {}
				(4.5,1) node[label=above:{3}] {};
		\draw [dotted] (1,1)--(2,1)
				(2.5,1)--(3.5,1);
		\draw	(4,1)--(4.5,1)
				(2,1)--(2.5,1);
		\draw	(3.5,1.05)--(3.95,1.05)
				(3.5,.95)--(3.95,.95)
				(2,1.05)--(1.55,1.05)
				(2,.95)--(1.55,.95);
		\draw 	(2.5,.2)--(3.5,.2)
				(2.5,1.8)--(3.5,1.8)
				(2,.3)--(2.5,.3)
				(2,1.7)--(2.5,1.7);
	\end{tikzpicture}
        \caption{}
    	\end{subfigure}
    	\hfill
    	\begin{subfigure}[t]{.49\textwidth}
    	\centering
	\begin{tikzpicture}
		\draw 	(1,0)--(1,2)
				(2,0)--(2,2)
				(2.5,0)--(2.5,2)
				(3.5,0)--(3.5,2);
		\draw 	(1,0) node[label=below:{1}] {} 
				(2,0)node[label=below:{1}] {} 
				(2.5,0)node[label=below:{1}] {} 
				(3.5,0)node[label=below:{1}] {} ;
		\draw 	(1.5,1) node[cross] {}
				(3,1) node[cross] {}
				(4,1) node[cross] {}
				(4.5,1) node[cross] {};
		\draw 	(1.5,1) node[label=above:{3}] {}
				(3,1) node[label=above:{3}] {}
				(4,1) node[label=above:{3}] {}
				(4.5,1) node[label=above:{3}] {};
		\draw [dotted] (1,1)--(2,1)
				(2.5,1)--(3.5,1);
		\draw	(4,1)--(4.5,1)
				(2,1)--(2.5,1);
		\draw	(3.5,1.05)--(3.95,1.05)
				(3.5,.95)--(3.95,.95)
				(2,1.05)--(1.55,1.05)
				(2,.95)--(1.55,.95);
		\draw 	(2.5,.95)--(3.5,.95)
				(2.5,1.05)--(3.5,1.05)
				(2,.95)--(2.5,.95)
				(2,1.05)--(2.5,1.05);
	\end{tikzpicture}
        \caption{}
    	\end{subfigure}
    	\hfill
    	\begin{subfigure}[t]{.49\textwidth}
    	\centering
	\begin{tikzpicture}
		\draw[dashed] 	(1.5,0)--(1.5,2)
				(3,0)--(3,2)
				(4,0)--(4,2)
				(4.5,0)--(4.5,2);
		\draw [dotted] (1,1)--(2,1)
				(2.5,1)--(3.5,1);
		\draw	(4,1)--(4.5,1)
				(2,1)--(2.5,1);
		\draw 	(1,1) node[label=below:{1}] {} 
				(2,1)node[label=below:{1}] {} 
				(2.5,1)node[label=below:{1}] {} 
				(3.5,1)node[label=below:{1}] {} ;
		\draw 	(1.5,2) node[label=above:{3}] {}
				(3,2) node[label=above:{3}] {}
				(4,2) node[label=above:{3}] {}
				(4.5,2) node[label=above:{3}] {};
		\draw	(3.5,1.05)--(4,1.05)
				(3.5,.95)--(4,.95)
				(2,1.05)--(1.5,1.05)
				(2,.95)--(1.5,.95);
		\draw 	(2.5,.95)--(3.5,.95)
				(2.5,1.05)--(3.5,1.05)
				(2,.95)--(2.5,.95)
				(2,1.05)--(2.5,1.05);
		\draw 	(1,1) node[circle,fill=white,draw=black] {}
				(2,1) node[circle,fill=white,draw=black] {}
				(2.5,1) node[circle,fill=white,draw=black] {}
				(3.5,1) node[circle,fill=white,draw] {};
		\draw 	(1,1) node[cross] {}
				(2,1) node[cross] {}
				(2.5,1) node[cross] {}
				(3.5,1) node[cross] {};
	\end{tikzpicture}
        \caption{}
    	\end{subfigure}
    	\hfill
    	\begin{subfigure}[t]{.49\textwidth}
    	\centering
	\begin{tikzpicture}
		\draw[dashed] 	(1.5,0)--(1.5,2)
				(3,0)--(3,2)
				(4,0)--(4,2)
				(4.5,0)--(4.5,2);
		\draw [dotted] (1,1)--(2,1)
				(2.5,1)--(3.5,1);
		\draw	(4,1)--(4.5,1)
				(2,1)--(2.5,1);
		\draw 	(1,1) node[label=below:{1}] {} 
				(2,1)node[label=below:{1}] {} 
				(2.5,1)node[label=below:{1}] {} 
				(3.5,1)node[label=below:{1}] {} ;
		\draw 	(1.5,2) node[label=above:{3}] {}
				(3,2) node[label=above:{3}] {}
				(4,2) node[label=above:{3}] {}
				(4.5,2) node[label=above:{3}] {};
		\draw	(1.5,1.8)--(3,1.8)
				(1.5,.2)--(3,.2)
				(3,1.6)--(4,1.6)
				(3,.4)--(4,.4);
		\draw 	(1,1) node[circle,fill=white,draw=black] {}
				(2,1) node[circle,fill=white,draw=black] {}
				(2.5,1) node[circle,fill=white,draw=black] {}
				(3.5,1) node[circle,fill=white,draw] {};
		\draw 	(1,1) node[cross] {}
				(2,1) node[cross] {}
				(2.5,1) node[cross] {}
				(3.5,1) node[cross] {};
	\end{tikzpicture}
        \caption{}
    	\end{subfigure}
	\hfill
 	\caption{Higgsing of the model with $\vec{l}_d=(3,3,3,3)$, $\vec{l}_s=(1,1,1,1)$, and rightmost orientifold plane $O3^-$. (a) represents the Coulomb branch brane configuration. (b-g) are one step phase transitions and (h) is the Higgs branch brane configuration.}
	\label{fig:SO4maxHiggs}
\end{figure}

The phase transition to the Higgs branch brane configuration is shown in figure \ref{fig:SO4maxHiggs}. Let us go over the transition step by step. First, the rightmost half D5-brane is pulled to the right of the neighboring half NS5-brane. The fact that no physical brane is created between them keeps the linking numbers unchanged, figure \ref{fig:SO4maxHiggs}(b). This is always the case when an $O3^-$ is connected to one and only one of the two fivebranes involved in the \emph{one step} phase transition. Note that the orientifold plane between the half D5-brane and the half NS5-brane has changed from $O3^+$ to $\widetilde{O3^-}$. 

Then, the same process is repeated for the second rightmost half D5-brane, a D3-brane (or two half D3-branes) needs to be created in this case in order to preserve the linking numbers, see figure  \ref{fig:SO4maxHiggs}(c). 

One can now pull the leftmost half D5-brane to the left of its neighboring half NS5-brane, figure \ref{fig:SO4maxHiggs}(d). No brane is created, as expected from the presence of $O3^-$ connected to the half NS5-brane. The same half D5-brane can be pulled one more step to the left. Brane creation is needed so the linking numbers are preserved, figure \ref{fig:SO4maxHiggs}(e).

Complete Higgsing can then be achieved: all half D3-branes are aligned at the origin, figure \ref{fig:SO4maxHiggs}(f). The next step is to change the perspective in the diagram. This is achieved by performing a rotation that takes directions $\vec{x}$ to $\vec{y}$ and directions $\vec{y}$ are taken to $-\vec{x}$. In the diagram now vertical directions correspond to $\vec{y}$  and directions perpendicular to the paper correspond to $\vec{x}$. The half D5-branes are represented with vertical dashed lines and the half NS5-branes are represented with circled crosses, figure \ref{fig:SO4maxHiggs}(g). In the final step one can realign and split the half D3-branes into half D3-brane segments that end in D5-branes, figure \ref{fig:SO4maxHiggs}(h). This is the Higgs branch brane configuration.

\subsubsection{Brane realization of the D-collapse}\label{sec:branecollapse}

A final remark: by moving some of the half NS5-branes via phase transitions with no brane creation/annihilation the partition $\lambda^t=(3,1)^t=(2,1,1)$ can be made manifest. The brane transition that we are looking after is:
\paragraph{Collapse transition:}\emph{If there is a half D5-brane connected to an $O3^-$ and can be pushed through a half NS5-brane so the NS5-brane is then connected to the $O3^-$ instead, then perform all such transitions.}\\

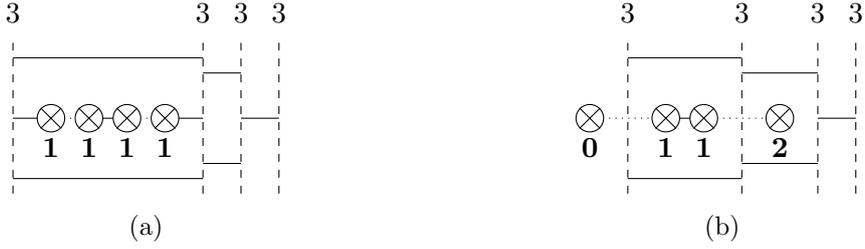
\begin{figure}[t]
	\centering
    	\begin{subfigure}[t]{.49\textwidth}
    	\centering
	\begin{tikzpicture}
		\draw[dashed] 	(1,0)--(1,2)
				(3.5,0)--(3.5,2)
				(4,0)--(4,2)
				(4.5,0)--(4.5,2);
		\draw [dotted] (1.5,1)--(2,1)
				(2.5,1)--(3,1);
		\draw	(4,1)--(4.5,1)
				(2,1)--(2.5,1)
				(1,1)--(1.5,1)
				(3,1)--(3.5,1);
		\draw 	(1.5,1) node[label=below:{$\mathbf{1}$}] {} 
				(2,1)node[label=below:{$\mathbf{1}$}] {} 
				(2.5,1)node[label=below:{$\mathbf{1}$}] {} 
				(3,1)node[label=below:{$\mathbf{1}$}] {} ;
		\draw 	(1,2) node[label=above:{3}] {}
				(3.5,2) node[label=above:{3}] {}
				(4,2) node[label=above:{3}] {}
				(4.5,2) node[label=above:{3}] {};
		\draw	(1,1.8)--(3.5,1.8)
				(1,.2)--(3.5,.2)
				(3.5,1.6)--(4,1.6)
				(3.5,.4)--(4,.4);
		\draw 	(1.5,1) node[circle,draw,fill=white] {}
				(2,1) node[circle,draw,fill=white] {}
				(2.5,1) node[circle,draw,fill=white] {}
				(3,1) node[circle,draw,fill=white] {};
		\draw 	(1.5,1) node[cross] {}
				(2,1) node[cross] {}
				(2.5,1) node[cross] {}
				(3,1) node[cross] {};
	\end{tikzpicture}
        \caption{}
    	\end{subfigure}
	\hfill
    	\begin{subfigure}[t]{.49\textwidth}
    	\centering
	\begin{tikzpicture}
		\draw[dashed] 	(1.5,0)--(1.5,2)
				(3,0)--(3,2)
				(4,0)--(4,2)
				(4.5,0)--(4.5,2);
		\draw [dotted] (1,1)--(2,1)
				(2.5,1)--(3.5,1);
		\draw	(4,1)--(4.5,1)
				(2,1)--(2.5,1);
		\draw 	(1,1) node[label=below:{$\mathbf{0}$}] {} 
				(2,1)node[label=below:{$\mathbf{1}$}] {} 
				(2.5,1)node[label=below:{$\mathbf{1}$}] {} 
				(3.5,1)node[label=below:{$\mathbf{2}$}] {} ;
		\draw 	(1.5,2) node[label=above:{3}] {}
				(3,2) node[label=above:{3}] {}
				(4,2) node[label=above:{3}] {}
				(4.5,2) node[label=above:{3}] {};
		\draw	(1.5,1.8)--(3,1.8)
				(1.5,.2)--(3,.2)
				(3,1.6)--(4,1.6)
				(3,.4)--(4,.4);
		\draw 	(1,1) node[circle,draw,fill=white] {}
				(2,1) node[circle,draw,fill=white] {}
				(2.5,1) node[circle,draw,fill=white] {}
				(3.5,1) node[circle,draw,fill=white] {};
		\draw 	(1,1) node[cross] {}
				(2,1) node[cross] {}
				(2.5,1) node[cross] {}
				(3.5,1) node[cross] {};
	\end{tikzpicture}
        \caption{}
    	\end{subfigure}
	\hfill
 	\caption{Brane realization of the D-collapse for the maximal nilpotent orbit of $\mathfrak{so}(4)$. (a) Higgs branch brane configuration of the model with $\vec{l}_d=(3,3,3,3)$, $\vec{l}_s=(1,1,1,1)$ and rightmost orientifold plane $O3^-$ before the \emph{collapse} transition is performed. The interval numbers of the half NS5-branes are shown in bold font under the corresponding branes, they form the partition $(\lambda')^t=(1,1,1,1)$. (b) After the  \emph{collapse} transition is performed, the interval numbers change to $\vec{k}_s=(0,1,1,2)$, this gives the partition $\lambda^t=(2,1,1)$. This shows the D-collapse of $\lambda'=(1,1,1,1)^t=(4)$ to $\lambda=(4)_D=(2,1,1)^t=(3,1)$.}
	\label{fig:SO4maxO3push}
\end{figure}

We call this the \emph{collapse} transition. In  figure \ref{fig:SO4maxHiggs}(h) all half D5-branes have already been pushed away of the $O3^-s$. The partition $\lambda^t=(2,1,1)$ can be seen in the following way: there are two half NS5-branes in the first interval from the left between half D5-branes, giving two parts with value 1; there is one half NS5-brane in the second interval between half D5-branes, giving one part of value 2. The leftmost half NS5-brane gives a part of value 0. We can define these parts as \emph{interval numbers} $\vec{k}_s$ for the half NS5-branes (see figure \ref{fig:SO4maxO3push}(b)): 
\begin{align}\label{eq:420}
\vec{k}_s:=(0,1,1,2)
\end{align}

A different set $\vec k_s'$ of \emph{inteval numbers} that is also relevant can be obtained from the same brane configuration. This is obtained by the inverse phase transition to the \emph{collapse transition}, i.e. by pulling the half D5-branes towards the $O3^-$ planes. The result of this transition is depicted in figure \ref{fig:SO4maxO3push}(a). The new interval numbers are:

\begin{align}\label{eq:421}
\vec{k}_s'=(1,1,1,1)
\end{align}	

The apparent ambiguity of the values of $\vec{k}_s$ and $\vec{k}_s'$ is actually in our advantage, since it allows one to see the brane realization of the D-collapse of the partition $\lambda'=(4)$ into partition $\lambda=(4)_D=(3,1)$. In order to see this let a set of interval numbers $\vec k_s$ define a partition $\lambda$ where $\lambda^t$ is the partition formed with all the integer numbers in $\vec k_s$. Then we have:

\begin{align}\label{eq:interval}
	\begin{aligned}
		\vec k_s'=(1,1,1,1)&\rightarrow\lambda'=(4)\\
		\vec k_s=(0,1,1,2)&\rightarrow\lambda=(3,1)
	\end{aligned}
\end{align}

The collapse transition changes the interval numbers of the half NS5-branes from $\vec k_s'$ to $\vec{k}_s$, furnishing a map from partition $\lambda'$ to partition $\lambda$. This is the physical realization of the D-collapse.

In general, the \emph{collapse} transition defined above, applied to the brane configurations that are discussed in this work, always realizes a D-collapse for partitions of the D-type.

Note that the prescription to obtain partition $\lambda'$ from the brane system has to be: perform the inverse transition to the collapse transition, i.e. pull the half D5-branes towards the $O3^-$s without brane creation/annihilation. However, for B-type, the inverse transition of the collapse transition always reduces the sum of the parts from $2n+1$ to $2n$. In order to obtain partition $\lambda'$, such that $\lambda'_B=\lambda$, an additional transition that moves half an NS5-brane with interval number $0$ back to the first interval needs to be performed after the \emph{inverse} collapse transition. For the C-type, the opposite effect takes place: the inverse of the collapse transition needs to be followed by a transition that reduces by one the interval number of the half NS5-brane with highest interval number. These additional transitions can be directly related to the augmented and reduced partitions defined in (\ref{eq:aug}).

\subsubsection{S-duality of the example}\label{sec:noCoulomb}

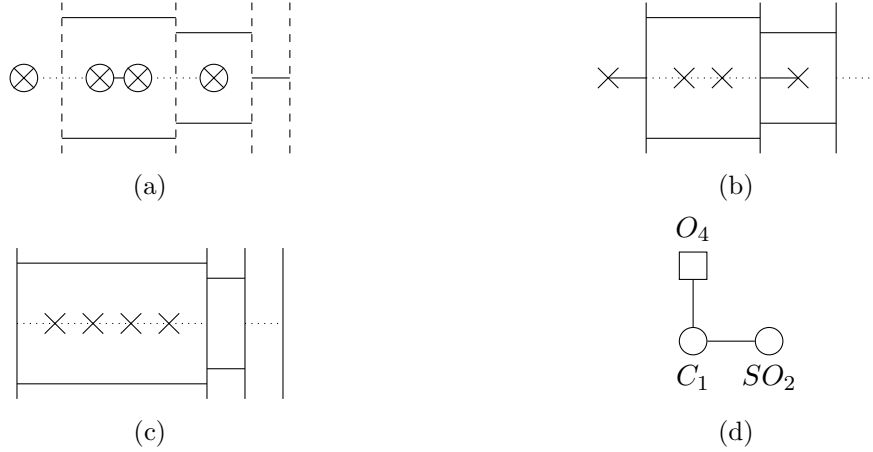
\begin{figure}[t]
	\centering
    	\begin{subfigure}[t]{.49\textwidth}
    	\centering
	\begin{tikzpicture}
		\draw[dashed] 	(1.5,0)--(1.5,2)
				(3,0)--(3,2)
				(4,0)--(4,2)
				(4.5,0)--(4.5,2);
		\draw [dotted] (1,1)--(2,1)
				(2.5,1)--(3.5,1);
		\draw	(4,1)--(4.5,1)
				(2,1)--(2.5,1);
		\draw	(1.5,1.8)--(3,1.8)
				(1.5,.2)--(3,.2)
				(3,1.6)--(4,1.6)
				(3,.4)--(4,.4);
		\draw 	(1,1) node[circ] {}
				(2,1) node[circ] {}
				(2.5,1) node[circ] {}
				(3.5,1) node[circ] {};
		\draw 	(1,1) node[cross] {}
				(2,1) node[cross] {}
				(2.5,1) node[cross] {}
				(3.5,1) node[cross] {};
	\end{tikzpicture}
        \caption{}
    	\end{subfigure}
	\hfill
    	\begin{subfigure}[t]{.49\textwidth}
    	\centering
	\begin{tikzpicture}
		\draw 	(1.5,0)--(1.5,2)
				(3,0)--(3,2)
				(4,0)--(4,2)
				(4.5,0)--(4.5,2);
		\draw 	(1,1) node[cross] {}
				(2,1) node[cross] {}
				(2.5,1) node[cross] {}
				(3.5,1) node[cross] {};
		\draw  (1,1)--(1.5,1)
				(3,1)--(3.5,1);
		\draw [dotted]	(4,1)--(4.5,1)
				(1.5,1)--(3,1);
		\draw	(1.5,1.8)--(3,1.8)
				(1.5,.2)--(3,.2)
				(3,1.6)--(4,1.6)
				(3,.4)--(4,.4);
	\end{tikzpicture}
        \caption{}
    	\end{subfigure}
	\hfill
    	\begin{subfigure}[t]{.49\textwidth}
    	\centering
	\begin{tikzpicture}
		\draw 	(1,0)--(1,2)
				(3.5,0)--(3.5,2)
				(4,0)--(4,2)
				(4.5,0)--(4.5,2);
		\draw 	(1.5,1) node[cross] {}
				(2,1) node[cross] {}
				(2.5,1) node[cross] {}
				(3,1) node[cross] {};
		\draw [dotted] (1.5,1)--(2,1)
				(2.5,1)--(3,1);
		\draw [dotted]	(4,1)--(4.5,1)
				(2,1)--(2.5,1)
				(1,1)--(1.5,1)
				(3,1)--(3.5,1);
		\draw	(1,1.8)--(3.5,1.8)
				(1,.2)--(3.5,.2)
				(3.5,1.6)--(4,1.6)
				(3.5,.4)--(4,.4);
	\end{tikzpicture}
        \caption{}
    	\end{subfigure}
    \hfill
	\begin{subfigure}[t]{.49\textwidth}
    \centering
	\begin{tikzpicture}[]
	\tikzstyle{gauge} = [circle,draw];
	\tikzstyle{flavour} = [regular polygon,regular polygon sides=4,draw];
	\node (g2) [gauge, label=below:{$C_1$}] {};
	\node (g3) [gauge,right of=g2, label=below:{$SO_2$}] {};
	\node (f2) [flavour,above of=g2, label=above:{$O_4$}] {};
	\draw (g3)--(g2)--(f2)
		;
	\end{tikzpicture}

        \caption{}
    \end{subfigure}
 	\caption{S-duality of the model with $\vec{l}_d=(3,3,3,3)$, $\vec{l}_s=(1,1,1,1)$ and rightmost orientifold plane $O3^-$. (a) Higgs branch brane configuration of the starting model. (b) Result of S-duality. (c) Result of performing the \emph{collapse} transition that pulls the half D5-branes away from $O3^-$ planes. This is necessary to read the corresponding quiver. (d) Quiver with a choice of $SO_2$ group in the rightmost node, this is related to the Lusztig's Canonical Quotient $\bar A(\mathcal O_{(3,1)})$ of the maximal orbit of $\gso (4)$ being trivial \cite{CHZ17}.}
	\label{fig:SO4maxMirror}
\end{figure}

After the transition to the Higgs branch brane configuration, figure \ref{fig:SO4maxMirror}(a), an S-duality transformation can be performed. This replaces the half D5-branes with half NS5-branes and vice versa. The O3-planes change accordingly: $\widetilde{O3^-}$ swaps with $O3^+$. $O3^-$ and $\widetilde{O3^+}$ do not change. The result is depicted in figure \ref{fig:SO4maxMirror}(b). 

The result is the Coulomb branch brane configuration of a model where $\vec{l}_d$ and $\vec{l}_s$ have swapped: $\vec{l}_d=(1,1,1,1)$, $\vec{l}_s=(3,3,3,3)$. Note that the O3-plane at the right end is still $O3^-$. A \emph{collapse} transition as defined in section \ref{sec:collapse} can be performed, figure \ref{fig:SO4maxMirror}(c), in order to read the quiver, figure \ref{fig:SO4maxMirror}(d). We have observed that the collapse transition is always necessary and are lead to propose the following prescription:
\begin{prescription}
	In order to read the quiver corresponding to a Coulomb branch brane configuration it is convenient to first perform the collapse transition.
\end{prescription}

Note that in order to obtain the closure of the maximal nilpotent orbit of $\mathfrak{so}(4)$, $\Or_{(3,1)}$, as the Coulomb branch of the resulting quiver, figure \ref{fig:SO4maxMirror}(d), the gauge node to the right of the quiver needs to be chosen $SO_2$ instead of $O_2$. The issue that different choices of $O_{2n}/SO_{2n}$ nodes result in different Coulomb branches was raised in \cite{CHMZ14}. In \cite{CHZ17} we provide a systematic study of this problem and identify the relation between the choice that has the closure of a nilpotent orbit $\bar{\mathcal O}_\lambda$ as its Coulomb branch and the Lusztig's Canonical Quotient $\bar A(\mathcal O_\lambda)$ of the nilpotent orbit. At this point one cannot claim that both quivers (the one in figure \ref{fig:SO4maximal}(b) and the one in figure \ref{fig:SO4maxMirror}(d)) satisfy $3d$ mirror symmetry, but rather that the former has $\M_H=\Or _{(3,1)}$ and the latter $\M_C=\Or _{(3,1)}$. In general, the results of the present paper should be understood as results on the Higgs branches of quivers with $O_n$ choices for all their orthogonal gauge group factors. All these Higgs branches are either closures of special nilpotent orbits of Lie algebras of types B, C and D \cite{BTX10,CDT13,CHMZ14} or transverse slices between such nilpotent orbit closures (see appendices \ref{app:DnHiggs}, \ref{app:AnHiggs} and \ref{app:AnAnHiggs} of the present note). The challenge to translate these results to the Coulomb branch still remains open: as discussed in \cite{CHZ17}, Coulomb branches of specific choices of $O_{2n}/SO_{2n}$ can be computed to be the closures of special nilpotent orbits of B and D types but only if they are \emph{normal}. A Coulomb branch construction for \emph{non-normal} closures of orbits\footnote{To find these constructions is a particularly interesting challenge since Nakajima has introduced a mathematical Coulomb branch description \cite{N15} that has been shown to always be normal \cite{BFN16}.} or for generic orbits of the C type is yet to be found.

\subsection{Example: Remaining nilpotent orbits of $\mathfrak{so(4)}$}\label{sec:5}

The next example to analyze is the model whose Higgs branch is: 
\begin{align}
	\M_H=\Or_{(2,2)}\subset \gso (4)
\end{align}

In order to do this, the prescription selects the following parameters for the brane model:

\begin{align}
	\begin{aligned}
		n_d&=4\\
		n_s&=4\\
		\vec{l}_d&=(3,3,3,3)
	\end{aligned}
\end{align}

and $\vec{l}_s$ is related to $d_{BV}(\lambda)=d_{BV}(2,2)=(2,2)$. Since $\vec{l}_s$ is a 4-tuple the first step is to pad the partition $d_{BV}(\lambda)$ with two zeroes, in order to obtain 

\begin{align}
d_{BV}(\lambda)=(2,2,0,0)
\end{align}

Since $\vec{l}_s$ corresponds to the linking numbers of the half NS5-branes ordered from left to right, the partition needs to be reversed:

\begin{align}
\vec{l}'_s=(0,0,2,2)
\end{align}

The apostrophe is added because this is not the right choice of linking numbers. As mentioned before, several sets of linking numbers can correspond to the same model. In the present example, the choice of linking numbers is determined because the Coulomb brane configuration is built in a the same fashion as in the previous example, following the steps:
\begin{enumerate}
	\item Set all the half NS5-branes.
	\item Set all the half D5-branes.
	\item Set the nature of the O3-planes, choosing for example the type of the rightmost one.
	\item Add physical D3-branes.
\end{enumerate}

In the case of $\mathfrak{so}(2n)$ orbits, after the orientifold planes are chosen, the half NS5-branes acquire linking numbers $(1,-1,1,-1,\dots,1,-1,1,2n-1)$, this is because the number of half NS5-branes is even, the number of half D5-branes is even and all half D5-branes are placed in the interval between the two rightmost half NS5-branes. 
The last step is number 4, to add physical (two halves) D3-branes, this increments or decreases the linking numbers of the half NS5-branes in jumps of 2. As a result of this, a brane system built following the previous steps, whose Higgs branch is a closure of a nilpotent orbit of $\mathfrak{so}(2n)$, can only have negative or positive odd integers as the elements of $\vec{l}_s$.

 The prescription to turn $\vec{l}'_s$ into an array of odd numbers is the following:
\begin{prescription}
	Since $d_{BV}(\lambda)\in \mathcal{P}_{+1}(2n)$, any even element in $\vec{l}'_s$ has even multiplicity. Let all even elements of $\vec{l}'_s$ be divided into distinct pairs $\{(\vec{l}'_s)_i,(\vec{l}'_s)_{i+1}\}$. Substitute these elements with $(\vec{l}_s)_i=(\vec{l}'_s)_i+1$ and $(\vec{l}_s)_{i+1}=(\vec{l}'_s)_{i+1}-1$.
\end{prescription}

This procedure applied to $\vec{l}_s'$ gives the result:
\begin{align}
\vec{l}_s=(1,-1,3,1)
\end{align}

\begin{figure}[t]
	\centering
    \begin{subfigure}[t]{.49\textwidth}
    \centering
	\begin{tikzpicture}
		\draw 	(1,0)--(1,2)
				(1.5,0)--(1.5,2)
				(2,0)--(2,2)
				(4.5,0)--(4.5,2);
		\draw 	(1,0) node[label=below:{1}] {} 
				(1.5,0)node[label=below:{-1}] {} 
				(2,0)node[label=below:{3}] {} 
				(4.5,0)node[label=below:{1}] {} ;
		\draw 	(2.5,1) node[cross] {}
				(3,1) node[cross] {}
				(3.5,1) node[cross] {}
				(4,1) node[cross] {};
		\draw 	(2.5,1) node[label=above:{3}] {}
				(3,1) node[label=above:{3}] {}
				(3.5,1) node[label=above:{3}] {}
				(4,1) node[label=above:{3}] {};
		\draw [dotted] (1,1)--(1.5,1)
				(2,1)--(4.5,1);
		\draw 	(2,.2)--(4.5,.2)
				(2,1.8)--(4.5,1.8);
	\end{tikzpicture}
        \caption{}
    \end{subfigure}
    \hfill
	\begin{subfigure}[t]{.49\textwidth}
    \centering
	\begin{tikzpicture}[]
	\tikzstyle{gauge} = [circle,draw];
	\tikzstyle{flavour} = [regular polygon,regular polygon sides=4,draw];
	\node (g3) [gauge,right of=g2,label=below:{$C_1$}] {};
	\node (f3) [flavour,above of=g3,label=above:{$O_4$}] {};
	\draw (g3)--(f3)
		;
	\end{tikzpicture}
	\caption{}
    \end{subfigure}
    \hfill
    \begin{subfigure}[t]{.49\textwidth}
    \centering
	\begin{tikzpicture}
		\draw 	(1,0)--(1,2)
				(1.5,0)--(1.5,2)
				(2.5,0)--(2.5,2)
				(4,0)--(4,2);
		\draw 	(1,0) node[label=below:{1}] {} 
				(1.5,0)node[label=below:{-1}] {} 
				(2.5,0)node[label=below:{3}] {} 
				(4,0)node[label=below:{1}] {} ;
		\draw 	(2,1) node[cross] {}
				(3,1) node[cross] {}
				(3.5,1) node[cross] {}
				(4.5,1) node[cross] {};
		\draw 	(2,1) node[label=above:{3}] {}
				(3,1) node[label=above:{3}] {}
				(3.5,1) node[label=above:{3}] {}
				(4.5,1) node[label=above:{3}] {};
		\draw [dotted] (1,1)--(1.5,1)
				(2.5,1)--(4,1);
		\draw	(2,1)--(2.5,1)
				(4,1)--(4.5,1);
		\draw 	(2.5,.2)--(4,.2)
				(2.5,1.8)--(4,1.8);
	\end{tikzpicture}
        \caption{}
    \end{subfigure}
    \hfill
    \begin{subfigure}[t]{.49\textwidth}
    \centering
	\begin{tikzpicture}
		\draw 	(1,0)--(1,2)
				(1.5,0)--(1.5,2)
				(3,0)--(3,2)
				(3.5,0)--(3.5,2);
		\draw 	(1,0) node[label=below:{1}] {} 
				(1.5,0)node[label=below:{-1}] {} 
				(3,0)node[label=below:{3}] {} 
				(3.5,0)node[label=below:{1}] {} ;
		\draw 	(2,1) node[cross] {}
				(2.5,1) node[cross] {}
				(4,1) node[cross] {}
				(4.5,1) node[cross] {};
		\draw 	(2,1) node[label=above:{3}] {}
				(2.5,1) node[label=above:{3}] {}
				(4,1) node[label=above:{3}] {}
				(4.5,1) node[label=above:{3}] {};
		\draw [dotted] (1,1)--(1.5,1)
				(3,1)--(3.5,1);
		\draw	(2,1)--(2.5,1)
				(4,1)--(4.5,1);
		\draw	(2.55,1.05)--(3,1.05)
				(2.55,.95)--(3,.95)
				(3.5,1.05)--(3.95,1.05)
				(3.5,.95)--(3.95,.95);
		\draw 	(3,.2)--(3.5,.2)
				(3,1.8)--(3.5,1.8);
	\end{tikzpicture}
        \caption{}
    \end{subfigure}
    \hfill
    \begin{subfigure}[t]{.49\textwidth}
    \centering
    \begin{tikzpicture}
		\draw [dashed] (2,0)--(2,2)
				(2.5,0)--(2.5,2)
				(4,0)--(4,2)
				(4.5,0)--(4.5,2);
		\draw [dotted] (1,1)--(1.5,1)
				(3,1)--(3.5,1);
		\draw	(2,1)--(2.5,1)
				(4,1)--(4.5,1);
		\draw	(2.5,1.8)--(4,1.8)
				(2.5,.2)--(4,.2);
		\draw 	(1,1) node[label=below:{1}] {} 
				(1.5,1)node[label=below:{-1}] {} 
				(3,1)node[label=below:{3}] {} 
				(3.5,1)node[label=below:{1}] {} ;
		\draw 	(2,2) node[label=above:{3}] {}
				(2.5,2) node[label=above:{3}] {}
				(4,2) node[label=above:{3}] {}
				(4.5,2) node[label=above:{3}] {};
		\draw	(1,1) node[circ]{}
				(1.5,1) node[circ]{}
				(3,1) node[circ]{}
				(3.5,1) node[circ]{};
		\draw	(1,1) node[cross]{}
				(1.5,1) node[cross]{}
				(3,1) node[cross]{}
				(3.5,1) node[cross]{};
    \end{tikzpicture}
    \caption{}
    \end{subfigure}
    \hfill
    \begin{subfigure}[t]{.49\textwidth}
    \centering
    \begin{tikzpicture}
		\draw [dashed] (2,0)--(2,2)
				(2.5,0)--(2.5,2)
				(4,0)--(4,2)
				(4.5,0)--(4.5,2);
		\draw	(2,1)--(2.5,1)
				(4,1)--(4.5,1);
		\draw	(2.5,1.8)--(4,1.8)
				(2.5,.2)--(4,.2);
		\draw 	(1.25,1.3) node[label=right:{0}] {} 
				(1.25,.7)node[label=right:{0}] {} 
				(3.25,1.4)node[label=right:{2}] {} 
				(3.25,.6)node[label=right:{2}] {} ;
		\draw 	(2,2) node[label=above:{3}] {}
				(2.5,2) node[label=above:{3}] {}
				(4,2) node[label=above:{3}] {}
				(4.5,2) node[label=above:{3}] {};
		\draw	(1.25,1.3) node[circ]{}
				(1.25,.7) node[circ]{}
				(3.25,1.4) node[circ]{}
				(3.25,.6) node[circ]{};
		\draw	(1.25,1.3) node[cross]{}
				(1.25,.7) node[cross]{}
				(3.25,1.4) node[cross]{}
				(3.25,.6) node[cross]{};
    \end{tikzpicture}
    \caption{}
    \end{subfigure}
    \hfill
 	\caption{Orthosymplectic model with $\M_H=\Or_{\lambda}\subset \mathfrak{so}(4)$ and $\lambda=(2^2)$. The parameters of the model are $n_s=n_d=4$, $\vec{l}_d=(3,3,3,3)$, $\vec{l}_s=(1,-1,3,1)$ and rightmost orientifold plane is of type $O3^-$. (a) Coulomb branch brane configuration of the model. (b) Quiver. (c-e) Phase transition to the Higgs branch brane configuration. The \emph{interval numbers} of the half NS5-branes in (e), $\vec{k}_s=(0,0,2,2)$, represent the reverse of the transpose partition $\lambda^t=(2,2,0,0)$. In (e) the half NS5-branes can coalesce into entire fivebranes in a supersymmetric way, and then be pulled away from the orientifold plane. This is a process of \emph{un-splitting} and the resulting un-split linking numbers correspond to $\vec{l}'_s=(0,0,2,2)$, the result is (f).}
	\label{fig:SO4minimal}
\end{figure}
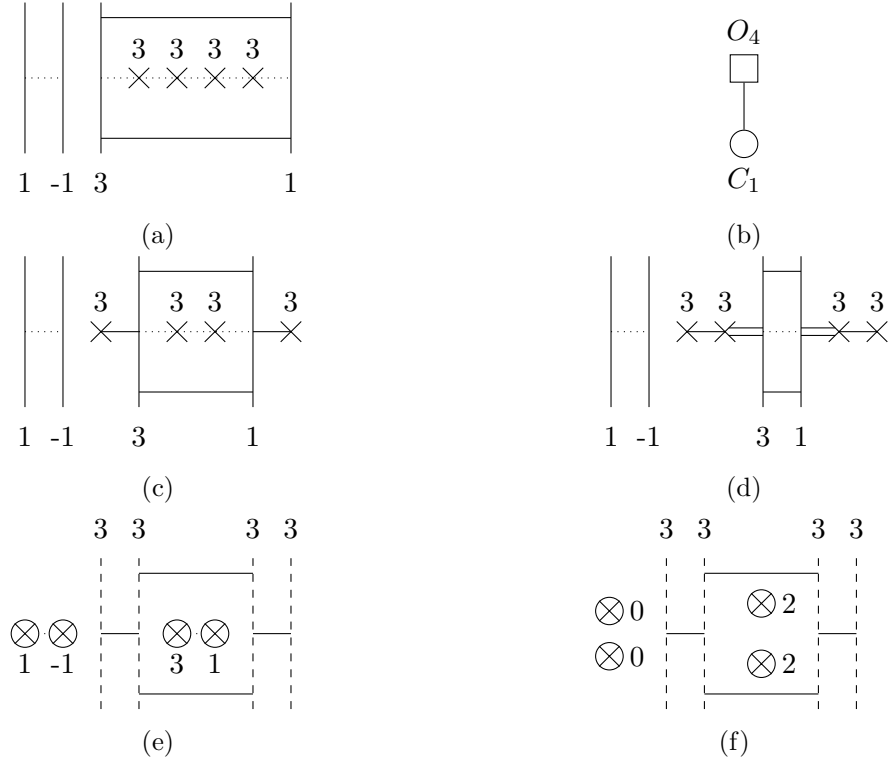

The parameters $n_d=4$, $n_s=4$, $\vec{l}_d=(3,3,3,3)$, $\vec{l}_s=(1,-1,3,1)$ and rightmost orientifold plane $O3^-$ specify the Coulomb branch brane configuration in figure \ref{fig:SO4minimal}(a). The quiver can then be read, figure \ref{fig:SO4minimal}(b). The phase transition to the Higgs branch brane configuration is also computed, figures \ref{fig:SO4minimal}(c-e).

\paragraph{Linking numbers and splitting mechanism}
Figure \ref{fig:SO4minimal}(f) shows how the splitting of the NS5-branes changes their linking numbers. The figure shows the result after the two pairs of half NS5-branes coalesce into entire NS5-branes, and then are pulled away from the orientifold plane as half NS5-branes and their images (physical NS5-branes). The linking numbers of the new half NS5-branes are $\vec{l}'_s=(0,0,2,2)$, which correspond to $d_{BV}(\lambda)=(2,2,0,0)$. This illustrates our previous statement that different sets of linking numbers correspond to the same brane system. In our case $\vec{l}_s=(1,-1,3,1)$ are the linking numbers when the half NS5-branes are stacked on the orientifold plane and $\vec{l}'_s=(0,0,2,2)$ are the linking numbers when the half NS5-branes are away from the orientifold plane.

The \emph{interval numbers} of the half NS5-branes in the Higgs branch brane configuration (figure \ref{fig:SO4minimal}(e)) are:

\begin{align}
	\vec{k}_s=(0,0,2,2)
\end{align}

As in the previous case, the elements of $\vec{k}_s$ define partition $\lambda^t=(2,2,0,0)$. This illustrates once more how the transpose partition of the defining partition $\lambda=(2,2)$ of the nilpotent orbit can be read directly from the position of the half NS5-branes in the Higgs branch brane configuration. The fact that this Higgs branch brane configuration has a trivial \emph{collapse} transition (there are no phase transitions that can be performed such that there is no brane creation or brane annihilation) means that there is no partition $\lambda'\in \mathcal{P}_{+1}(4)$, with $\lambda' \neq \lambda$, such that $(\lambda')_D=\lambda$. Therefore, branes are shown to encode important information about the combinatorics of $\mathcal{P}_{+1}(2n)$.

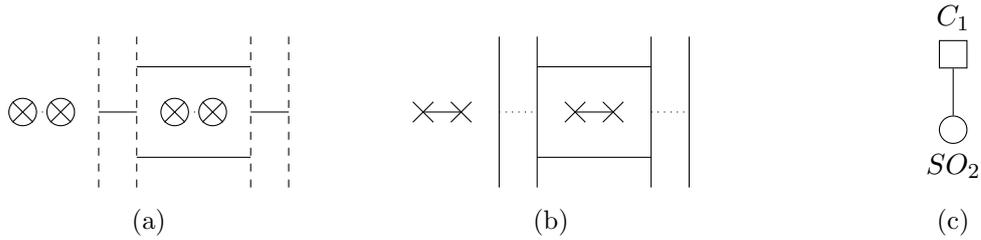
\begin{figure}[t]
	\centering
	\begin{subfigure}[t]{.30\textwidth}
    	\centering
	\begin{tikzpicture}
		\draw[dashed] 	(2,0)--(2,2)
				(2.5,0)--(2.5,2)
				(4,0)--(4,2)
				(4.5,0)--(4.5,2);
		\draw [dotted] (1,1)--(1.5,1)
				(3,1)--(3.5,1);
		\draw	(4,1)--(4.5,1)
				(2,1)--(2.5,1);
		\draw	(2.5,1.6)--(4,1.6)
				(2.5,.4)--(4,.4);
		\draw 	(1,1) node[circ] {}
				(1.5,1) node[circ] {}
				(3,1) node[circ] {}
				(3.5,1) node[circ] {};
		\draw 	(1,1) node[cross] {}
				(1.5,1) node[cross] {}
				(3,1) node[cross] {}
				(3.5,1) node[cross] {};
	\end{tikzpicture}
        \caption{}
    	\end{subfigure}
	\hfill
	\begin{subfigure}[t]{.30\textwidth}
    	\centering
	\begin{tikzpicture}
		\draw 	(2,0)--(2,2)
				(2.5,0)--(2.5,2)
				(4,0)--(4,2)
				(4.5,0)--(4.5,2);
		\draw 	(1,1) node[cross] {}
				(1.5,1) node[cross] {}
				(3,1) node[cross] {}
				(3.5,1) node[cross] {};
		\draw  (1,1)--(1.5,1)
				(3,1)--(3.5,1);
		\draw[dotted]	(4,1)--(4.5,1)
				(2,1)--(2.5,1);
		\draw	(2.5,1.6)--(4,1.6)
				(2.5,.4)--(4,.4);
	\end{tikzpicture}
        \caption{}
    	\end{subfigure}
	\hfill
	\begin{subfigure}[t]{.30\textwidth}
    \centering
	\begin{tikzpicture}[]
	\tikzstyle{gauge} = [circle,draw];
	\tikzstyle{flavour} = [regular polygon,regular polygon sides=4,draw];
	\node (g2) [gauge,label=below:{$SO_2$}] {};
	\node (f2) [flavour,above of=g2,label=above:{$C_1$}] {};
	\draw (g2)--(f2)
		;
	\end{tikzpicture}

        \caption{}
    \end{subfigure}
	\caption{Model with $\M_C=\Or_{(2,2)}^I\subset \mathfrak{so}(4)$. (a) is the Higgs branch configuration of the model with $\M_H=\Or_{(2,2)}^{I}\cup \Or_{(2,2)}^{II}$. (b) is the Coulomb branch brane configuration that results after performing S-duality. (c) is the quiver read from (b) with a choice of $SO_2$ gauge node instead of $O_2$. This choice is related to the fact that Lusztig's Canonical Quotient $\bar A(\mathcal O_{(2,2)})$ is trivial \cite{CHZ17}.}
	\label{fig:SO4minMirror}
\end{figure}

An S-duality transformation can be performed on the Higgs branch brane configuration in order to obtain a quiver with $\M_C=\Or_{(2,2)}^I$, the result is depicted on figure \ref{fig:SO4minMirror}. Note that once more a gauge group $SO_2$ has to be chosen \cite{CHZ17} and that even with this choice the Coulomb branch is not isomorphic to the Higgs branch counterpart $\M_H=\Or_{(2,2)}=\Or_{(2,2)}^{I}\cup \Or_{(2,2)}^{II}$, but to a single one of the two orbit closures under the adjoint action of $SO(4)$ on the $\gso (4)$ algebra.

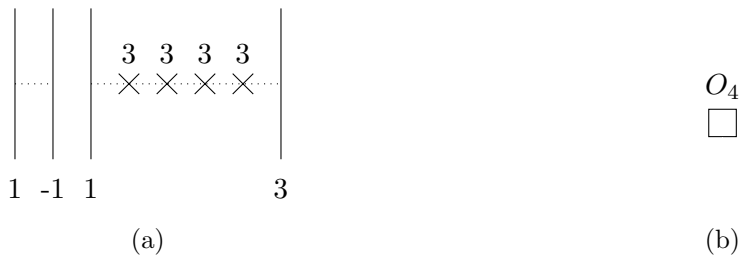
\begin{figure}[t]
	\centering
    \begin{subfigure}[t]{.49\textwidth}
    \centering
	\begin{tikzpicture}
		\draw 	(1,0)--(1,2)
				(1.5,0)--(1.5,2)
				(2,0)--(2,2)
				(4.5,0)--(4.5,2);
		\draw 	(1,0) node[label=below:{1}] {} 
				(1.5,0)node[label=below:{-1}] {} 
				(2,0)node[label=below:{1}] {} 
				(4.5,0)node[label=below:{3}] {} ;
		\draw 	(2.5,1) node[cross] {}
				(3,1) node[cross] {}
				(3.5,1) node[cross] {}
				(4,1) node[cross] {};
		\draw 	(2.5,1) node[label=above:{3}] {}
				(3,1) node[label=above:{3}] {}
				(3.5,1) node[label=above:{3}] {}
				(4,1) node[label=above:{3}] {};
		\draw [dotted] (1,1)--(1.5,1)
				(2,1)--(4.5,1);
	\end{tikzpicture}
        \caption{}
    \end{subfigure}
    \hfill
	\begin{subfigure}[t]{.49\textwidth}
    \centering
	\begin{tikzpicture}[]
	\tikzstyle{gauge} = [circle,draw];
	\tikzstyle{flavour} = [regular polygon,regular polygon sides=4, draw];
	\node (g3) [] {};
	\node (f3) [flavour,above of=g3,label=above:{$O_4$}] {};
	\draw[white] (g3)--(f3)
		;
	\end{tikzpicture}
	\caption{}
    \end{subfigure}
    \hfill
    \hfill
 	\caption{Orthosymplectic model with $n_s=n_d=4$, $\vec{l}_d=(3,3,3,3)$, $\vec{l}_s=(1,-1,1,3)$ and rightmost orientifold plane $O3^-$. The quiver is trivial and can be read from (a), and it is depicted in (b).}
	\label{fig:SO4trivial}
\end{figure}

\paragraph{Trivial orbit}
For $\lambda=(1,1,1,1)$, the dual partition is $d_{BV}(\lambda)=(3,1)$ and therefore $\vec{l}'_s=(0,0,1,3)$. After using the prescription above to turn it into an array of odd numbers the result is:
\begin{align}
	\vec{l}_s=(1,-1,1,3)
\end{align}
The Coulomb branch brane configuration and quiver are depicted in figure \ref{fig:SO4trivial}. We say that the Higgs branch corresponds to the closure of the trivial orbit, i.e. a single point.

\subsection{Kraft-Procesi transition for the nilpotent orbits of $\mathfrak{so}(4)$}\label{sec:6}

This section contains the first three examples of new results introduced by the present paper.

\subsubsection{$A_1$ transition}

The transition between the different models whose Higgs branches are the closures of the different nilpotent orbits of $\mathfrak{so}(4)$ can be analyzed
in an analogous manner to the $\mathfrak{sl}(n)$ case \cite{CH16}. The starting point is the model with Higgs branch:
\begin{align}
	\M_H=\Or_{(3,1)}
\end{align}

The corresponding Higgs branch brane configuration is depicted in figure \ref{fig:SO4maxHiggs}(h). To find the transition one should focus on the rightmost physical (entire) D3-brane and treat the other one as an spectator. The question of \emph{what is the moduli space $S$ generated by the motion of this brane along the directions spanned by the two half D3-branes?} is addressed. One way to answer this question is to analyze the local subsystem around the D3-brane, depicted in figure \ref{fig:SO4maxInstNew}(a). This represents the Higgs branch of the quiver with gauge group $G=O(2)$ and flavor symmetry $F=Sp(1)$, figure \ref{fig:SO4maxInstNew}(d). The Higgs branch of this quiver is:
\begin{align}
	S=A_1
\end{align}

\begin{figure}[t]
	\centering
    	\begin{subfigure}[t]{.49\textwidth}
    	\centering
	\begin{tikzpicture}
		\draw[dashed] 	(1.5,0)--(1.5,2)
				(3,0)--(3,2);
		\draw [dotted] (1,1)--(2,1)
				(2.5,1)--(3.5,1);
		\draw	(2,1)--(2.5,1);
		\draw	(1.5,1.8)--(3,1.8)
				(1.5,.2)--(3,.2);
		\draw 	(1,1) node[circ] {}
				(2,1) node[circ] {}
				(2.5,1) node[circ] {}
				(3.5,1) node[circ] {};
		\draw 	(1,1) node[cross] {}
				(2,1) node[cross] {}
				(2.5,1) node[cross] {}
				(3.5,1) node[cross] {};
	\end{tikzpicture}
        \caption{}
    	\end{subfigure}
	\hfill
    	\begin{subfigure}[t]{.49\textwidth}
    	\centering
	\begin{tikzpicture}
		\draw[dashed] 	(2,0)--(2,2)
				(2.5,0)--(2.5,2);
		\draw [dotted] (1,1)--(1.5,1)
				(3,1)--(3.5,1);
		\draw	(2,1)--(2.5,1);
		\draw	(2.5,1.05)--(3,1.05)
				(2.5,.95)--(3,.95);
		\draw	(1.5,1.05)--(2,1.05)
				(1.5,.95)--(2,.95);
		\draw	(2,1.8)--(2.5,1.8)
				(2,.2)--(2.5,.2);
		\draw 	(1,1) node[circ] {}
				(1.5,1) node[circ] {}
				(3,1) node[circ] {}
				(3.5,1) node[circ] {};
		\draw 	(1,1) node[cross] {}
				(1.5,1) node[cross] {}
				(3,1) node[cross] {}
				(3.5,1) node[cross] {};
	\end{tikzpicture}
        \caption{}
    	\end{subfigure}
	\hfill
    	\begin{subfigure}[t]{.49\textwidth}
    	\centering
	\begin{tikzpicture}
		\draw 	(1,0)--(1,2)
				(1.5,0)--(1.5,2)
				(3,0)--(3,2)
				(3.5,0)--(3.5,2);
		\draw 	(2,1) node[cross] {}
				(2.5,1) node[cross] {};
		\draw [dotted] (1,1)--(1.5,1)
				(3,1)--(3.5,1);
		\draw	(2,1)--(2.5,1);
		\draw	(1.5,1.8)--(3,1.8)
				(1.5,.2)--(3,.2);
	\end{tikzpicture}
        \caption{}
    	\end{subfigure}
	\hfill
	\begin{subfigure}[t]{.49\textwidth}
    \centering
	\begin{tikzpicture}[]
	\tikzstyle{gauge} = [circle,draw];
	\tikzstyle{flavour} = [regular polygon,regular polygon sides=4,draw];
	\node (g2) [gauge,label=below:{$O_2$}] {};
	\node (f2) [flavour,above of=g2,label=above:{$C_1$}] {};
	\draw (g2)--(f2)
		;
	\end{tikzpicture}
	\caption{}
    \end{subfigure}
 	\caption{Transverse slice $S=A_1$ that can be removed via Kraft-Procesi transition. (a) Local brane configuration obtained from model in figure \ref{fig:SO4maxHiggs}(h). (b) Phase transition to get closer to the Coulomb branch brane configuration . (c) Coulomb branch brane configuration after aligning the D3-branes in (b), performing a rotation and giving the D3-branes a generic position along the $\vec x$ direction. (d) Quiver corresponding to brane system (c).}
	\label{fig:SO4maxInstNew}
\end{figure}
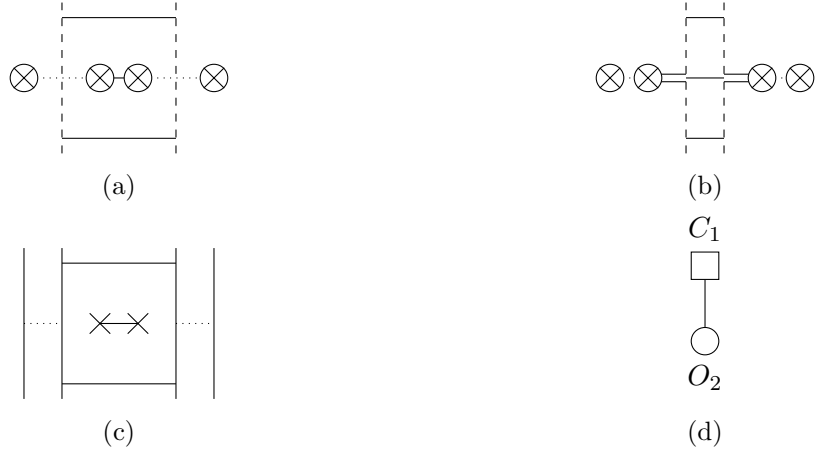

\begin{figure}[t]
	\centering
    	\begin{subfigure}[t]{.49\textwidth}
    	\centering
	\begin{tikzpicture}
		\draw[dashed] 	(1.5,0)--(1.5,2)
				(3,0)--(3,2)
				(4,0)--(4,2)
				(4.5,0)--(4.5,2);
		\draw [dotted] (1,1)--(2,1)
				(2.5,1)--(3.5,1);
		\draw	(4,1)--(4.5,1)
				(2,1)--(2.5,1);
		\draw	(1.5,1.8)--(3,1.8)
				(1.5,.2)--(3,.2)
				(3,1.6)--(4,1.6)
				(3,.4)--(4,.4);
		\draw 	(1,1) node[circ] {}
				(2,1) node[circ] {}
				(2.5,1) node[circ] {}
				(3.5,1) node[circ] {};
		\draw 	(1,1) node[cross] {}
				(2,1) node[cross] {}
				(2.5,1) node[cross] {}
				(3.5,1) node[cross] {};
	\end{tikzpicture}
        \caption{}
    	\end{subfigure}
	\hfill
	\begin{subfigure}[t]{.49\textwidth}
    	\centering
	\begin{tikzpicture}
		\draw[dashed] 	(1.5,0)--(1.5,2)
				(3,0)--(3,2)
				(4,0)--(4,2)
				(4.5,0)--(4.5,2);
		\draw [dotted] (1,1)--(2,1)
				(2.5,1)--(3.5,1);
		\draw	(4,1)--(4.5,1)
				(2,1)--(2.5,1);
		\draw	(1.5,1.05)--(3,1.05)
				(1.5,.95)--(3,.95)
				(3,1.6)--(4,1.6)
				(3,.4)--(4,.4);
		\draw 	(1,1) node[circ] {}
				(2,1) node[circ] {}
				(2.5,1) node[circ] {}
				(3.5,1) node[circ] {};
		\draw 	(1,1) node[cross] {}
				(2,1) node[cross] {}
				(2.5,1) node[cross] {}
				(3.5,1) node[cross] {};
	\end{tikzpicture}
        \caption{}
    	\end{subfigure}
	\hfill
	\begin{subfigure}[t]{.49\textwidth}
    	\centering
	\begin{tikzpicture}
		\draw[dashed] 	(1.5,0)--(1.5,2)
				(3,0)--(3,2)
				(4,0)--(4,2)
				(4.5,0)--(4.5,2);
		\draw [dotted] (1,1)--(2,1)
				(2.5,1)--(3.5,1);
		\draw	(4,1)--(4.5,1)
				(2,1)--(2.5,1);
		\draw	(1.5,1.05)--(1.95,1.05)
				(1.5,.95)--(1.95,.95)
				(3,1.05)--(2.55,1.05)
				(3,.95)--(2.55,.95)
				(3,1.6)--(4,1.6)
				(3,.4)--(4,.4);
		\draw 	(1,1) node[circ] {}
				(2,1) node[circ] {}
				(2.5,1) node[circ] {}
				(3.5,1) node[circ] {};
		\draw 	(1,1) node[cross] {}
				(2,1) node[cross] {}
				(2.5,1) node[cross] {}
				(3.5,1) node[cross] {};
	\end{tikzpicture}
        \caption{}
    	\end{subfigure}
	\hfill
	\begin{subfigure}[t]{.49\textwidth}
    	\centering
	\begin{tikzpicture}
		\draw[dashed] 	(2,0)--(2,2)
				(2.5,0)--(2.5,2)
				(4,0)--(4,2)
				(4.5,0)--(4.5,2);
		\draw [dotted] (1,1)--(1.5,1)
				(3,1)--(3.5,1);
		\draw	(4,1)--(4.5,1)
				(2,1)--(2.5,1);
		\draw	(2.5,1.6)--(4,1.6)
				(2.5,.4)--(4,.4);
		\draw 	(1,1) node[circ] {}
				(1.5,1) node[circ] {}
				(3,1) node[circ] {}
				(3.5,1) node[circ] {};
		\draw 	(1,1) node[cross] {}
				(1.5,1) node[cross] {}
				(3,1) node[cross] {}
				(3.5,1) node[cross] {};
	\end{tikzpicture}
        \caption{}
    	\end{subfigure}
	\hfill
 	\caption{$A_1$ Kraft-Procesi transition starting from the model with $\vec{l}_d=(3,3,3,3)$, $\vec{l}_s=(1,1,1,1)$ and rightmost orientifold plane $O3^-$. (a) Higgs branch brane configuration of the starting model. Its Higgs branch is $\M_H=\Or_{(3,1)}$. (b) The physical D3-brane is aligned with the half NS5-branes. (c) The D3-brane is split in three segments, the middle one is \emph{removed} by taking it to infinity in the directions spanned by the NS5-branes. (d) Two phase transitions are performed to obtain the Higgs branch brane configuration of the resulting model, with \emph{interval numbers} $\vec{k}_s=(0,0,2,2)$. It's Higgs branch is the closure $\Or_{(2,2)}^{I}\cup \Or_{(2,2)}^{II}\subset\mathfrak{so}(4)$.}
	\label{fig:SO4maxKP}
\end{figure}
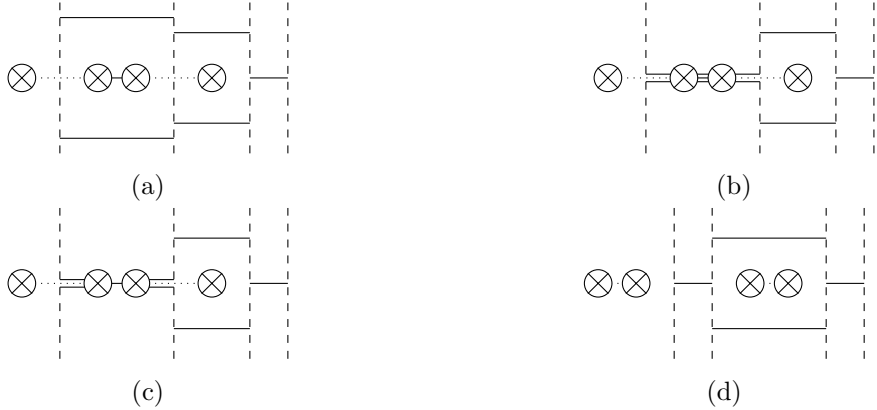

The slice $S\subseteq \Or_{(3,1)}$ can be \emph{removed} by partial Higgsing. This process is the generalization of the same process already discussed for the $\mathfrak{sl}(n)$ case (section \ref{sec:2} and \cite{CH16}). Let figure \ref{fig:SO4maxKP} illustrate it step by step. The first step is to align the half D3-branes with the orientifold plane, figure \ref{fig:SO4maxKP}(b). Then, the D3-brane is split in three segments. The middle segment can acquire nonzero positions in direction $\vec{x}$ spanned by the half NS5-branes. The limit when these positions go to infinity is considered. Therefore the D3-brane is effectively removed from the system, figure \ref{fig:SO4maxKP}(c). Two phase transitions are then performed to remove the fixed D3-brane segments, figure \ref{fig:SO4maxKP}(d).

The resulting model has new linking numbers: $\vec{l}_d=(3,3,3,3)$ (do not change) and  $\vec{l}_s=(1,-1,3,1)$ (change). As explained before, a partition $\lambda=(2,2)$ can be read from the position of the half NS5-branes in the Higgs branch brane configuration (i.e. its \emph{interval numbers}) after the \emph{collapse} partition has been performed\footnote{In this case the collapse transition is trivial.}:
\begin{align}
	\vec{k}_s=(0,0,2,2)
\end{align}
These correspond to $\lambda^t$. This model can be recognized as one of the models discussed above, in section \ref{sec:5}. Its Higgs branch brane configuration is depicted in figure \ref{fig:SO4minimal}(e), whose Higgs branch is:
\begin{align}
	\M_H=\Or_{(2,2)}=\Or_{(2,2)}^{I}\cup \Or_{(2,2)}^{II}\subset \gso (4)
\end{align}

Note that Kraft and Procesi \cite{KP82} compute the slice that is transverse to each orbit $\mathcal O_{(2,2)}^I$ and $\mathcal O_{(2,2)}^{II}$ separately, and they find them both to be $A_1$ singularities. Since both orbits are isomorphic, it is expected that both transverse slices are identical. In general, when we deal with transverse slices $S$ to orbits with \emph{very even partitions} $\lambda\in \mathcal P_{+1}(4n)$ there is a single brane system whose Higgs branch is the union $\M_H=\Or_{(\lambda)}^{I}\cup \Or_{(\lambda)}^{II}$. The moduli space $S$ generated by the D3-branes that are removed in the KP (Kraft-Procesi) transition to this system is isomorphic to the slices found by Kraft and Procesi that were transverse to each single orbit separately.

This concludes the first example of a Kraft-Procesi transition for an orthogonal algebra. The initial variety is the closure of the maximal nilpotent orbit of $\mathfrak{so}(4)$, $\Or_{(3,1)}$. The singularity $S=A_1$ is found to be the slice of $\Or_{(3,1)}$ transverse to the subregular orbits $\mathcal{O}_{(2^2)}^I,\mathcal O_{(2,2)}^{II}\subset \Or_{(3,1)}$. Removing the transverse slice corresponds to performing a Higgs mechanism in the brane system.

This shows how the theory developed in \cite{CH16} for Kraft-Procesi transitions of $\mathfrak{sl}(n)$ can be extended to the orthogonal and symplectic algebras by introducing O3-planes. The remaining part of this section shows the remaining KP transitions for $\mathfrak{so}(4)$. Section \ref{sec:sigmarho} provides an overview on the general way of building brane systems for effective gauge theories whose Higgs or Coulomb branch are closures of nilpotent orbits of orthogonal and symplectic algebras. Section \ref{sec:so5interlude} gives the KP transitions of $\mathfrak{so}(5)$ and $\mathfrak{sp}(2)$ as further examples. Section \ref{sec:11} covers the general definition of all possible Kraft-Procesi transitions in the brane systems for orthogonal and symplectic groups.

\subsubsection{$A_1\cup A_1$ transition}
A further KP transition can be performed in the resulting model of figure \ref{fig:SO4minimal}(e), with Higgs branch $\M_H=\Or_{(2,2)}^{I}\cup \Or_{(2,2)}^{II}$. This time there is a single D3-brane in the brane configuration. The KP transition removes such brane, resulting in a Higgs branch with zero dimension, i.e. the trivial nilpotent orbit.

\begin{figure}[t]
	\centering
	\begin{subfigure}[t]{.49\textwidth}
    	\centering
	\begin{tikzpicture}
		\draw[dashed] 	(2,0)--(2,2)
				(2.5,0)--(2.5,2)
				(4,0)--(4,2)
				(4.5,0)--(4.5,2);
		\draw [dotted] (1,1)--(1.5,1)
				(3,1)--(3.5,1);
		\draw	(4,1)--(4.5,1)
				(2,1)--(2.5,1);
		\draw	(2.5,1.6)--(4,1.6)
				(2.5,.4)--(4,.4);
		\draw 	(1,1) node[circ] {}
				(1.5,1) node[circ] {}
				(3,1) node[circ] {}
				(3.5,1) node[circ] {};
		\draw 	(1,1) node[cross] {}
				(1.5,1) node[cross] {}
				(3,1) node[cross] {}
				(3.5,1) node[cross] {};
	\end{tikzpicture}
        \caption{}
    	\end{subfigure}
	\hfill
	\begin{subfigure}[t]{.49\textwidth}
    	\centering
	\begin{tikzpicture}
		\draw[dashed] 	(2,0)--(2,2)
				(2.5,0)--(2.5,2)
				(4,0)--(4,2)
				(4.5,0)--(4.5,2);
		\draw [dotted] (1,1)--(1.5,1)
				(3,1)--(3.5,1);
		\draw	(4,1)--(4.5,1)
				(2,1)--(2.5,1);
		\draw	(2.5,1.05)--(4,1.05)
				(2.5,.95)--(4,.95);
		\draw 	(1,1) node[circ] {}
				(1.5,1) node[circ] {}
				(3,1) node[circ] {}
				(3.5,1) node[circ] {};
		\draw 	(1,1) node[cross] {}
				(1.5,1) node[cross] {}
				(3,1) node[cross] {}
				(3.5,1) node[cross] {};
	\end{tikzpicture}
        \caption{}
    	\end{subfigure}
	\hfill
	\begin{subfigure}[t]{.49\textwidth}
    	\centering
	\begin{tikzpicture}
		\draw[dashed] 	(2,0)--(2,2)
				(2.5,0)--(2.5,2)
				(4,0)--(4,2)
				(4.5,0)--(4.5,2);
		\draw [dotted] (1,1)--(1.5,1)
				(3,1)--(3.5,1);
		\draw	(4,1)--(4.5,1)
				(2,1)--(2.5,1);
		\draw	(2.5,1.05)--(2.95,1.05)
				(2.5,.95)--(2.95,.95)
				(4,1.05)--(3.55,1.05)
				(4,.95)--(3.55,.95);
		\draw 	(1,1) node[circ] {}
				(1.5,1) node[circ] {}
				(3,1) node[circ] {}
				(3.5,1) node[circ] {};
		\draw 	(1,1) node[cross] {}
				(1.5,1) node[cross] {}
				(3,1) node[cross] {}
				(3.5,1) node[cross] {};
	\end{tikzpicture}
        \caption{}
    	\end{subfigure}
	\hfill
	\begin{subfigure}[t]{.49\textwidth}
    	\centering
	\begin{tikzpicture}
		\draw[dashed] 	(2,0)--(2,2)
				(3,0)--(3,2)
				(3.5,0)--(3.5,2)
				(4.5,0)--(4.5,2);
		\draw [dotted] (1,1)--(1.5,1)
				(2.5,1)--(4,1);
		\draw	(4,1)--(4.5,1)
				(2,1)--(2.5,1);
		\draw 	(1,1) node[circ] {}
				(1.5,1) node[circ] {}
				(2.5,1) node[circ] {}
				(4,1) node[circ] {};
		\draw 	(1,1) node[cross] {}
				(1.5,1) node[cross] {}
				(2.5,1) node[cross] {}
				(4,1) node[cross] {};
	\end{tikzpicture}
        \caption{}
    	\end{subfigure}
	\hfill
	\begin{subfigure}[t]{.49\textwidth}
    	\centering
	\begin{tikzpicture}
		\draw[dashed] 	(2.5,0)--(2.5,2)
				(3,0)--(3,2)
				(3.5,0)--(3.5,2)
				(4,0)--(4,2);
		\draw [dotted] (1,1)--(1.5,1)
				(2,1)--(4.5,1);
		\draw 	(1,1) node[circ] {}
				(1.5,1) node[circ] {}
				(2,1) node[circ] {}
				(4.5,1) node[circ] {};
		\draw 	(1,1) node[cross] {}
				(1.5,1) node[cross] {}
				(2,1) node[cross] {}
				(4.5,1) node[cross] {};
	\end{tikzpicture}
        \caption{}
    	\end{subfigure}
	\hfill
 	\caption{$A_1\cup A_1$ KP transition from minimal of $\mathfrak{so(4)}$ to trivial. (a) Higgs branch brane configuration of the model with $\M_H=\Or_{(2,2)}^{I}\cup \Or_{(2,2)}^{II}$. (b) Critical point where the D3-brane coincides with the orientifold plane. (c) The D3-brane splits in three segments, the middle one is taken to infinity along the $\vec{x}$ directions. The result is a new model with a different Higgs branch. (d) Two phase transitions are performed to remove the remaining fixed D3-brane segments. (e) The collapse transition is performed. The interval numbers of the half NS5-branes are $\vec{k}_s=(0,0,0,4)$, corresponding to the partition of the trivial orbit $\lambda=(1,1,1,1)$.}
	\label{fig:SO4minKP}
\end{figure}
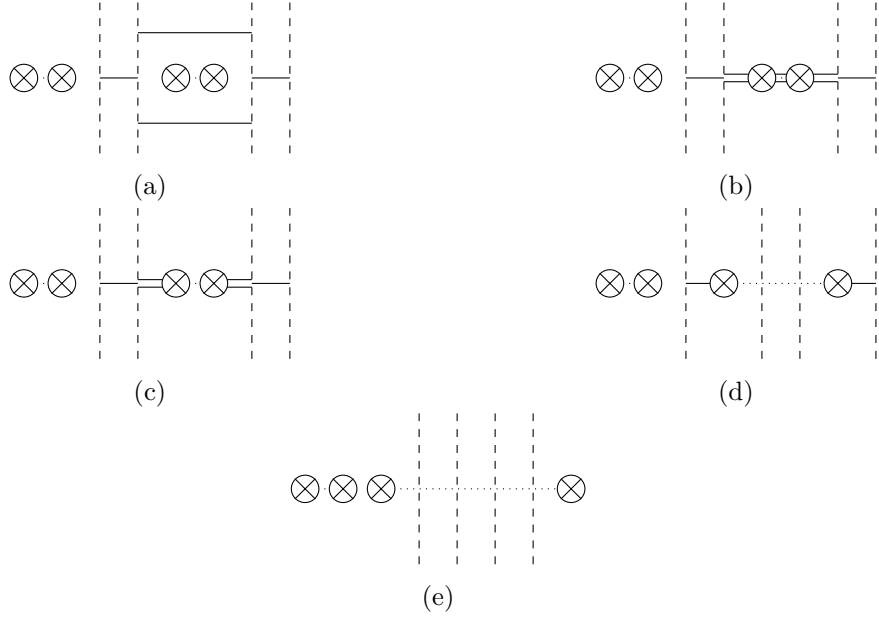

The transition is depicted step by step in figure \ref{fig:SO4minKP}. The interval numbers of the half NS5-branes after the collapse transition,  figure \ref{fig:SO4minKP}(e), are: 

\begin{align}
	\vec{k}_s=(0,0,0,4)
\end{align}
The elements of $\vec{k}_s$ form partition: $\lambda^t=(4)$, its transpose is indeed the defining partition for the trivial orbit $\lambda=(1,1,1,1)$.

In this case the trivial orbit $\mathcal{O}_{(1^4)}\subseteq {V}=\Or_{(2^2)}^{I}\cup \Or_{(2^2)}^{II}$ is a single point, i.e. the origin of ${V}$. The slice $S\subseteq {V}$ transverse to the origin is the variety itself, $S={V}$.

The closure of the minimal nilpotent orbit $\Or_{(2,2)}^{I}\cup \Or_{(2,2)}^{II}$ under the action of the group $O(4)$ is the union of two $A_1$ singularities that intersect at their singular points, since $\Or_{(2,2)}^I=A_1$ and $\Or_{(2,2)}^{II}=A_1$. Hence:

\begin{align}
S=A_1\cup A_1
\end{align}

This is indeed the Higgs branch of the quiver gauge theory of the corresponding model, figure \ref{fig:SO4minimal}(b), see for example \cite{FH16}.

\subsubsection{Hasse diagram for $\mathfrak{so}(4)$}

\begin{figure}[t]
	\centering
\begin{subfigure}[t]{.45\textwidth}
    \centering
	\begin{tikzpicture}
	\tikzstyle{hasse} = [circle, fill,inner sep=2pt]
		\node [hasse] (1) [] {};
		\node [hasse] (2) [below of=1] {};
		\node [hasse] (3) [below of=2] {};
		\draw (1) edge [] node[label=left:$A_1$] {} (2)
			(2) edge [] node[label=left:$A_1\cup A_1$] {} (3);
		\node (e1) [right of=1] {};
		\node (d1) [right of=e1] {$(1,1,1,1)$};
		\node (c1) [right of=d1] {};
		\node (b1) [right of=c1] {$(3,3,3,3)$};
		\node (e2) [right of=2] {};
		\node (d2) [right of=e2] {$(1,-1,3,1)$};
		\node (c2) [right of=d2] {};
		\node (b2) [right of=c2] {$(3,3,3,3)$};
		\node (e3) [right of=3] {};
		\node (d3) [right of=e3] {$(1,-1,1,3)$};
		\node (c3) [right of=d3] {};
		\node (b3) [right of=c3] {$(3,3,3,3)$};
		\node (d) [above of=d1] {$\vec{l}_s$};
		\node (b) [above of=b1] {$\vec{l}_d$};
	\end{tikzpicture}
	\caption{}
 	\end{subfigure}
 	\hfill
 	\begin{subfigure}[t]{.45\textwidth}
 	\centering
 	\begin{tikzpicture}
		\tikzstyle{hasse} = [circle, fill,inner sep=2pt]
		\node [hasse] (1) [] {};
		\node [hasse] (2) [below of=1] {};
		\node [hasse] (3) [below of=2] {};
		\draw (1) edge [] node[label=left:$A_1$] {} (2)
			(2) edge [] node[label=left:$A_1\cup A_1$] {} (3);
		\node (e1) [right of=1] {};
		\node (d1) [right of=e1] {$(3,1)$};
		\node (c1) [right of=d1] {};
		\node (b1) [right of=c1] {$2$};
		\node (e2) [right of=2] {};
		\node (d2) [right of=e2] {$(2^4)$};
		\node (c2) [right of=d2] {};
		\node (b2) [right of=c2] {$1$};
		\node (e3) [right of=3] {};
		\node (d3) [right of=e3] {$(1^4)$};
		\node (c3) [right of=d3] {};
		\node (b3) [right of=c3] {$0$};
		\node (d) [above of=d1] {$\lambda$};
		\node (b) [above of=b1] {$dim$};
	\end{tikzpicture}
	\caption{}
 	\end{subfigure}
 	\hfill
	\caption{Hasse diagram for the models whose Higgs branch is the closure of a nilpotent orbit of $\mathfrak{so}(4)$ under the adjoint action of the group $O(4)$. (a) Represents the brane configurations, where the linking numbers $\vec{l}_s$ and $\vec{l}_d$ are provided for each orbit and the rightmost orientifold plane is always of type $O3^-$. (b) Depicts the information of the orbits: $\lambda$ is the corresponding partition and $dim$ is the quaternionic dimension of the closure of the orbit. With respect to the brane configurations $\lambda^t$ is the partition formed by the elements of $\vec{k}_s$, where $\vec{k}_s$ are the interval numbers of the half NS5-branes after the \emph{collapse} transition of the Higgs branch brane configuration. \emph{dim} is the number of physical D3-branes that generate the Higgs branch in each model.}
	\label{fig:HasseSO4}
\end{figure}
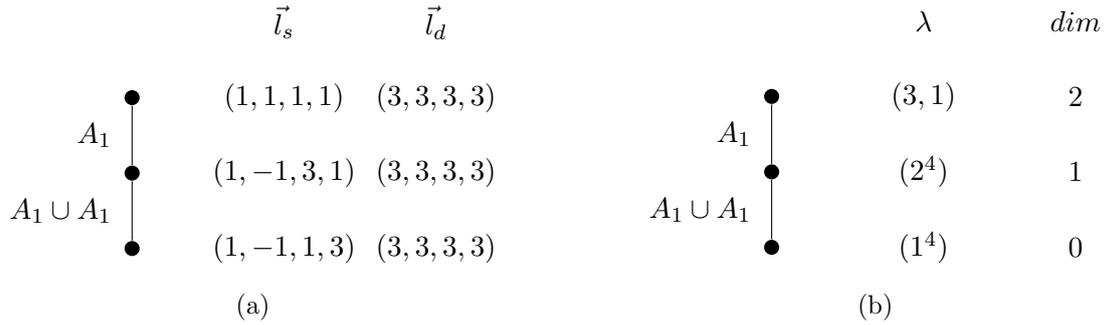

\begin{table}[t]
	\centering
	\begin{tabular}{ c c}
	  
		\raisebox{-.5\height}{\begin{tikzpicture}[node distance=80pt]
		\node at (0,0)[]{\large{$\mathfrak{so}(4)$}};
		\node at (0,-0.5) [hasse] (1) [] {};
		\node [hasse] (2) [below of=1] {};
		\node [hasse] (3) [below of=2] {};
		\draw (1) edge [] node[label=left:$A_1$] {} (2)
			(2) edge [] node[label=left:$A_1\cup A_1$] {} (3);
	\end{tikzpicture}}
	&
	\fontsize{10}{11}\selectfont
	\begin{tabular}{ c c c  c}
	\toprule
	\textbf{Partition}  & \textbf{Branes} & $\M_H$ & $\M_C$ \\ 
	\midrule \addlinespace[3ex]
$\mathbf{3,1}$ & \raisebox{-.4\height}{\begin{tikzpicture}[scale=1, every node/.style={transform shape}]
		\draw[dashed] 	(1.5,0)--(1.5,2)
				(3,0)--(3,2)
				(4,0)--(4,2)
				(4.5,0)--(4.5,2);
		\draw [dotted] (1,1)--(2,1)
				(2.5,1)--(3.5,1);
		\draw	(4,1)--(4.5,1)
				(2,1)--(2.5,1);
		\draw	(1.5,1.8)--(3,1.8)
				(1.5,.2)--(3,.2)
				(3,1.6)--(4,1.6)
				(3,.4)--(4,.4);
		\draw 	(1,1) node[circ,draw] {}
				(2,1) node[circ,draw] {}
				(2.5,1) node[circ,draw] {}
				(3.5,1) node[circ,draw] {};
		\draw 	(1,1) node[cross] {}
				(2,1) node[cross] {}
				(2.5,1) node[cross] {}
				(3.5,1) node[cross] {};
		\end{tikzpicture}}
& \raisebox{-.4\height}{\begin{tikzpicture}[scale=1, every node/.style={transform shape},node distance=20]
	\tikzstyle{gauge} = [circle, draw];
	\tikzstyle{flavour} = [regular polygon,regular polygon sides=4,draw];
	\node (g3) [gauge,label=below:{$O_2$}] {};
	\node (g4) [gauge,right of=g3,label=below:{$C_1$}] {};
	\node (f4) [flavour,above of=g4,label=above:{$O_4$}] {};
	\draw (g3)--(g4)--(f4);
		\end{tikzpicture}} & \raisebox{-.4\height}{\begin{tikzpicture}[scale=1, every node/.style={transform shape},node distance=20]
			\tikzstyle{gauge} = [circle, draw];
			\tikzstyle{flavour} = [regular polygon,regular polygon sides=4,draw];
			\node (g1) [gauge, label=below:{$C_1$}]{};
			\node (g2) [gauge, right of=g1, label=below:{$SO_2$}]{};
			\node (f1) [flavour,above of=g1,label=above:{$O_4$}] {};
			\draw (f1)--(g1)--(g2);
		\end{tikzpicture}}
 \\
  \addlinespace[5ex]
 $\mathbf{2,2}$ & 
		\raisebox{-.4\height}{\begin{tikzpicture}[scale=1, every node/.style={transform shape}]
		\draw[dashed] 	(2,0)--(2,2)
				(2.5,0)--(2.5,2)
				(4,0)--(4,2)
				(4.5,0)--(4.5,2);
		\draw [dotted] (1,1)--(1.5,1)
				(3,1)--(3.5,1);
		\draw	(4,1)--(4.5,1)
				(2,1)--(2.5,1);
		\draw	(2.5,1.6)--(4,1.6)
				(2.5,.4)--(4,.4);
		\draw 	(1,1) node[circ,draw] {}
				(1.5,1) node[circ,draw] {}
				(3,1) node[circ,draw] {}
				(3.5,1) node[circ,draw] {};
		\draw 	(1,1) node[cross] {}
				(1.5,1) node[cross] {}
				(3,1) node[cross] {}
				(3.5,1) node[cross] {};
		\end{tikzpicture}}
 & \raisebox{-.4\height}{\begin{tikzpicture}[scale=1, every node/.style={transform shape},node distance=20]
	\tikzstyle{gauge} = [circle, draw];
	\tikzstyle{flavour} = [regular polygon,regular polygon sides=4,draw];
	\node (g4) [gauge,right of=g3,label=below:{$C_1$}] {};
	\node (f4) [flavour,above of=g4,label=above:{$O_4$}] {};
	\draw (g4)--(f4);
		\end{tikzpicture}}  & \raisebox{-.4\height}{\begin{tikzpicture}[scale=1, every node/.style={transform shape},node distance=20]
			\tikzstyle{gauge} = [circle, draw];
			\tikzstyle{flavour} = [regular polygon,regular polygon sides=4,draw];
			
			\node (g1) [gauge, label=below:{$SO_2$}]{};
			\node (f1) [flavour,above of=g1,label=above:{$C_1$}] {};
			\node (g0) [gauge, left of=g1, label=below:{$C_0$}]{};
			\draw (f1)--(g1)--(g0);
		\end{tikzpicture}}\\
  \addlinespace[5ex]
 $\mathbf{1,1,1,1}$  & 
		\raisebox{-.4\height}{\begin{tikzpicture}[scale=1, every node/.style={transform shape}]
		\draw[dashed] 	(2.5,0)--(2.5,2)
				(3,0)--(3,2)
				(3.5,0)--(3.5,2)
				(4,0)--(4,2);
		\draw [dotted] (1,1)--(1.5,1)
				(2,1)--(4.5,1);
		\draw 	(1,1) node[circ,draw] {}
				(1.5,1) node[circ,draw] {}
				(2,1) node[circ,draw] {}
				(4.5,1) node[circ,draw] {};
		\draw 	(1,1) node[cross] {}
				(1.5,1) node[cross] {}
				(2,1) node[cross] {}
				(4.5,1) node[cross] {};
		\end{tikzpicture}}
 & \raisebox{-.4\height}{\begin{tikzpicture}[scale=1, every node/.style={transform shape},node distance=20]
			\tikzstyle{gauge} = [circle, draw];
			\tikzstyle{flavour} = [regular polygon,regular polygon sides=4, draw];
			\node (g6) [] {};
			\node (f6) [flavour,above of=g6,label=above:{$O_4$}] {};
		\end{tikzpicture}} & \raisebox{-.4\height}{\begin{tikzpicture}[scale=1, every node/.style={transform shape},node distance=20]
			\tikzstyle{gauge} = [circle,draw];
			\tikzstyle{flavour} = [regular polygon,regular polygon sides=4,draw];
			\node (g1) [gauge, label=below:{$O_0$}]{};
			\node (g0) [gauge, left of=g1, label=below:{$C_0$}]{};
			\draw (g1)--(g0);
		\end{tikzpicture}} \\ \addlinespace[3ex]
	\bottomrule
	\end{tabular}
	\end{tabular}
		\caption{Summary of Kraft-Procesi transitions for the nilpotent orbits of $\mathfrak{so}(4)$. The brane systems in the middle depict models whose Higgs branch is the closure of the nilpotent orbit labelled by each \emph{Partition}. The column $\M_H$ contains the quivers of such models, with $\M_H=\Or_{\lambda}$. The column $\M_C$ depicts the quivers of the models obtained after performing S-duality transformation on the brane systems, they have $\M_C=\Or_\lambda$. The $C_0$ and $O_0$ gauge nodes have been left in this column to highlight the effect that the transition has on these quivers: the flavor nodes move along a fixed structure of gauge nodes, while the rank of the gauge nodes decreases.}
		\label{tab:SO4summary}
\end{table}

A Hasse diagram for the two different Kraft-Procesi transitions can be sketched, figure \ref{fig:HasseSO4}. There is a difference with the Hasse diagram in \cite{KP82}. In the present case there are only three different nilpotent orbits under $O(4)$, corresponding to the three different partitions in $\mathcal{P}_{+}(4)=\{(3,1),(2^2),(1^4)\}$. In the Hasse diagram  of \cite{KP82} the orbits under $SO(4)$ are considered, finding two distinct orbits corresponding to the very even partition $\lambda=(2,2)$.

\subsubsection{Summary}

Table \ref{tab:SO4summary} summarizes the different brane systems and the corresponding $3d$ $\mathcal{N}=4$ quiver gauge theories that are related by KP transitions on the algebra $\mathfrak{so}(4)$. Note that the Coulomb branch of the quiver in the second row of column $\M_C$ is isomorphic to closure of one of the orbits $\M_C=\Or_{(2,2)}^I=A_1$ under the adjoint action of $SO(4)$, while the Higgs branch of the quiver in the second row of column $\M_H$ is $\M_H=\Or_{(2,2)}^I\cup\Or_{(2,2)}^{II}=A_1\cup A_1$. This is a general result of these brane systems: for very even partitions  $\lambda \in \mathcal P_{+1}(4n)$ the KP transitions take us to an orthosymplectic quiver with Higgs branch  $\M_H=\Or_{\lambda}^I\cup\Or_{\lambda}^{II}$
while the quiver \emph{candidate for mirror symmetry} has Coulomb branch isomorphic to the closure of a single orbit $\M_C=\Or_{\lambda}^I$.

\section{$T_\rho(G)$ Theories for Nilpotent Orbits and their Brane Configurations}\label{sec:sigmarho}

After discussing the initial examples in the previous section, the following paragraphs aim to give a complete characterization of brane systems and their corresponding orthosymplectic quivers such that their Higgs branch is the closure of a \emph{special} nilpotent orbit with partition $\lambda$ of either $\mathfrak{so}(n)$ or $\mathfrak{sp}(n)$ algebra:
\begin{align}
	\M_H=\Or_{\lambda}
\end{align}

In order to find the relevant theories one can start with the  orthosymplectic quiver whose Higgs branch is the closure of the maximal nilpotent orbit suggested in \cite{GW09} (see figure \ref{fig:nilpotentconequivers}). After performing KP transitions one finds that all resulting theories belong to the family of quiver gauge theories whose IR superconformal limits are known as $T_{\rho}(G)$ theories \cite{GW09}.

In particular, if one chooses $\rho=d_{BV}(\lam)$ where $d_{BV}$ is the Barbasch-Vogan map defined above in section \ref{sec:BV}, the Higgs branch of the theory is $\M_H=\Or_{\lambda}\subset \mathfrak g$, with $\mathfrak g = Lie(G)$.  This result was already computed in \cite{KP79} for the $\gsl (n)$ case. For $\gso (2n)$ and $\gsp (n)$ this was found in \cite{BTX10} (see its \emph{Appendix B}). Finally \cite{CDT13} completed the picture, including the $\gso (2n+1)$ case and introducing the Barbasch-Vogan map to provide a unified description for all classical Lie groups.

This section reviews the general description on how to obtain the right set of parameters that define the brane system with $\M_H=\Or_{\lambda}$. Note that a consequence of the Barbasch-Vogan map is that a partition $\lambda$ that is \emph{non-special} does not produce a brane system/orthosymplectic quiver with the desired Higgs branch $\M_H=\Or_{\lambda}$, but rather 
\begin{align}
	\M_H=\Or_{(d_{LS})^{2}(\lambda)}
\end{align}
where $d_{LS}$ is the Lusztig-Spaltenstein map (defined in section \ref{sec:LS}).

An analogous analysis can be attempted for Coulomb branches, the \emph{candidate mirror theory} $T^{d_{BV}(\lambda)}(G^\vee)$ is then the IR fixed point of the brane system obtained after performing an S-duality transformation on the system for $T_{d_{BV}(\lambda)}(G)$. However, as explained at the end of section \ref{sec:noCoulomb}, there are still some challenges that remain to be tackled concerning the computation of certain Coulomb branches. A first step in this direction is \cite{CHZ17}.

\subsection{Type $B_n$ orbits}

Let the partition $\lambda\in \mathcal{P}_{+1}(2n+1)$ correspond to a special nilpotent orbit in the algebra $\mathfrak{so}({2n+1})$ under the adjoint action of the group $O({2n+1})$. The brane system whose Higgs branch is the closure of the nilpotent orbit is determined by the following parameters:
\begin{itemize}
	\item $n_s=n_d=2n+1$
	\item $\vec{l}_d=(2n,2n,\dots, 2n)$
	\item $\vec{l}_s=Even(d_{BV}(\lambda))$
	\item Rightmost orientifold plane: $O3^-$
\end{itemize}
where all half fivebranes are stacked at the orientifold plane. The map $Even()$ takes a partition $\mu=(\mu_1,\mu_2,\dots,\mu_m)$ with $m$ number of parts, and turns it into an array $\vec{l}_s$ of $2n+1$ even integer numbers. The map takes the following steps:
\begin{enumerate}
	\item Pad partition $\mu$ with $2n+1-m$ parts of zero value $\mu_{m+i}=0$. Then reverse the order in the array to obtain array: 
	\begin{align}
	\vec{l}'_s=(0,\dots,0,\mu_m,\dots,\mu_1)
	\end{align}
	\item Change all odd elements in $\vec{l}'_s$ with even elements with the following prescription:
	\begin{prescription}
	Since we are considering partitions of the form $\mu=d_{BV}(\lambda)\in \mathcal{P}_{-1}(2n)$ every odd element in $\vec{l}'_s$ has even multiplicity. Let all odd elements of $\vec{l}'_s$ be divided into distinct pairs $\{(\vec{l}'_s)_i,(\vec{l}'_s)_{i+1}\}$. Substitute these elements with $(\vec{l}_s)_i=(\vec{l}'_s)_i+1$ and $(\vec{l}_s)_{i+1}=(\vec{l}'_s)_{i+1}-1$.
	\end{prescription}
	The resulting array is denoted:
	\begin{align}
		\vec{l}_s=Even(\mu)
	\end{align}
\end{enumerate}

\subsection{Type $C_n$ orbits}

Let the partition $\lambda\in \mathcal{P}_{-1}(2n)$ correspond to a special nilpotent orbit in the algebra $\mathfrak{sp}({n})$ under the adjoint action of the group $Sp({n})$. The brane system whose Higgs branch is the closure of the nilpotent orbit is determined by the following parameters:
\begin{itemize}
	\item $n_s=n_d=2n+1$
	\item $\vec{l}_d=(2n+1,2n-1,2n+1,\dots,2n-1, 2n+1)$
	\item $\vec{l}_s=Odd(d_{BV}(\lambda))$
	\item Rightmost orientifold plane: $\widetilde{O3^+}$
\end{itemize}
where all half fivebranes are stacked at the orientifold plane. The map $Odd()$ takes a partition $\mu=(\mu_1,\mu_2,\dots,\mu_m)$ with $m$ number of parts, and turns it into an array $\vec{l}_s$ of $2n+1$ odd integer numbers. The map takes similar steps as the $Even()$ map:
\begin{enumerate}
	\item Pad partition $\mu$ with $2n+1-m$ parts of zero value $\mu_{m+i}=0$. Then reverse the order in the array to obtain array: 
	\begin{align}
	\vec{l}'_s=(0,\dots,0,\mu_m,\dots,\mu_1)
	\end{align}
	\item Change all even parts in $\vec{l}'_s$ with odd parts with the following prescription: 
	\begin{prescription}
	Since partitions of the form $\mu=d_{BV}(\lambda)\in \mathcal{P}_{+1}(2n+1)$ are being considered, every even element in $\vec{l}'_s$ has even multiplicity (Note that 0 also has even multiplicity, since $m=|\mu|$ is always odd). Let all even elements of $\vec{l}'_s$ be divided into distinct pairs $\{(\vec{l}'_s)_i,(\vec{l}'_s)_{i+1}\}$. Substitute these elements with $(\vec{l}_s)_i=(\vec{l}_s)_i+1$ and $(\vec{l}_s)_{i+1}=(\vec{l}_s)_{i+1}-1$.	
	\end{prescription}
	The resulting array is denoted:
	\begin{align}
		\vec{l}_s=Odd(\mu)
	\end{align}
\end{enumerate}
		
\subsection{Type $D_n$ orbits}

Let the partition $\lambda\in \mathcal{P}_{+1}(2n)$ correspond to a special nilpotent orbit in the algebra $\mathfrak{so}({2n})$ under the adjoint action of the group $O({2n})$. The brane system whose Higgs branch is the closure of the nilpotent orbit is determined by the following parameters:
\begin{itemize}
	\item $n_s=n_d=2n$
	\item $\vec{l}_d=(2n-1,2n-1,\dots,2n-1)$
	\item $\vec{l}_s=Odd(d_{BV}(\lambda))$
	\item Rightmost orientifold plane: $O3^-$
\end{itemize}
where all half fivebranes are stacked at the orientifold plane. The map $Odd()$ is defined as above. Note that this time it is also the case that every even element in $\vec{l}'_s$ has even multiplicity since $d_{BV}(\lambda)\in\mathcal{P}_{+1}(2n)$ (the element 0 also has even multiplicity since $d_{BV}(\lambda)$ always has an even number of nonzero parts).

\section{$SO(5)\simeq Sp(2)$ Interlude}\label{sec:so5interlude}
\subsection{Example: branes for maximal orbits of $\mathfrak{so}({5})$ and $\mathfrak{sp}({2})$}\label{sec:8}

\subsubsection{Partitions}\label{sec:BVexample}
Let us illustrate the previous prescriptions with the specific cases of nilpotent orbits of $\gso (5)$ and $\gsp (2)$. The corresponding groups are Langlands dual to each other:
\begin{align}
	\begin{aligned}
		O(5)^\vee&=Sp(2)\\
		Sp(2)^\vee &=O(5)
	\end{aligned}
\end{align}

For $\gso (5)$ let us order the relevant set of partitions:
\begin{align}
	\mathcal{P}_{+1}(5)=\left\lbrace\begin{array}{c}
		(5)\\
		(3,1^2)\\
		(2^2,1)\\
		(1^5)\\
		\end{array}\right\rbrace
\end{align}
They correspond to four different nilpotent orbits in the algebra\footnote{The partitions here are expressed in \emph{exponential notation}, i.e. $(3,1^2)$ denotes partition $(3,1,1)$.}. Let us compare it with the partitions for nilpotent orbits of $\gsp (2)$:
\begin{align}
	\mathcal{P}_{-1}(4)=\left\lbrace\begin{array}{c}
		(4)\\
		(2^2)\\
		(2,1^2)\\
		(1^4)\\
		\end{array}\right\rbrace
\end{align}
They also correspond to four different orbits of the algebra.

\paragraph{Transpose map} Let us perform the transpose map on each partition set:
\begin{align}
	\mathcal{P}_{+1}(5)^t=\left\lbrace\begin{array}{c}
		(1^5)\\
		(3,1^2)\\
		(3,2)\\
		(5)\\
		\end{array}\right\rbrace
\end{align}

\begin{align}
	\mathcal{P}_{-1}(4)^t=\left\lbrace\begin{array}{c}
		(1^4)\\
		(2^2)\\
		(3,1)\\
		(4)\\
		\end{array}\right\rbrace
\end{align}

\paragraph{B- and C-collapse}

From the set $\mathcal{P}_{+1}(5)^t$ we see that partition $\lambda=(3,2)$ does not belong to $\mathcal{P}_{+1}(5)$, since there is an even part $\lambda_2=2$ with odd multiplicity; its B-collapse is defined in section \ref{sec:collapse} and gives:
\begin{align}
	(3,2)_B=(3,1^2)
\end{align}

From the set $\mathcal{P}_{-1}(4)^t$ we see that partition $\lambda=(3,1)$ does not belong to $\mathcal{P}_{-1}(4)$, since there are odd parts $\lambda_1=3$ and $\lambda_2=1$ with odd multiplicity; its C-collapse, as defined in section \ref{sec:collapse}, gives:
\begin{align}
	(3,1)_C=(2^2)
\end{align}

All other partitions in  $\mathcal{P}_{+1}(5)^t$ and $\mathcal{P}_{-1}(4)^t$ are mapped to themselves under the B-collapse and the C-collapse, respectively. We get:
\begin{align}
	\mathcal{P}_{+1}(5)^t_{\ B}=\left\lbrace\begin{array}{c}
		(1^5)\\
		(3,1^2)\\
		(3,1^2)\\
		(5)\\
		\end{array}\right\rbrace
\end{align}

\begin{align}
	\mathcal{P}_{-1}(4)^t_{\ C}=\left\lbrace\begin{array}{c}
		(1^4)\\
		(2^2)\\
		(2^2)\\
		(4)\\
		\end{array}\right\rbrace
\end{align}

\paragraph{Lusztig-Spaltenstein map} As mentioned above, the Lusztig-Spaltenstein map $d_{LS}$ is the composition of the transpose map with the X-collapse:
\begin{align}
	d_{LS}(\mathcal{P}_{+1}(5))=\mathcal{P}_{+1}(5)^t_{\ B}=\left\lbrace\begin{array}{c}
		(1^5)\\
		(3,1^2)\\
		(3,1^2)\\
		(5)\\
		\end{array}\right\rbrace
\end{align}

\begin{align}
	d_{LS}(\mathcal{P}_{-1}(4))=\mathcal{P}_{-1}(4)^t_{\ C}=\left\lbrace\begin{array}{c}
		(1^4)\\
		(2^2)\\
		(2^2)\\
		(4)\\
		\end{array}\right\rbrace
\end{align}

At this point one can notice that there is one partition $\lambda=(2^2,1)\in\mathcal{P}_{+1}(5)$ such that $\lambda\not\in d_{LS}(\mathcal{P}_{+1}(5))$, this is equivalent to the fact that $\lambda=(2^2,1)$ is a \emph{non-special} partition, corresponding to a \emph{non-special} nilpotent orbit of $\gso (5)$. The remaining partitions of $\mathcal{P}_{+1}(5)$ are all \emph{special}. The same is true for partition $\lambda=(2,1^2)\in\mathcal{P}_{-1}(4)$. Remember that a partition $\lambda$ is defined to be special in the previous section if and only if $(d_{LS})^2(\lambda)=\lambda$, this is an equivalent definition: A partition $\lambda \in \mathcal P_{+1}(n)$ is special if and only if $\lambda \in d_{LS}(\mathcal P_{+1}(n))$ (and similarly for $\lambda\in \mathcal P_{-1}(2n)$). In \cite{CM93} one can find a quick way to check whether partition $\lambda$ is special, following:
\begin{prescription} There are three possible cases to determine whether $\lambda$ is \emph{special}:
\begin{itemize}
	\item If $\lambda\in\mathcal{P}_{+1}(2n+1)$ then $\lambda$ is special if and only if $\lambda^t\in\mathcal{P}_{+1}(2n+1)$
	\item If $\lambda\in\mathcal{P}_{-1}(2n)$ then $\lambda$ is special if and only if $\lambda^t\in\mathcal{P}_{-1}(2n)$
	\item If $\lambda\in\mathcal{P}_{+1}(2n)$ then $\lambda$ is special if and only if $\lambda^t\in\mathcal{P}_{-1}(2n)$
\end{itemize}
\end{prescription}
\paragraph{Barbasch-Vogan map}
Since we have kept $\mathcal{P}_{+1}(5)$ and $\mathcal{P}_{-1}(4)$ as ordered sets, the Barbasch-Vogan map described in section \ref{sec:BV} takes a simple form:
\begin{align}\label{eq:1}
	d_{BV}(\mathcal{P}_{+1}(5))=d_{LS}(\mathcal{P}_{-1}(4))
\end{align}
so:
\begin{align}
	\begin{aligned}
		d_{BV}(5)&=(1^4)\\
		d_{BV}(3,1^2)&=(2^2)\\
		d_{BV}(2^2,1)&=(2^2)\\
		d_{BV}(1^5)&=(4)\\
	\end{aligned}
\end{align}
Similarly:
\begin{align}\label{eq:2}
	d_{BV}(\mathcal{P}_{-1}(4))=d_{LS}(\mathcal{P}_{+1}(5))
\end{align}
gives:
\begin{align}
	\begin{aligned}
		d_{BV}(4)&=(1^5)\\
		d_{BV}(2^2)&=(3,1^2)\\
		d_{BV}(2,1^2)&=(3,1^2)\\
		d_{BV}(1^4)&=(5)\\
	\end{aligned}
\end{align}

Identities like (\ref{eq:1}) and (\ref{eq:2}) are straightforward here because the natural partial ordering becomes a total ordering for partition sets $\mathcal P_{-1}(4)$ and $\mathcal P_{+1}(5)$. This helps to illustrate the nature of the Barbasch-Vogan map, that reverses the partial ordering of the set and collapses to partitions of the dual group.

\subsubsection{Maximal orbit of $\gso (5)$}\label{sec:maxO5}

Let us focus on $\lambda=(5)$. According to prescription in section \ref{sec:sigmarho}, the brane system with
\begin{align}
	\M_H=\Or_{(5)}
\end{align}
is determined by:
\begin{align}
	\begin{aligned}
		n_s&=5\\
		n_d&=5\\
		\vec{l}_d&=(4,4,4,4,4)\\
		\vec{l}_s&=(0,2,0,2,0)\\
		\text{rightmost}\ O3&=O3^-
	\end{aligned}
\end{align}

Remember that
\begin{align}
	\begin{aligned}
		\vec{l}_s=Even(d_{BV}(\lambda))=Even(d_{BV}(5))=Even(1^4)=(0,2,0,2,0)
	\end{aligned}
\end{align}

\begin{figure}[t]
	\centering
    \begin{subfigure}[t]{.49\textwidth}
    \centering
	\begin{tikzpicture}
		\draw 	(1,0)--(1,2)
				(1.5,0)--(1.5,2)
				(2,0)--(2,2)
				(2.5,0)--(2.5,2)
				(5.5,0)--(5.5,2);
		\draw 	(3,1) node[cross] {}
				(3.5,1) node[cross] {}
				(4,1) node[cross] {}
				(4.5,1) node[cross] {}
				(5,1) node[cross] {};
		\draw [dotted] (.5,1)--(1,1)
				(1.5,1)--(2,1)
				(2.5,1)--(5.5,1);
		\draw	(1,1)--(1.5,1)
				(2,1)--(2.5,1);
		\draw	(1.5,1.3)--(2,1.3)
				(1.5,.7)--(2,.7)
				(2,1.5)--(2.5,1.5)
				(2,.5)--(2.5,.5)
				(2.5,1.3)--(5.5,1.3)
				(2.5,.7)--(5.5,.7)
				(2.5,1.7)--(5.5,1.7)
				(2.5,.3)--(5.5,.3);
		\end{tikzpicture}
        \caption{}
    \end{subfigure}
    \hfill
	\begin{subfigure}[t]{.49\textwidth}
    \centering
	\begin{tikzpicture}[]
	\tikzstyle{gauge} = [circle, draw];
	\tikzstyle{flavour} = [regular polygon,regular polygon sides=4,draw];
	\node (g1) [gauge,label=below:{$O_1$}] {};
	\node (g2) [gauge,right of=g1,label=below:{$C_1$}] {};
	\node (g3) [gauge,right of=g2,label=below:{$O_3$}] {};
	\node (g4) [gauge,right of=g3,label=below:{$C_2$}] {};
	\node (f4) [flavour,above of=g4,label=above:{$O_5$}] {};
	\draw (g1)--(g2)--(g3)--(g4)--(f4)
		;
	\end{tikzpicture}

        \caption{}
    \end{subfigure}
 	\caption{Orthosymplectic model with $n_s=n_d=5$, $\vec{l}_d=(4,4,4,4,4)$, $\vec{l}_s=(0,2,0,2,0)$ and rightmost orientifold plane $O3^-$. (a) is the Coulomb branch brane configuration. (b) is the quiver that can be read from the brane system. The Higgs branch of this model is the closure of the maximal nilpotent orbit of $\gso (5)$, corresponding to partition $\lambda=(5)$: $\M_H=\Or_{(5)}$.}
	\label{fig:SO5maximal}
\end{figure}
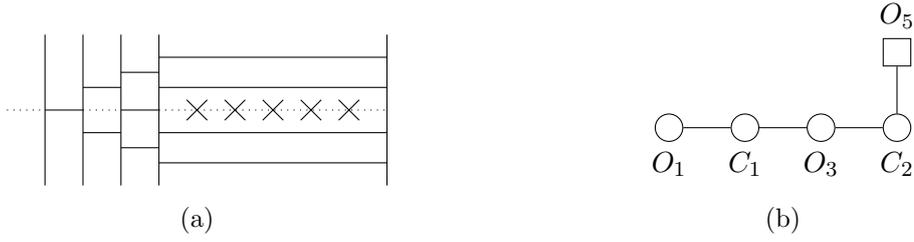

Now one can draw the brane configuration and the quiver for the Higgs branch being the closure of the nilpotent orbit, figure \ref{fig:SO5maximal}. This can be done following steps analogous to the ones described in section \ref{sec:so4interlude} for the $\gso (4)$ examples.

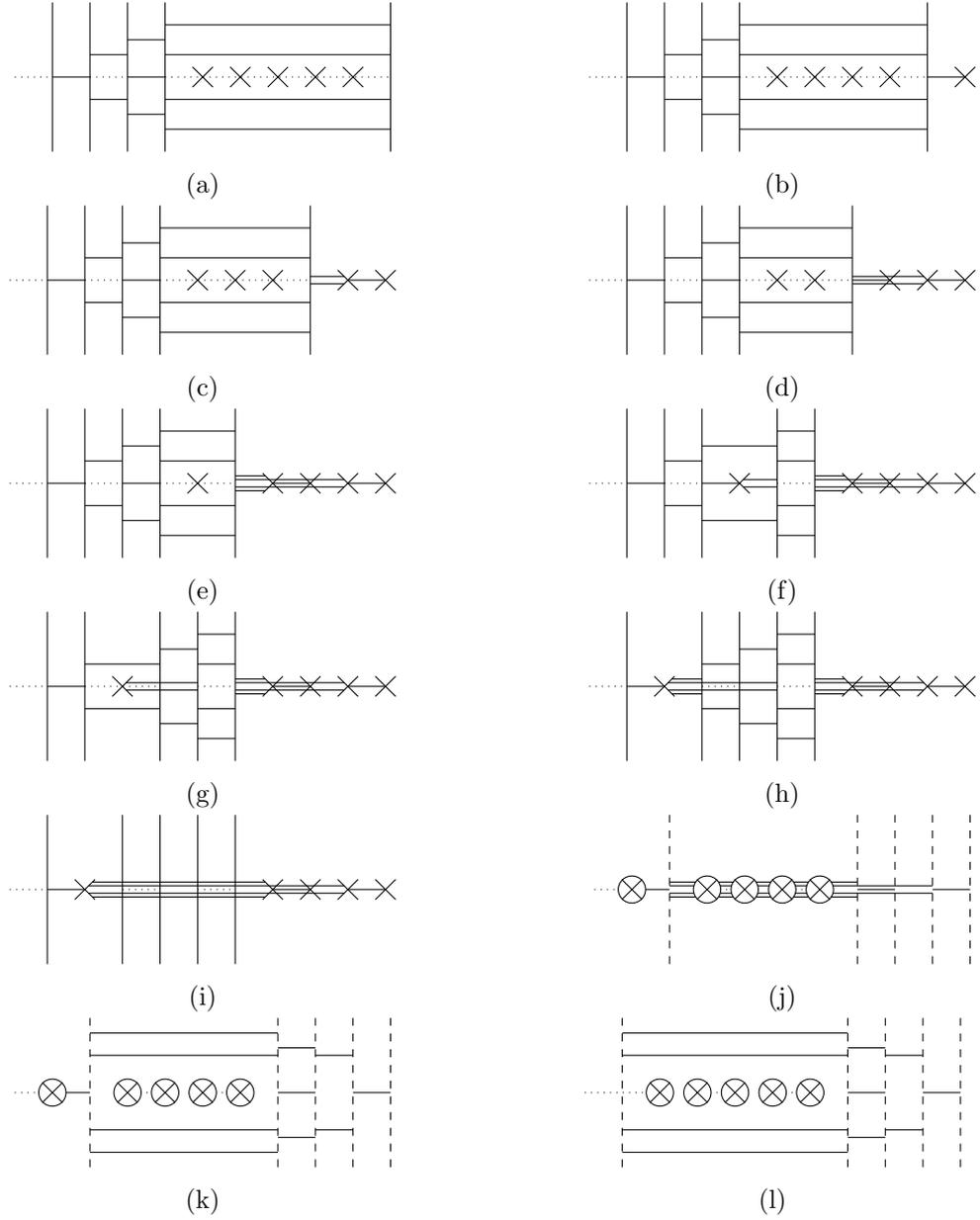
\begin{figure}[t]
	\centering
    \begin{subfigure}[t]{.49\textwidth}
    \centering
	\begin{tikzpicture}
		\draw 	(1,0)--(1,2)
				(1.5,0)--(1.5,2)
				(2,0)--(2,2)
				(2.5,0)--(2.5,2)
				(5.5,0)--(5.5,2);
		\draw 	(3,1) node[cross] {}
				(3.5,1) node[cross] {}
				(4,1) node[cross] {}
				(4.5,1) node[cross] {}
				(5,1) node[cross] {};
		\draw [dotted] (.5,1)--(1,1)
				(1.5,1)--(2,1)
				(2.5,1)--(5.5,1);
		\draw	(1,1)--(1.5,1)
				(2,1)--(2.5,1);
		\draw	(1.5,1.3)--(2,1.3)
				(1.5,.7)--(2,.7)
				(2,1.5)--(2.5,1.5)
				(2,.5)--(2.5,.5)
				(2.5,1.3)--(5.5,1.3)
				(2.5,.7)--(5.5,.7)
				(2.5,1.7)--(5.5,1.7)
				(2.5,.3)--(5.5,.3);
		\end{tikzpicture}
        \caption{}
    \end{subfigure}
    \hfill
    \begin{subfigure}[t]{.49\textwidth}
    \centering
	\begin{tikzpicture}
		\draw 	(1,0)--(1,2)
				(1.5,0)--(1.5,2)
				(2,0)--(2,2)
				(2.5,0)--(2.5,2)
				(5,0)--(5,2);
		\draw 	(3,1) node[cross] {}
				(3.5,1) node[cross] {}
				(4,1) node[cross] {}
				(4.5,1) node[cross] {}
				(5.5,1) node[cross] {};
		\draw [dotted] (.5,1)--(1,1)
				(1.5,1)--(2,1)
				(2.5,1)--(5,1);
		\draw	(1,1)--(1.5,1)
				(2,1)--(2.5,1)
				(5,1)--(5.5,1);
		\draw	(1.5,1.3)--(2,1.3)
				(1.5,.7)--(2,.7)
				(2,1.5)--(2.5,1.5)
				(2,.5)--(2.5,.5)
				(2.5,1.3)--(5,1.3)
				(2.5,.7)--(5,.7)
				(2.5,1.7)--(5,1.7)
				(2.5,.3)--(5,.3);
		\end{tikzpicture}
        \caption{}
    \end{subfigure}
    \hfill
    \begin{subfigure}[t]{.49\textwidth}
    \centering
	\begin{tikzpicture}
		\draw 	(1,0)--(1,2)
				(1.5,0)--(1.5,2)
				(2,0)--(2,2)
				(2.5,0)--(2.5,2)
				(4.5,0)--(4.5,2);
		\draw 	(3,1) node[cross] {}
				(3.5,1) node[cross] {}
				(4,1) node[cross] {}
				(5,1) node[cross] {}
				(5.5,1) node[cross] {};
		\draw [dotted] (.5,1)--(1,1)
				(1.5,1)--(2,1)
				(2.5,1)--(4.5,1);
		\draw	(1,1)--(1.5,1)
				(2,1)--(2.5,1)
				(5,1)--(5.5,1);
		\draw	(1.5,1.3)--(2,1.3)
				(1.5,.7)--(2,.7)
				(2,1.5)--(2.5,1.5)
				(2,.5)--(2.5,.5)
				(2.5,1.3)--(4.5,1.3)
				(2.5,.7)--(4.5,.7)
				(2.5,1.7)--(4.5,1.7)
				(2.5,.3)--(4.5,.3)
				(4.5,1.05)--(4.95,1.05)
				(4.5,.95)--(4.95,.95);
		\end{tikzpicture}
        \caption{}
    \end{subfigure}
    \hfill
    \begin{subfigure}[t]{.49\textwidth}
    \centering
	\begin{tikzpicture}
		\draw 	(1,0)--(1,2)
				(1.5,0)--(1.5,2)
				(2,0)--(2,2)
				(2.5,0)--(2.5,2)
				(4,0)--(4,2);
		\draw 	(3,1) node[cross] {}
				(3.5,1) node[cross] {}
				(4.5,1) node[cross] {}
				(5,1) node[cross] {}
				(5.5,1) node[cross] {};
		\draw [dotted] (.5,1)--(1,1)
				(1.5,1)--(2,1)
				(2.5,1)--(4,1);
		\draw	(1,1)--(1.5,1)
				(2,1)--(2.5,1)
				(4,1)--(4.5,1)
				(5,1)--(5.5,1);
		\draw	(1.5,1.3)--(2,1.3)
				(1.5,.7)--(2,.7)
				(2,1.5)--(2.5,1.5)
				(2,.5)--(2.5,.5)
				(2.5,1.3)--(4,1.3)
				(2.5,.7)--(4,.7)
				(2.5,1.7)--(4,1.7)
				(2.5,.3)--(4,.3)
				(4,1.05)--(4.95,1.05)
				(4,.95)--(4.95,.95);
		\end{tikzpicture}
        \caption{}
    \end{subfigure}
    \hfill
    \begin{subfigure}[t]{.49\textwidth}
    \centering
	\begin{tikzpicture}
		\draw 	(1,0)--(1,2)
				(1.5,0)--(1.5,2)
				(2,0)--(2,2)
				(2.5,0)--(2.5,2)
				(3.5,0)--(3.5,2);
		\draw 	(3,1) node[cross] {}
				(4,1) node[cross] {}
				(4.5,1) node[cross] {}
				(5,1) node[cross] {}
				(5.5,1) node[cross] {};
		\draw [dotted] (.5,1)--(1,1)
				(1.5,1)--(2,1)
				(2.5,1)--(3.5,1);
		\draw	(1,1)--(1.5,1)
				(2,1)--(2.5,1)
				(4,1)--(4.5,1)
				(5,1)--(5.5,1);
		\draw	(1.5,1.3)--(2,1.3)
				(1.5,.7)--(2,.7)
				(2,1.5)--(2.5,1.5)
				(2,.5)--(2.5,.5)
				(2.5,1.3)--(3.5,1.3)
				(2.5,.7)--(3.5,.7)
				(2.5,1.7)--(3.5,1.7)
				(2.5,.3)--(3.5,.3)
				(3.5,1.05)--(4.95,1.05)
				(3.5,.95)--(4.95,.95)
				(3.5,1.1)--(3.9,1.1)
				(3.5,.9)--(3.9,.9);
		\end{tikzpicture}
        \caption{}
    \end{subfigure}
    \hfill
    \begin{subfigure}[t]{.49\textwidth}
    \centering
	\begin{tikzpicture}
		\draw 	(1,0)--(1,2)
				(1.5,0)--(1.5,2)
				(2,0)--(2,2)
				(3,0)--(3,2)
				(3.5,0)--(3.5,2);
		\draw 	(2.5,1) node[cross] {}
				(4,1) node[cross] {}
				(4.5,1) node[cross] {}
				(5,1) node[cross] {}
				(5.5,1) node[cross] {};
		\draw [dotted] (.5,1)--(1,1)
				(1.5,1)--(2,1)
				(3,1)--(3.5,1);
		\draw	(1,1)--(1.5,1)
				(2,1)--(2.5,1)
				(4,1)--(4.5,1)
				(5,1)--(5.5,1);
		\draw	(1.5,1.3)--(2,1.3)
				(1.5,.7)--(2,.7)
				(2,1.5)--(3,1.5)
				(2,.5)--(3,.5)
				(3,1.3)--(3.5,1.3)
				(3,.7)--(3.5,.7)
				(3,1.7)--(3.5,1.7)
				(3,.3)--(3.5,.3)
				(3.5,1.05)--(4.95,1.05)
				(3.5,.95)--(4.95,.95)
				(3.5,1.1)--(3.9,1.1)
				(3.5,.9)--(3.9,.9)
				(2.55,1.05)--(3,1.05)
				(2.55,.95)--(3,.95);
		\end{tikzpicture}
        \caption{}
    \end{subfigure}
    \hfill
    \begin{subfigure}[t]{.49\textwidth}
    \centering
	\begin{tikzpicture}
		\draw 	(1,0)--(1,2)
				(1.5,0)--(1.5,2)
				(2.5,0)--(2.5,2)
				(3,0)--(3,2)
				(3.5,0)--(3.5,2);
		\draw 	(2,1) node[cross] {}
				(4,1) node[cross] {}
				(4.5,1) node[cross] {}
				(5,1) node[cross] {}
				(5.5,1) node[cross] {};
		\draw [dotted] (.5,1)--(1,1)
				(1.5,1)--(2.5,1)
				(3,1)--(3.5,1);
		\draw	(1,1)--(1.5,1)
				(4,1)--(4.5,1)
				(5,1)--(5.5,1);
		\draw	(1.5,1.3)--(2.5,1.3)
				(1.5,.7)--(2.5,.7)
				(2.5,1.5)--(3,1.5)
				(2.5,.5)--(3,.5)
				(3,1.3)--(3.5,1.3)
				(3,.7)--(3.5,.7)
				(3,1.7)--(3.5,1.7)
				(3,.3)--(3.5,.3)
				(3.5,1.05)--(4.95,1.05)
				(3.5,.95)--(4.95,.95)
				(3.5,1.1)--(3.9,1.1)
				(3.5,.9)--(3.9,.9)
				(2.05,1.05)--(3,1.05)
				(2.05,.95)--(3,.95);
		\end{tikzpicture}
        \caption{}
    \end{subfigure}
    \hfill
    \begin{subfigure}[t]{.49\textwidth}
    \centering
	\begin{tikzpicture}
		\draw 	(1,0)--(1,2)
				(2,0)--(2,2)
				(2.5,0)--(2.5,2)
				(3,0)--(3,2)
				(3.5,0)--(3.5,2);
		\draw 	(1.5,1) node[cross] {}
				(4,1) node[cross] {}
				(4.5,1) node[cross] {}
				(5,1) node[cross] {}
				(5.5,1) node[cross] {};
		\draw [dotted] (.5,1)--(1,1)
				(2,1)--(2.5,1)
				(3,1)--(3.5,1);
		\draw	(1,1)--(1.5,1)
				(4,1)--(4.5,1)
				(5,1)--(5.5,1);
		\draw	(2,1.3)--(2.5,1.3)
				(2,.7)--(2.5,.7)
				(2.5,1.5)--(3,1.5)
				(2.5,.5)--(3,.5)
				(3,1.3)--(3.5,1.3)
				(3,.7)--(3.5,.7)
				(3,1.7)--(3.5,1.7)
				(3,.3)--(3.5,.3)
				(3.5,1.05)--(4.95,1.05)
				(3.5,.95)--(4.95,.95)
				(3.5,1.1)--(3.9,1.1)
				(3.5,.9)--(3.9,.9)
				(1.55,1.05)--(3,1.05)
				(1.55,.95)--(3,.95)
				(1.6,1.1)--(2,1.1)
				(1.6,.9)--(2,.9);
		\end{tikzpicture}
        \caption{}
    \end{subfigure}
    \hfill
    \begin{subfigure}[t]{.49\textwidth}
    \centering
	\begin{tikzpicture}
		\draw 	(1,0)--(1,2)
				(2,0)--(2,2)
				(2.5,0)--(2.5,2)
				(3,0)--(3,2)
				(3.5,0)--(3.5,2);
		\draw 	(1.5,1) node[cross] {}
				(4,1) node[cross] {}
				(4.5,1) node[cross] {}
				(5,1) node[cross] {}
				(5.5,1) node[cross] {};
		\draw [dotted] (.5,1)--(1,1)
				(2,1)--(2.5,1)
				(3,1)--(3.5,1);
		\draw	(1,1)--(1.5,1)
				(4,1)--(4.5,1)
				(5,1)--(5.5,1);
		\draw	(1.55,1.05)--(4.95,1.05)
				(1.55,.95)--(4.95,.95)
				(1.6,1.1)--(3.9,1.1)
				(1.6,.9)--(3.9,.9);
		\end{tikzpicture}
        \caption{}
    \end{subfigure}
    \hfill
    \begin{subfigure}[t]{.49\textwidth}
    \centering
	\begin{tikzpicture}
		\draw[dashed] 	(1.5,0)--(1.5,2)
				(4,0)--(4,2)
				(4.5,0)--(4.5,2)
				(5,0)--(5,2)
				(5.5,0)--(5.5,2);
		\draw [dotted] (.5,1)--(1,1)
				(2,1)--(2.5,1)
				(3,1)--(3.5,1);
		\draw	(1,1)--(1.5,1)
				(4,1)--(4.5,1)
				(5,1)--(5.5,1);
		\draw	(1.5,1.05)--(5,1.05)
				(1.5,.95)--(5,.95)
				(1.5,1.1)--(4,1.1)
				(1.5,.9)--(4,.9);
		\draw 	(1,1) node[circ]{}
				(2,1)node[circ]{}
				(2.5,1)node[circ]{}
				(3,1)node[circ]{}
				(3.5,1)node[circ]{};
		\draw 	(1,1) node[cross]{}
				(2,1)node[cross]{}
				(2.5,1)node[cross]{}
				(3,1)node[cross]{}
				(3.5,1)node[cross]{};
		\end{tikzpicture}
        \caption{}
    \end{subfigure}
    \hfill
    \begin{subfigure}[t]{.49\textwidth}
    \centering
	\begin{tikzpicture}
		\draw[dashed] 	(1.5,0)--(1.5,2)
				(4,0)--(4,2)
				(4.5,0)--(4.5,2)
				(5,0)--(5,2)
				(5.5,0)--(5.5,2);
		\draw [dotted] (.5,1)--(1,1)
				(2,1)--(2.5,1)
				(3,1)--(3.5,1);
		\draw	(1,1)--(1.5,1)
				(4,1)--(4.5,1)
				(5,1)--(5.5,1);
		\draw	(1.5,1.5)--(4,1.5)
				(4,1.6)--(4.5,1.6)
				(4.5,1.5)--(5,1.5)
				(1.5,.5)--(4,.5)
				(4,.4)--(4.5,.4)
				(4.5,.5)--(5,.5)
				(1.5,1.8)--(4,1.8)
				(1.5,.2)--(4,.2);
		\draw 	(1,1) node[circ]{}
				(2,1)node[circ]{}
				(2.5,1)node[circ]{}
				(3,1)node[circ]{}
				(3.5,1)node[circ]{};
		\draw 	(1,1) node[cross]{}
				(2,1)node[cross]{}
				(2.5,1)node[cross]{}
				(3,1)node[cross]{}
				(3.5,1)node[cross]{};
		\end{tikzpicture}
        \caption{}
    \end{subfigure}
    \hfill
    \begin{subfigure}[t]{.49\textwidth}
    \centering
	\begin{tikzpicture}
		\draw[dashed] 	(1,0)--(1,2)
				(4,0)--(4,2)
				(4.5,0)--(4.5,2)
				(5,0)--(5,2)
				(5.5,0)--(5.5,2);
		\draw [dotted] (.5,1)--(1.5,1)
				(2,1)--(2.5,1)
				(3,1)--(3.5,1);
		\draw	(4,1)--(4.5,1)
				(5,1)--(5.5,1);
		\draw	(1,1.5)--(4,1.5)
				(4,1.6)--(4.5,1.6)
				(4.5,1.5)--(5,1.5)
				(1,.5)--(4,.5)
				(4,.4)--(4.5,.4)
				(4.5,.5)--(5,.5)
				(1,1.8)--(4,1.8)
				(1,.2)--(4,.2);
		\draw 	(1.5,1) node[circ]{}
				(2,1)node[circ]{}
				(2.5,1)node[circ]{}
				(3,1)node[circ]{}
				(3.5,1)node[circ]{};
		\draw 	(1.5,1) node[cross]{}
				(2,1)node[cross]{}
				(2.5,1)node[cross]{}
				(3,1)node[cross]{}
				(3.5,1)node[cross]{};
		\end{tikzpicture}
        \caption{}
    \end{subfigure}
    \hfill
 	\caption{Higgsing of the orthosymplectic model with $n_s=n_d=5$, $\vec{l}_d=(4,4,4,4,4)$, $\vec{l}_s=(0,2,0,2,0)$ and rightmost orientifold plane $O3^-$. (a) is the Coulomb branch. (b-j) are one step phase transitions where the linking numbers of the fivebranes are kept constant by creating or not creating D3-branes after each transition. (k) Higgs branch before the \emph{collapse} transition; the half NS5-branes have interval numbers $\vec k_s'=(0,1,1,1,1)$. (l) Higgs branch after the \emph{collapse} transition, the half NS5-branes have interval numbers $\vec k_s=(1,1,1,1,1)$.}
	\label{fig:SO5maxHiggs}
\end{figure}

One can also obtain the Higgs branch brane configuration. The transition step by step is depicted in figure \ref{fig:SO5maxHiggs}. Note that once more, since all the half fivebranes are kept at the orientifold plane at all times, the creation/annihilation of D3-branes is only restricted by the need of preserving the linking numbers of the fivebranes in each step.

The interval numbers of the half NS5-branes in figure \ref{fig:SO5maxHiggs}(l), after performing the collapse transition, are $\vec{k}_s=(1,1,1,1,1)$. This defines a partition $\lambda$ as in equation (\ref{eq:interval}), were $\lambda^t$ is formed by all the elements in $\vec{k}_s=(1,1,1,1,1)$. Hence, one recovers the partition $\lambda=(5)$ directly from the Higgs branch brane configuration:

\begin{align}
	\begin{aligned}
		\vec k_s=(1,1,1,1,1)&\rightarrow\lambda=(5)	
	\end{aligned}
\end{align}

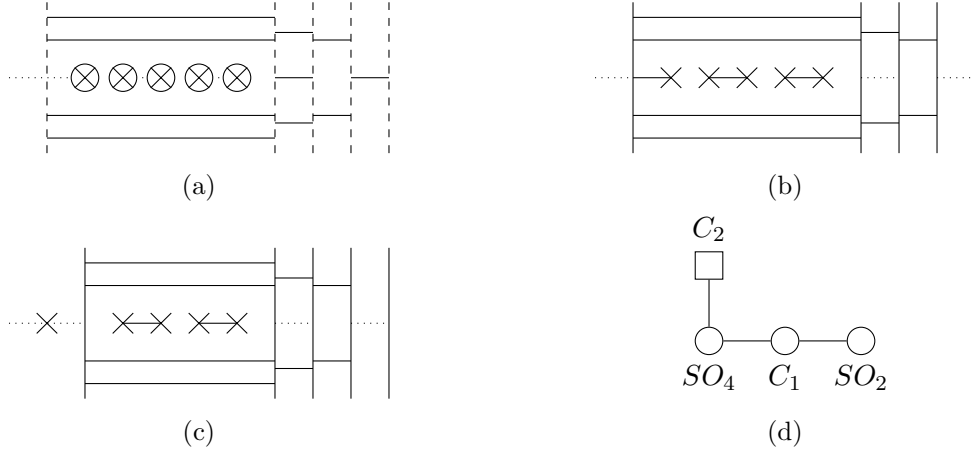
\begin{figure}[t]
	\centering
    \begin{subfigure}[t]{.49\textwidth}
    \centering
	\begin{tikzpicture}
		\draw[dashed] 	(1,0)--(1,2)
				(4,0)--(4,2)
				(4.5,0)--(4.5,2)
				(5,0)--(5,2)
				(5.5,0)--(5.5,2);
		\draw [dotted] (.5,1)--(1.5,1)
				(2,1)--(2.5,1)
				(3,1)--(3.5,1);
		\draw	(4,1)--(4.5,1)
				(5,1)--(5.5,1);
		\draw	(1,1.5)--(4,1.5)
				(4,1.6)--(4.5,1.6)
				(4.5,1.5)--(5,1.5)
				(1,.5)--(4,.5)
				(4,.4)--(4.5,.4)
				(4.5,.5)--(5,.5)
				(1,1.8)--(4,1.8)
				(1,.2)--(4,.2);
		\draw 	(1.5,1) node[circ]{}
				(2,1)node[circ]{}
				(2.5,1)node[circ]{}
				(3,1)node[circ]{}
				(3.5,1)node[circ]{};
		\draw 	(1.5,1) node[cross]{}
				(2,1)node[cross]{}
				(2.5,1)node[cross]{}
				(3,1)node[cross]{}
				(3.5,1)node[cross]{};
		\end{tikzpicture}
        \caption{}
    \end{subfigure}
    \hfill
    \begin{subfigure}[t]{.49\textwidth}
    \centering
	\begin{tikzpicture}
		\draw 	(1.5,1) node[cross]{}
				(2,1)node[cross]{}
				(2.5,1)node[cross]{}
				(3,1)node[cross]{}
				(3.5,1)node[cross]{};
		\draw 	(1,0)--(1,2)
				(4,0)--(4,2)
				(4.5,0)--(4.5,2)
				(5,0)--(5,2)
				(5.5,0)--(5.5,2);
		\draw (2,1)--(2.5,1)
				(3,1)--(3.5,1);
		\draw	(1,1)--(1.5,1);
		\draw[dotted](.5,1)--(1,1)
				(4,1)--(4.5,1)
				(5,1)--(5.5,1);
		\draw	(1,1.5)--(4,1.5)
				(4,1.6)--(4.5,1.6)
				(4.5,1.5)--(5,1.5)
				(1,.5)--(4,.5)
				(4,.4)--(4.5,.4)
				(4.5,.5)--(5,.5)
				(1,1.8)--(4,1.8)
				(1,.2)--(4,.2);
		\end{tikzpicture}
        \caption{}
    \end{subfigure}
    \hfill
    \begin{subfigure}[t]{.49\textwidth}
    \centering
	\begin{tikzpicture}
		\draw 	(1,1) node[cross]{}
				(2,1)node[cross]{}
				(2.5,1)node[cross]{}
				(3,1)node[cross]{}
				(3.5,1)node[cross]{};
		\draw 	(1.5,0)--(1.5,2)
				(4,0)--(4,2)
				(4.5,0)--(4.5,2)
				(5,0)--(5,2)
				(5.5,0)--(5.5,2);
		\draw (2,1)--(2.5,1)
				(3,1)--(3.5,1);
		\draw[dotted](.5,1)--(1.5,1)
				(4,1)--(4.5,1)
				(5,1)--(5.5,1);
		\draw	(1.5,1.5)--(4,1.5)
				(4,1.6)--(4.5,1.6)
				(4.5,1.5)--(5,1.5)
				(1.5,.5)--(4,.5)
				(4,.4)--(4.5,.4)
				(4.5,.5)--(5,.5)
				(1.5,1.8)--(4,1.8)
				(1.5,.2)--(4,.2);
		\end{tikzpicture}
        \caption{}
    \end{subfigure}
    \hfill
	\begin{subfigure}[t]{.49\textwidth}
    \centering
	\begin{tikzpicture}[]
	\tikzstyle{gauge} = [circle, draw];
	\tikzstyle{flavour} = [regular polygon,regular polygon sides=4,draw];
	\node (g1) [gauge,label=below:{$SO_4$}] {};
	\node (g2) [gauge,right of=g1,label=below:{$C_1$}] {};
	\node (g3) [gauge,right of=g2,label=below:{$SO_2$}] {};
	\node (f1) [flavour,above of=g1,label=above:{$C_2$}] {};
	\draw (f1)--(g1)--(g2)--(g3)
		;
	\end{tikzpicture}

        \caption{}
    \end{subfigure}
 	\caption{Result of S-duality on the model with $n_s=n_d=5$, $\vec{l}_d=(4,4,4,4,4)$, $\vec{l}_s=(0,2,0,2,0)$ and rightmost orientifold plane $O3^-$. (a) Higgs branch of the initial model. (b) Resulting model after performing S-duality on (a). (c) Resulting model after performing the \emph{collapse} transition (i.e. pull all half D5-branes away from $O3^-$ planes without brane creation) on (b), this is necessary to read the quiver. (d) Quiver obtained from (c) with the choice of $SO_4$ and $SO_2$ as gauge nodes. This choice is necessary to obtain a Coulomb branch $\M_C=\Or_{(5)}\subset \gso (5)$, \cite{CHZ17}.}
	\label{fig:SO5maxMirror}
\end{figure}

S-duality can be performed in order to obtain a quiver with $\M_C=\Or_{(5)}\subset \gso (5)$, figure \ref{fig:SO5maxMirror}.

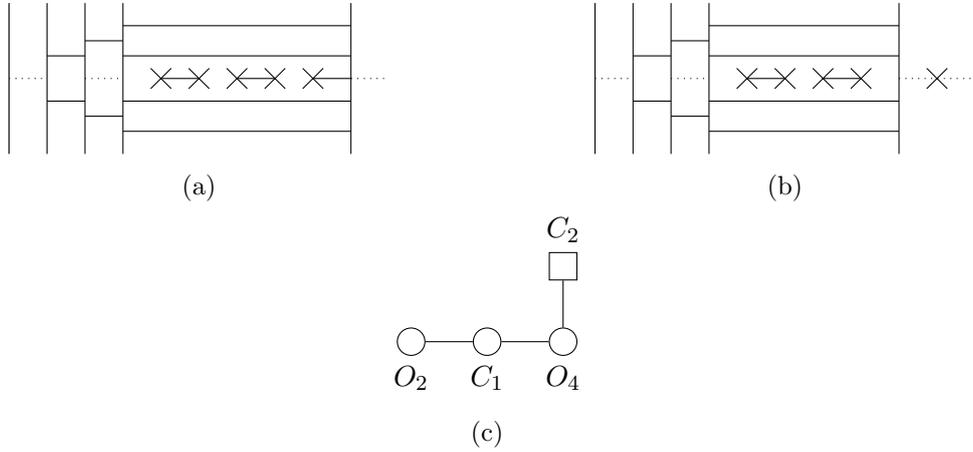
\begin{figure}[t]
	\centering
	\begin{subfigure}[t]{.49\textwidth}
		\centering
		\begin{tikzpicture}
			\draw 	(1,0)--(1,2)
					(1.5,0)--(1.5,2)
					(2,0)--(2,2)
					(2.5,0)--(2.5,2)
					(5.5,0)--(5.5,2);
			\draw 	(3,1) node[cross]{}
					(3.5,1)node[cross]{}
					(4,1)node[cross]{}
					(4.5,1)node[cross]{}
					(5,1)node[cross]{};
			\draw	[dotted](1,1)--(1.5,1)
					(2,1)--(2.5,1)
					(5.5,1)--(6,1);
			\draw	(3,1)--(3.5,1)
					(4,1)--(4.5,1)
					(5,1)--(5.5,1);
			\draw	(1.5,1.3)--(2,1.3)
					(1.5,.7)--(2,.7)
					(2,1.5)--(2.5,1.5)
					(2,.5)--(2.5,.5)
					(2.5,1.3)--(5.5,1.3)
					(2.5,.7)--(5.5,.7)
					(2.5,1.7)--(5.5,1.7)
					(2.5,.3)--(5.5,.3);
		\end{tikzpicture}
		\caption{}
	\end{subfigure}
	\hfill
	\begin{subfigure}[t]{.49\textwidth}
		\centering
		\begin{tikzpicture}
			\draw 	(1,0)--(1,2)
					(1.5,0)--(1.5,2)
					(2,0)--(2,2)
					(2.5,0)--(2.5,2)
					(5,0)--(5,2);
			\draw 	(3,1) node[cross]{}
					(3.5,1)node[cross]{}
					(4,1)node[cross]{}
					(4.5,1)node[cross]{}
					(5.5,1)node[cross]{};
			\draw	[dotted](1,1)--(1.5,1)
					(2,1)--(2.5,1)
					(5,1)--(6,1);
			\draw	(3,1)--(3.5,1)
					(4,1)--(4.5,1);
			\draw	(1.5,1.3)--(2,1.3)
					(1.5,.7)--(2,.7)
					(2,1.5)--(2.5,1.5)
					(2,.5)--(2.5,.5)
					(2.5,1.3)--(5,1.3)
					(2.5,.7)--(5,.7)
					(2.5,1.7)--(5,1.7)
					(2.5,.3)--(5,.3);
		\end{tikzpicture}
		\caption{}
	\end{subfigure}
	\hfill
	\begin{subfigure}[t]{.49\textwidth}
	\centering
		\begin{tikzpicture}[]
			\tikzstyle{gauge} = [circle, draw];
			\tikzstyle{flavour} = [regular polygon,regular polygon sides=4,draw];
			\node (g1) [gauge, label=below:{$O_2$}]{};
			\node (g2) [gauge, right of=g1, label=below:{$C_1$}]{};
			\node (g3) [gauge,right of=g2, label=below:{$O_4$}]{};
			\node (f3) [flavour,above of=g3,label=above:{$C_2$}] {};
			\draw (g1)--(g2)--(g3)--(f3);
		\end{tikzpicture}
		\caption{}
	\end{subfigure}
	\hfill
	\caption{Orthosymplectic model with $n_s=n_d=5$, $\vec{l}_d=(5,3,5,3,5)$, $\vec{l}_s=(1,1,1,1,1)$ and rightmost orientifold plane $\widetilde{O3^+}$. (a) Coulomb branch brane configuration. (b) Resulting model after the \emph{collapse} transition has been performed in (a), in order to being able to read the quiver. (c) Quiver that can be read from the brane system (b). The Higgs branch of this model is the closure of the maximal nilpotent orbit of $\gsp (2)$, corresponding to partition $\lambda=(4)$: $\M_H=\Or_{(4)}$.}
	\label{fig:Sp4maximal}
\end{figure}

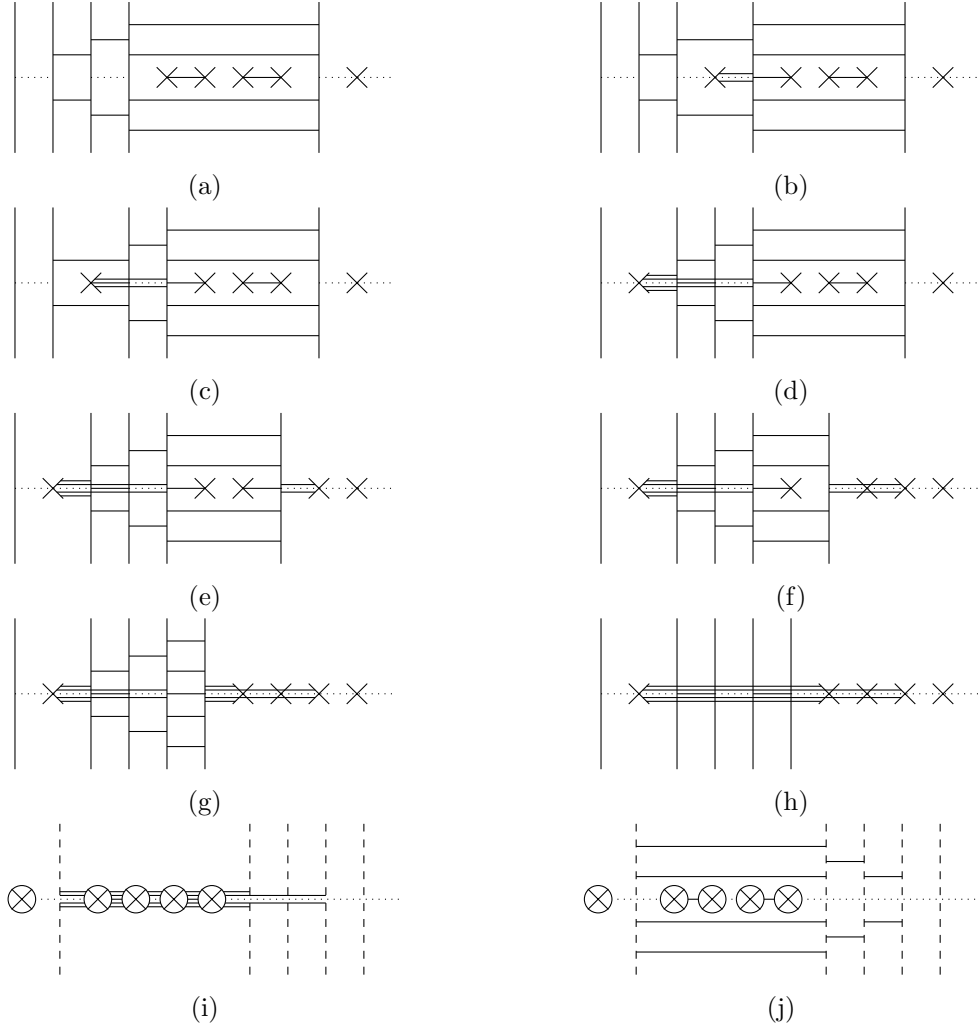
\begin{figure}[t]
	\centering
	\begin{subfigure}[t]{.49\textwidth}
		\centering
		\begin{tikzpicture}
			\draw 	(1,0)--(1,2)
					(1.5,0)--(1.5,2)
					(2,0)--(2,2)
					(2.5,0)--(2.5,2)
					(5,0)--(5,2);
			\draw 	(3,1) node[cross]{}
					(3.5,1)node[cross]{}
					(4,1)node[cross]{}
					(4.5,1)node[cross]{}
					(5.5,1)node[cross]{};
			\draw	[dotted](1,1)--(1.5,1)
					(2,1)--(2.5,1)
					(5,1)--(6,1);
			\draw	(3,1)--(3.5,1)
					(4,1)--(4.5,1);
			\draw	(1.5,1.3)--(2,1.3)
					(1.5,.7)--(2,.7)
					(2,1.5)--(2.5,1.5)
					(2,.5)--(2.5,.5)
					(2.5,1.3)--(5,1.3)
					(2.5,.7)--(5,.7)
					(2.5,1.7)--(5,1.7)
					(2.5,.3)--(5,.3);
		\end{tikzpicture}
		\caption{}
	\end{subfigure}
	\hfill
	\begin{subfigure}[t]{.49\textwidth}
		\centering
		\begin{tikzpicture}
			\draw 	(1,0)--(1,2)
					(1.5,0)--(1.5,2)
					(2,0)--(2,2)
					(3,0)--(3,2)
					(5,0)--(5,2);
			\draw 	(2.5,1) node[cross]{}
					(3.5,1)node[cross]{}
					(4,1)node[cross]{}
					(4.5,1)node[cross]{}
					(5.5,1)node[cross]{};
			\draw	[dotted](1,1)--(1.5,1)
					(2,1)--(3,1)
					(5,1)--(6,1);
			\draw	(3,1)--(3.5,1)
					(4,1)--(4.5,1);
			\draw	(1.5,1.3)--(2,1.3)
					(1.5,.7)--(2,.7)
					(2,1.5)--(3,1.5)
					(2,.5)--(3,.5)
					(3,1.3)--(5,1.3)
					(3,.7)--(5,.7)
					(3,1.7)--(5,1.7)
					(3,.3)--(5,.3);
			\draw	(2.55,1.05)--(3,1.05)
					(2.55,.95)--(3,.95);
		\end{tikzpicture}
		\caption{}
	\end{subfigure}
	\hfill
	\begin{subfigure}[t]{.49\textwidth}
		\centering
		\begin{tikzpicture}
			\draw 	(1,0)--(1,2)
					(1.5,0)--(1.5,2)
					(2.5,0)--(2.5,2)
					(3,0)--(3,2)
					(5,0)--(5,2);
			\draw 	(2,1) node[cross]{}
					(3.5,1)node[cross]{}
					(4,1)node[cross]{}
					(4.5,1)node[cross]{}
					(5.5,1)node[cross]{};
			\draw	[dotted](1,1)--(1.5,1)
					(2.5,1)--(3,1)
					(5,1)--(6,1);
			\draw	(2,1)--(2.5,1)
					(3,1)--(3.5,1)
					(4,1)--(4.5,1);
			\draw	(1.5,1.3)--(2.5,1.3)
					(1.5,.7)--(2.5,.7)
					(2.5,1.5)--(3,1.5)
					(2.5,.5)--(3,.5)
					(3,1.3)--(5,1.3)
					(3,.7)--(5,.7)
					(3,1.7)--(5,1.7)
					(3,.3)--(5,.3);
			\draw	(2.05,1.05)--(3,1.05)
					(2.05,.95)--(3,.95);
		\end{tikzpicture}
		\caption{}
	\end{subfigure}
	\hfill
	\begin{subfigure}[t]{.49\textwidth}
		\centering
		\begin{tikzpicture}
			\draw 	(1,0)--(1,2)
					(2,0)--(2,2)
					(2.5,0)--(2.5,2)
					(3,0)--(3,2)
					(5,0)--(5,2);
			\draw 	(1.5,1) node[cross]{}
					(3.5,1)node[cross]{}
					(4,1)node[cross]{}
					(4.5,1)node[cross]{}
					(5.5,1)node[cross]{};
			\draw	[dotted](1,1)--(2,1)
					(2.5,1)--(3,1)
					(5,1)--(6,1);
			\draw	(2,1)--(2.5,1)
					(3,1)--(3.5,1)
					(4,1)--(4.5,1);
			\draw	(2,1.3)--(2.5,1.3)
					(2,.7)--(2.5,.7)
					(2.5,1.5)--(3,1.5)
					(2.5,.5)--(3,.5)
					(3,1.3)--(5,1.3)
					(3,.7)--(5,.7)
					(3,1.7)--(5,1.7)
					(3,.3)--(5,.3);
			\draw	(1.55,1.05)--(3,1.05)
					(1.55,.95)--(3,.95)
					(1.6,1.1)--(2,1.1)
					(1.6,.9)--(2,.9);
		\end{tikzpicture}
		\caption{}
	\end{subfigure}
	\hfill
	\begin{subfigure}[t]{.49\textwidth}
		\centering
		\begin{tikzpicture}
			\draw 	(1,0)--(1,2)
					(2,0)--(2,2)
					(2.5,0)--(2.5,2)
					(3,0)--(3,2)
					(4.5,0)--(4.5,2);
			\draw 	(1.5,1) node[cross]{}
					(3.5,1)node[cross]{}
					(4,1)node[cross]{}
					(5,1)node[cross]{}
					(5.5,1)node[cross]{};
			\draw	[dotted](1,1)--(2,1)
					(2.5,1)--(3,1)
					(4.5,1)--(6,1);
			\draw	(2,1)--(2.5,1)
					(3,1)--(3.5,1)
					(4,1)--(4.5,1);
			\draw	(2,1.3)--(2.5,1.3)
					(2,.7)--(2.5,.7)
					(2.5,1.5)--(3,1.5)
					(2.5,.5)--(3,.5)
					(3,1.3)--(4.5,1.3)
					(3,.7)--(4.5,.7)
					(3,1.7)--(4.5,1.7)
					(3,.3)--(4.5,.3);
			\draw	(1.55,1.05)--(3,1.05)
					(1.55,.95)--(3,.95)
					(1.6,1.1)--(2,1.1)
					(1.6,.9)--(2,.9)
					(4.5,1.05)--(4.95,1.05)
					(4.5,.95)--(4.95,.95);
		\end{tikzpicture}
		\caption{}
	\end{subfigure}
	\hfill
	\begin{subfigure}[t]{.49\textwidth}
		\centering
		\begin{tikzpicture}
			\draw 	(1,0)--(1,2)
					(2,0)--(2,2)
					(2.5,0)--(2.5,2)
					(3,0)--(3,2)
					(4,0)--(4,2);
			\draw 	(1.5,1) node[cross]{}
					(3.5,1)node[cross]{}
					(4.5,1)node[cross]{}
					(5,1)node[cross]{}
					(5.5,1)node[cross]{};
			\draw	[dotted](1,1)--(2,1)
					(2.5,1)--(3,1)
					(4,1)--(6,1);
			\draw	(2,1)--(2.5,1)
					(3,1)--(3.5,1);
			\draw	(2,1.3)--(2.5,1.3)
					(2,.7)--(2.5,.7)
					(2.5,1.5)--(3,1.5)
					(2.5,.5)--(3,.5)
					(3,1.3)--(4,1.3)
					(3,.7)--(4,.7)
					(3,1.7)--(4,1.7)
					(3,.3)--(4,.3);
			\draw	(1.55,1.05)--(3,1.05)
					(1.55,.95)--(3,.95)
					(1.6,1.1)--(2,1.1)
					(1.6,.9)--(2,.9)
					(4,1.05)--(4.95,1.05)
					(4,.95)--(4.95,.95);
		\end{tikzpicture}
		\caption{}
	\end{subfigure}
	\hfill
	\begin{subfigure}[t]{.49\textwidth}
		\centering
		\begin{tikzpicture}
			\draw 	(1,0)--(1,2)
					(2,0)--(2,2)
					(2.5,0)--(2.5,2)
					(3,0)--(3,2)
					(3.5,0)--(3.5,2);
			\draw 	(1.5,1) node[cross]{}
					(4,1)node[cross]{}
					(4.5,1)node[cross]{}
					(5,1)node[cross]{}
					(5.5,1)node[cross]{};
			\draw	[dotted](1,1)--(2,1)
					(2.5,1)--(3,1)
					(3.5,1)--(6,1);
			\draw	(2,1)--(2.5,1)
					(3,1)--(3.5,1);
			\draw	(2,1.3)--(2.5,1.3)
					(2,.7)--(2.5,.7)
					(2.5,1.5)--(3,1.5)
					(2.5,.5)--(3,.5)
					(3,1.3)--(3.5,1.3)
					(3,.7)--(3.5,.7)
					(3,1.7)--(3.5,1.7)
					(3,.3)--(3.5,.3);
			\draw	(1.55,1.05)--(3,1.05)
					(1.55,.95)--(3,.95)
					(1.6,1.1)--(2,1.1)
					(1.6,.9)--(2,.9)
					(3.5,1.05)--(4.95,1.05)
					(3.5,.95)--(4.95,.95)
					(3.5,1.1)--(3.9,1.1)
					(3.5,.9)--(3.9,.9);
		\end{tikzpicture}
		\caption{}
	\end{subfigure}
	\hfill
	\begin{subfigure}[t]{.49\textwidth}
		\centering
		\begin{tikzpicture}
			\draw 	(1,0)--(1,2)
					(2,0)--(2,2)
					(2.5,0)--(2.5,2)
					(3,0)--(3,2)
					(3.5,0)--(3.5,2);
			\draw 	(1.5,1) node[cross]{}
					(4,1)node[cross]{}
					(4.5,1)node[cross]{}
					(5,1)node[cross]{}
					(5.5,1)node[cross]{};
			\draw	[dotted](1,1)--(2,1)
					(2.5,1)--(3,1)
					(3.5,1)--(6,1);
			\draw	(2,1)--(2.5,1)
					(3,1)--(3.5,1);
			\draw	(1.55,1.05)--(4.95,1.05)
					(1.55,.95)--(4.95,.95)
					(1.6,1.1)--(3.9,1.1)
					(1.6,.9)--(3.9,.9);
		\end{tikzpicture}
		\caption{}
	\end{subfigure}
	\hfill
	\begin{subfigure}[t]{.49\textwidth}
		\centering
		\begin{tikzpicture}
			\draw 	[dashed](1.5,0)--(1.5,2)
					(4,0)--(4,2)
					(4.5,0)--(4.5,2)
					(5,0)--(5,2)
					(5.5,0)--(5.5,2);
			\draw	[dotted](1,1)--(2,1)
					(2.5,1)--(3,1)
					(3.5,1)--(6,1);
			\draw	(2,1)--(2.5,1)
					(3,1)--(3.5,1);
			\draw	(1.5,1.05)--(5,1.05)
					(1.5,.95)--(5,.95)
					(1.5,1.1)--(4,1.1)
					(1.5,.9)--(4,.9);
			\draw 	(1,1) node[circ]{}
					(2,1) node[circ]{}
					(2.5,1) node[circ]{}
					(3,1) node[circ]{}
					(3.5,1) node[circ]{};
			\draw 	(1,1) node[cross]{}
					(2,1) node[cross]{}
					(2.5,1) node[cross]{}
					(3,1) node[cross]{}
					(3.5,1) node[cross]{};
		\end{tikzpicture}
		\caption{}
	\end{subfigure}
	\hfill
	\begin{subfigure}[t]{.49\textwidth}
		\centering
		\begin{tikzpicture}
			\draw 	[dashed](1.5,0)--(1.5,2)
					(4,0)--(4,2)
					(4.5,0)--(4.5,2)
					(5,0)--(5,2)
					(5.5,0)--(5.5,2);
			\draw	[dotted](1,1)--(2,1)
					(2.5,1)--(3,1)
					(3.5,1)--(6,1);
			\draw	(2,1)--(2.5,1)
					(3,1)--(3.5,1);
			\draw	(1.5,1.7)--(4,1.7)
					(1.5,.3)--(4,.3)
					(1.5,1.3)--(4,1.3)
					(1.5,.7)--(4,.7)
					(4,1.5)--(4.5,1.5)
					(4,.5)--(4.5,.5)
					(4.5,1.3)--(5,1.3)
					(4.5,.7)--(5,.7);
			\draw 	(1,1) node[circ]{}
					(2,1) node[circ]{}
					(2.5,1) node[circ]{}
					(3,1) node[circ]{}
					(3.5,1) node[circ]{};
			\draw 	(1,1) node[cross]{}
					(2,1) node[cross]{}
					(2.5,1) node[cross]{}
					(3,1) node[cross]{}
					(3.5,1) node[cross]{};
		\end{tikzpicture}
		\caption{}
	\end{subfigure}
	\hfill
	\caption{Higgsing of the maximal model of $\mathfrak{sp}(2)$. The parameters defining this model are $n_s=n_d=5$, $\vec{l}_d=(5,3,5,3,5)$, $\vec{l}_s=(1,1,1,1,1)$ and rightmost orientifold plane $\widetilde{O3^+}$. The interval numbers of the half NS5-branes in (j), after performing a collapse transition, are $\vec{k}_s=(0,1,1,1,1)$. This gives $\lambda^t=(1^4)$, hence $\lambda=(4)$ can be obtained in this way.}
	\label{fig:Sp4maxHiggs}
\end{figure}

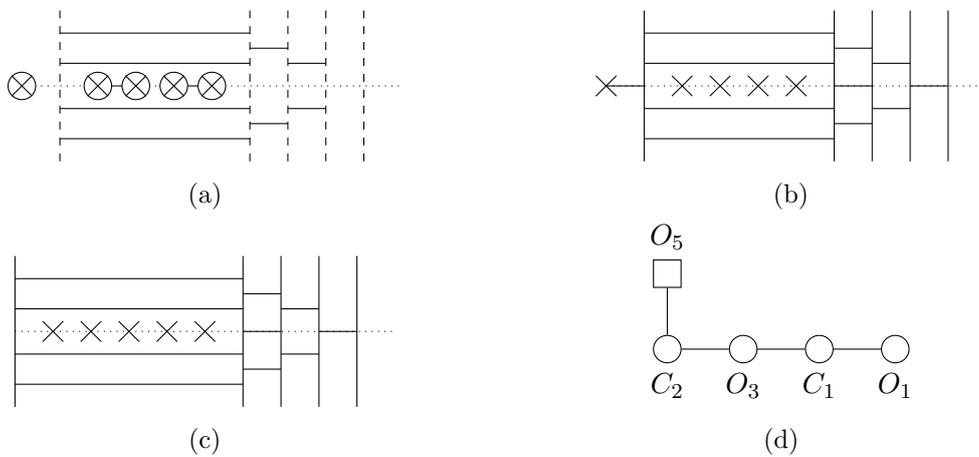
\begin{figure}[t]
	\centering
	\begin{subfigure}[t]{.49\textwidth}
		\centering
		\begin{tikzpicture}
			\draw 	[dashed](1.5,0)--(1.5,2)
					(4,0)--(4,2)
					(4.5,0)--(4.5,2)
					(5,0)--(5,2)
					(5.5,0)--(5.5,2);
			\draw	[dotted](1,1)--(2,1)
					(2.5,1)--(3,1)
					(3.5,1)--(6,1);
			\draw	(2,1)--(2.5,1)
					(3,1)--(3.5,1);
			\draw	(1.5,1.7)--(4,1.7)
					(1.5,.3)--(4,.3)
					(1.5,1.3)--(4,1.3)
					(1.5,.7)--(4,.7)
					(4,1.5)--(4.5,1.5)
					(4,.5)--(4.5,.5)
					(4.5,1.3)--(5,1.3)
					(4.5,.7)--(5,.7);
			\draw 	(1,1) node[circ]{}
					(2,1) node[circ]{}
					(2.5,1) node[circ]{}
					(3,1) node[circ]{}
					(3.5,1) node[circ]{};
			\draw 	(1,1) node[cross]{}
					(2,1) node[cross]{}
					(2.5,1) node[cross]{}
					(3,1) node[cross]{}
					(3.5,1) node[cross]{};
		\end{tikzpicture}
		\caption{}
	\end{subfigure}
	\hfill
	\begin{subfigure}[t]{.49\textwidth}
		\centering
		\begin{tikzpicture}
			\draw 	(1,1) node[cross]{}
					(2,1) node[cross]{}
					(2.5,1) node[cross]{}
					(3,1) node[cross]{}
					(3.5,1) node[cross]{};
			\draw 	(1.5,0)--(1.5,2)
					(4,0)--(4,2)
					(4.5,0)--(4.5,2)
					(5,0)--(5,2)
					(5.5,0)--(5.5,2);
			\draw	[dotted](1,1)--(2,1)
					(2.5,1)--(3,1)
					(3.5,1)--(6,1);
			\draw	[dotted](2,1)--(2.5,1)
					(3,1)--(3.5,1);
			\draw	(1,1)--(1.5,1)
					(4,1)--(4.5,1)
					(5,1)--(5.5,1);
			\draw	(1.5,1.7)--(4,1.7)
					(1.5,.3)--(4,.3)
					(1.5,1.3)--(4,1.3)
					(1.5,.7)--(4,.7)
					(4,1.5)--(4.5,1.5)
					(4,.5)--(4.5,.5)
					(4.5,1.3)--(5,1.3)
					(4.5,.7)--(5,.7);
		\end{tikzpicture}
		\caption{}
	\end{subfigure}
	\hfill
	\begin{subfigure}[t]{.49\textwidth}
		\centering
		\begin{tikzpicture}
			\draw 	(1.5,1) node[cross]{}
					(2,1) node[cross]{}
					(2.5,1) node[cross]{}
					(3,1) node[cross]{}
					(3.5,1) node[cross]{};
			\draw 	(1,0)--(1,2)
					(4,0)--(4,2)
					(4.5,0)--(4.5,2)
					(5,0)--(5,2)
					(5.5,0)--(5.5,2);
			\draw	[dotted](1,1)--(2,1)
					(2.5,1)--(3,1)
					(3.5,1)--(6,1);
			\draw	[dotted](2,1)--(2.5,1)
					(3,1)--(3.5,1);
			\draw	(4,1)--(4.5,1)
					(5,1)--(5.5,1);
			\draw	(1,1.7)--(4,1.7)
					(1,.3)--(4,.3)
					(1,1.3)--(4,1.3)
					(1,.7)--(4,.7)
					(4,1.5)--(4.5,1.5)
					(4,.5)--(4.5,.5)
					(4.5,1.3)--(5,1.3)
					(4.5,.7)--(5,.7);
		\end{tikzpicture}
		\caption{}
	\end{subfigure}
	\hfill
	\begin{subfigure}[t]{.49\textwidth}
	\centering
		\begin{tikzpicture}[]
			\tikzstyle{gauge} = [circle, draw];
			\tikzstyle{flavour} = [regular polygon,regular polygon sides=4,draw];
			\node (g1) [gauge, label=below:{$C_2$}]{};
			\node (g2) [gauge, right of=g1, label=below:{$O_3$}]{};
			\node (g3) [gauge,right of=g2, label=below:{$C_1$}]{};
			\node (g4) [gauge,right of=g3, label=below:{$O_1$}]{};
			\node (f1) [flavour,above of=g1,label=above:{$O_5$}] {};
			\draw (f1)--(g1)--(g2)--(g3)--(g4);
		\end{tikzpicture}
		\caption{}
	\end{subfigure}
	\hfill
	\caption{S-duality transformation of the maximal model of $\mathfrak{sp}(2)$. The parameters of the initial model are $n_s=n_d=5$, $\vec{l}_d=(5,3,5,3,5)$, $\vec{l}_d=(1,1,1,1,1)$ and rightmost O3-plane $\widetilde{O3^+}$. (a) Higgs branch of the initial model. (b) Resulting model after performing S-duality on (a). (c) Resulting model after performing the collapse transition on (b). (d) Quiver obtained from (c). The Coulomb branch of the resulting model (d) is predicted to be the closure of the maximal nilpotent orbit of $\mathfrak{sp}(2)$, corresponding to partition $\lambda=(4)$.}
	\label{fig:Sp4maxMirror}
\end{figure}

\afterpage{\clearpage}

\subsubsection{Maximal orbit of $\gsp (2)$}
Let us now focus on $\lambda=(4)$, corresponding to the maximal nilpotent orbit of $\gsp (2)$. Let us construct the model with:
\begin{align}
	\M_H=\Or_{(4)}
\end{align}
Following prescription in section \ref{sec:sigmarho} the parameters of the brane model are:

\begin{align}
	\begin{aligned}
		n_s&=5\\
		n_d&=5\\
		\vec{l}_d&=(5,3,5,3,5)\\
		\vec{l}_s&=(1,1,1,1,1)\\
		\text{rightmost}\ O3&=\widetilde{O3^+}
	\end{aligned}
\end{align}

Remember that:
\begin{align}
	\begin{aligned}
		\vec{l}_s=Odd(d_{BV}(\lambda))=Odd(d_{BV}(4))=Odd(1^5)=(1,1,1,1,1)
	\end{aligned}
\end{align}

Now one can draw the Coulomb branch brane configuration and the quiver for the model, figure \ref{fig:Sp4maximal}. One can also obtain the Higgs branch brane configuration. The transition step by step is depicted in figure \ref{fig:Sp4maxHiggs}. An S-duality transformation can be performed to obtain the candidate mirror quiver, figure \ref{fig:Sp4maxMirror}. The Coulomb branch of this quiver has been predicted to be $\M_C=\Or_{(4)}\subset \gsp (2)$, \cite{GW09} but no Hilbert series has been computed yet utilizing the \emph{monopole formula} \cite{CHZ13}, and so the question on whether the gauge nodes have to be chosen $O_n$ or $SO_n$ is yet unanswered.

\subsection{Kraft-Procesi transitions for the nilpotent orbits of $\gso (5)$}\label{sec:9}

This and the next are the last two sections with examples, before giving a general description of all possible Kraft-Procesi transitions that can take place in B, C and D type brane configurations with O3-planes. So far, we have discussed examples of brane configurations whose Higgs branches are closures of nilpotent orbits of $\mathfrak{so}({2n})$, $\mathfrak{so}({2n+1})$ and $\mathfrak{sp}(n)$. We have also discussed the KP transitions that take place among the brane configurations related to nilpotent orbits of $\mathfrak{so}(4)$. In the following paragraphs we introduce the KP transitions that take place in $\gso (5)$. In the next section we do the same for $\gsp (2)$.

\subsubsection{$A_3$ transition}

Let us start with the model from section \ref{sec:maxO5} whose Higgs branch is $\M_H=\Or_{(5)}$, i.e. the closure of the maximal nilpotent orbit of $\gso (5)$. Its Higgs branch is depicted again in figure \ref{fig:SO5maxKP}(a). In this figure one can see that there is only a single way of performing a Kraft-Procesi transition, by \emph{Higgsing away} a D3-brane from the leftmost interval between half D5-branes. Such transition is depicted step by step in figures \ref{fig:SO5maxKP}(b-e), and the resulting model can be found in figure \ref{fig:SO5maxKP}(f). The two questions that need answering are: \emph{What is the moduli space generated by the remaining D3-branes after the transition?} and \emph{What is the moduli space generated by the physical D3-brane that has been removed?}. The answer to the first question determines a subvariety $O\subset\Or_{(5)}$ which is an orbit under the adjoint action of $O(5)$. The answer to the second question determines the slice $S\subset \Or_{(5)}$ which is transverse to $O$.

\begin{figure}[t]
	\centering
    \begin{subfigure}[t]{.49\textwidth}
    \centering
	\begin{tikzpicture}
		\draw[dashed] 	(1,0)--(1,2)
				(4,0)--(4,2)
				(4.5,0)--(4.5,2)
				(5,0)--(5,2)
				(5.5,0)--(5.5,2);
		\draw [dotted] (.5,1)--(1.5,1)
				(2,1)--(2.5,1)
				(3,1)--(3.5,1);
		\draw	(4,1)--(4.5,1)
				(5,1)--(5.5,1);
		\draw	(1,1.5)--(4,1.5)
				(4,1.6)--(4.5,1.6)
				(4.5,1.5)--(5,1.5)
				(1,.5)--(4,.5)
				(4,.4)--(4.5,.4)
				(4.5,.5)--(5,.5)
				(1,1.8)--(4,1.8)
				(1,.2)--(4,.2);
		\draw 	(1.5,1) node[circ]{}
				(2,1)node[circ]{}
				(2.5,1)node[circ]{}
				(3,1)node[circ]{}
				(3.5,1)node[circ]{};
		\draw 	(1.5,1) node[cross]{}
				(2,1)node[cross]{}
				(2.5,1)node[cross]{}
				(3,1)node[cross]{}
				(3.5,1)node[cross]{};
		\end{tikzpicture}
        \caption{}
    \end{subfigure}
    \hfill
    \begin{subfigure}[t]{.49\textwidth}
   		\centering
		\begin{tikzpicture}
		\draw[dashed] 	(1,0)--(1,2)
				(4,0)--(4,2)
				(4.5,0)--(4.5,2)
				(5,0)--(5,2)
				(5.5,0)--(5.5,2);
		\draw [dotted] (.5,1)--(1.5,1)
				(2,1)--(2.5,1)
				(3,1)--(3.5,1);
		\draw	(4,1)--(4.5,1)
				(5,1)--(5.5,1);
		\draw	(1,1.05)--(4,1.05)
				(4,1.6)--(4.5,1.6)
				(4.5,1.5)--(5,1.5)
				(1,.95)--(4,.95)
				(4,.4)--(4.5,.4)
				(4.5,.5)--(5,.5)
				(1,1.8)--(4,1.8)
				(1,.2)--(4,.2);
		\draw 	(1.5,1) node[circ]{}
				(2,1)node[circ]{}
				(2.5,1)node[circ]{}
				(3,1)node[circ]{}
				(3.5,1)node[circ]{};
		\draw 	(1.5,1) node[cross]{}
				(2,1)node[cross]{}
				(2.5,1)node[cross]{}
				(3,1)node[cross]{}
				(3.5,1)node[cross]{};
		\end{tikzpicture}
        \caption{}
    \end{subfigure}
    \hfill
    \begin{subfigure}[t]{.49\textwidth}
   		\centering
		\begin{tikzpicture}
		\draw[dashed] 	(1.5,0)--(1.5,2)
				(4,0)--(4,2)
				(4.5,0)--(4.5,2)
				(5,0)--(5,2)
				(5.5,0)--(5.5,2);
		\draw [dotted] (.5,1)--(1,1)
				(2,1)--(2.5,1)
				(3,1)--(3.5,1);
		\draw	(1,1)--(1.5,1)
				(4,1)--(4.5,1)
				(5,1)--(5.5,1);
		\draw	(1.5,1.05)--(4,1.05)
				(4,1.6)--(4.5,1.6)
				(4.5,1.5)--(5,1.5)
				(1.5,.95)--(4,.95)
				(4,.4)--(4.5,.4)
				(4.5,.5)--(5,.5)
				(1.5,1.8)--(4,1.8)
				(1.5,.2)--(4,.2);
		\draw 	(1,1) node[circ]{}
				(2,1)node[circ]{}
				(2.5,1)node[circ]{}
				(3,1)node[circ]{}
				(3.5,1)node[circ]{};
		\draw 	(1,1) node[cross]{}
				(2,1)node[cross]{}
				(2.5,1)node[cross]{}
				(3,1)node[cross]{}
				(3.5,1)node[cross]{};
		\end{tikzpicture}
        \caption{}
    \end{subfigure}
    \hfill
    \begin{subfigure}[t]{.49\textwidth}
   		\centering
		\begin{tikzpicture}
		\draw[dashed] 	(1.5,0)--(1.5,2)
				(4,0)--(4,2)
				(4.5,0)--(4.5,2)
				(5,0)--(5,2)
				(5.5,0)--(5.5,2);
		\draw [dotted] (.5,1)--(1,1)
				(2,1)--(2.5,1)
				(3,1)--(3.5,1);
		\draw	(1,1)--(1.5,1)
				(4,1)--(4.5,1)
				(5,1)--(5.5,1);
		\draw	(1.5,1.05)--(1.95,1.05)
				(3.55,1.05)--(4,1.05)
				(4,1.6)--(4.5,1.6)
				(4.5,1.5)--(5,1.5)
				(1.5,.95)--(1.95,.95)
				(3.55,.95)--(4,.95)
				(4,.4)--(4.5,.4)
				(4.5,.5)--(5,.5)
				(1.5,1.8)--(4,1.8)
				(1.5,.2)--(4,.2);
		\draw 	(1,1) node[circ]{}
				(2,1)node[circ]{}
				(2.5,1)node[circ]{}
				(3,1)node[circ]{}
				(3.5,1)node[circ]{};
		\draw 	(1,1) node[cross]{}
				(2,1)node[cross]{}
				(2.5,1)node[cross]{}
				(3,1)node[cross]{}
				(3.5,1)node[cross]{};
		\end{tikzpicture}
        \caption{}
    \end{subfigure}
    \hfill
    \begin{subfigure}[t]{.49\textwidth}
   		\centering
		\begin{tikzpicture}
		\draw[dashed] 	(2,0)--(2,2)
				(3.5,0)--(3.5,2)
				(4.5,0)--(4.5,2)
				(5,0)--(5,2)
				(5.5,0)--(5.5,2);
		\draw [dotted] (.5,1)--(1,1)
				(1.5,1)--(2.5,1)
				(3,1)--(4,1);
		\draw	(1,1)--(1.5,1)
				(4,1)--(4.5,1)
				(5,1)--(5.5,1);
		\draw	(3.5,1.6)--(4.5,1.6)
				(4.5,1.5)--(5,1.5)
				(3.5,.4)--(4.5,.4)
				(4.5,.5)--(5,.5)
				(2,1.8)--(3.5,1.8)
				(2,.2)--(3.5,.2);
		\draw 	(1,1) node[circ]{}
				(1.5,1)node[circ]{}
				(2.5,1)node[circ]{}
				(3,1)node[circ]{}
				(4,1)node[circ]{};
		\draw 	(1,1) node[cross]{}
				(1.5,1)node[cross]{}
				(2.5,1)node[cross]{}
				(3,1)node[cross]{}
				(4,1)node[cross]{};
		\end{tikzpicture}
        \caption{}
    \end{subfigure}
    \hfill
    \begin{subfigure}[t]{.49\textwidth}
   		\centering
		\begin{tikzpicture}
		\draw[dashed] 	(2,0)--(2,2)
				(3.5,0)--(3.5,2)
				(4,0)--(4,2)
				(5,0)--(5,2)
				(5.5,0)--(5.5,2);
		\draw [dotted] (.5,1)--(1,1)
				(1.5,1)--(2.5,1)
				(3,1)--(4.5,1);
		\draw	(1,1)--(1.5,1)
				(5,1)--(5.5,1);
		\draw	(3.5,1.6)--(4,1.6)
				(4,1.5)--(5,1.5)
				(3.5,.4)--(4,.4)
				(4,.5)--(5,.5)
				(2,1.8)--(3.5,1.8)
				(2,.2)--(3.5,.2);
		\draw 	(1,1) node[circ]{}
				(1.5,1)node[circ]{}
				(2.5,1)node[circ]{}
				(3,1)node[circ]{}
				(4.5,1)node[circ]{};
		\draw 	(1,1) node[cross]{}
				(1.5,1)node[cross]{}
				(2.5,1)node[cross]{}
				(3,1)node[cross]{}
				(4.5,1)node[cross]{};
		\end{tikzpicture}
        \caption{}
    \end{subfigure}
    \hfill
 	\caption{$A_3$ Kraft-Procesi transition.  The starting model has $n_s=n_d=5$, $\vec{l}_d=(4,4,4,4,4)$, $\vec{l}_s=(0,2,0,2,0)$ and rightmost O3-plane $O3^-$. (a) Higgs branch of the initial model. (b) We focus on one of the leftmost physical D3-branes, considering all the rest as spectators. We align this brane with the half NS5-branes at the orientifold plane. Some vector multiplets  in the effective gauge theory become massless at this position. The D3-brane is split into six different segments, four of them only end on half NS5-branes. These four different segments can have their position $\vec x_i$ along the half NS5-branes to be non-zero. In order to perform the Kraft-Procesi transition one would like to take them all to infinity, effectively removing them from the system. However, if the leftmost of these segments, with position $\vec x_1$, is removed, the resulting model is no longer in a sypersymmetric configuration (see details of this phenomenon in section \ref{sec:311transition}). (c) In order to avoid this, the leftmost half NS5-brane can be pushed through its neighboring half D5-brane first. This transition fixes the $\vec x_1$ position of the D3-brane segment to zero, since it now ends on two half fivebranes of different type. (d) The KP transition takes place. The three remaining D3-brane segments are taken to infinity along their $\vec x_i$ directions, with $i=2,3,4$. (e) The fixed D3-brane segments are annihilated via phase transitions. (f) The \emph{collapse} transition is performed. Now the interval numbers of the half NS5-branes are $\vec{k}_s=(0,0,1,1,3)$, and the Higgs branch of the resulting model is identified as the closure of the nilpotent orbit of $\gso (5)$ corresponding to partition $\lambda=(3,1^2)^t=(3,1^2)$.}
	\label{fig:SO5maxKP}
\end{figure}
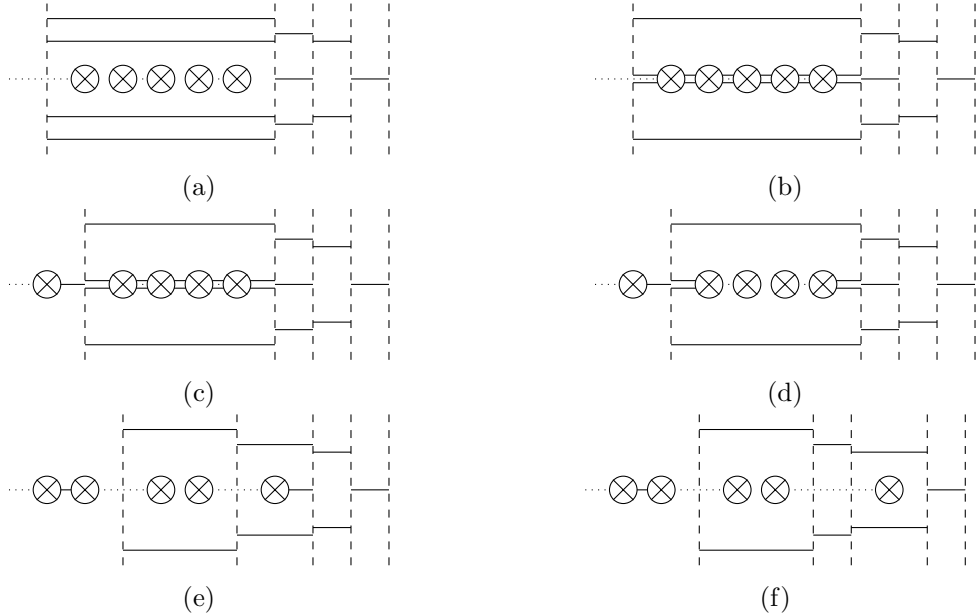

\begin{figure}[t]
	\centering
    \begin{subfigure}[t]{.49\textwidth}
   		\centering
		\begin{tikzpicture}
		\draw[dashed] 	(2,0)--(2,2)
				(3.5,0)--(3.5,2)
				(4,0)--(4,2)
				(5,0)--(5,2)
				(5.5,0)--(5.5,2);
		\draw [dotted] (.5,1)--(1,1)
				(1.5,1)--(2.5,1)
				(3,1)--(4.5,1);
		\draw	(1,1)--(1.5,1)
				(5,1)--(5.5,1);
		\draw	(3.5,1.6)--(4,1.6)
				(4,1.5)--(5,1.5)
				(3.5,.4)--(4,.4)
				(4,.5)--(5,.5)
				(2,1.8)--(3.5,1.8)
				(2,.2)--(3.5,.2);
		\draw 	(1,1) node[circ]{}
				(1.5,1)node[circ]{}
				(2.5,1)node[circ]{}
				(3,1)node[circ]{}
				(4.5,1)node[circ]{};
		\draw 	(1,1) node[cross]{}
				(1.5,1)node[cross]{}
				(2.5,1)node[cross]{}
				(3,1)node[cross]{}
				(4.5,1)node[cross]{};
		\end{tikzpicture}
        \caption{}
    \end{subfigure}
    \hfill
    \begin{subfigure}[t]{.49\textwidth}
   		\centering
		\begin{tikzpicture}
		\draw[dashed] 	(2.5,0)--(2.5,2)
				(3.5,0)--(3.5,2)
				(4,0)--(4,2)
				(5,0)--(5,2)
				(5.5,0)--(5.5,2);
		\draw [dotted] (.5,1)--(1,1)
				(1.5,1)--(2,1)
				(3,1)--(4.5,1);
		\draw	(1,1)--(1.5,1)
				(2,1)--(2.5,1)
				(5,1)--(5.5,1);
		\draw	(3.5,1.6)--(4,1.6)
				(4,1.5)--(5,1.5)
				(3.5,.4)--(4,.4)
				(4,.5)--(5,.5)
				(2.5,1.8)--(3.5,1.8)
				(2.5,.2)--(3.5,.2);
		\draw 	(1,1) node[circ]{}
				(1.5,1)node[circ]{}
				(2,1)node[circ]{}
				(3,1)node[circ]{}
				(4.5,1)node[circ]{};
		\draw 	(1,1) node[cross]{}
				(1.5,1)node[cross]{}
				(2,1)node[cross]{}
				(3,1)node[cross]{}
				(4.5,1)node[cross]{};
		\end{tikzpicture}
        \caption{}
    \end{subfigure}
    \hfill
    \begin{subfigure}[t]{.49\textwidth}
   		\centering
		\begin{tikzpicture}
		\draw[dashed] 	(3,0)--(3,2)
				(3.5,0)--(3.5,2)
				(4,0)--(4,2)
				(4.5,0)--(4.5,2)
				(5.5,0)--(5.5,2);
		\draw [dotted] (.5,1)--(1,1)
				(1.5,1)--(2,1)
				(2.5,1)--(5,1);
		\draw	(1,1)--(1.5,1)
				(2,1)--(2.5,1)
				(5,1)--(5.5,1);
		\draw	(3.5,1.6)--(4,1.6)
				(4,1.5)--(4.5,1.5)
				(3.5,.4)--(4,.4)
				(4,.5)--(4.5,.5)
				(3,1.8)--(3.5,1.8)
				(3,.2)--(3.5,.2)
				(2.55,1.05)--(3,1.05)
				(2.55,.95)--(3,.95)
				(4.5,1.05)--(4.95,1.05)
				(4.5,.95)--(4.95,.95);
		\draw 	(1,1) node[circ]{}
				(1.5,1)node[circ]{}
				(2,1)node[circ]{}
				(2.5,1)node[circ]{}
				(5,1)node[circ]{};
		\draw 	(1,1) node[cross]{}
				(1.5,1)node[cross]{}
				(2,1)node[cross]{}
				(2.5,1)node[cross]{}
				(5,1)node[cross]{};
		\end{tikzpicture}
        \caption{}
    \end{subfigure}
    \hfill
    \begin{subfigure}[t]{.49\textwidth}
   		\centering
		\begin{tikzpicture}
		\draw[dashed] 	(3,0)--(3,2)
				(3.5,0)--(3.5,2)
				(4,0)--(4,2)
				(4.5,0)--(4.5,2)
				(5.5,0)--(5.5,2);
		\draw [dotted] (.5,1)--(1,1)
				(1.5,1)--(2,1)
				(2.5,1)--(5,1);
		\draw	(1,1)--(1.5,1)
				(2,1)--(2.5,1)
				(5,1)--(5.5,1);
		\draw	(2.55,1.05)--(4.95,1.05)
				(2.55,.95)--(4.95,.95);
		\draw 	(1,1) node[circ]{}
				(1.5,1)node[circ]{}
				(2,1)node[circ]{}
				(2.5,1)node[circ]{}
				(5,1)node[circ]{};
		\draw 	(1,1) node[cross]{}
				(1.5,1)node[cross]{}
				(2,1)node[cross]{}
				(2.5,1)node[cross]{}
				(5,1)node[cross]{};
		\end{tikzpicture}
        \caption{}
    \end{subfigure}
    \hfill
    \begin{subfigure}[t]{.49\textwidth}
   		\centering
		\begin{tikzpicture}
			\draw 	(1,0)--(1,2)
					(1.5,0)--(1.5,2)
					(2,0)--(2,2)
					(2.5,0)--(2.5,2)
					(5,0)--(5,2);
			\draw 	(3,1)node[cross]{}
				(3.5,1)node[cross]{}
				(4,1)node[cross]{}
				(4.5,1)node[cross]{}
				(5.5,1)node[cross]{};
			\draw [dotted] (.5,1)--(1,1)
					(1.5,1)--(2,1)
					(2.5,1)--(5,1);
			\draw	(1,1)--(1.5,1)
					(2,1)--(2.5,1)
					(5,1)--(5.5,1);
			\draw	(2.5,1.05)--(5,1.05)
					(2.5,.95)--(5,.95);
		\end{tikzpicture}
        \caption{}
    \end{subfigure}
    \hfill
    \begin{subfigure}[t]{.49\textwidth}
   		\centering
		\begin{tikzpicture}
			\draw 	(1,0)--(1,2)
					(1.5,0)--(1.5,2)
					(2,0)--(2,2)
					(2.5,0)--(2.5,2)
					(5,0)--(5,2);
			\draw 	(3,1)node[cross]{}
				(3.5,1)node[cross]{}
				(4,1)node[cross]{}
				(4.5,1)node[cross]{}
				(5.5,1)node[cross]{};
			\draw [dotted] (.5,1)--(1,1)
					(1.5,1)--(2,1)
					(2.5,1)--(5,1);
			\draw	(1,1)--(1.5,1)
					(2,1)--(2.5,1)
					(5,1)--(5.5,1);
			\draw	(2.5,1.5)--(5,1.5)
					(2.5,.5)--(5,.5);
		\end{tikzpicture}
        \caption{}
    \end{subfigure}
    \hfill
    \begin{subfigure}[t]{.49\textwidth}
   		\centering
		\begin{tikzpicture}
			\draw 	(1,0)--(1,2)
					(1.5,0)--(1.5,2)
					(2,0)--(2,2)
					(2.5,0)--(2.5,2)
					(5.5,0)--(5.5,2);
			\draw 	(3,1)node[cross]{}
				(3.5,1)node[cross]{}
				(4,1)node[cross]{}
				(4.5,1)node[cross]{}
				(5,1)node[cross]{};
			\draw [dotted] (.5,1)--(1,1)
					(1.5,1)--(2,1)
					(2.5,1)--(5.5,1);
			\draw	(1,1)--(1.5,1)
					(2,1)--(2.5,1);
			\draw	(2.5,1.5)--(5.5,1.5)
					(2.5,.5)--(5.5,.5);
		\end{tikzpicture}
        \caption{}
    \end{subfigure}
    \hfill
	\begin{subfigure}[t]{.49\textwidth}
    \centering
	\begin{tikzpicture}[]
	\tikzstyle{gauge} = [circle, draw];
	\tikzstyle{flavour} = [regular polygon,regular polygon sides=4,draw];
	\node (g3) [gauge,right of=g2,label=below:{$O_1$}] {};
	\node (g4) [gauge,right of=g3,label=below:{$C_1$}] {};
	\node (f4) [flavour,above of=g4,label=above:{$O_5$}] {};
	\draw(g3)--(g4)--(f4)
		;
	\end{tikzpicture}
        \caption{}
    \end{subfigure}
 	\caption{Phase transition to the Coulomb branch brane configuration. In order to read the quiver for the model resulting after the $A_3$ KP transition in figure \ref{fig:SO5maxKP}(f). (a) Higgs branch of the model. (b-e) Phase transitions. (f) Coulomb branch of the model before the \emph{collapse} transition. (g) Coulomb branch of the model after the \emph{collapse} transition. (h) Quiver, it can be identified with the model whose Higgs branch is the closure of the next to minimal nilpotent orbit of $\gso (5)$, correspondig to partition $\lambda=(3,1^2)$: $\M_H=\Or_{(3,1^2)}$.}
	\label{fig:SO5subreg}
\end{figure}
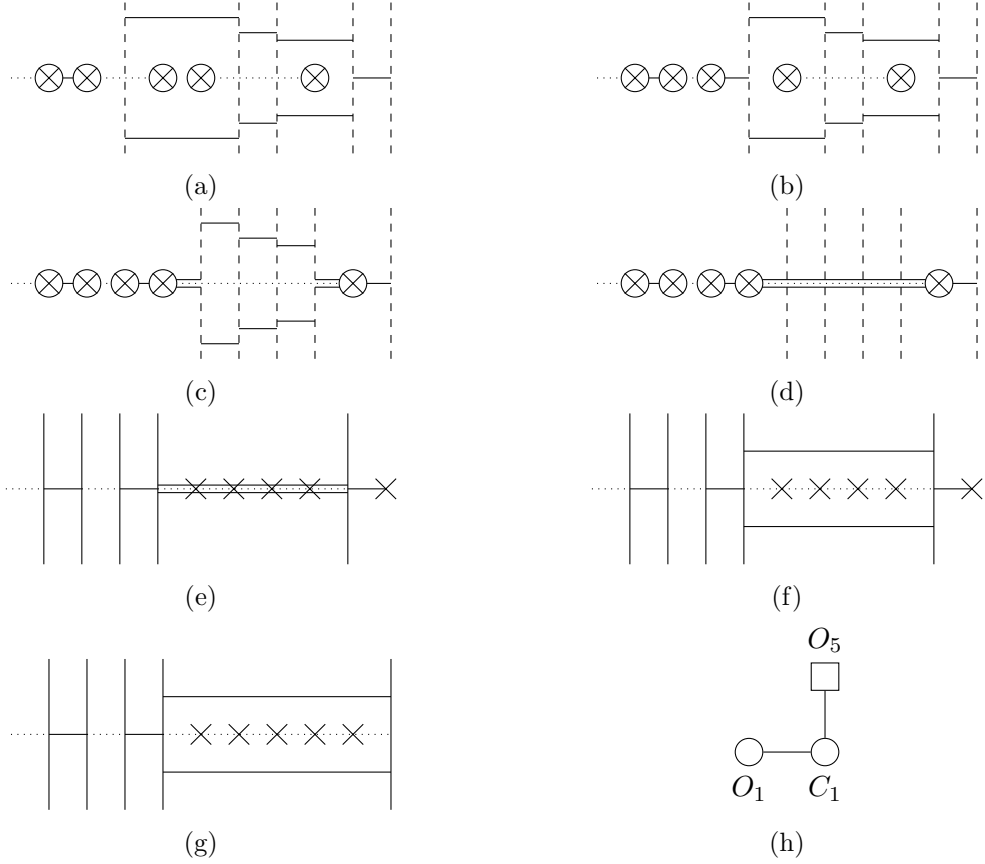

To answer the first question one realizes that the resulting model in figure \ref{fig:SO5maxKP}(f) is in a Higgs branch phase. A phase transition to the Coulomb branch phase is performed in order to read the corresponding quiver. This transition is depicted step by step in figure \ref{fig:SO5subreg}. Note that even before computing the Coulomb branch brane configuration in figure \ref{fig:SO5subreg} one can establish which are the parameters of the new model:

\begin{align}
	\begin{aligned}
		n_s&=5\\
		n_d&=5\\
		\vec{l}_d&=(4,4,4,4,4)\\
		\vec{l}_s&=(0,0,0,2,2)\\
		\text{rightmost}\ O3&= O3^- 
	\end{aligned}
\end{align}

The linking numbers of the half NS5-branes can be expressed as $\vec{l}_s=Even(2^2)$ and $(2^2)=d_{BV}(3,1^2)$. Therefore, according with section \ref{sec:sigmarho} this would be the model with:

\begin{align}
	\M_H=\Or_{(3,1^2)}\subset \gso (5)
\end{align}

Note that actually we have not defined an inverse map of $Even()$. This is not necessary and instead one can exploit the position of the half NS5-branes as is described by their \emph{interval numbers}. This is indeed the first example in which the \emph{interval numbers} of the half NS5-branes become truly useful. Once more, let a set of interval numbers $\vec k _s$ define a partition $\lambda$ such that the parts of $\lambda^t$ are the elements of $\vec k_s$. In the present case figure \ref{fig:SO5maxKP}(f) represents the Higgs branch brane configuration after the collapse transition, the interval numbers of the half NS5-branes are:
\begin{align}
	\vec{k}_s=(0,0,1,1,3)\rightarrow \lambda=(3,1^2)
\end{align}
Since $\lambda^t=(3,1^2)$, which univocally determines $\lambda=(3,1^2)$. 

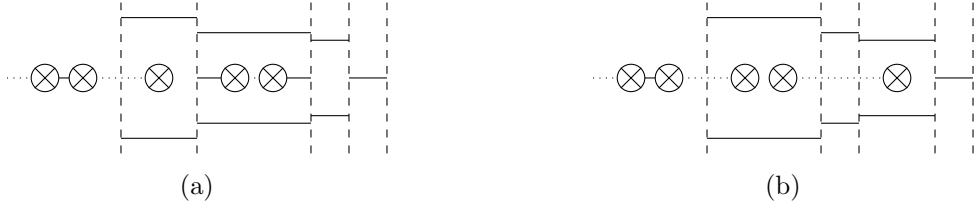
\begin{figure}[t]
	\centering
    \begin{subfigure}[t]{.49\textwidth}
   		\centering
		\begin{tikzpicture}
		\draw[dashed] 	(2,0)--(2,2)
				(3,0)--(3,2)
				(4.5,0)--(4.5,2)
				(5,0)--(5,2)
				(5.5,0)--(5.5,2);
		\draw [dotted] (.5,1)--(1,1)
				(1.5,1)--(2.5,1)
				(3.5,1)--(4,1);
		\draw	(1,1)--(1.5,1)
				(3,1)--(3.5,1)
				(4,1)--(4.5,1)
				(5,1)--(5.5,1);
		\draw	(3,1.6)--(4.5,1.6)
				(4.5,1.5)--(5,1.5)
				(3,.4)--(4.5,.4)
				(4.5,.5)--(5,.5)
				(2,1.8)--(3,1.8)
				(2,.2)--(3,.2);
		\draw 	(1,1) node[circ]{}
				(1.5,1)node[circ]{}
				(2.5,1)node[circ]{}
				(3.5,1)node[circ]{}
				(4,1)node[circ]{};
		\draw 	(1,1) node[cross]{}
				(1.5,1)node[cross]{}
				(2.5,1)node[cross]{}
				(3.5,1)node[cross]{}
				(4,1)node[cross]{};
		\end{tikzpicture}
        \caption{}
    \end{subfigure}
    \hfill
    \begin{subfigure}[t]{.49\textwidth}
   		\centering
		\begin{tikzpicture}
		\draw[dashed] 	(2,0)--(2,2)
				(3.5,0)--(3.5,2)
				(4,0)--(4,2)
				(5,0)--(5,2)
				(5.5,0)--(5.5,2);
		\draw [dotted] (.5,1)--(1,1)
				(1.5,1)--(2.5,1)
				(3,1)--(4.5,1);
		\draw	(1,1)--(1.5,1)
				(5,1)--(5.5,1);
		\draw	(3.5,1.6)--(4,1.6)
				(4,1.5)--(5,1.5)
				(3.5,.4)--(4,.4)
				(4,.5)--(5,.5)
				(2,1.8)--(3.5,1.8)
				(2,.2)--(3.5,.2);
		\draw 	(1,1) node[circ]{}
				(1.5,1)node[circ]{}
				(2.5,1)node[circ]{}
				(3,1)node[circ]{}
				(4.5,1)node[circ]{};
		\draw 	(1,1) node[cross]{}
				(1.5,1)node[cross]{}
				(2.5,1)node[cross]{}
				(3,1)node[cross]{}
				(4.5,1)node[cross]{};
		\end{tikzpicture}
        \caption{}
    \end{subfigure}
 	\caption{Brane realization of the B-collapse of partitions $(3,2)_B=(3,1^2)$. (a) Has interval numbers for the half NS5-branes $\vec k'=(0,0,1,2,2)$ corresponding to partition $\lambda'=(3,2)$. (b) Is obtained after performing the \emph{collapse} transition on (a). The interval numbers of (b) are $\vec k=(0,0,1,1,3)$, corresponding to partition $\lambda=(3,1^2)=\lambda'_B$.}
	\label{fig:Bcollapse}
\end{figure}

\paragraph{Brane realization of the B-collapse.} Analogous to the discussion in section \ref{sec:branecollapse}, we can see how the \emph{collapse} transition in the branes realizes the B-collapse of different partitions. In this case, the B-collapse of $\lambda'=(3,2)$ into $\lambda=(3,1^2)$. Starting with the brane system in figure \ref{fig:SO5maxKP}(f), the inverse transition of the collapse transition (i.e., the phase transition that pulls all possible half NS5-branes away from $O3^-$ planes without any brane creation/annihilation) rearranges the position of the half NS5-branes, so they acquire interval numbers $\vec k''=(0,0,0,2,2)$. However, $\vec k''$ cannot produce a partition of $\mathcal P(5)$, since the sum of all its elements is 4. Hence, the rightmost half NS5-brane with interval number $0$ must be pushed through its neighboring half D5-brane, in order for its interval number to be $1$. This phase transition does not involve any brane creation/annihilation. This results in the brane system in figure \ref{fig:Bcollapse}(a) with interval numbers: 
\begin{align}
	\vec{k}'_s=(0,0,1,2,2)\rightarrow \lambda'=(3,2)
\end{align}
The \emph{collapse} transition on brane system in figure \ref{fig:Bcollapse}(a) gives brane system on figure \ref{fig:Bcollapse}(b), mapping partition $\lambda'=(3,2)$ to $\lambda'_B=\lambda=(3,1^2)$. Hence, giving a physical realization to the B-collapse between partitions. This result is general and can be applied to all the brane systems of B-type in the present work.

\begin{figure}[t]
	\centering
    \begin{subfigure}[t]{.49\textwidth}
   		\centering
		\begin{tikzpicture}
		\draw[dashed] 	(2,0)--(2,2)
				(3.5,0)--(3.5,2)
				(4,0)--(4,2)
				(5,0)--(5,2)
				(5.5,0)--(5.5,2);
		\draw [dotted] (.5,1)--(1,1)
				(1.5,1)--(2.5,1)
				(3,1)--(4.5,1);
		\draw	(1,1)--(1.5,1)
				(5,1)--(5.5,1);
		\draw	(3.5,1.6)--(4,1.6)
				(4,1.5)--(5,1.5)
				(3.5,.4)--(4,.4)
				(4,.5)--(5,.5)
				(2,1.8)--(3.5,1.8)
				(2,.2)--(3.5,.2);
		\draw 	(1,1) node[circ]{}
				(1.5,1)node[circ]{}
				(2.5,1)node[circ]{}
				(3,1)node[circ]{}
				(4.5,1)node[circ]{};
		\draw 	(1,1) node[cross]{}
				(1.5,1)node[cross]{}
				(2.5,1)node[cross]{}
				(3,1)node[cross]{}
				(4.5,1)node[cross]{};
		\end{tikzpicture}
        \caption{}
    \end{subfigure}
    \hfill
    \begin{subfigure}[t]{.49\textwidth}
   		\centering
		\begin{tikzpicture}
		\draw 	(1,1) node[cross]{}
				(1.5,1)node[cross]{}
				(2.5,1)node[cross]{}
				(3,1)node[cross]{}
				(4.5,1)node[cross]{};
		\draw	(2,0)--(2,2)
				(3.5,0)--(3.5,2)
				(4,0)--(4,2)
				(5,0)--(5,2)
				(5.5,0)--(5.5,2);
		\draw [dotted] (.5,1)--(2,1)
				(3.5,1)--(4,1)
				(5,1)--(5.5,1);
		\draw	(2,1)--(2.5,1)
				(3,1)--(3.5,1)
				(4,1)--(4.5,1);
		\draw	(3.5,1.6)--(4,1.6)
				(4,1.5)--(5,1.5)
				(3.5,.4)--(4,.4)
				(4,.5)--(5,.5)
				(2,1.8)--(3.5,1.8)
				(2,.2)--(3.5,.2);
		\end{tikzpicture}
        \caption{}
    \end{subfigure}
    \hfill
    \begin{subfigure}[t]{.49\textwidth}
   		\centering
		\begin{tikzpicture}
		\draw 	(1,1) node[cross]{}
				(1.5,1)node[cross]{}
				(2,1)node[cross]{}
				(3.5,1)node[cross]{}
				(4,1)node[cross]{};
		\draw	(2.5,0)--(2.5,2)
				(3,0)--(3,2)
				(4.5,0)--(4.5,2)
				(5,0)--(5,2)
				(5.5,0)--(5.5,2);
		\draw [dotted] (.5,1)--(2.5,1)
				(3,1)--(4.5,1)
				(5,1)--(5.5,1);
		\draw	(3,1.6)--(4.5,1.6)
				(4.5,1.5)--(5,1.5)
				(3,.4)--(4.5,.4)
				(4.5,.5)--(5,.5)
				(2.5,1.8)--(3,1.8)
				(2.5,.2)--(3,.2);
		\end{tikzpicture}
        \caption{}
    \end{subfigure}
    \hfill
	\begin{subfigure}[t]{.49\textwidth}
    \centering
	\begin{tikzpicture}[]
	\tikzstyle{gauge} = [circle, draw];
	\tikzstyle{flavour} = [regular polygon,regular polygon sides=4,draw];
	\node (g2) [gauge,label=below:{$O_2$}] {};
	\node (g3) [gauge,right of=g2,label=below:{$C_1$}] {};
	\node (g4) [gauge,right of=g3,label=below:{$SO_2$}] {};
	\node (f3) [flavour,above of=g3,label=above:{$O_2$}] {};
	\draw(g2)--(g3)--(g4)
			(f3)--(g3)
		;
	\end{tikzpicture}
        \caption{}
    \end{subfigure}
 	\caption{S-duality on the model for the next to minimal orbit of $\gso (5)$. The parameters of the initial model are $n_s=n_d=5$, $\vec{l}_d=(4,4,4,4,4)$, $\vec{l}_s=(0,0,0,2,2)$ and rightmost orientifold plane $O3^-$. (a) Higgs branch of the initial model. (b) Resulting model after performing S-duality on (a). (c) Resulting model after performing the collapse transition on (b). (d) Quiver obtained from (c), note that one of the gauge nodes has to be $SO_2$ instead of $O_2$ for the Coulomb branch of the resulting model to be $\M_C=\Or_{(3,1^2)}\subset \gso (5)$ \cite{CHMZ14,CHZ17}. This is related to the fact that the Lusztig's Canonical Quotient is $\bar A (\mathcal O_{(3,1^2)})=\mathbb Z_2$. }
	\label{fig:SO5subregMirror}
\end{figure}
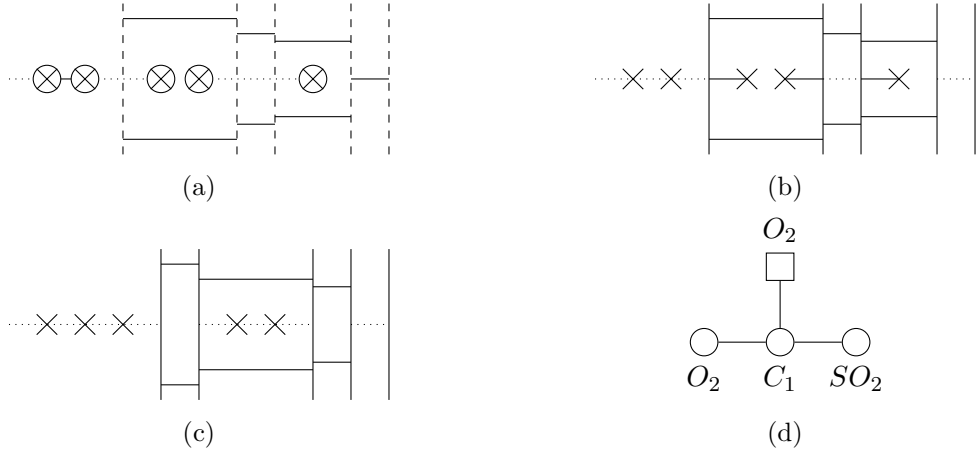

\paragraph{S-duality.} An S-duality transformation can be performed on the brane configuration in figure \ref{fig:SO5subreg}(a) in order to obtain a model with $\M_C=\Or_{(3,1^2)}\subset \gso (5)$. This is depicted in figure \ref{fig:SO5subregMirror}. 

\begin{figure}[t]
	\centering
    	\begin{subfigure}[t]{.49\textwidth}
    	\centering
	\begin{tikzpicture}	
		\draw[dashed] 	(1,0)--(1,2)
				(4,0)--(4,2);
		\draw [dotted] (.5,1)--(1.5,1)
				(2,1)--(2.5,1)
				(3,1)--(3.5,1);
		\draw	(4,1)--(4.5,1);
		\draw	(1,1.8)--(4,1.8)
				(1,.2)--(4,.2);
		\draw 	(1.5,1) node[circ]{}
				(2,1)node[circ]{}
				(2.5,1)node[circ]{}
				(3,1)node[circ]{}
				(3.5,1)node[circ]{};
		\draw 	(1.5,1) node[cross]{}
				(2,1)node[cross]{}
				(2.5,1)node[cross]{}
				(3,1)node[cross]{}
				(3.5,1)node[cross]{};
	\end{tikzpicture}
        \caption{}
    	\end{subfigure}
	\hfill
    	\begin{subfigure}[t]{.49\textwidth}
    	\centering
	\begin{tikzpicture}
			\draw[dashed] 	(2,0)--(2,2)
					(3.5,0)--(3.5,2);
			\draw [dotted] (.5,1)--(1,1)
					(1.5,1)--(2.5,1)
					(3,1)--(4,1);
			\draw	(1,1)--(1.5,1);
			\draw	(4,1)--(4.5,1);
			\draw	(1.5,1.05)--(2,1.05)
					(1.5,.95)--(2,.95);
			\draw	(3.5,1.05)--(4,1.05)
					(3.5,.95)--(4,.95);
			\draw	(2,1.8)--(3.5,1.8)
					(2,.2)--(3.5,.2);
			\draw 	(1,1) node[circ]{}
					(1.5,1)node[circ]{}
					(2.5,1)node[circ]{}
					(3,1)node[circ]{}
					(4,1)node[circ]{};
			\draw 	(1,1) node[cross]{}
					(1.5,1)node[cross]{}
					(2.5,1)node[cross]{}
					(3,1)node[cross]{}
					(4,1)node[cross]{};
	\end{tikzpicture}
        \caption{}
    	\end{subfigure}
	\hfill
    	\begin{subfigure}[t]{.49\textwidth}
    	\centering
	\begin{tikzpicture}
				\draw 	(1,0)--(1,2)
						(1.5,0)--(1.5,2)
						(2.5,0)--(2.5,2)
						(3,0)--(3,2)
						(4,0)--(4,2);
				\draw [dotted] (.5,1)--(1,1)
						(1.5,1)--(2.5,1)
						(3,1)--(4,1);
				\draw	(1,1)--(1.5,1);
				\draw	(4,1)--(4.5,1);
				\draw	(1.5,1.5)--(2.5,1.5)
						(1.5,.5)--(2.5,.5);
				\draw	(3,1.5)--(4,1.5)
						(3,.5)--(4,.5);
				\draw	(2.5,1.8)--(3,1.8)
						(2.5,.2)--(3,.2);
				\draw 	(2,1) node[cross]{}
						(3.5,1)node[cross]{};
	\end{tikzpicture}
        \caption{}
    	\end{subfigure}
	\hfill
	\begin{subfigure}[t]{.49\textwidth}
    \centering
	\begin{tikzpicture}[]
	\tikzstyle{gauge} = [circle,draw];
	\tikzstyle{flavour} = [regular polygon,regular polygon sides=4,draw];
	\node (g1) [gauge,label=below:{$O_1$}] {};
	\node (g2) [gauge,right of=g1,label=below:{$C_1$}] {};
	\node (g3) [gauge,right of=g2,label=below:{$O_2$}] {};
	\node (g4) [gauge,right of=g3,label=below:{$C_1$}] {};
	\node (f2) [flavour,above of=g2,label=above:{$O_1$}] {};
	\node (f4) [flavour,above of=g4,label=above:{$O_2$}] {};
	\draw (g1)--(g2)--(f2)
			(g2)--(g3)--(g4)--(f4)
		;
	\end{tikzpicture}
	\caption{}
    \end{subfigure}
 	\caption{Transverse slice $S=A_3$ that can be removed via Kraft-Procesi transition. (a) Local brane configuration. (b) Phase transition towards the Coulomb branch brane configuration, two D3-branes are created. (c) Phase transition that obtains the Coulomb branch brane configuration after aligning all D3-branes at the origin and then splitting them so they only end at half NS5-branes, a rotation that takes $\vec x$ to $\vec y$ and $\vec y$ to $-\vec x$ has been performed. (d) Quiver from (c), the Higgs branch of this quiver is $\M_H=A_3$, see the computation in appendix \ref{app:AnHiggs}.}
	\label{fig:SO5maxInstNew}
\end{figure}
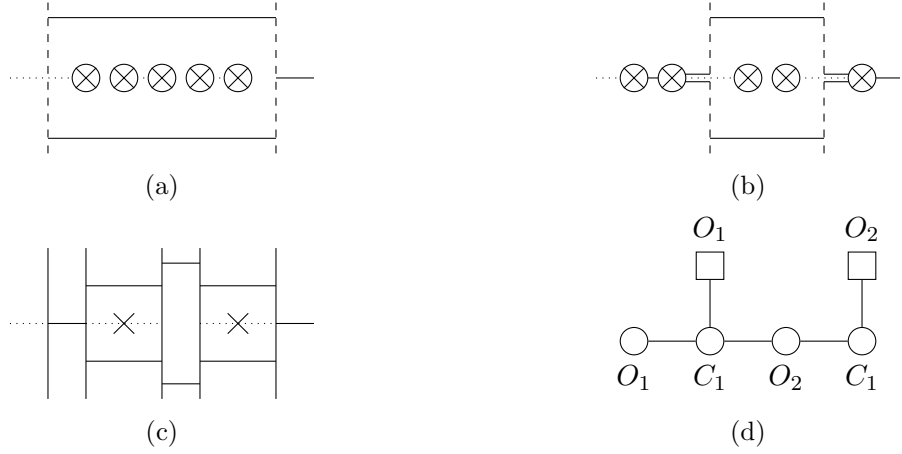

\paragraph{Transverse slice.} In order to answer the second question about the \emph{transverse slice} $S$ one focusses on the local subsystem containing the physical D3-brane that is \emph{Higgsed away} in the Kraft-Procesi transition. This system is depicted in figure \ref{fig:SO5maxInstNew}(a). Performing a phase transition, figure \ref{fig:SO5maxInstNew}(b-c), one can find the Coulomb branch brane configuration of such local system and hence the corresponding quiver, figure \ref{fig:SO5maxInstNew}(d). Note that, in order to write the quiver, the semi-infinite $\widetilde{O3^-}$ to the right is considered to contribute with half a flavor to the rightmost gauge node. In order to see this one can pull the rightmost half D5-brane to the right. This creates a physical D3-brane that is stacked on top of a $O3^-$ plane, thus giving a $O(2)$ flavor node. The Higgs branch of this quiver is the slice $S\subseteq \Or_{(5)}$ transverse to $\mathcal O_{(3,1^2)}$. In order to compute this Higgs branch, its Hilbert Series can be obtained utilizing the methods described in \cite{HK16}. This computation is carried out explicitly in the appendix \ref{app:AnHiggs}, the result is the variety:
\begin{align}
	S=A_3
\end{align}
Therefore, one says that there exists an $A_3$ KP transition from the closure of the maximal nilpotent orbit of $\gso (5)$, corresponding to partition $(5)$, to the \emph{subregular} nilpotent orbit, corresponding to partition $(3,1^2)$. This is a physical realization of the Brieskorn-Slodowy result \cite{B70,Sl80}: \emph{The slice in the closure of the maximal nilpotent orbit of type $B_n$ transverse to the subregular orbit with parition $(2n-1,1^2)$ is the surface singularity $A_{2n-1}$.}

\subsubsection{$\Or_{(3,1^2)}$ transition}\label{sec:311transition}

Let us take the model we just found as a starting point for a new KP transition. It has Higgs branch:

\begin{align}
	\M_H=\Or_{(3,1^2)}\subset \gso (5)
\end{align}

Consider the Higgs branch brane configuration depicted in figure \ref{fig:SO5maxKP}(f). We ask the question again: \emph{what is the minimal singularity that can be removed?} The answer is that all three physical D3-branes need to be removed, if one tries to remove just one or two of them this would break the supersymmetric configuration of the system. 

\begin{figure}[t]
	\centering\begin{subfigure}[t]{.49\textwidth}
    	\centering
	\begin{tikzpicture}
				\draw 	(1.5,0)--(1.5,2);
				\draw [dotted] (1.5,1)--(2.5,1);
				\draw	(0.5,1)--(1.5,1);
				\draw	(0.55,1.05)--(1.5,1.05)
						(0.55,.95)--(1.5,.95);
				\draw 	(0.5,1) node[cross]{};
				\node at (1,1)[label=above:$N$]{};
	\end{tikzpicture}
        \caption{}
    	\end{subfigure}
	\hfill
	\centering\begin{subfigure}[t]{.49\textwidth}
    	\centering
	\begin{tikzpicture}
				\draw 	(.5,0)--(.5,2);
				\draw [dotted] (.5,1)--(2.5,1);
				\draw	(0.5,1.05)--(1.45,1.05)
						(0.5,.95)--(1.45,.95);
				\draw 	(1.5,1) node[cross]{};
				\node at (1,1)[label=above:$\tilde N$]{};
	\end{tikzpicture}
        \caption{}
    	\end{subfigure}
    	\caption{Supersymmetric Configuration. (a) Shows a system with $N$ D3-branes. A phase transition takes it to the system (b), with $\tilde N=-N$ D3-branes. The only possibility for both numbers to be positive is $N=\tilde N=0$.}
    	\label{fig:susybreak}
\end{figure}
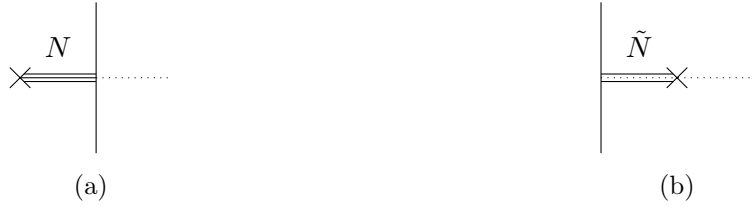

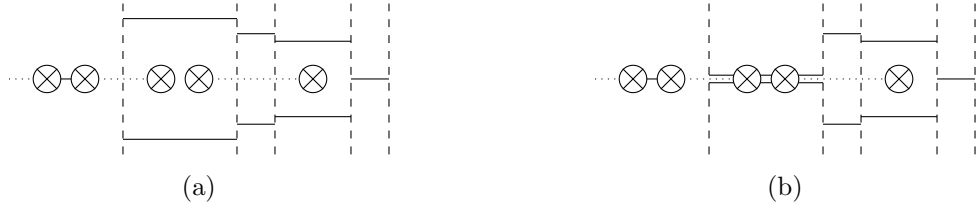
\begin{figure}[t]
	\centering
    \begin{subfigure}[t]{.49\textwidth}
   		\centering
		\begin{tikzpicture}
		\draw[dashed] 	(2,0)--(2,2)
				(3.5,0)--(3.5,2)
				(4,0)--(4,2)
				(5,0)--(5,2)
				(5.5,0)--(5.5,2);
		\draw [dotted] (.5,1)--(1,1)
				(1.5,1)--(2.5,1)
				(3,1)--(4.5,1);
		\draw	(1,1)--(1.5,1)
				(5,1)--(5.5,1);
		\draw	(3.5,1.6)--(4,1.6)
				(4,1.5)--(5,1.5)
				(3.5,.4)--(4,.4)
				(4,.5)--(5,.5)
				(2,1.8)--(3.5,1.8)
				(2,.2)--(3.5,.2);
		\draw 	(1,1) node[circ]{}
				(1.5,1)node[circ]{}
				(2.5,1)node[circ]{}
				(3,1)node[circ]{}
				(4.5,1)node[circ]{};
		\draw 	(1,1) node[cross]{}
				(1.5,1)node[cross]{}
				(2.5,1)node[cross]{}
				(3,1)node[cross]{}
				(4.5,1)node[cross]{};
		\end{tikzpicture}
        \caption{}
    \end{subfigure}
    \hfill
    \begin{subfigure}[t]{.49\textwidth}
   		\centering
		\begin{tikzpicture}
		\draw[dashed] 	(2,0)--(2,2)
				(3.5,0)--(3.5,2)
				(4,0)--(4,2)
				(5,0)--(5,2)
				(5.5,0)--(5.5,2);
		\draw [dotted] (.5,1)--(1,1)
				(1.5,1)--(2.5,1)
				(3,1)--(4.5,1);
		\draw	(1,1)--(1.5,1)
				(5,1)--(5.5,1);
		\draw	(3.5,1.6)--(4,1.6)
				(4,1.5)--(5,1.5)
				(3.5,.4)--(4,.4)
				(4,.5)--(5,.5)
				(2,1.05)--(3.5,1.05)
				(2,.95)--(3.5,.95);
		\draw 	(1,1) node[circ]{}
				(1.5,1)node[circ]{}
				(2.5,1)node[circ]{}
				(3,1)node[circ]{}
				(4.5,1)node[circ]{};
		\draw 	(1,1) node[cross]{}
				(1.5,1)node[cross]{}
				(2.5,1)node[cross]{}
				(3,1)node[cross]{}
				(4.5,1)node[cross]{};
		\end{tikzpicture}
        \caption{}
    \end{subfigure}
 	\caption{\emph{Naive} attempt at KP transition on the model with $\vec{k}_s=(0,0,1,1,3)$. (a) Higgs branch of the model. (b) After splitting the D3-brane in three segments, the middle one can acquire non zero position $\vec{x}$ along the directions spanned by the half NS5-branes. However, when the limit is taken where this position is infinite and the segment is effectively removed from the system, the resulting system would not be in a supersymmetric configuration. This is because there would be a subset of the system identical to the configuration in figure \ref{fig:susybreak}(a), with $N=1$.}
	\label{fig:SO5subregObs}
\end{figure}

Let us illustrate this phenomenon by trying to remove the leftmost D3-brane via a \emph{naive} KP transition. This attempt is illustrated in figure \ref{fig:SO5subregObs}. The obstruction to the transition stems from a fact discussed in \cite{FH00}: a brane configuration like the one in figure \ref{fig:susybreak}(a), that only consists on a left to right sequence of $O3^-$, half D5-brane, $\widetilde{O3^-}$, half NS5-brane, $\widetilde{O3^+}$, with no D3-branes ending on the half D5-brane from the left and no D3-branes ending on the half NS5-brane from the right, cannot have a non zero number $N$ of D3-branes in between the two different half fivebranes. If this were the case, the linking number of the half D5-brane would be $2N+1$, and the half NS5-brane would have linking number $-2N+1$. The phase transition that interchanges them would see $\tilde{N}$ branes in between, figure \ref{fig:susybreak}(b); the new linking number for the half D5-brane would be $-2\tilde{N}+1$ and the new linking number for the half NS5-brane would be $2\tilde{N}+1$, so one needs:

\begin{align}
	2N+1=-2\tilde{N}+1
\end{align}

hence
\begin{align}
	N=-\tilde{N}
\end{align}

Therefore, the only solution for both $N$ and $\widetilde{N}$ to be positive is $N=\tilde{N}=0$. The removal of a single D3-brane by partial Higgsing can result in a system where some of the supersymmetry is broken. This is the case if the resulting system has either the brane configuration in figure \ref{fig:susybreak}(a) or the brane configuration in figure \ref{fig:susybreak}(b) as a subsystem, and $N = -\tilde{N}\not = 0$. We define the Kraft-Procesi transition to preserve all the supersymmetry of the system. Hence, a Kraft-Procesi transition never removes this type of D3-branes via partial Higgsing\footnote{Note that in \cite{KP82} transitions that from our point of view would break supersymmetry are allowed. These are always transitions either to or from a \emph{non-special} orbit. This difference is discussed in more detail in sections \ref{sec:11} and \ref{sec:12}.}.

\begin{figure}[t]
	\centering
    \begin{subfigure}[t]{.49\textwidth}
   		\centering
		\begin{tikzpicture}
		\draw[dashed] 	(2,0)--(2,2)
				(3.5,0)--(3.5,2)
				(4,0)--(4,2)
				(5,0)--(5,2)
				(5.5,0)--(5.5,2);
		\draw [dotted] (.5,1)--(1,1)
				(1.5,1)--(2.5,1)
				(3,1)--(4.5,1);
		\draw	(1,1)--(1.5,1)
				(5,1)--(5.5,1);
		\draw	(3.5,1.6)--(4,1.6)
				(4,1.5)--(5,1.5)
				(3.5,.4)--(4,.4)
				(4,.5)--(5,.5)
				(2,1.8)--(3.5,1.8)
				(2,.2)--(3.5,.2);
		\draw 	(1,1) node[circ]{}
				(1.5,1)node[circ]{}
				(2.5,1)node[circ]{}
				(3,1)node[circ]{}
				(4.5,1)node[circ]{};
		\draw 	(1,1) node[cross]{}
				(1.5,1)node[cross]{}
				(2.5,1)node[cross]{}
				(3,1)node[cross]{}
				(4.5,1)node[cross]{};
		\end{tikzpicture}
        \caption{}
    \end{subfigure}
    \hfill
    \begin{subfigure}[t]{.49\textwidth}
   		\centering
		\begin{tikzpicture}
		\draw[dashed] 	(2,0)--(2,2)
				(3.5,0)--(3.5,2)
				(4,0)--(4,2)
				(5,0)--(5,2)
				(5.5,0)--(5.5,2);
		\draw [dotted] (.5,1)--(1,1)
				(1.5,1)--(2.5,1)
				(3,1)--(4.5,1);
		\draw	(1,1)--(1.5,1)
				(5,1)--(5.5,1);
		\draw	(2,1.05)--(5,1.05)
				(2,.95)--(5,.95);
		\draw 	(1,1) node[circ]{}
				(1.5,1)node[circ]{}
				(2.5,1)node[circ]{}
				(3,1)node[circ]{}
				(4.5,1)node[circ]{};
		\draw 	(1,1) node[cross]{}
				(1.5,1)node[cross]{}
				(2.5,1)node[cross]{}
				(3,1)node[cross]{}
				(4.5,1)node[cross]{};
		\end{tikzpicture}
        \caption{}
    \end{subfigure}
    \hfill
    \begin{subfigure}[t]{.49\textwidth}
   		\centering
		\begin{tikzpicture}
		\draw[dashed] 	(2.5,0)--(2.5,2)
				(3.5,0)--(3.5,2)
				(4,0)--(4,2)
				(5,0)--(5,2)
				(5.5,0)--(5.5,2);
		\draw [dotted] (.5,1)--(1,1)
				(1.5,1)--(2,1)
				(3,1)--(4.5,1);
		\draw	(1,1)--(1.5,1)
				(2,1)--(2.5,1)
				(5,1)--(5.5,1);
		\draw	(2.5,1.05)--(5,1.05)
				(2.5,.95)--(5,.95);
		\draw 	(1,1) node[circ]{}
				(1.5,1)node[circ]{}
				(2,1)node[circ]{}
				(3,1)node[circ]{}
				(4.5,1)node[circ]{};
		\draw 	(1,1) node[cross]{}
				(1.5,1)node[cross]{}
				(2,1)node[cross]{}
				(3,1)node[cross]{}
				(4.5,1)node[cross]{};
		\end{tikzpicture}
        \caption{}
    \end{subfigure}
    \hfill
    \begin{subfigure}[t]{.49\textwidth}
   		\centering
		\begin{tikzpicture}
		\draw[dashed] 	(2.5,0)--(2.5,2)
				(3.5,0)--(3.5,2)
				(4,0)--(4,2)
				(5,0)--(5,2)
				(5.5,0)--(5.5,2);
		\draw [dotted] (.5,1)--(1,1)
				(1.5,1)--(2,1)
				(3,1)--(4.5,1);
		\draw	(1,1)--(1.5,1)
				(2,1)--(2.5,1)
				(5,1)--(5.5,1);
		\draw	(2.5,1.05)--(2.95,1.05)
				(2.5,.95)--(2.95,.95)
				(4.55,1.05)--(5,1.05)
				(4.55,.95)--(5,.95);
		\draw 	(1,1) node[circ]{}
				(1.5,1)node[circ]{}
				(2,1)node[circ]{}
				(3,1)node[circ]{}
				(4.5,1)node[circ]{};
		\draw 	(1,1) node[cross]{}
				(1.5,1)node[cross]{}
				(2,1)node[cross]{}
				(3,1)node[cross]{}
				(4.5,1)node[cross]{};
		\end{tikzpicture}
        \caption{}
    \end{subfigure}
    \hfill
    \begin{subfigure}[t]{.49\textwidth}
   		\centering
		\begin{tikzpicture}
		\draw[dashed] 	(3,0)--(3,2)
				(3.5,0)--(3.5,2)
				(4,0)--(4,2)
				(4.5,0)--(4.5,2)
				(5.5,0)--(5.5,2);
		\draw [dotted] (.5,1)--(1,1)
				(1.5,1)--(2,1)
				(2.5,1)--(5,1);
		\draw	(1,1)--(1.5,1)
				(2,1)--(2.5,1)
				(5,1)--(5.5,1);
		\draw 	(1,1) node[circ]{}
				(1.5,1)node[circ]{}
				(2,1)node[circ]{}
				(2.5,1)node[circ]{}
				(5,1)node[circ]{};
		\draw 	(1,1) node[cross]{}
				(1.5,1)node[cross]{}
				(2,1)node[cross]{}
				(2.5,1)node[cross]{}
				(5,1)node[cross]{};
		\end{tikzpicture}
        \caption{}
    \end{subfigure}
    \hfill
    \begin{subfigure}[t]{.49\textwidth}
   		\centering
		\begin{tikzpicture}
		\draw[dashed] 	(3,0)--(3,2)
				(3.5,0)--(3.5,2)
				(4,0)--(4,2)
				(4.5,0)--(4.5,2)
				(5,0)--(5,2);
		\draw [dotted] (.5,1)--(1,1)
				(1.5,1)--(2,1)
				(2.5,1)--(5.5,1);
		\draw	(1,1)--(1.5,1)
				(2,1)--(2.5,1);
		\draw 	(1,1) node[circ]{}
				(1.5,1)node[circ]{}
				(2,1)node[circ]{}
				(2.5,1)node[circ]{}
				(5.5,1)node[circ]{};
		\draw 	(1,1) node[cross]{}
				(1.5,1)node[cross]{}
				(2,1)node[cross]{}
				(2.5,1)node[cross]{}
				(5.5,1)node[cross]{};
		\end{tikzpicture}
        \caption{}
    \end{subfigure}
    \hfill
 	\caption{$\Or_{(3,1^2)}$ KP transition. The starting model is defined by parameters $n_s=n_d=5$, $\vec{l}_d=(4,4,4,4,4)$, $\vec{l}_s=(0,0,0,2,2)$ and rightmost O3-plane $O3^-$. (a) Higgs branch of the starting model. (b) The D3-branes align with the half NS5-branes, stacked at the orientifold plane. The D3-branes are split into four segments: two in the middle only ending on half NS5-branes and two fixed segments ending on both half a D3-brane and half an NS5-brane.  (c) We perform a phase transition that annihilates the leftmost fixed D3-brane segment and fixes the position of the leftmost mobile D3-brane segment, in order to preserve the supersymmetric configuration. (d) The remaining mobile D3-brane segment acquires a non zero position $\vec{x}$ along the direction spanned by the half NS5-branes. One takes the limit were $\vec x$ goes to infinity, effectively removing the brane from the system. (e) The fixed D3-brane segments are annihilated via phase transitions. (f) Result after performing the collapse transition on (e), the Higgs branch of this system is the trivial orbit; the interval numbers of the half NS5-branes are $\vec{k}_s=(0,0,0,0,5)$ giving the trivial partition $\lambda=(5)^t=(1^5)$.}
	\label{fig:SO5subregKP}
\end{figure}
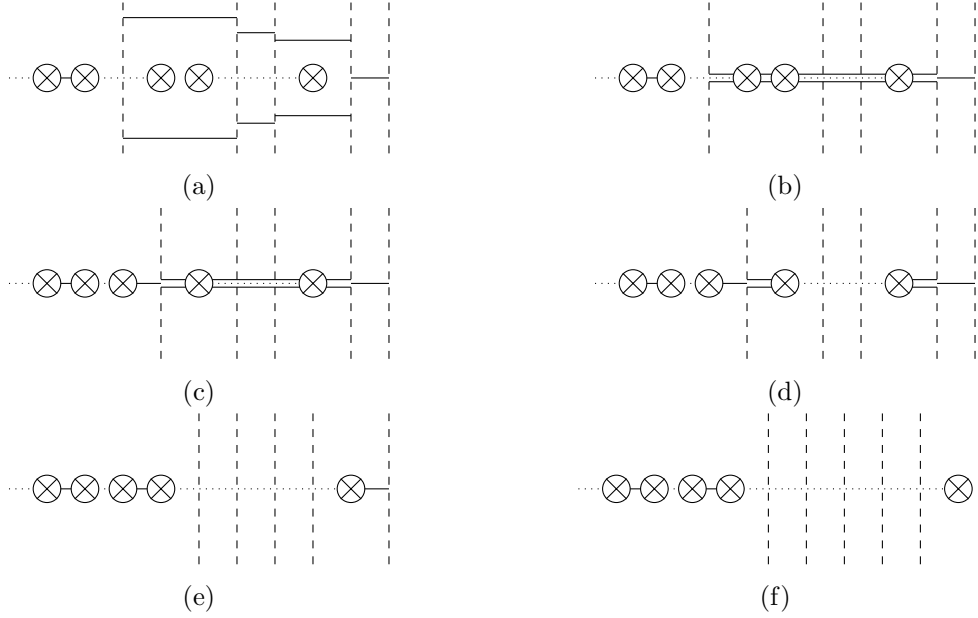

The only possible KP transition in the case of the system in figure \ref{fig:SO5subregObs}(a) is a KP transition that removes the three D3-branes, as shown in figure \ref{fig:SO5subregKP}. This removes the whole variety $\M_H$, the resulting Higgs branch is the trivial orbit $\M'_H=\Or_{(1^5)}$, and the transverse slice to the trivial orbit is $S=\M_H=\Or_{(3,1^2)}$.

\subsubsection{Hasse diagram for $\gso (5)$}

The Hasse diagram with the KP transitions found for the $\gso (5)$ models is depicted in figure \ref{fig:HasseSO5}. Note that this diagram differs with the one in \cite{KP82} in that the non-special orbit corresponding to partition $\lambda=(2^2,1)$ does not appear. Instead, one finds a transition from the next to minimal orbit with partition $\lambda=(3,1^2)$ to the trivial orbit. In general, one does not have brane systems with O3-planes and orthosymplcetic quivers that correspond to closures of non-special orbits, like the one labelled by partition $\lambda=(2^2,1^{2n-3})$.

\subsubsection{Summary}

A summary of all the models we have found, with the brane systems, the quivers and the KP transitions is depicted in table \ref{tab:SO5summary}.

\begin{figure}[t]
	\centering
\begin{subfigure}[t]{.45\textwidth}
    \centering
	\begin{tikzpicture}
		\node [hasse] (1) [] {};
		\node [hasse] (2) [below of=1] {};
		\node [hasse] (3) [below of=2] {};
		\draw (1) edge [] node[label=left:$A_3$] {} (2)
			(2) edge [] node[label=left:$\Or_{(3,1^2)}$] {} (3);
		\node (e1) [right of=1] {};
		\node (d1) [right of=e1] {$(0,2,0,2,0)$};
		\node (c1) [right of=d1] {};
		\node (m1) [right of=c1] {};
		\node (b1) [right of=m1] {$(4,4,4,4,4)$};
		\node (e2) [right of=2] {};
		\node (d2) [right of=e2] {$(0,0,0,2,2)$};
		\node (c2) [right of=d2] {};
		\node (m2) [right of=c2] {};
		\node (b2) [right of=m2] {$(4,4,4,4,4)$};
		\node (e3) [right of=3] {};
		\node (d3) [right of=e3] {$(0,0,0,0,4)$};
		\node (c3) [right of=d3] {};
		\node (m3) [right of=c3] {};
		\node (b3) [right of=m3] {$(4,4,4,4,4)$};
		\node (d) [above of=d1] {$\vec l_s$};
		\node (b) [above of=b1] {$\vec l_d$};
	\end{tikzpicture}
	\caption{}
 	\end{subfigure}
 	\hfill
 	\begin{subfigure}[t]{.45\textwidth}
 	\centering
 	\begin{tikzpicture}
		\node [hasse] (1) [] {};
		\node [hasse] (2) [below of=1] {};
		\node [hasse] (3) [below of=2] {};
		\draw (1) edge [] node[label=left:$A_3$] {} (2)
			(2) edge [] node[label=left:$\Or_{(3,1^2)}$] {} (3);
		\node (e1) [right of=1] {};
		\node (d1) [right of=e1] {$(5)$};
		\node (c1) [right of=d1] {};
		\node (b1) [right of=c1] {$4$};
		\node (e2) [right of=2] {};
		\node (d2) [right of=e2] {$(3,1^2)$};
		\node (c2) [right of=d2] {};
		\node (b2) [right of=c2] {$3$};
		\node (e3) [right of=3] {};
		\node (d3) [right of=e3] {$(1^5)$};
		\node (c3) [right of=d3] {};
		\node (b3) [right of=c3] {$0$};
		\node (d) [above of=d1] {$\lambda$};
		\node (b) [above of=b1] {$dim$};
	\end{tikzpicture}
	\caption{}
 	\end{subfigure}
 	\hfill
	\caption{Hasse diagram for the models whose Higgs branch is the closure of a special nilpotent orbit of $\gso (5)$ under the adjoint action of the group $O(5)$. (a) represents the brane configurations, where the linking numbers $\vec{l}_s$ and $\vec{l}_d$ are provided for each orbit and the rightmost orientifold plane is always of type $O3^-$. (b) depicts the information of the orbits. $\lambda$ is the partition obtained from the interval numbers $\vec{k}_s$ as discussed above. \emph{dim} is the number of physical D3-branes that generate the Higgs branch in each model.}
	\label{fig:HasseSO5}	
\end{figure}

\begin{table}[t]
	\centering
	\begin{tabular}{ c c}
	  
		\raisebox{-.5\height}{\begin{tikzpicture}[node distance=80pt]
		\node at (0,0)[]{\large{$\mathfrak{so}(5)$}};
		\node at (0,-0.5) [hasse] (1) [] {};
		\node [hasse] (2) [below of=1] {};
		\node [hasse] (3) [below of=2] {};
		\draw (1) edge [] node[label=left:$A_3$] {} (2)
			(2) edge [] node[label=left:$\Or_{(3,1^2)}$] {} (3);
	\end{tikzpicture}}
	&
	\fontsize{10}{11}\selectfont
	\begin{tabular}{ c c c  c}
	\toprule
	\textbf{Partition}  & \textbf{Branes} & $\M_H$  & $\M_C$  \\ 
	\midrule \addlinespace[3ex]
$\mathbf{5}$ & \raisebox{-.4\height}{\begin{tikzpicture}[scale=.8, every node/.style={transform shape}]
		\draw[dashed] 	(1,0)--(1,2)
				(4,0)--(4,2)
				(4.5,0)--(4.5,2)
				(5,0)--(5,2)
				(5.5,0)--(5.5,2);
		\draw [dotted] (.5,1)--(1.5,1)
				(2,1)--(2.5,1)
				(3,1)--(3.5,1);
		\draw	(4,1)--(4.5,1)
				(5,1)--(5.5,1);
		\draw	(1,1.5)--(4,1.5)
				(4,1.6)--(4.5,1.6)
				(4.5,1.5)--(5,1.5)
				(1,.5)--(4,.5)
				(4,.4)--(4.5,.4)
				(4.5,.5)--(5,.5)
				(1,1.8)--(4,1.8)
				(1,.2)--(4,.2);
		\draw 	(1.5,1) node[circ]{}
				(2,1)node[circ]{}
				(2.5,1)node[circ]{}
				(3,1)node[circ]{}
				(3.5,1)node[circ]{};
		\draw 	(1.5,1) node[cross]{}
				(2,1)node[cross]{}
				(2.5,1)node[cross]{}
				(3,1)node[cross]{}
				(3.5,1)node[cross]{};
		\end{tikzpicture}}
& \raisebox{-.4\height}{\begin{tikzpicture}[scale=1, every node/.style={transform shape},node distance=20]
	\tikzstyle{gauge} = [circle, draw];
	\tikzstyle{flavour} = [regular polygon,regular polygon sides=4,draw];
	\node (g1) [gauge,label=below:{$O_1$}] {};
	\node (g2) [gauge,right of=g1,label=below:{$C_1$}] {};
	\node (g3) [gauge,right of=g2,label=below:{$O_3$}] {};
	\node (g4) [gauge,right of=g3,label=below:{$C_2$}] {};
	\node (f4) [flavour,above of=g4,label=above:{$O_5$}] {};
	\draw (g1)--(g2)--(g3)--(g4)--(f4);
		\end{tikzpicture}} & \raisebox{-.4\height}{\begin{tikzpicture}[scale=1, every node/.style={transform shape},node distance=20]
			\tikzstyle{gauge} = [circle, draw];
			\tikzstyle{flavour} = [regular polygon,regular polygon sides=4,draw];
	\node (g2) [gauge,label=below:{$SO_4$}] {};
	\node (g3) [gauge,right of=g2,label=below:{$C_1$}] {};
	\node (g4) [gauge,right of=g3,label=below:{$SO_2$}] {};
	\node (f2) [flavour,above of=g2,label=above:{$C_2$}] {};
	\draw(g2)--(g3)--(g4)
			(f2)--(g2)
		;
		\end{tikzpicture}}
 \\
  \addlinespace[5ex]
 $\mathbf{3,1^2}$ & 
		\raisebox{-.4\height}{\begin{tikzpicture}[scale=.8, every node/.style={transform shape}]
		\draw[dashed] 	(2,0)--(2,2)
				(3.5,0)--(3.5,2)
				(4,0)--(4,2)
				(5,0)--(5,2)
				(5.5,0)--(5.5,2);
		\draw [dotted] (.5,1)--(1,1)
				(1.5,1)--(2.5,1)
				(3,1)--(4.5,1);
		\draw	(1,1)--(1.5,1)
				(5,1)--(5.5,1);
		\draw	(3.5,1.6)--(4,1.6)
				(4,1.5)--(5,1.5)
				(3.5,.4)--(4,.4)
				(4,.5)--(5,.5)
				(2,1.8)--(3.5,1.8)
				(2,.2)--(3.5,.2);
		\draw 	(1,1) node[circ]{}
				(1.5,1)node[circ]{}
				(2.5,1)node[circ]{}
				(3,1)node[circ]{}
				(4.5,1)node[circ]{};
		\draw 	(1,1) node[cross]{}
				(1.5,1)node[cross]{}
				(2.5,1)node[cross]{}
				(3,1)node[cross]{}
				(4.5,1)node[cross]{};
		\end{tikzpicture}}
 & \raisebox{-.4\height}{\begin{tikzpicture}[scale=1, every node/.style={transform shape},node distance=20]
	\tikzstyle{gauge} = [circle, draw];
	\tikzstyle{flavour} = [regular polygon,regular polygon sides=4,draw];
	\node (g3) [gauge,label=below:{$O_1$}] {};
	\node (g4) [gauge,right of=g3,label=below:{$C_1$}] {};
	\node (f4) [flavour,above of=g4,label=above:{$O_5$}] {};
	\draw (g3)--(g4)--(f4)
		;
		\end{tikzpicture}}  & \raisebox{-.4\height}{\begin{tikzpicture}[scale=1, every node/.style={transform shape},node distance=20]
			\tikzstyle{gauge} = [circle, draw];
			\tikzstyle{flavour} = [regular polygon,regular polygon sides=4,draw];
	\node (g2) [gauge,label=below:{$O_2$}] {};
	\node (g3) [gauge,right of=g2,label=below:{$C_1$}] {};
	\node (g4) [gauge,right of=g3,label=below:{$SO_2$}] {};
	\node (f3) [flavour,above of=g3,label=above:{$O_2$}] {};
	\draw(g2)--(g3)--(g4)
			(f3)--(g3)
		;
		\end{tikzpicture}}\\
  \addlinespace[5ex]
 $\mathbf{1^5}$  & 
		\raisebox{-.4\height}{\begin{tikzpicture}[scale=.8, every node/.style={transform shape}]
		\draw [dotted] (.5,1)--(1,1)
				(1.5,1)--(2,1)
				(2.5,1)--(5.5,1);
		\draw	(1,1)--(1.5,1)
				(2,1)--(2.5,1);
		\draw[dashed] 	(3,0)--(3,2)
				(3.5,0)--(3.5,2)
				(4,0)--(4,2)
				(4.5,0)--(4.5,2)
				(5,0)--(5,2);
		\draw 	(1,1) node[circ]{}
				(1.5,1)node[circ]{}
				(2,1)node[circ]{}
				(2.5,1)node[circ]{}
				(5.5,1)node[circ]{};
		\draw 	(1,1) node[cross]{}
				(1.5,1)node[cross]{}
				(2,1)node[cross]{}
				(2.5,1)node[cross]{}
				(5.5,1)node[cross]{};
		\end{tikzpicture}}
 & \raisebox{-.4\height}{\begin{tikzpicture}[scale=1, every node/.style={transform shape},node distance=20]
			\tikzstyle{gauge} = [circle, draw];
			\tikzstyle{flavour} = [regular polygon,regular polygon sides=4, draw];
			\node (g6) [] {};
			\node (f6) [flavour,above of=g6,label=above:{$O_5$}] {};
		\end{tikzpicture}} & \raisebox{-.4\height}{\begin{tikzpicture}[scale=1, every node/.style={transform shape},node distance=20]
			\tikzstyle{gauge} = [circle,draw];
			\tikzstyle{flavour} = [regular polygon,regular polygon sides=4,draw];
	\node (g2) [gauge,label=below:{$O_0$}] {};
	\node (g3) [gauge,right of=g2,label=below:{$C_0$}] {};
	\node (g4) [gauge,right of=g3,label=below:{$O_0$}] {};
	\draw(g2)--(g3)--(g4)
		;
		\end{tikzpicture}} \\ \addlinespace[3ex]
	\bottomrule
	\end{tabular}
	\end{tabular}
		\caption{Summary of Kraft-Procesi transitions for the nilpotent orbits of $\mathfrak{so}(5)$. The brane systems in the middle depict models whose Higgs branch is the closure of the nilpotent orbit labelled by each \emph{Partition}. The column $\M_H$   contains the quivers of such models, with $\M_H=\Or_{\lambda}$. The column $\M_C$  depicts the quivers of the models obtained after performing S-duality transformation on the Brane systems, they have $\M_C=\Or_\lambda$. The $C_0$ and $O_0$ gauge nodes have been left in this column to highlight the effect that the transition has on these quivers: the flavor nodes move along a fixed structure of gauge nodes, while the rank of the gauge nodes decreases.}
		\label{tab:SO5summary}
\end{table}

\subsection{Kraft-Procesi transitions for the nilpotent orbits of $\gsp (2)$}\label{sec:10}

The KP transitions among closures of special nilpotent orbits of $\gsp (2)$ are the last set of examples.

\subsubsection{$D_3$ transition}

\begin{figure}[t]
	\centering
	\begin{subfigure}[t]{.49\textwidth}
		\centering
		\begin{tikzpicture}
			\draw 	[dashed](1.5,0)--(1.5,2)
					(4,0)--(4,2)
					(4.5,0)--(4.5,2)
					(5,0)--(5,2)
					(5.5,0)--(5.5,2);
			\draw	[dotted](1,1)--(2,1)
					(2.5,1)--(3,1)
					(3.5,1)--(6,1);
			\draw	(2,1)--(2.5,1)
					(3,1)--(3.5,1);
			\draw	(1.5,1.7)--(4,1.7)
					(1.5,.3)--(4,.3)
					(1.5,1.3)--(4,1.3)
					(1.5,.7)--(4,.7)
					(4,1.5)--(4.5,1.5)
					(4,.5)--(4.5,.5)
					(4.5,1.3)--(5,1.3)
					(4.5,.7)--(5,.7);
			\draw 	(1,1) node[circ]{}
					(2,1) node[circ]{}
					(2.5,1) node[circ]{}
					(3,1) node[circ]{}
					(3.5,1) node[circ]{};
			\draw 	(1,1) node[cross]{}
					(2,1) node[cross]{}
					(2.5,1) node[cross]{}
					(3,1) node[cross]{}
					(3.5,1) node[cross]{};
		\end{tikzpicture}
		\caption{}
	\end{subfigure}
	\hfill
	\begin{subfigure}[t]{.49\textwidth}
		\centering
		\begin{tikzpicture}
			\draw 	[dashed](1.5,0)--(1.5,2)
					(4,0)--(4,2)
					(4.5,0)--(4.5,2)
					(5,0)--(5,2)
					(5.5,0)--(5.5,2);
			\draw	[dotted](1,1)--(2,1)
					(2.5,1)--(3,1)
					(3.5,1)--(6,1);
			\draw	(2,1)--(2.5,1)
					(3,1)--(3.5,1);
			\draw	(1.5,1.7)--(4,1.7)
					(1.5,.3)--(4,.3)
					(1.5,1.05)--(4,1.05)
					(1.5,.95)--(4,.95)
					(4,1.5)--(4.5,1.5)
					(4,.5)--(4.5,.5)
					(4.5,1.3)--(5,1.3)
					(4.5,.7)--(5,.7);
			\draw 	(1,1) node[circ]{}
					(2,1) node[circ]{}
					(2.5,1) node[circ]{}
					(3,1) node[circ]{}
					(3.5,1) node[circ]{};
			\draw 	(1,1) node[cross]{}
					(2,1) node[cross]{}
					(2.5,1) node[cross]{}
					(3,1) node[cross]{}
					(3.5,1) node[cross]{};
		\end{tikzpicture}
		\caption{}
	\end{subfigure}
	\hfill
	\begin{subfigure}[t]{.49\textwidth}
		\centering
		\begin{tikzpicture}
			\draw 	[dashed](1.5,0)--(1.5,2)
					(4,0)--(4,2)
					(4.5,0)--(4.5,2)
					(5,0)--(5,2)
					(5.5,0)--(5.5,2);
			\draw	[dotted](1,1)--(2,1)
					(2.5,1)--(3,1)
					(3.5,1)--(6,1);
			\draw	(2,1)--(2.5,1)
					(3,1)--(3.5,1);
			\draw	(1.5,1.7)--(4,1.7)
					(1.5,.3)--(4,.3)
					(1.5,1.05)--(1.95,1.05)
					(3.55,1.05)--(4,1.05)
					(1.5,.95)--(1.95,.95) 
					(3.55,.95)--(4,.95)
					(4,1.5)--(4.5,1.5)
					(4,.5)--(4.5,.5)
					(4.5,1.3)--(5,1.3)
					(4.5,.7)--(5,.7);
			\draw 	(1,1) node[circ]{}
					(2,1) node[circ]{}
					(2.5,1) node[circ]{}
					(3,1) node[circ]{}
					(3.5,1) node[circ]{};
			\draw 	(1,1) node[cross]{}
					(2,1) node[cross]{}
					(2.5,1) node[cross]{}
					(3,1) node[cross]{}
					(3.5,1) node[cross]{};
		\end{tikzpicture}
		\caption{}
	\end{subfigure}
	\hfill
	\begin{subfigure}[t]{.49\textwidth}
		\centering
		\begin{tikzpicture}
			\draw 	[dashed](2,0)--(2,2)
					(3.5,0)--(3.5,2)
					(4.5,0)--(4.5,2)
					(5,0)--(5,2)
					(5.5,0)--(5.5,2);
			\draw	[dotted](1,1)--(1.5,1)
					(2.5,1)--(3,1)
					(4,1)--(6,1);
			\draw	(2,1)--(2.5,1)
					(3,1)--(3.5,1);
			\draw	(2,1.7)--(3.5,1.7)
					(2,.3)--(3.5,.3)
					(3.5,1.5)--(4.5,1.5)
					(3.5,.5)--(4.5,.5)
					(4.5,1.3)--(5,1.3)
					(4.5,.7)--(5,.7);
			\draw 	(1,1) node[circ]{}
					(1.5,1) node[circ]{}
					(2.5,1) node[circ]{}
					(3,1) node[circ]{}
					(4,1) node[circ]{};
			\draw 	(1,1) node[cross]{}
					(1.5,1) node[cross]{}
					(2.5,1) node[cross]{}
					(3,1) node[cross]{}
					(4,1) node[cross]{};
		\end{tikzpicture}
		\caption{}
	\end{subfigure}
	\hfill
	\begin{subfigure}[t]{.49\textwidth}
		\centering
		\begin{tikzpicture}
			\draw 	[dashed](2.5,0)--(2.5,2)
					(3,0)--(3,2)
					(4.5,0)--(4.5,2)
					(5,0)--(5,2)
					(5.5,0)--(5.5,2);
			\draw	[dotted](1,1)--(1.5,1)
					(2,1)--(3.5,1)
					(4,1)--(6,1);
			\draw	(2.5,1.7)--(3,1.7)
					(2.5,.3)--(3,.3)
					(3,1.5)--(4.5,1.5)
					(3,.5)--(4.5,.5)
					(4.5,1.3)--(5,1.3)
					(4.5,.7)--(5,.7);
			\draw 	(1,1) node[circ]{}
					(1.5,1) node[circ]{}
					(2,1) node[circ]{}
					(3.5,1) node[circ]{}
					(4,1) node[circ]{};
			\draw 	(1,1) node[cross]{}
					(1.5,1) node[cross]{}
					(2,1) node[cross]{}
					(3.5,1) node[cross]{}
					(4,1) node[cross]{};
		\end{tikzpicture}
		\caption{}
	\end{subfigure}
	\hfill
	\caption{$D_3$ Kraft-Procesi transition.  The starting model has $n_s=n_d=5$, $\vec{l}_d=(5,3,5,3,5)$, $\vec{l}_s=(1,1,1,1,1)$ and rightmost O3-plane $\widetilde{O3^+}$. (a) Higgs branch brane configuration of the initial model with $\M_H=\Or_{(4)}\subset \mathfrak{sp}(2)$. (b) The half D3-branes that are involved in the KP transition are aligned with the half NS5-branes. (c) The D3-brane is split into five segments. The three segments in the middle are taken to infinity in the $\vec{x}$ directions spanned by the half NS5-branes, effectively removing them from the system. (d) Phase transitions are performed that annihilate the fixed segments of D3-brane. (e) Result after performing the \emph{collapse} transition on (d), the resulting interval numbers of the half NS5-branes are $\vec{k}_s=(0,0,0,2,2)$, and the resulting Higgs branch is identified as the closure of the nilpotent orbit corresponding to partition $\lambda=(2^2)^t=(2^2)$: $\M_H'=\Or_{(2^2)}\subset \mathfrak {sp}(2)$.}
	\label{fig:Sp4maxKP}
\end{figure}
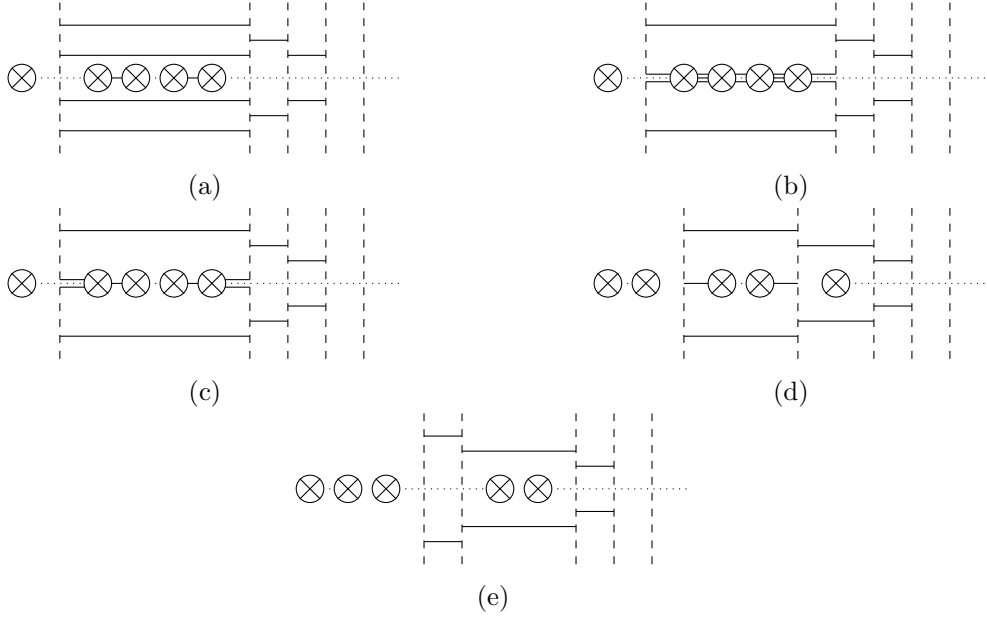

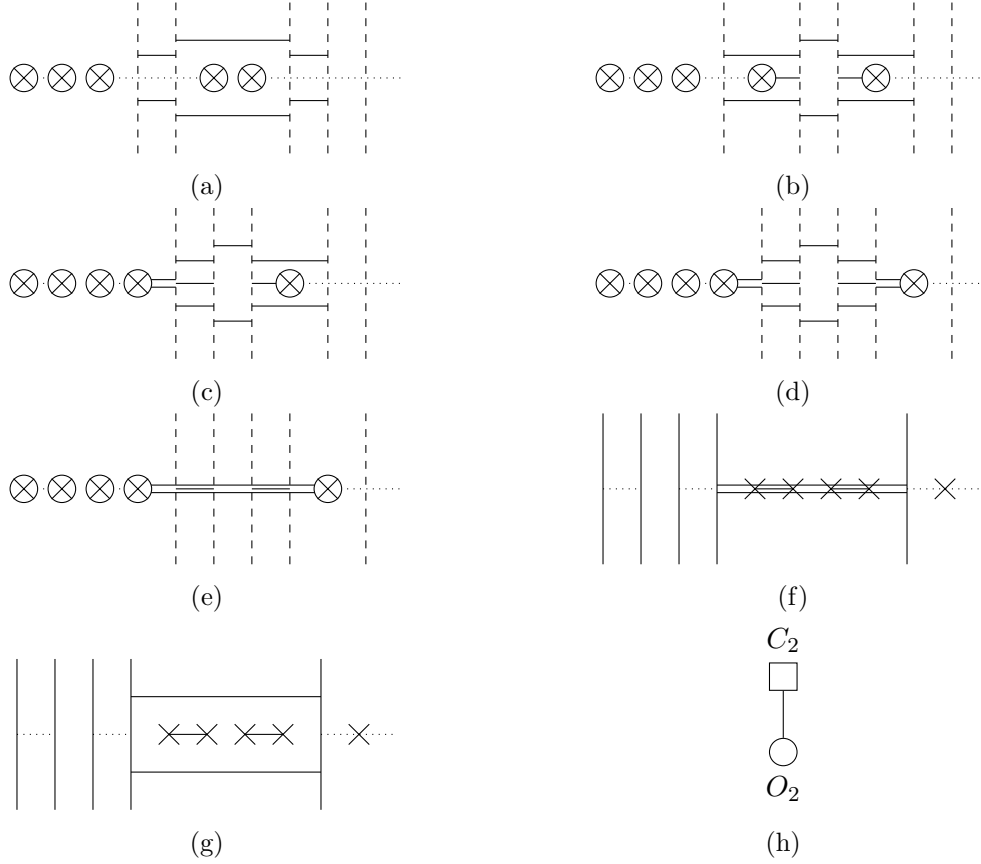
\begin{figure}[t]
	\centering
	\begin{subfigure}[t]{.49\textwidth}
		\centering
		\begin{tikzpicture}
			\draw 	[dashed](2.5,0)--(2.5,2)
					(3,0)--(3,2)
					(4.5,0)--(4.5,2)
					(5,0)--(5,2)
					(5.5,0)--(5.5,2);
			\draw	[dotted](1,1)--(1.5,1)
					(2,1)--(3.5,1)
					(4,1)--(6,1);
			\draw	(2.5,1.3)--(3,1.3)
					(2.5,.7)--(3,.7)
					(3,1.5)--(4.5,1.5)
					(3,.5)--(4.5,.5)
					(4.5,1.3)--(5,1.3)
					(4.5,.7)--(5,.7);
			\draw 	(1,1) node[circ]{}
					(1.5,1) node[circ]{}
					(2,1) node[circ]{}
					(3.5,1) node[circ]{}
					(4,1) node[circ]{};
			\draw 	(1,1) node[cross]{}
					(1.5,1) node[cross]{}
					(2,1) node[cross]{}
					(3.5,1) node[cross]{}
					(4,1) node[cross]{};
		\end{tikzpicture}
		\caption{}
	\end{subfigure}
	\hfill
	\begin{subfigure}[t]{.49\textwidth}
		\centering
		\begin{tikzpicture}
			\draw 	[dashed](2.5,0)--(2.5,2)
					(3.5,0)--(3.5,2)
					(4,0)--(4,2)
					(5,0)--(5,2)
					(5.5,0)--(5.5,2);
			\draw	[dotted](1,1)--(1.5,1)
					(2,1)--(3,1)
					(4.5,1)--(6,1);
			\draw	(3,1)--(3.5,1)
					(4,1)--(4.5,1);
			\draw	(2.5,1.3)--(3.5,1.3)
					(2.5,.7)--(3.5,.7)
					(3.5,1.5)--(4,1.5)
					(3.5,.5)--(4,.5)
					(4,1.3)--(5,1.3)
					(4,.7)--(5,.7);
			\draw 	(1,1) node[circ]{}
					(1.5,1) node[circ]{}
					(2,1) node[circ]{}
					(3,1) node[circ]{}
					(4.5,1) node[circ]{};
			\draw 	(1,1) node[cross]{}
					(1.5,1) node[cross]{}
					(2,1) node[cross]{}
					(3,1) node[cross]{}
					(4.5,1) node[cross]{};
		\end{tikzpicture}
		\caption{}
	\end{subfigure}
	\hfill
	\begin{subfigure}[t]{.49\textwidth}
		\centering
		\begin{tikzpicture}
			\draw 	[dashed](3,0)--(3,2)
					(3.5,0)--(3.5,2)
					(4,0)--(4,2)
					(5,0)--(5,2)
					(5.5,0)--(5.5,2);
			\draw	[dotted](1,1)--(1.5,1)
					(2,1)--(2.5,1)
					(4.5,1)--(6,1);
			\draw	(3,1)--(3.5,1)
					(4,1)--(4.5,1);
			\draw	(3,1.3)--(3.5,1.3)
					(3,.7)--(3.5,.7)
					(3.5,1.5)--(4,1.5)
					(3.5,.5)--(4,.5)
					(4,1.3)--(5,1.3)
					(4,.7)--(5,.7)
					(2.55,1.05)--(3,1.05)
					(2.55,.95)--(3,.95);
			\draw 	(1,1) node[circ]{}
					(1.5,1) node[circ]{}
					(2,1) node[circ]{}
					(2.5,1) node[circ]{}
					(4.5,1) node[circ]{};
			\draw 	(1,1) node[cross]{}
					(1.5,1) node[cross]{}
					(2,1) node[cross]{}
					(2.5,1) node[cross]{}
					(4.5,1) node[cross]{};
		\end{tikzpicture}
		\caption{}
	\end{subfigure}
	\hfill
	\begin{subfigure}[t]{.49\textwidth}
		\centering
		\begin{tikzpicture}
			\draw 	[dashed](3,0)--(3,2)
					(3.5,0)--(3.5,2)
					(4,0)--(4,2)
					(4.5,0)--(4.5,2)
					(5.5,0)--(5.5,2);
			\draw	[dotted](1,1)--(1.5,1)
					(2,1)--(2.5,1)
					(5,1)--(6,1);
			\draw	(3,1)--(3.5,1)
					(4,1)--(4.5,1);
			\draw	(3,1.3)--(3.5,1.3)
					(3,.7)--(3.5,.7)
					(3.5,1.5)--(4,1.5)
					(3.5,.5)--(4,.5)
					(4,1.3)--(4.5,1.3)
					(4,.7)--(4.5,.7)
					(2.55,1.05)--(3,1.05)
					(2.55,.95)--(3,.95)
					(4.5,1.05)--(4.95,1.05)
					(4.5,.95)--(4.95,.95);
			\draw 	(1,1) node[circ]{}
					(1.5,1) node[circ]{}
					(2,1) node[circ]{}
					(2.5,1) node[circ]{}
					(5,1) node[circ]{};
			\draw 	(1,1) node[cross]{}
					(1.5,1) node[cross]{}
					(2,1) node[cross]{}
					(2.5,1) node[cross]{}
					(5,1) node[cross]{};
		\end{tikzpicture}
		\caption{}
	\end{subfigure}
	\hfill
	\begin{subfigure}[t]{.49\textwidth}
		\centering
		\begin{tikzpicture}
			\draw 	[dashed](3,0)--(3,2)
					(3.5,0)--(3.5,2)
					(4,0)--(4,2)
					(4.5,0)--(4.5,2)
					(5.5,0)--(5.5,2);
			\draw	[dotted](1,1)--(1.5,1)
					(2,1)--(2.5,1)
					(5,1)--(6,1);
			\draw	(3,1)--(3.5,1)
					(4,1)--(4.5,1);
			\draw	(2.55,1.05)--(4.95,1.05)
					(2.55,.95)--(4.95,.95);
			\draw 	(1,1) node[circ]{}
					(1.5,1) node[circ]{}
					(2,1) node[circ]{}
					(2.5,1) node[circ]{}
					(5,1) node[circ]{};
			\draw 	(1,1) node[cross]{}
					(1.5,1) node[cross]{}
					(2,1) node[cross]{}
					(2.5,1) node[cross]{}
					(5,1) node[cross]{};
		\end{tikzpicture}
		\caption{}
	\end{subfigure}
	\hfill
	\begin{subfigure}[t]{.49\textwidth}
		\centering
		\begin{tikzpicture}
			\draw 	(1,0)--(1,2)
					(1.5,0)--(1.5,2)
					(2,0)--(2,2)
					(2.5,0)--(2.5,2)
					(5,0)--(5,2);
			\draw 	(3,1) node[cross]{}
					(3.5,1) node[cross]{}
					(4,1) node[cross]{}
					(4.5,1) node[cross]{}
					(5.5,1) node[cross]{};
			\draw	[dotted](1,1)--(1.5,1)
					(2,1)--(2.5,1)
					(5,1)--(6,1);
			\draw	(3,1)--(3.5,1)
					(4,1)--(4.5,1);
			\draw	(2.5,1.05)--(5,1.05)
					(2.5,.95)--(5,.95);
		\end{tikzpicture}
		\caption{}
	\end{subfigure}
	\hfill
	\begin{subfigure}[t]{.49\textwidth}
		\centering
		\begin{tikzpicture}
			\draw 	(1,0)--(1,2)
					(1.5,0)--(1.5,2)
					(2,0)--(2,2)
					(2.5,0)--(2.5,2)
					(5,0)--(5,2);
			\draw 	(3,1) node[cross]{}
					(3.5,1) node[cross]{}
					(4,1) node[cross]{}
					(4.5,1) node[cross]{}
					(5.5,1) node[cross]{};
			\draw	[dotted](1,1)--(1.5,1)
					(2,1)--(2.5,1)
					(5,1)--(6,1);
			\draw	(3,1)--(3.5,1)
					(4,1)--(4.5,1);
			\draw	(2.5,1.5)--(5,1.5)
					(2.5,.5)--(5,.5);
		\end{tikzpicture}
		\caption{}
	\end{subfigure}
	\hfill
	\begin{subfigure}[t]{.49\textwidth}
	\centering
		\begin{tikzpicture}[]
			\tikzstyle{gauge} = [circle, draw];
			\tikzstyle{flavour} = [regular polygon,regular polygon sides=4,draw];
			\node (g1) [gauge, label=below:{$O_2$}]{};
			\node (f1) [flavour,above of=g1,label=above:{$C_2$}] {};
			\draw (f1)--(g1);
		\end{tikzpicture}
		\caption{}
	\end{subfigure}
	\hfill
	\caption{Quiver whose Higgs branch is $\M_H=\Or_{(2^2)}\subset\gsp (2)$, the closure of the \emph{subregular} nilpotent orbit of $\mathfrak{sp}(2)$. The interval numbers of the half NS5-branes are $\vec{k}_s=(0,0,0,2,2)$, corresponding to partition $\lambda^t=(2^2)$. (a) Higgs branch. (b-f) One step phase transitions to the Coulomb branch. (g) Coulomb branch. (h) Quiver.}
	\label{fig:Sp4subreg}
\end{figure}

This section investigates the KP transitions starting from the model whose Higgs branch is the closure of the maximal nilpotent orbit of $\gsp (2)$. Its Higgs branch brane configuration is depicted again in figure \ref{fig:Sp4maxKP}(a). There is only one possible KP transition allowed. This KP transition can be performed in an analogous way to the ones performed in the previous examples so far, figures \ref{fig:Sp4maxKP}(b-d), in order to obtain a new model, figure \ref{fig:Sp4maxKP}(e). As before, one can perform phase transitions to the Coulomb branch brane configuration of the new model in order to read the corresponding quiver, figure \ref{fig:Sp4subreg}. The parameters of the new model are:

\begin{align}
	\begin{aligned}
		n_s&=5\\
		n_d&=5\\
		\vec{l}_d&=(5,3,5,3,5)\\
		\vec{l}_s&=(1,-1,1,1,3)\\
		\text{rightmost}\ O3&= \widetilde{O3^+} 
	\end{aligned}
\end{align}

The interval numbers of the half NS5-branes $\vec{k}_s$ in the Higgs branch after the collapse transition are:
\begin{align}
	\vec{k}_s=(0,0,0,2,2)\rightarrow \lambda=(2^2)
\end{align}
This indicates that the Higgs branch of the new model is:
\begin{align}
	\M_H=\Or_{(2^2)}
\end{align}
where $\Or_{(2^2)}$ is the closure of the subregular nilpotent orbit of $\gsp (2)$. Note that indeed:
\begin{align}
	Odd(d_{BV}(\lambda))=Odd(d_{BV}(2^2))=Odd(3,1^2)=(1,-1,1,1,3)=\vec{l}_s
\end{align}
as expected.

\paragraph{Brane realization of the C-collapse.} As happened before for the algebras of type D and B, the \emph{collapse} transition in this case realizes the C-collapse of partition $\lambda'=(3,1)$ into partition $\lambda'_C=\lambda=(2^2)$. Performing the inverse transition of the collapse transition on brane system in figure \ref{fig:Sp4subreg}(a) gives interval numbers $\vec k''=(0,0,1,1,3)$. The sum of all elements of $\vec k''$ gives $5$, hence it cannot form a partition in $\mathcal P (4)$. The rightmost half NS5-brane with interval number $3$ can be pushed through its half D5-brane neighbor to the left without brane creation/annihilation. This changes its interval number to $2$, resulting in the system depicted in figure \ref{fig:Sp4Collapse}(a). The interval numbers for this system are:
\begin{align}
	\vec{k}'_s=(0,0,1,1,2)\rightarrow \lambda'=(3,1)
\end{align}
The \emph{collapse} transition takes this system to the brane configuration in figure \ref{fig:Sp4Collapse}(b), with interval numbers $\vec k=(0,0,0,2,2)$, corresponding to partition $\lambda=\lambda'_C=(2^2)$. This result is general and can be applied to all brane systems of C-type on this work.

\paragraph{S-duality.} S-duality can be performed to find a candidate quiver with $\M_C=\Or_{(2^2)}\subset \gsp (2)$, figure \ref{fig:Sp4subregMirror}.

\begin{figure}[t]
	\centering
    	\begin{subfigure}[t]{.45\textwidth}
    	\centering
	\begin{tikzpicture}	
			\draw 	[dashed](1.5,0)--(1.5,2)
					(4,0)--(4,2);
			\draw	[dotted](1,1)--(2,1)
					(2.5,1)--(3,1)
					(3.5,1)--(4.5,1);
			\draw	(2,1)--(2.5,1)
					(3,1)--(3.5,1);
			\draw	(1.5,1.7)--(4,1.7)
					(1.5,.3)--(4,.3);
			\draw 	(1,1) node[circ]{}
					(2,1) node[circ]{}
					(2.5,1) node[circ]{}
					(3,1) node[circ]{}
					(3.5,1) node[circ]{};
			\draw 	(1,1) node[cross]{}
					(2,1) node[cross]{}
					(2.5,1) node[cross]{}
					(3,1) node[cross]{}
					(3.5,1) node[cross]{};
	\end{tikzpicture}
        \caption{}
    	\end{subfigure}
	\hfill
    	\begin{subfigure}[t]{.45\textwidth}
    	\centering
	\begin{tikzpicture}
			\draw 	[dashed](2,0)--(2,2)
					(3.5,0)--(3.5,2);
			\draw	[dotted](1,1)--(1.5,1)
					(2.5,1)--(3,1)
					(4,1)--(4.5,1);
			\draw	(2,1)--(2.5,1)
					(3,1)--(3.5,1);
			\draw	(1.5,1.05)--(2,1.05)
					(1.5,.95)--(2,.95);
			\draw	(2,1.7)--(3.5,1.7)
					(2,.3)--(3.5,.3);
			\draw	(3.5,1.05)--(4,1.05)
					(3.5,.95)--(4,.95);
			\draw 	(1,1) node[circ]{}
					(1.5,1) node[circ]{}
					(2.5,1) node[circ]{}
					(3,1) node[circ]{}
					(4,1) node[circ]{};
			\draw 	(1,1) node[cross]{}
					(1.5,1) node[cross]{}
					(2.5,1) node[cross]{}
					(3,1) node[cross]{}
					(4,1) node[cross]{};
	\end{tikzpicture}
        \caption{}
    	\end{subfigure}
		\hfill
\begin{subfigure}[t]{.45\textwidth}
    	\centering
	\begin{tikzpicture}
			\draw 	(1,0)--(1,2)
					(1.5,0)--(1.5,2)
					(2.5,0)--(2.5,2)
					(3,0)--(3,2)
					(4,0)--(4,2);
			\draw	[dotted](1,1)--(1.5,1)
					(2.5,1)--(3,1)
					(4,1)--(4.5,1);
			\draw	(2,1)--(2.5,1)
					(3,1)--(3.5,1);
			\draw	(1.5,1.5)--(2.5,1.5)
					(1.5,.5)--(2.5,.5);
			\draw	(2.5,1.7)--(3,1.7)
					(2.5,.3)--(3,.3);
			\draw	(3,1.5)--(4,1.5)
					(3,.5)--(4,.5);
			\draw 	(2,1) node[cross]{}
					(3.5,1) node[cross]{};
	\end{tikzpicture}
        \caption{}
    	\end{subfigure}
		\hfill
\begin{subfigure}[t]{.45\textwidth}
    	\centering
	\begin{tikzpicture}
			\draw 	(1,0)--(1,2)
					(1.5,0)--(1.5,2)
					(2,0)--(2,2)
					(3.5,0)--(3.5,2)
					(4,0)--(4,2);
			\draw	[dotted](1,1)--(1.5,1)
					(2,1)--(3.5,1)
					(4,1)--(4.5,1);
			\draw	(1.5,1.5)--(2,1.5)
					(1.5,.5)--(2,.5);
			\draw	(2,1.7)--(3.5,1.7)
					(2,.3)--(3.5,.3);
			\draw	(3.5,1.5)--(4,1.5)
					(3.5,.5)--(4,.5);
			\draw 	(2.5,1) node[cross]{}
					(3,1) node[cross]{};
	\end{tikzpicture}
        \caption{}
    	\end{subfigure}
		\hfill
	\begin{subfigure}[t]{.45\textwidth}
    \centering
	\begin{tikzpicture}
		\tikzstyle{gauge} = [circle, draw];
	\tikzstyle{flavour} = [regular polygon,regular polygon sides=4, draw];
	\node (g2) [gauge,label=below:{$O_2$}] {};
	\node (g3) [gauge, right of=g2,label=below:{$C_1$}] {};
	\node (g4) [gauge, right of=g3,label=below:{$O_2$}] {};
	\node (f3) [flavour,above of=g3,label=above:{$O_{2}$}] {};
	\draw (g2)--(g3)--(g4)
			(g3)--(f3)
		;
	\end{tikzpicture}
	\caption{}
    \end{subfigure}
 	\caption{Transverse slice $S=D_3$ that can be removed via Kraft-Procesi transition. (a) Local brane configuration. (b) Phase transition towards the Coulomb branch brane configuration of the local system. (c) Coulomb branch brane configuration. (d) A \emph{collapse} transition is performed in (c) in order to read the quiver. (e) Quiver, its Higgs branch is $\M_H=D_3$, see the computation in appendix \ref{app:DnHiggs}.}
	\label{fig:Sp2maxInst}
\end{figure}
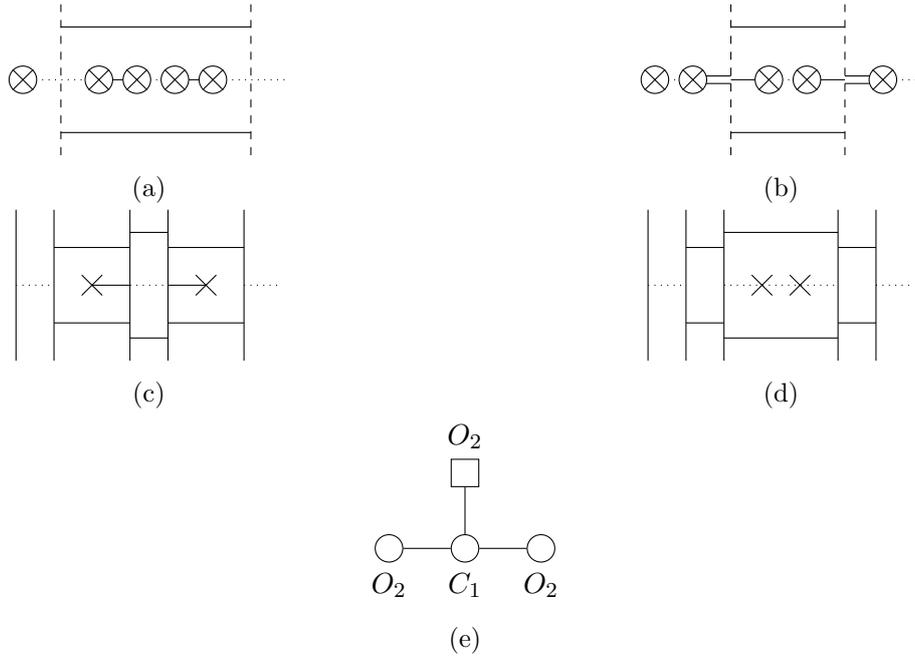

\paragraph{Transverse slice.} In order to find the slice $S\subseteq \Or_{(4)}$ transverse to $\mathcal{O}_{(2^2)}$ that has been \emph{removed} and that gives name to the KP transition, one focuses on the local brane system around the D3-brane that is \emph{Higgsed away} during the transition, figure \ref{fig:Sp2maxInst}(a). A phase transition can be performed, figure \ref{fig:Sp2maxInst}(b-d), in order to find the quiver for which this local brane system constitutes the Higgs branch, figure \ref{fig:Sp2maxInst}(e). The Higgs branch of this quiver is a surface Kleinian singularity of the type $\M_H=D_3$ (see appendix \ref{app:DnHiggs} for the computation of the corresponding Hilbert series). Therefore:

\begin{align}
	S=D_3
\end{align}

Note that this coincides with the mathematical result by Brieskorn and Slodowy \cite{B70,Sl80}: \emph{the slice of the closure of the maximal nilpotent orbit of $\gsp (n)$, $S\subseteq \Or_{(2n)}$, transverse to the subregular nilpotent orbit $\mathcal O_{(2n-2,2)}$ is $S=D_{n+1}$}. We say that there is a $D_3$ KP transition between the closures of orbits corresponding to partitions $(4)$ and $(2^2)$ in $\gsp (2)$.

\begin{figure}[t]
	\centering
	\begin{subfigure}[t]{.49\textwidth}
		\centering
		\begin{tikzpicture}
			\draw 	[dashed](2,0)--(2,2)
					(3.5,0)--(3.5,2)
					(4.5,0)--(4.5,2)
					(5,0)--(5,2)
					(5.5,0)--(5.5,2);
			\draw	[dotted](1,1)--(1.5,1)
					(2.5,1)--(3,1)
					(4,1)--(6,1);
			\draw	(2,1)--(2.5,1)
					(3,1)--(3.5,1);
			\draw	(2,1.3)--(3.5,1.3)
					(2,.7)--(3.5,.7)
					(3.5,1.5)--(4.5,1.5)
					(3.5,.5)--(4.5,.5)
					(4.5,1.3)--(5,1.3)
					(4.5,.7)--(5,.7);
			\draw 	(1,1) node[circ]{}
					(1.5,1) node[circ]{}
					(2.5,1) node[circ]{}
					(3,1) node[circ]{}
					(4,1) node[circ]{};
			\draw 	(1,1) node[cross]{}
					(1.5,1) node[cross]{}
					(2.5,1) node[cross]{}
					(3,1) node[cross]{}
					(4,1) node[cross]{};
		\end{tikzpicture}
		\caption{}
	\end{subfigure}
	\hfill
	\begin{subfigure}[t]{.49\textwidth}
		\centering
		\begin{tikzpicture}
			\draw 	[dashed](2.5,0)--(2.5,2)
					(3,0)--(3,2)
					(4.5,0)--(4.5,2)
					(5,0)--(5,2)
					(5.5,0)--(5.5,2);
			\draw	[dotted](1,1)--(1.5,1)
					(2,1)--(3.5,1)
					(4,1)--(6,1);
			\draw	(2.5,1.3)--(3,1.3)
					(2.5,.7)--(3,.7)
					(3,1.5)--(4.5,1.5)
					(3,.5)--(4.5,.5)
					(4.5,1.3)--(5,1.3)
					(4.5,.7)--(5,.7);
			\draw 	(1,1) node[circ]{}
					(1.5,1) node[circ]{}
					(2,1) node[circ]{}
					(3.5,1) node[circ]{}
					(4,1) node[circ]{};
			\draw 	(1,1) node[cross]{}
					(1.5,1) node[cross]{}
					(2,1) node[cross]{}
					(3.5,1) node[cross]{}
					(4,1) node[cross]{};
		\end{tikzpicture}
		\caption{}
	\end{subfigure}
	\caption{Brane realization of the C-collapse of partitions $(3,1)_C=(2^2)$. (a) Has interval numbers for the half NS5-branes $\vec k'=(0,0,1,1,2)$, corresponding to partition $\lambda'=(3,1)$. (b) Is obtained after performing the \emph{collapse} transition on (a). The interval numbers of (b) are $\vec k=(0,0,0,2,2)$, corresponding to partition $\lambda=(2^2)=\lambda'_C$.}
	\label{fig:Sp4Collapse}
\end{figure}
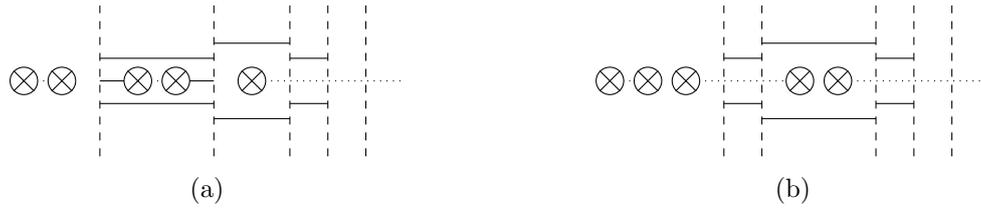

\begin{figure}[t]
	\centering
	\begin{subfigure}[t]{.49\textwidth}
		\centering
		\begin{tikzpicture}
			\draw 	[dashed](2.5,0)--(2.5,2)
					(3,0)--(3,2)
					(4.5,0)--(4.5,2)
					(5,0)--(5,2)
					(5.5,0)--(5.5,2);
			\draw	[dotted](1,1)--(1.5,1)
					(2,1)--(3.5,1)
					(4,1)--(6,1);
			\draw	(2.5,1.3)--(3,1.3)
					(2.5,.7)--(3,.7)
					(3,1.5)--(4.5,1.5)
					(3,.5)--(4.5,.5)
					(4.5,1.3)--(5,1.3)
					(4.5,.7)--(5,.7);
			\draw 	(1,1) node[circ]{}
					(1.5,1) node[circ]{}
					(2,1) node[circ]{}
					(3.5,1) node[circ]{}
					(4,1) node[circ]{};
			\draw 	(1,1) node[cross]{}
					(1.5,1) node[cross]{}
					(2,1) node[cross]{}
					(3.5,1) node[cross]{}
					(4,1) node[cross]{};
		\end{tikzpicture}
		\caption{}
	\end{subfigure}
	\hfill
	\begin{subfigure}[t]{.49\textwidth}
		\centering
		\begin{tikzpicture}
			\draw 	(1,1) node[cross]{}
					(1.5,1) node[cross]{}
					(2,1) node[cross]{}
					(3.5,1) node[cross]{}
					(4,1) node[cross]{};
			\draw 	(2.5,0)--(2.5,2)
					(3,0)--(3,2)
					(4.5,0)--(4.5,2)
					(5,0)--(5,2)
					(5.5,0)--(5.5,2);
			\draw	(1,1)--(1.5,1)
					(2,1)--(2.5,1)
					(3,1)--(3.5,1)
					(4,1)--(4.5,1)
					(5,1)--(5.5,1);
			\draw	[dotted] (2.5,1)--(3,1)
					(4.5,1)--(5,1)
					(5.5,1)--(6,1);
			\draw	(2.5,1.3)--(3,1.3)
					(2.5,.7)--(3,.7)
					(3,1.5)--(4.5,1.5)
					(3,.5)--(4.5,.5)
					(4.5,1.3)--(5,1.3)
					(4.5,.7)--(5,.7);
		\end{tikzpicture}
		\caption{}
	\end{subfigure}
	\hfill
	\begin{subfigure}[t]{.49\textwidth}
		\centering
		\begin{tikzpicture}
			\draw 	(1,1) node[cross]{}
					(1.5,1) node[cross]{}
					(2.5,1) node[cross]{}
					(3,1) node[cross]{}
					(4.5,1) node[cross]{};
			\draw 	(2,0)--(2,2)
					(3.5,0)--(3.5,2)
					(4,0)--(4,2)
					(5,0)--(5,2)
					(5.5,0)--(5.5,2);
			\draw	(1,1)--(1.5,1)
					(5,1)--(5.5,1);
			\draw	[dotted] (2,1)--(3.5,1)
					(4,1)--(5,1)
					(5.5,1)--(6,1);
			\draw	(2,1.3)--(3.5,1.3)
					(2,.7)--(3.5,.7)
					(3.5,1.5)--(4,1.5)
					(3.5,.5)--(4,.5)
					(4,1.3)--(5,1.3)
					(4,.7)--(5,.7);
		\end{tikzpicture}
		\caption{}
	\end{subfigure}
	\hfill
	\begin{subfigure}[t]{.49\textwidth}
	\centering
		\begin{tikzpicture}[]
			\tikzstyle{gauge} = [circle, draw];
			\tikzstyle{flavour} = [regular polygon,regular polygon sides=4,draw];
			\node (g1) [gauge, label=below:{$C_1$}]{};
			\node (g2) [gauge, right of=g1, label=below:{$O_2$}]{};
			\node (g3) [gauge, right of=g2, label=below:{$C_1$}]{};
			\node (g4) [gauge, right of=g3, label=below:{$O_1$}]{};
			\node (f1) [flavour,above of=g1,label=above:{$O_2$}] {};
			\node (f3) [flavour,above of=g3,label=above:{$O_1$}] {};
			\draw (f1)--(g1)--(g2)--(g3)--(g4)
					(f3)--(g3);
		\end{tikzpicture}
		\caption{}
	\end{subfigure}
	\hfill
	\caption{S-duality transformation of the model in figure \ref{fig:Sp4subreg}(h). (a) Higgs branch of the initial model. (b) Result of performing S-duality on (a). (c) Result of performing the \emph{collapse} transition on (b). (d) Quiver read from (c). The Coulomb branch of this quiver is predicted to be $\M_C=\Or_{(2^2)}$.}
	\label{fig:Sp4subregMirror}
\end{figure}
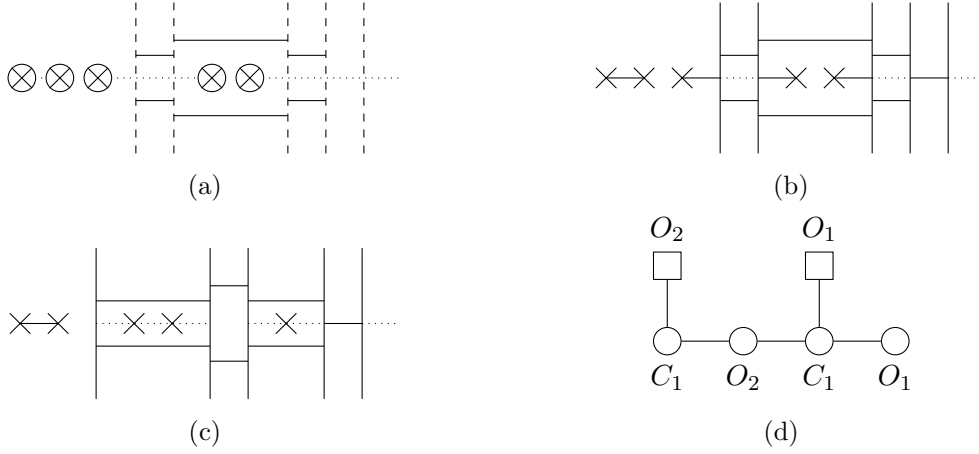

\subsubsection{$\Or_{(2^2)}$ transition}

Let us once more take the newly found model as a starting point for a KP transition. It has Higgs branch:

\begin{align}
	\M_H=\Or_{(2^2)}\subset \gsp (2)
\end{align}

Consider the Higgs branch brane configuration depicted in figure \ref{fig:Sp4maxKP}(e). We ask the question again: \emph{what is the minimal singularity that can be removed?} The answer is that all three physical D3-branes need to be removed, as it happened in the case of the KP transition from the $\mathcal O_{(3,1^2)}$ orbit of $\gso(5)$. In this case, one can start the transition by thinking that only a single D3-brane has been removed, figure \ref{fig:Sp4subregObstruction}. However, when the fixed D3-branes are annihilated in order to reach a new Higgs branch, all other D3-branes are also annihilated and one finds the trivial orbit as the resulting Higgs branch.
 
\begin{figure}[ht]
	\centering
	\begin{subfigure}[t]{.49\textwidth}
		\centering
		\begin{tikzpicture}
			\draw 	[dashed](2.5,0)--(2.5,2)
					(3,0)--(3,2)
					(4.5,0)--(4.5,2)
					(5,0)--(5,2)
					(5.5,0)--(5.5,2);
			\draw	[dotted](1,1)--(1.5,1)
					(2,1)--(3.5,1)
					(4,1)--(6,1);
			\draw	(2.5,1.3)--(3,1.3)
					(2.5,.7)--(3,.7)
					(3,1.5)--(4.5,1.5)
					(3,.5)--(4.5,.5)
					(4.5,1.3)--(5,1.3)
					(4.5,.7)--(5,.7);
			\draw 	(1,1) node[circ]{}
					(1.5,1) node[circ]{}
					(2,1) node[circ]{}
					(3.5,1) node[circ]{}
					(4,1) node[circ]{};
			\draw 	(1,1) node[cross]{}
					(1.5,1) node[cross]{}
					(2,1) node[cross]{}
					(3.5,1) node[cross]{}
					(4,1) node[cross]{};
		\end{tikzpicture}
		\caption{}
	\end{subfigure}
	\hfill
	\begin{subfigure}[t]{.49\textwidth}
		\centering
		\begin{tikzpicture}
			\draw 	[dashed](2.5,0)--(2.5,2)
					(3.5,0)--(3.5,2)
					(4,0)--(4,2)
					(5,0)--(5,2)
					(5.5,0)--(5.5,2);
			\draw	[dotted](1,1)--(1.5,1)
					(2,1)--(3,1)
					(4.5,1)--(6,1);
			\draw	(3,1)--(3.5,1)
					(4,1)--(4.5,1);
			\draw	(2.5,1.3)--(3.5,1.3)
					(2.5,.7)--(3.5,.7)
					(3.5,1.5)--(4,1.5)
					(3.5,.5)--(4,.5)
					(4,1.3)--(5,1.3)
					(4,.7)--(5,.7);
			\draw 	(1,1) node[circ]{}
					(1.5,1) node[circ]{}
					(2,1) node[circ]{}
					(3,1) node[circ]{}
					(4.5,1) node[circ]{};
			\draw 	(1,1) node[cross]{}
					(1.5,1) node[cross]{}
					(2,1) node[cross]{}
					(3,1) node[cross]{}
					(4.5,1) node[cross]{};
		\end{tikzpicture}
		\caption{}
	\end{subfigure}
	\hfill
	\begin{subfigure}[t]{.49\textwidth}
		\centering
		\begin{tikzpicture}
			\draw 	[dashed](2.5,0)--(2.5,2)
					(3.5,0)--(3.5,2)
					(4,0)--(4,2)
					(5,0)--(5,2)
					(5.5,0)--(5.5,2);
			\draw	[dotted](1,1)--(1.5,1)
					(2,1)--(3,1)
					(4.5,1)--(6,1);
			\draw	(3,1)--(3.5,1)
					(4,1)--(4.5,1);
			\draw	(2.5,1.05)--(5,1.05)
					(2.5,.95)--(5,.95);
			\draw 	(1,1) node[circ]{}
					(1.5,1) node[circ]{}
					(2,1) node[circ]{}
					(3,1) node[circ]{}
					(4.5,1) node[circ]{};
			\draw 	(1,1) node[cross]{}
					(1.5,1) node[cross]{}
					(2,1) node[cross]{}
					(3,1) node[cross]{}
					(4.5,1) node[cross]{};
		\end{tikzpicture}
		\caption{}
	\end{subfigure}
	\hfill
	\begin{subfigure}[t]{.49\textwidth}
		\centering
		\begin{tikzpicture}
			\draw 	[dashed](2.5,0)--(2.5,2)
					(3.5,0)--(3.5,2)
					(4,0)--(4,2)
					(5,0)--(5,2)
					(5.5,0)--(5.5,2);
			\draw	[dotted](1,1)--(1.5,1)
					(2,1)--(3,1)
					(4.5,1)--(6,1);
			\draw	(3,1)--(3.5,1)
					(4,1)--(4.5,1);
			\draw	(2.5,1.05)--(2.95,1.05)
					(2.5,.95)--(2.95,.95)
					(4.55,1.05)--(5,1.05)
					(4.55,.95)--(5,.95);
			\draw 	(1,1) node[circ]{}
					(1.5,1) node[circ]{}
					(2,1) node[circ]{}
					(3,1) node[circ]{}
					(4.5,1) node[circ]{};
			\draw 	(1,1) node[cross]{}
					(1.5,1) node[cross]{}
					(2,1) node[cross]{}
					(3,1) node[cross]{}
					(4.5,1) node[cross]{};
		\end{tikzpicture}
		\caption{}
	\end{subfigure}
	\hfill
	\begin{subfigure}[t]{.49\textwidth}
		\centering
		\begin{tikzpicture}
			\draw 	[dashed](3,0)--(3,2)
					(3.5,0)--(3.5,2)
					(4,0)--(4,2)
					(4.5,0)--(4.5,2)
					(5.5,0)--(5.5,2);
			\draw	[dotted](1,1)--(1.5,1)
					(2,1)--(2.5,1)
					(5,1)--(6,1);
			\draw	(3,1)--(3.5,1)
					(4,1)--(4.5,1);
			\draw 	(1,1) node[circ]{}
					(1.5,1) node[circ]{}
					(2,1) node[circ]{}
					(2.5,1) node[circ]{}
					(5,1) node[circ]{};
			\draw 	(1,1) node[cross]{}
					(1.5,1) node[cross]{}
					(2,1) node[cross]{}
					(2.5,1) node[cross]{}
					(5,1) node[cross]{};
		\end{tikzpicture}
		\caption{}
	\end{subfigure}
	\hfill
	\caption{$\Or_{(2^2)}$ KP transition from the closure of the subregular nilpotent orbit to the trivial orbit of $\gsp (2)$. (a) Higgs branch brane configuration of the initial system. (b) Phase transition that moves two of the half NS5-branes without brane creation/annihilation. (c) The three D3-branes are taken to the origin. They realign into a single D3-brane and then split into three new segments: the segment in the middle ends on both half NS5-branes. (d) This middle segment can acquire non zero $\vec x$ position along the directions spanned by the half NS5-branes. The limit is taken where this position goes to infinity, effectively removing the threebrane from the system. (e) The fixed segments of D3-branes from are annihilated via phase transitions. The resulting Higgs branch is the trivial orbit.}
	\label{fig:Sp4subregKP}
\end{figure}
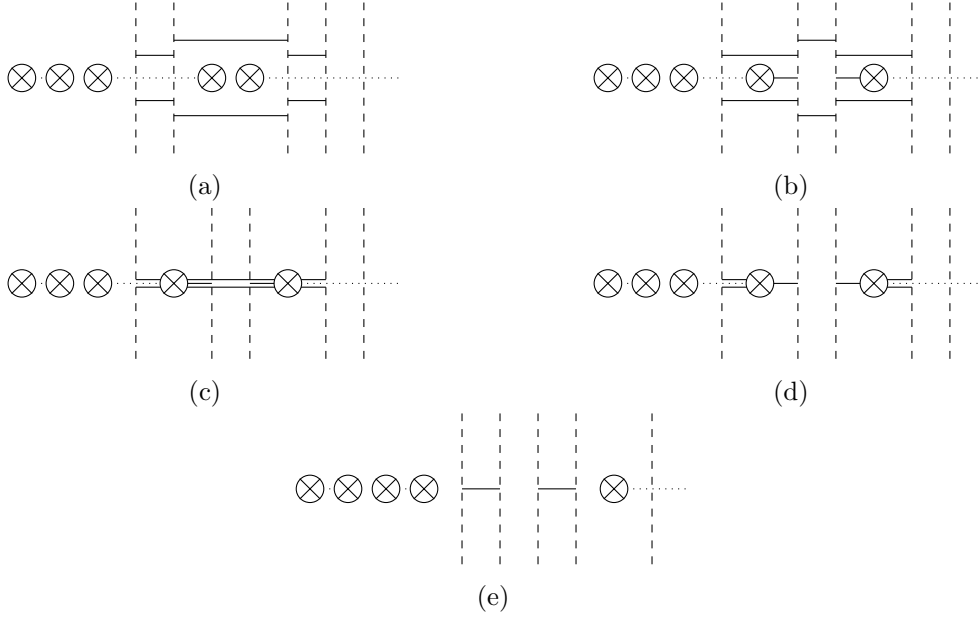

Alternatively, one can rearrange the half NS5-branes and \emph{remove} all the D3-branes from the beginning of the KP transition. This process step by step is depicted in figure \ref{fig:Sp4subregKP}. Note that if one tries to remove the rightmost D3-brane in figure \ref{fig:Sp4maxKP}(e) after giving two half NS5-branes interval number equal to $1$, one finds that the resulting system breaks the supersymmetric configuration, as it is the case in section \ref{sec:311transition}.

\begin{figure}[t]
	\centering
	\begin{subfigure}[t]{.49\textwidth}
		\centering
		\begin{tikzpicture}
			\draw 	[dashed](2.5,0)--(2.5,2)
					(3,0)--(3,2)
					(4.5,0)--(4.5,2)
					(5,0)--(5,2)
					(5.5,0)--(5.5,2);
			\draw	[dotted](1,1)--(1.5,1)
					(2,1)--(3.5,1)
					(4,1)--(6,1);
			\draw	(2.5,1.3)--(3,1.3)
					(2.5,.7)--(3,.7)
					(3,1.5)--(4.5,1.5)
					(3,.5)--(4.5,.5)
					(4.5,1.3)--(5,1.3)
					(4.5,.7)--(5,.7);
			\draw 	(1,1) node[circ]{}
					(1.5,1) node[circ]{}
					(2,1) node[circ]{}
					(3.5,1) node[circ]{}
					(4,1) node[circ]{};
			\draw 	(1,1) node[cross]{}
					(1.5,1) node[cross]{}
					(2,1) node[cross]{}
					(3.5,1) node[cross]{}
					(4,1) node[cross]{};
		\end{tikzpicture}
		\caption{}
	\end{subfigure}
	\hfill
	\begin{subfigure}[t]{.49\textwidth}
		\centering
		\begin{tikzpicture}
			\draw 	[dashed](2.5,0)--(2.5,2)
					(3,0)--(3,2)
					(4.5,0)--(4.5,2)
					(5,0)--(5,2)
					(5.5,0)--(5.5,2);
			\draw	[dotted](1,1)--(1.5,1)
					(2,1)--(3.5,1)
					(4,1)--(6,1);
			\draw	(2.5,1.3)--(3,1.3)
					(2.5,.7)--(3,.7)
					(3,1.05)--(4.5,1.05)
					(3,.95)--(4.5,.95)
					(4.5,1.3)--(5,1.3)
					(4.5,.7)--(5,.7);
			\draw 	(1,1) node[circ]{}
					(1.5,1) node[circ]{}
					(2,1) node[circ]{}
					(3.5,1) node[circ]{}
					(4,1) node[circ]{};
			\draw 	(1,1) node[cross]{}
					(1.5,1) node[cross]{}
					(2,1) node[cross]{}
					(3.5,1) node[cross]{}
					(4,1) node[cross]{};
		\end{tikzpicture}
		\caption{}
	\end{subfigure}
	\hfill
	\begin{subfigure}[t]{.49\textwidth}
		\centering
		\begin{tikzpicture}
			\draw 	[dashed](2.5,0)--(2.5,2)
					(3,0)--(3,2)
					(4.5,0)--(4.5,2)
					(5,0)--(5,2)
					(5.5,0)--(5.5,2);
			\draw	[dotted](1,1)--(1.5,1)
					(2,1)--(3.5,1)
					(4,1)--(6,1);
			\draw	(2.5,1.3)--(3,1.3)
					(2.5,.7)--(3,.7)
					(3,1.05)--(3.5,1.05)
					(3,.95)--(3.5,.95)
					(4,1.05)--(4.5,1.05)
					(4,.95)--(4.5,.95)
					(4.5,1.3)--(5,1.3)
					(4.5,.7)--(5,.7);
			\draw 	(1,1) node[circ]{}
					(1.5,1) node[circ]{}
					(2,1) node[circ]{}
					(3.5,1) node[circ]{}
					(4,1) node[circ]{};
			\draw 	(1,1) node[cross]{}
					(1.5,1) node[cross]{}
					(2,1) node[cross]{}
					(3.5,1) node[cross]{}
					(4,1) node[cross]{};
		\end{tikzpicture}
		\caption{}
	\end{subfigure}
	\hfill
	\begin{subfigure}[t]{.49\textwidth}
		\centering
		\begin{tikzpicture}
			\draw 	[dashed](2.5,0)--(2.5,2)
					(3.5,0)--(3.5,2)
					(4,0)--(4,2)
					(5,0)--(5,2)
					(5.5,0)--(5.5,2);
			\draw	[dotted](1,1)--(1.5,1)
					(2,1)--(3,1)
					(4.5,1)--(6,1);
			\draw	(3,1)--(3.5,1)
					(4,1)--(4.5,1);
			\draw	(2.5,1.05)--(2.95,1.05)
					(2.5,.95)--(2.95,.95)
					(4.55,1.05)--(5,1.05)
					(4.55,.95)--(5,.95);
			\draw 	(1,1) node[circ]{}
					(1.5,1) node[circ]{}
					(2,1) node[circ]{}
					(3,1) node[circ]{}
					(4.5,1) node[circ]{};
			\draw 	(1,1) node[cross]{}
					(1.5,1) node[cross]{}
					(2,1) node[cross]{}
					(3,1) node[cross]{}
					(4.5,1) node[cross]{};
		\end{tikzpicture}
		\caption{}
	\end{subfigure}
	\hfill
	\begin{subfigure}[t]{.49\textwidth}
		\centering
		\begin{tikzpicture}
			\draw 	[dashed](3,0)--(3,2)
					(3.5,0)--(3.5,2)
					(4,0)--(4,2)
					(4.5,0)--(4.5,2)
					(5.5,0)--(5.5,2);
			\draw	[dotted](1,1)--(1.5,1)
					(2,1)--(2.5,1)
					(5,1)--(6,1);
			\draw	(3,1)--(3.5,1)
					(4,1)--(4.5,1);
			\draw 	(1,1) node[circ]{}
					(1.5,1) node[circ]{}
					(2,1) node[circ]{}
					(2.5,1) node[circ]{}
					(5,1) node[circ]{};
			\draw 	(1,1) node[cross]{}
					(1.5,1) node[cross]{}
					(2,1) node[cross]{}
					(2.5,1) node[cross]{}
					(5,1) node[cross]{};
		\end{tikzpicture}
		\caption{}
	\end{subfigure}
	\hfill
	\caption{$\Or_{(2^2)}$ KP transition from the closure of the subregular nilpotent orbit to the trivial orbit of $\gsp (2)$. (a) Higgs branch brane configuration of the initial system. (b) One focusses on the D3-brane in the middle interval. This brane is taken to the origin and split into three segments. (c) The middle segment can acquire non zero $\vec x$ position along the directions spanned by the half NS5-branes. The limit is taken where this position goes to infinity, effectively removing the threebrane from the system. (d-e) An attempt to remove the fixed segments of D3-branes from (c). During these phase transitions the remaining D3-branes that could have generated a new Higgs branch are also annihilated. The resulting Higgs branch is the trivial orbit.}
	\label{fig:Sp4subregObstruction}
\end{figure}
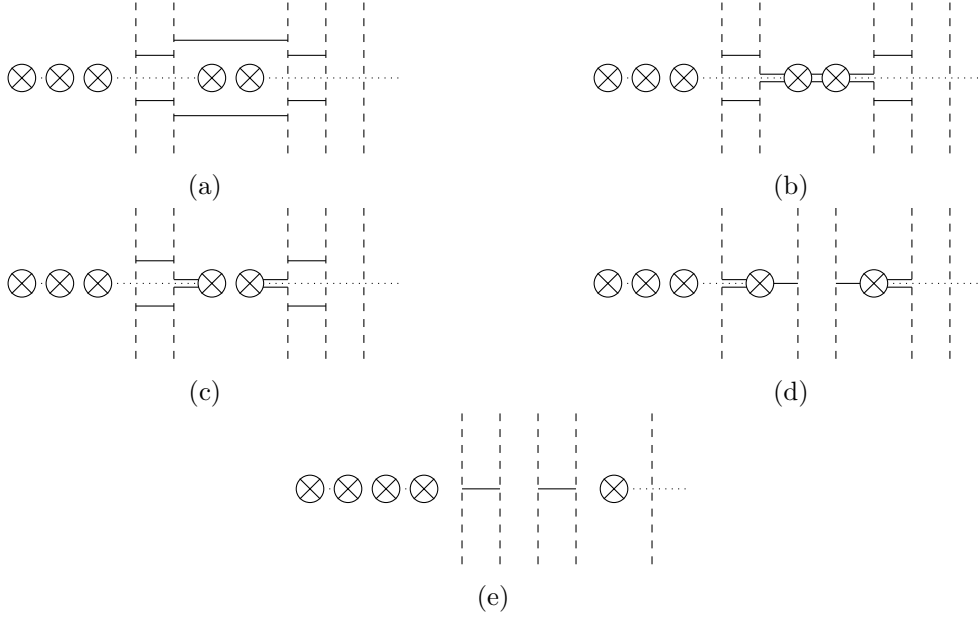

\subsubsection{Hasse diagram for $\gsp (2)$}

The corresponding Hasse diagram can be found in figure \ref{fig:HasseSp4}. Note that only the special orbits appear in this diagram. Once again, this diagram differs with the one found in \cite{KP82} in that the non-special orbit corresponding to partition $\lambda=(2,1^2)$ does not appear and  a transition from the next to minimal orbit with partition $\lambda=(2^2)$ to the trivial orbit has been found instead.

\begin{figure}[t]
	\centering
\begin{subfigure}[t]{.45\textwidth}
    \centering
	\begin{tikzpicture}
		\node [hasse] (1) [] {};
		\node [hasse] (2) [below of=1] {};
		\node [hasse] (3) [below of=2] {};
		\draw (1) edge [] node[label=left:$A_3$] {} (2)
			(2) edge [] node[label=left:$\Or_{(2^2)}$] {} (3);
		\node (e1) [right of=1] {};
		\node (d1) [right of=e1] {$(1,1,1,1,1)$};
		\node (c1) [right of=d1] {};
		\node (m1) [right of=c1] {};
		\node (b1) [right of=m1] {$(5,3,5,3,5)$};
		\node (e2) [right of=2] {};
		\node (d2) [right of=e2] {$(1,-1,1,1,3)$};
		\node (c2) [right of=d2] {};
		\node (m2) [right of=c2] {};
		\node (b2) [right of=m2] {$(5,3,5,3,5)$};
		\node (e3) [right of=3] {};
		\node (d3) [right of=e3] {$(1,-1,1,-1,5)$};
		\node (c3) [right of=d3] {};
		\node (m3) [right of=c3] {};
		\node (b3) [right of=m3] {$(5,3,5,3,5)$};
		\node (d) [above of=d1] {$\vec l_s$};
		\node (b) [above of=b1] {$\vec l_d$};
	\end{tikzpicture}
	\caption{}
 	\end{subfigure}
 	\hfill
 	\begin{subfigure}[t]{.45\textwidth}
 	\centering
 	\begin{tikzpicture}
		\node [hasse] (1) [] {};
		\node [hasse] (2) [below of=1] {};
		\node [hasse] (3) [below of=2] {};
		\draw (1) edge [] node[label=left:$A_3$] {} (2)
			(2) edge [] node[label=left:$\Or_{(2^2)}$] {} (3);
		\node (e1) [right of=1] {};
		\node (d1) [right of=e1] {$(4)$};
		\node (c1) [right of=d1] {};
		\node (b1) [right of=c1] {$4$};
		\node (e2) [right of=2] {};
		\node (d2) [right of=e2] {$(2^2)$};
		\node (c2) [right of=d2] {};
		\node (b2) [right of=c2] {$3$};
		\node (e3) [right of=3] {};
		\node (d3) [right of=e3] {$(1^4)$};
		\node (c3) [right of=d3] {};
		\node (b3) [right of=c3] {$0$};
		\node (d) [above of=d1] {$\lambda$};
		\node (b) [above of=b1] {$dim$};
	\end{tikzpicture}
	\caption{}
 	\end{subfigure}
 	\hfill
	\caption{Hasse diagram for the models whose Higgs branch is the closure of a special nilpotent orbit of $\gsp (2)$ under the adjoint action of the group $Sp(2)$. (a) represents the brane configurations, where the linking numbers $\vec{l}_s$ and $\vec{l}_d$ are provided for each orbit and the rightmost orientifold plane is always of type $\widetilde{O3^+}$. (b) depicts the information of the orbits. $\lambda$ is the partition obtained from the interval numbers $\vec{k}_s$ as discussed above. \emph{dim} is the number of physical D3-branes that generate the Higgs branch in each model.}
	\label{fig:HasseSp4}
\end{figure}

\begin{table}[t]
	\centering
	\begin{tabular}{ c c}
	  
		\raisebox{-.5\height}{\begin{tikzpicture}[node distance=80pt]
		\node at (0,0)[]{\large{$\mathfrak{sp}(2)$}};
		\node at (0,-0.5) [hasse] (1) [] {};
		\node [hasse] (2) [below of=1] {};
		\node [hasse] (3) [below of=2] {};
		\draw (1) edge [] node[label=left:$D_3$] {} (2)
			(2) edge [] node[label=left:$\Or_{(2^2)}$] {} (3);
	\end{tikzpicture}}
	&
	\fontsize{10}{11}\selectfont
	\begin{tabular}{ c c c  c}
	\toprule
	\textbf{Partition}  & \textbf{Branes} & $\M_H$  & $\M_C$  \\ 
	\midrule \addlinespace[3ex]
$\mathbf{4}$ & \raisebox{-.4\height}{\begin{tikzpicture}[scale=.8, every node/.style={transform shape}]	
			\draw 	[dashed](1.5,0)--(1.5,2)
					(4,0)--(4,2)
					(4.5,0)--(4.5,2)
					(5,0)--(5,2)
					(5.5,0)--(5.5,2);
			\draw	[dotted](1,1)--(2,1)
					(2.5,1)--(3,1)
					(3.5,1)--(6,1);
			\draw	(2,1)--(2.5,1)
					(3,1)--(3.5,1);
			\draw	(1.5,1.7)--(4,1.7)
					(1.5,.3)--(4,.3)
					(1.5,1.3)--(4,1.3)
					(1.5,.7)--(4,.7)
					(4,1.5)--(4.5,1.5)
					(4,.5)--(4.5,.5)
					(4.5,1.3)--(5,1.3)
					(4.5,.7)--(5,.7);
			\draw 	(1,1) node[circ]{}
					(2,1) node[circ]{}
					(2.5,1) node[circ]{}
					(3,1) node[circ]{}
					(3.5,1) node[circ]{};
			\draw 	(1,1) node[cross]{}
					(2,1) node[cross]{}
					(2.5,1) node[cross]{}
					(3,1) node[cross]{}
					(3.5,1) node[cross]{};
	\end{tikzpicture}}
& \raisebox{-.4\height}{\begin{tikzpicture}[scale=1, every node/.style={transform shape},node distance=20]
	\tikzstyle{gauge} = [circle, draw];
	\tikzstyle{flavour} = [regular polygon,regular polygon sides=4,draw];
	\node (g2) [gauge,label=below:{$O_2$}] {};
	\node (g3) [gauge,right of=g2,label=below:{$C_1$}] {};
	\node (g4) [gauge,right of=g3,label=below:{$O_4$}] {};
	\node (f4) [flavour,above of=g4,label=above:{$C_2$}] {};
	\draw (g2)--(g3)--(g4)--(f4);
		\end{tikzpicture}} & \raisebox{-.4\height}{\begin{tikzpicture}[scale=1, every node/.style={transform shape},node distance=20]
			\tikzstyle{gauge} = [circle, draw];
			\tikzstyle{flavour} = [regular polygon,regular polygon sides=4,draw];
			\node (g1) [gauge, label=below:{$C_2$}]{};
			\node (g2) [gauge, right of=g1, label=below:{$O_3$}]{};
			\node (g3) [gauge,right of=g2, label=below:{$C_1$}]{};
			\node (g4) [gauge,right of=g3, label=below:{$O_1$}]{};
			\node (f1) [flavour,above of=g1,label=above:{$O_5$}] {};
			\draw (f1)--(g1)--(g2)--(g3)--(g4);		\end{tikzpicture}}
 \\
  \addlinespace[5ex]
 $\mathbf{2^2}$ & 
		\raisebox{-.4\height}{\begin{tikzpicture}[scale=.8, every node/.style={transform shape}]	
			\draw 	[dashed](2.5,0)--(2.5,2)
					(3,0)--(3,2)
					(4.5,0)--(4.5,2)
					(5,0)--(5,2)
					(5.5,0)--(5.5,2);
			\draw	[dotted](1,1)--(1.5,1)
					(2,1)--(3.5,1)
					(4,1)--(6,1);
			\draw	(2.5,1.3)--(3,1.3)
					(2.5,.7)--(3,.7)
					(3,1.5)--(4.5,1.5)
					(3,.5)--(4.5,.5)
					(4.5,1.3)--(5,1.3)
					(4.5,.7)--(5,.7);
			\draw 	(1,1) node[circ]{}
					(1.5,1) node[circ]{}
					(2,1) node[circ]{}
					(3.5,1) node[circ]{}
					(4,1) node[circ]{};
			\draw 	(1,1) node[cross]{}
					(1.5,1) node[cross]{}
					(2,1) node[cross]{}
					(3.5,1) node[cross]{}
					(4,1) node[cross]{};
	\end{tikzpicture}}
 & \raisebox{-.4\height}{\begin{tikzpicture}[scale=1, every node/.style={transform shape},node distance=20]
	\tikzstyle{gauge} = [circle, draw];
	\tikzstyle{flavour} = [regular polygon,regular polygon sides=4,draw];
	\node (g4) [gauge,label=below:{$O_2$}] {};
	\node (f4) [flavour,above of=g4,label=above:{$C_2$}] {};
	\draw (g4)--(f4);
		\end{tikzpicture}}  & \raisebox{-.4\height}{\begin{tikzpicture}[scale=1, every node/.style={transform shape},node distance=20]
			\tikzstyle{gauge} = [circle, draw];
			\tikzstyle{flavour} = [regular polygon,regular polygon sides=4,draw];
			\node (g1) [gauge, label=below:{$C_1$}]{};
			\node (g2) [gauge, right of=g1, label=below:{$O_2$}]{};
			\node (g3) [gauge, right of=g2, label=below:{$C_1$}]{};
			\node (g4) [gauge, right of=g3, label=below:{$O_1$}]{};
			\node (f1) [flavour,above of=g1,label=above:{$O_2$}] {};
			\node (f3) [flavour,above of=g3,label=above:{$O_1$}] {};
			\draw (f1)--(g1)--(g2)--(g3)--(g4)
					(f3)--(g3);
		\end{tikzpicture}}\\
  \addlinespace[5ex]
 $\mathbf{1^4}$  & 
		\raisebox{-.4\height}{\begin{tikzpicture}[scale=.8, every node/.style={transform shape}]	
			\draw 	[dashed](3,0)--(3,2)
					(3.5,0)--(3.5,2)
					(4,0)--(4,2)
					(4.5,0)--(4.5,2)
					(5.5,0)--(5.5,2);
			\draw	[dotted](1,1)--(1.5,1)
					(2,1)--(2.5,1)
					(5,1)--(6,1);
			\draw	(3,1)--(3.5,1)
					(4,1)--(4.5,1);
			\draw 	(1,1) node[circ]{}
					(1.5,1) node[circ]{}
					(2,1) node[circ]{}
					(2.5,1) node[circ]{}
					(5,1) node[circ]{};
			\draw 	(1,1) node[cross]{}
					(1.5,1) node[cross]{}
					(2,1) node[cross]{}
					(2.5,1) node[cross]{}
					(5,1) node[cross]{};
			\end{tikzpicture}}
 & \raisebox{-.4\height}{\begin{tikzpicture}[scale=1, every node/.style={transform shape},node distance=20]
			\tikzstyle{gauge} = [circle, draw];
			\tikzstyle{flavour} = [regular polygon,regular polygon sides=4, draw];
			\node (g6) [] {};
			\node (f6) [flavour,above of=g6,label=above:{$C_2$}] {};
		\end{tikzpicture}} & \raisebox{-.4\height}{\begin{tikzpicture}[scale=1, every node/.style={transform shape},node distance=20]
			\tikzstyle{gauge} = [circle,draw];
			\tikzstyle{flavour} = [regular polygon,regular polygon sides=4,draw];
	\node (g1) [gauge,label=below:{$C_0$}] {};
	\node (g2) [gauge,right of=g1,label=below:{$O_0$}] {};
	\node (g3) [gauge,right of=g2,label=below:{$C_0$}] {};
	\node (g4) [gauge,right of=g3,label=below:{$O_0$}] {};
	\draw(g1)--(g2)--(g3)--(g4)
		;
		\end{tikzpicture}} \\ \addlinespace[3ex]
	\bottomrule
	\end{tabular}
	\end{tabular}
		\caption{Summary of Kraft-Procesi transitions for the nilpotent orbits of $\mathfrak{sp}(2)$. The brane systems in the middle depict models whose Higgs branch is the closure of the nilpotent orbit labelled by each \emph{Partition}. The column $\M_H$   contains the quivers of such models, with $\M_H=\Or_{\lambda}$. The column $\M_C$  depicts the quivers of the models obtained after performing S-duality transformation on the Brane systems, they are predicted to have $\M_C=\Or_\lambda$. The $C_0$ and $O_0$ gauge nodes have been left in this column to highlight the effect that the transition has on these quivers: the flavor nodes move along a fixed structure of gauge nodes, while the rank of the gauge nodes decreases.}
		\label{tab:Sp2summary}
\end{table}

\subsubsection{Summary}

The quivers and brane configurations that have been found in this section for the nilpotent orbits of $\gsp (2)$ are summarized in table \ref{tab:Sp2summary}.

\section{Kraft-Procesi transitions for orthogonal and symplectic groups}\label{sec:11}

This section collects all KP transitions that can take place in the Type IIB superstring  brane configurations with O3-planes described above.  As in the previous examples, each transition can be given a label. The label is that of the hyperk\"ahler singularity $S\subseteq  \Or$, where $ \Or$ is the Higgs branch of the initial system and $S$ is a slice transverse to the orbit $\mathcal O'$, such that $\Or '\subset \Or$; the Higgs branch of the resulting system after the KP transition is $\bar {\mathcal O}'$. 

One of the main results of \cite{KP82} is that the hyperk\"ahler singularities $S$ obtained in all possible transitions between neighboring\footnote{By \emph{neighboring} we mean orbits that are connected by a link in the corresponding Hasse diagram.} nilpotent orbits of $\gso(n)$ or $\gsp(n)$ can be classified in two different types: \emph{surface singularities} and \emph{minimal singularities}. \emph{Surface singularities} are varieties of the type $A_{2k-1}$, $D_k$ or $A_{2k-1}\cup A_{2k-1}$, where the first two are the Kleinian surface singularities of the form:
\begin{align}
	A_{2k-1}&:=\mathbb C^2/\mathbb Z_{2k}\\
	D_{k}&:= \mathbb C^2/Dic_{k-2}
\end{align}
with $Dic_{k-2}$ the \emph{binary  dihedral group} (also known as \emph{dicyclic group}) of order $|Dic_{k-2}|=4(k-2)$. The third one is the union of two $A_{2k-1}$ singularities that meet at the singular point. In fact, \cite{B70,Sl80} already established which singularity should be found in the transition from the closure of the maximal nilpotent orbit to the subregular nilpotent orbit depending on the algebra:

\begin{align}\label{eq:subreg}
	\begin{aligned}
	 	\mathfrak{so}({2n+1})&\rightarrow S=A_{2n-1}\\
	 	\mathfrak{so}({2n}) &\rightarrow S=D_{n}\\
	 	\mathfrak{sp}({n})&\rightarrow S=D_{n+1}
	\end{aligned}
\end{align}

\emph{Minimal singularities} are closures of minimal nilpotent orbits of $\mathfrak{so}(2n+1)$, $\mathfrak{sp}(n)$ and $\mathfrak{so}(2n)$. They receive the labels $b_n$, $c_n$ and $d_n$ respectively. They also have a description in terms of transverse slices: they are transverse to the trivial nilpotent orbit in the transition from the closure of the minimal orbit to the trivial orbit. 

\subsection{$D_n$ transitions}\label{sec:Dntransitions}

\begin{figure}[t]
		\centering
		\begin{tikzpicture}
		\draw[dashed] 	(1.5,-1)--(1.5,3)
				(4.5,-1)--(4.5,3);
		
		\draw[dashed] (5.5,-1)--(5.5,3)
				(6,-1)--(6,3)
				(6.5,-1)--(6.5,3);
				
		\fill (7,1) circle [radius=1pt];
		\fill (7.25,1) circle [radius=1pt];
		\fill (7.5,1) circle [radius=1pt];
		
		\draw[dashed] 
				(8,-1)--(8,3)
				(8.5,-1)--(8.5,3)
				(9,-1)--(9,3)
				(9.5,-1)--(9.5,3)
				(10,-1)--(10,3);
		
		\draw (1.5,2.8)--(4.5,2.8);
		\draw (1.5,2.6)--(4.5,2.6);
		\draw (1.5,2)--(4.5,2);
		
		\fill (3,2.3) circle [radius=.5pt];
		\fill (3,2.45) circle [radius=.5pt];
		\fill (3,2.15) circle [radius=.5pt];
		
		\draw (1.5,-.8)--(4.5,-.8);
		\draw (1.5,-.6)--(4.5,-.6);
		\draw (1.5,0)--(4.5,0);
		
		\fill (3,-.3) circle [radius=.5pt];
		\fill (3,-.45) circle [radius=.5pt];
		\fill (3,-.15) circle [radius=.5pt];
		
		\draw (4.5,2.8)--(5.5,2.8);
		\draw (4.5,2.6)--(5.5,2.6);
		\draw (4.5,2)--(5.5,2);
		
		\fill (5,2.3) circle [radius=.5pt];
		\fill (5,2.45) circle [radius=.5pt];
		\fill (5,2.15) circle [radius=.5pt];
		
		\draw (4.5,-.8)--(5.5,-.8);
		\draw (4.5,-.6)--(5.5,-.6);
		\draw (4.5,0)--(5.5,0);
		
		\fill (5,-.3) circle [radius=.5pt];
		\fill (5,-.45) circle [radius=.5pt];
		\fill (5,-.15) circle [radius=.5pt];
		
		\draw (5.5,2.6)--(6,2.6);
		\draw (5.5,2)--(6,2);
		
		\fill (5.75,2.3) circle [radius=.5pt];
		\fill (5.75,2.45) circle [radius=.5pt];
		\fill (5.75,2.15) circle [radius=.5pt];
		
		\draw (5.5,-.6)--(6,-.6);
		\draw (5.5,0)--(6,0);
		
		\fill (5.75,-.3) circle [radius=.5pt];
		\fill (5.75,-.45) circle [radius=.5pt];
		\fill (5.75,-.15) circle [radius=.5pt];
		
		\draw (6,2.6)--(6.7,2.6);
		\draw (6,2)--(6.7,2);
		
		\fill (6.25,2.3) circle [radius=.5pt];
		\fill (6.25,2.45) circle [radius=.5pt];
		\fill (6.25,2.15) circle [radius=.5pt];
		
		\draw (6,-.6)--(6.7,-.6);
		\draw (6,0)--(6.7,0);
		
		\fill (6.25,-.3) circle [radius=.5pt];
		\fill (6.25,-.45) circle [radius=.5pt];
		\fill (6.25,-.15) circle [radius=.5pt];

		\draw (7.8,2)--(9,2);
		\draw (7.8,0)--(9,0);
		
		\draw[dotted] (1,1)--(2,1);
		\draw[dotted] (4,1)--(5,1);
		
		\draw (2,1)--(2.5,1);
		\draw (3.5,1)--(4,1);
		\draw (5.5,1)--(6,1);
		\draw (6.5,1)--(6.7,1);
		\draw (7.8,1)--(8,1);
		\draw (8.5,1)--(9,1);
		\draw (9.5,1)--(10,1);

		\draw (1.5,2.7)--(6.5,2.7);
		\draw (1.5,-.7)--(6.5,-.7);
		
		\draw (1.5,1.9)--(6.7,1.9);
		\draw (1.5,.1)--(6.7,.1);
		
		\draw (2,1) node[circle,fill=white,draw=black] {};
		\draw (2.5,1) node[circle,fill=white,draw=black] {};
		\fill (3,1) circle [radius=.5pt];
		\fill (3.15,1) circle [radius=.5pt];
		\fill (2.85,1) circle [radius=.5pt];
		\draw (3.5,1) node[circle,fill=white,draw=black] {};
		\draw (4,1) node[circle,fill=white,draw=black] {};
				
		\draw (2,1) node[cross] {};
		\draw (2.5,1) node[cross] {};
		\draw (3.5,1) node[cross] {};
		\draw (4,1) node[cross] {};
		
		\draw [decorate,decoration={brace,amplitude=5pt}](1.8,1.3) -- (4.2,1.3);
		\draw node at (3,1.6) {\footnotesize $2n-2$};

		\draw (1,1) node[circle,fill=white,draw=black] {};
		\draw (1,1) node[cross] {};
		\draw (5,1) node[circle,fill=white,draw=black] {};
		\draw (5,1) node[cross] {};

		\draw [decorate,decoration={brace,amplitude=4pt}](1.45,1.85) -- (1.45,2.85);
		\draw node at (0.7,2.35) {\footnotesize $n-1$};
		
		\draw [decorate,decoration={brace,mirror,amplitude=4pt}](1.45,.15) -- (1.45,-.85);
		\draw node at (0.7,-.35) {\footnotesize $n-1$};
		\end{tikzpicture}
	\caption{Higgs branch brane configuration for $\M_H$ the closure of the maximal nilpotent orbit of $\mathfrak{so}(2n)$. In the two leftmost intervals between half D5-branes there are $n-1$ D3-branes. The number of D3-branes then decreases by one on each next pair of intervals to the right, until it reaches zero.}
	\label{fig:SubregularDn}
\end{figure}
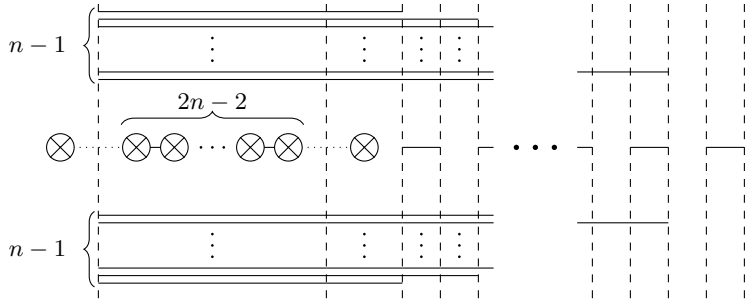

Based in equation (\ref{eq:subreg}), there are at least two different brane systems in which a $D_n$ KP transition can take place. The first case is the system with $\M_H=\Or _{(2n-1,1)}\subset \gso (2n)$, figure \ref{fig:SubregularDn}. The second case is the system with $\M_H=\Or _{(2n-2)}\subset \gsp (n-1)$ (figure \ref{fig:SubregularCn} depicts the system with $\M_H=\Or _{(2n)}\subset \gsp (n)$). In both cases the KP transition  removes a single physical D3-brane. The moduli space $S$ generated by the single D3-brane is $S=D_n$. $S$ is the transverse slice that gives name to the transition.

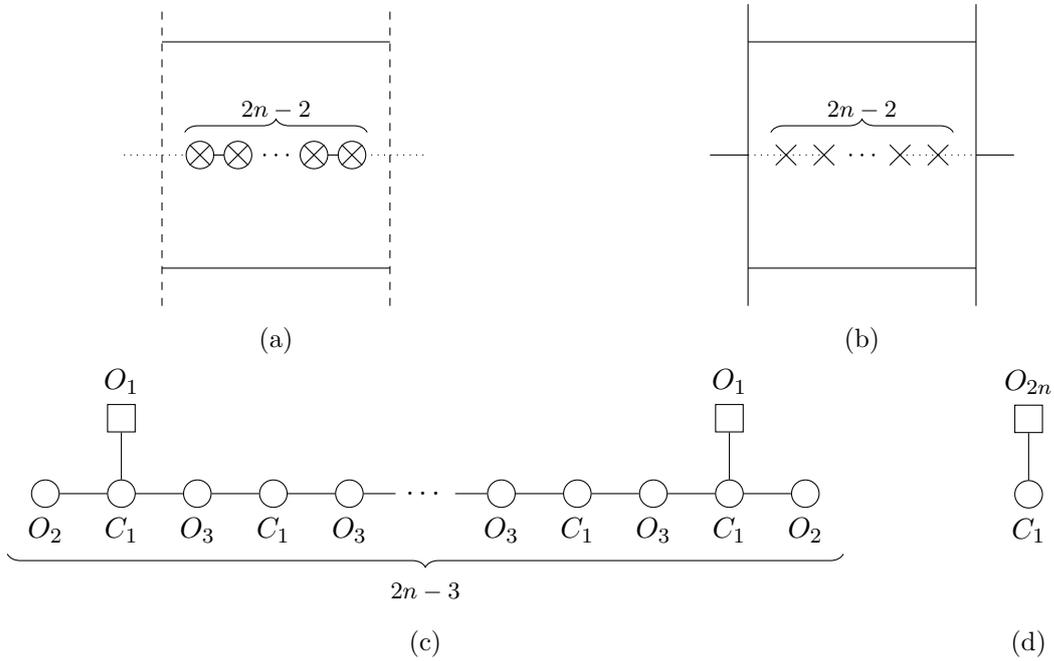
\begin{figure}[t]
		\centering
	\begin{subfigure}[t]{.49\textwidth}
		\centering
		\begin{tikzpicture}
				\draw[dashed] 	(1.5,-1)--(1.5,3)
				(4.5,-1)--(4.5,3);
		
		\draw (1.5,2.5)--(4.5,2.5);
		\draw (1.5,-.5)--(4.5,-.5);
		
		\draw[dotted] (1,1)--(2,1);
		\draw[dotted] (4,1)--(5,1);
		
		\draw (2,1)--(2.5,1);
		\draw (3.5,1)--(4,1);
		
		\draw (2,1) node[circle,fill=white,draw=black] {};
		\draw (2.5,1) node[circle,fill=white,draw=black] {};
		\fill (3,1) circle [radius=.5pt];
		\fill (3.15,1) circle [radius=.5pt];
		\fill (2.85,1) circle [radius=.5pt];
		\draw (3.5,1) node[circle,fill=white,draw=black] {};
		\draw (4,1) node[circle,fill=white,draw=black] {};
				
		\draw (2,1) node[cross] {};
		\draw (2.5,1) node[cross] {};
		\draw (3.5,1) node[cross] {};
		\draw (4,1) node[cross] {};
		
		\draw [decorate,decoration={brace,amplitude=5pt}](1.8,1.3) -- (4.2,1.3);
		\draw node at (3,1.6) {\footnotesize $2n-2$};
		
		\end{tikzpicture}
		\caption{}
	\end{subfigure}	
	\hfill
	\begin{subfigure}[t]{.49\textwidth}
		\centering
		\begin{tikzpicture}		\draw 	(1.5,-1)--(1.5,3)
				(4.5,-1)--(4.5,3);
		
		\draw (1.5,2.5)--(4.5,2.5);
		\draw (1.5,-.5)--(4.5,-.5);
		
		\draw[] (1,1)--(1.5,1);
		\draw[] (4.5,1)--(5,1);
		
		\draw [dotted] (1.5,1)--(2.5,1);
		\draw [dotted] (3.5,1)--(4.5,1);
		
		\fill (3,1) circle [radius=.5pt];
		\fill (3.15,1) circle [radius=.5pt];
		\fill (2.85,1) circle [radius=.5pt];
				
		\draw (2,1) node[cross] {};
		\draw (2.5,1) node[cross] {};
		\draw (3.5,1) node[cross] {};
		\draw (4,1) node[cross] {};
		
		\draw [decorate,decoration={brace,amplitude=5pt}](1.8,1.3) -- (4.2,1.3);
		\draw node at (3,1.6) {\footnotesize $2n-2$};
		\end{tikzpicture}
		\caption{}
	\end{subfigure}
	\hfill
	\begin{subfigure}[t]{.75\textwidth}
		  \centering
	\begin{tikzpicture}
	\tikzstyle{gauge} = [circle, draw];
	\tikzstyle{flavour} = [regular polygon,regular polygon sides=4, draw];
	\node (g1) [gauge,label=below:{$O_2$}] {};
	\node (g2) [gauge, right of=g1,label=below:{$C_1$}] {};
	\node (g3) [gauge, right of=g2,label=below:{$O_3$}] {};
	\node (g4) [gauge, right of=g3,label=below:{$C_1$}] {};
	\node (g5) [gauge, right of=g4,label=below:{$O_3$}] {};
	\node (gd) [right of=g5] {$\dots$};
	\node (g7) [gauge, right of=gd,label=below:{$O_3$}] {};
	\node (g8) [gauge, right of=g7,label=below:{$C_1$}] {};
	\node (g9) [gauge, right of=g8,label=below:{$O_3$}] {};
	\node (g10) [gauge, right of=g9,label=below:{$C_1$}] {};
	\node (g11) [gauge, right of=g10,label=below:{$O_2$}] {};	\node (f2) [flavour,above of=g2,label=above:{$O_1$}] {};
	\node (f10) [flavour,above of=g10,label=above:{$O_1$}] {};
	\draw (g1)--(g2)--(g3)--(g4)--(g5)--(gd)--(g7)--(g8)--(g9)--(g10)--(g11)
			(g2)--(f2)
			(g10)--(f10)
		;
	\draw [decorate,decoration={brace,mirror,amplitude=5pt}](-.5,-.8) -- (10.5,-.8) node [black,midway,xshift=-0.6cm] { };
		\draw node at (5,-1.3) {\footnotesize $2n-3$};
	\end{tikzpicture}
		\caption{}
	\end{subfigure}
	\hfill
	\begin{subfigure}[t]{.20\textwidth}
		\centering
		\raisebox{.3\height}{
		\begin{tikzpicture}
		\tikzstyle{gauge} = [circle,draw];
			\tikzstyle{flavour} = [regular polygon,regular polygon sides=4,draw];
			\node (g1) [gauge, label=below:{$C_1$}]{};
			\node (f1) [flavour, above of=g1, label=above:{$O_{2n}$}]{};
			\draw (g1)--(f1);
		\end{tikzpicture}
		}
		\caption{}
	\end{subfigure}
	\caption{$D_n$ subregular singularity. (a) Brane subsystem that represents the moduli space generated by a single D3-brane in the leftmost interval of figure \ref{fig:SubregularDn}. (b) After performing S-duality on (a) the moduli space generated by this single D3-brane can be understood as a Coulomb branch. (c) Quiver for brane system (a), its Higgs branch is the Kleinian singularity $\M_H=D_n$. (d) Quiver corresponding to brane system (b), its Coulomb branch is $\M_C=D_n$. The $\widetilde{O3^-}$ planes contribute to the quiver as half flavors. Note that \citep{FH00} already discusses  quivers (c) and (d) and their relation via S-duality.}
	\label{fig:Dn}
\end{figure}

Let us focus on the first case. The moduli space $\M_H$ of the brane configuration in figure \ref{fig:SubregularDn} is the closure of the maximal nilpotent orbit of $\gso (2n)$. The only possible KP transition consists on removing one D3-brane out of the D3-branes in the leftmost interval between half D5-branes. In order to establish the nature of the transition we compute the moduli space generated by this single brane. The local subsystem of branes is depicted in figure \ref{fig:Dn}(a). This corresponds to the Higgs branch of a $3d$ $\mathcal N=4$ quiver theory with $2n-3$ alternating gauge nodes: $O_2\times C_1\times O_3 \times C_1 \times O_3 \cdots O_3 \times C_1 \times  O_2 $ with a flavor node $O_1$ attached to the first $C_1$ gauge node and another flavor $O_1$ attached to the last $C_1$ gauge node, see figure \ref{fig:Dn}(c). Appendix \ref{app:DnHiggs} computes the Hilbert series for these Higgs branches for the lower values of $n$ and shows that this variety is indeed the Kleinian singularity $S=D_n$. Alternatively, S-duality can be performed, see figure \ref{fig:Dn}(b), to see this moduli space as the Coulomb branch of a $3d$ $\mathcal N=4$ theory with gauge group $Sp(1)$ and $O(2n)$ flavor symmetry, see the quiver in figure \ref{fig:Dn}(d). The Hilbert series of this Coulomb branch was computed in \cite{CHZ13}, employing the \emph{monopole formula}, and is recorded in appendix \ref{sec:monopole} for completeness.

Let us now consider the second case, the $D_{n+1}$ KP transition (with $n>1$) from the closure of the maximal  nilpotent orbit of $\gsp (n)$ to its subregular orbit. The Higgs branch brane configuration for the closure of the maximal orbit is depicted in figure \ref{fig:SubregularCn}. The KP transition removes one of the leftmost D3-branes. Figure \ref{fig:Dn2}(a) depicts the subsystem of one of these branes. As in the previous case, the question: \emph{what is the moduli space generated by this threebrane?} can be answered by obtaining the quiver for which this moduli space is a Higgs branch, figure \ref{fig:Dn2}(c). Notice that this quiver is the same as in the previous case, producing a similar Higgs branch, in this case $\M_H=D_{n+1}$. Alternatively an S-duality can be performed, figure \ref{fig:Dn2}(b) and the quiver for which the moduli space is a Coulomb branch obtained, figure \ref{fig:Dn2}(d). The Coulomb branch of such quiver is again the variety $\M_C=D_{n+1}$ \cite{CHZ13}.

\begin{figure}[t]
		\centering
		\begin{tikzpicture}
		\draw[dashed] 	(1.5,-1)--(1.5,3)
				(4.5,-1)--(4.5,3);
		
		\draw[dashed] (5.5,-1)--(5.5,3)
				(6,-1)--(6,3)
				(6.5,-1)--(6.5,3);
				
		\fill (7,1) circle [radius=1pt];
		\fill (7.25,1) circle [radius=1pt];
		\fill (7.5,1) circle [radius=1pt];
		
		\draw[dashed] 
				(8,-1)--(8,3)
				(8.5,-1)--(8.5,3)
				(9,-1)--(9,3)
				(9.5,-1)--(9.5,3);
		
		\draw (1.5,2.8)--(4.5,2.8);
		\draw (1.5,2.6)--(4.5,2.6);
		\draw (1.5,2)--(4.5,2);
		
		\fill (3,2.3) circle [radius=.5pt];
		\fill (3,2.45) circle [radius=.5pt];
		\fill (3,2.15) circle [radius=.5pt];
		
		\draw (1.5,-.8)--(4.5,-.8);
		\draw (1.5,-.6)--(4.5,-.6);
		\draw (1.5,0)--(4.5,0);
		
		\fill (3,-.3) circle [radius=.5pt];
		\fill (3,-.45) circle [radius=.5pt];
		\fill (3,-.15) circle [radius=.5pt];
		
		\draw (4.5,2.6)--(5.5,2.6);
		\draw (4.5,2)--(5.5,2);
		
		\fill (5,2.3) circle [radius=.5pt];
		\fill (5,2.45) circle [radius=.5pt];
		\fill (5,2.15) circle [radius=.5pt];
		
		\draw (4.5,-.6)--(5.5,-.6);
		\draw (4.5,0)--(5.5,0);
		
		\fill (5,-.3) circle [radius=.5pt];
		\fill (5,-.45) circle [radius=.5pt];
		\fill (5,-.15) circle [radius=.5pt];
		
		\draw (5.5,2.6)--(6,2.6);
		\draw (5.5,2)--(6,2);
		
		\fill (5.75,2.3) circle [radius=.5pt];
		\fill (5.75,2.45) circle [radius=.5pt];
		\fill (5.75,2.15) circle [radius=.5pt];
		
		\draw (5.5,-.6)--(6,-.6);
		\draw (5.5,0)--(6,0);
		
		\fill (5.75,-.3) circle [radius=.5pt];
		\fill (5.75,-.45) circle [radius=.5pt];
		\fill (5.75,-.15) circle [radius=.5pt];
		
		\draw (6,2.6)--(6.7,2.6);
		\draw (6,2)--(6.7,2);
		
		\fill (6.25,2.3) circle [radius=.5pt];
		\fill (6.25,2.45) circle [radius=.5pt];
		\fill (6.25,2.15) circle [radius=.5pt];
		
		\draw (6,-.6)--(6.7,-.6);
		\draw (6,0)--(6.7,0);
		
		\fill (6.25,-.3) circle [radius=.5pt];
		\fill (6.25,-.45) circle [radius=.5pt];
		\fill (6.25,-.15) circle [radius=.5pt];

		\draw (7.8,1.9)--(9,1.9);
		\draw (7.8,0.1)--(9,0.1);
		\draw (7.8,2)--(8,2);
		\draw (7.8,0)--(8,0.);
		
		\draw[dotted] (1,1)--(2,1);
		\draw[dotted] (4,1)--(6.7,1);
		\draw[dotted] (7.8,1)--(11,1);
		
		\draw (2,1)--(2.5,1);
		\draw (3.5,1)--(4,1);
		
		\draw (1.5,2.7)--(6,2.7);
		\draw (1.5,-.7)--(6,-.7);
		
		\draw (1.5,1.9)--(6.7,1.9);
		\draw (1.5,.1)--(6.7,.1);
		
		\draw (2,1) node[circle,fill=white,draw=black] {};
		\draw (2.5,1) node[circle,fill=white,draw=black] {};
		\fill (3,1) circle [radius=.5pt];
		\fill (3.15,1) circle [radius=.5pt];
		\fill (2.85,1) circle [radius=.5pt];
		\draw (3.5,1) node[circle,fill=white,draw=black] {};
		\draw (4,1) node[circle,fill=white,draw=black] {};
				
		\draw (2,1) node[cross] {};
		\draw (2.5,1) node[cross] {};
		\draw (3.5,1) node[cross] {};
		\draw (4,1) node[cross] {};
		
		\draw [decorate,decoration={brace,amplitude=5pt}](1.8,1.3) -- (4.2,1.3);
		\draw node at (3,1.6) {\footnotesize $2n$};

		\draw (1,1) node[circle,fill=white,draw=black] {};
		\draw (1,1) node[cross] {};

		\draw [decorate,decoration={brace,amplitude=4pt}](1.45,1.85) -- (1.45,2.85);
		\draw node at (0.7,2.35) {\footnotesize $n$};
		
		\draw [decorate,decoration={brace,mirror,amplitude=4pt}](1.45,.15) -- (1.45,-.85);
		\draw node at (0.7,-.35) {\footnotesize $n$};
		\end{tikzpicture}
	\caption{Higgs branch brane configuration for $\M_H$ the closure of the maximal nilpotent orbit of $\mathfrak{sp}(n)$. The leftmost interval between half D5-branes has $n$ D3-branes. The number of D3-branes then decreases by one on each next pair of intervals to the right, until it reaches zero.}
	\label{fig:SubregularCn}
\end{figure}
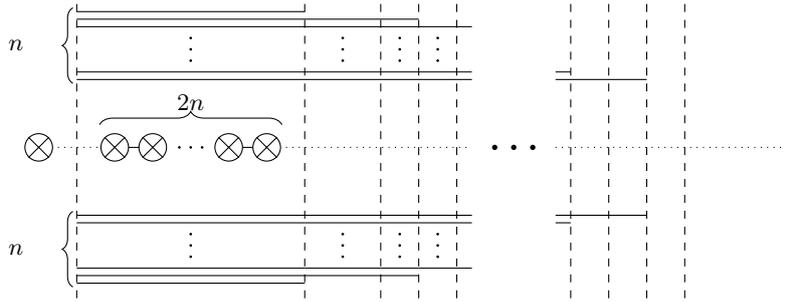

\begin{figure}[t]
		\centering
	\begin{subfigure}[t]{.49\textwidth}
		\centering
		\begin{tikzpicture}
				\draw[dashed] 	(1.5,-1)--(1.5,3)
				(4.5,-1)--(4.5,3);
		
		\draw (1.5,2.5)--(4.5,2.5);
		\draw (1.5,-.5)--(4.5,-.5);
		
		\draw[dotted] (1,1)--(2,1);
		\draw[dotted] (4,1)--(5,1);
		
		\draw (2,1)--(2.5,1);
		\draw (3.5,1)--(4,1);
		
		\draw (2,1) node[circle,fill=white,draw=black] {};
		\draw (2.5,1) node[circle,fill=white,draw=black] {};
		\fill (3,1) circle [radius=.5pt];
		\fill (3.15,1) circle [radius=.5pt];
		\fill (2.85,1) circle [radius=.5pt];
		\draw (3.5,1) node[circle,fill=white,draw=black] {};
		\draw (4,1) node[circle,fill=white,draw=black] {};
				
		\draw (2,1) node[cross] {};
		\draw (2.5,1) node[cross] {};
		\draw (3.5,1) node[cross] {};
		\draw (4,1) node[cross] {};
		
		\draw [decorate,decoration={brace,amplitude=5pt}](1.8,1.3) -- (4.2,1.3);
		\draw node at (3,1.6) {\footnotesize $2n$};
		\end{tikzpicture}
		\caption{}
	\end{subfigure}
	\hfill
	\begin{subfigure}[t]{.49\textwidth}
		\centering
		\begin{tikzpicture}		\draw 	(1.5,-1)--(1.5,3)
				(4.5,-1)--(4.5,3);
		
		\draw (1.5,2.5)--(4.5,2.5);
		\draw (1.5,-.5)--(4.5,-.5);
		
		\draw[] (1,1)--(1.5,1);
		\draw[] (4.5,1)--(5,1);
		
		\draw [dotted] (1.5,1)--(2.5,1);
		\draw [dotted] (3.5,1)--(4.5,1);
		
		\fill (3,1) circle [radius=.5pt];
		\fill (3.15,1) circle [radius=.5pt];
		\fill (2.85,1) circle [radius=.5pt];
				
		\draw (2,1) node[cross] {};
		\draw (2.5,1) node[cross] {};
		\draw (3.5,1) node[cross] {};
		\draw (4,1) node[cross] {};
		
		\draw [decorate,decoration={brace,amplitude=5pt}](1.8,1.3) -- (4.2,1.3);
		\draw node at (3,1.6) {\footnotesize $2n$};
		\end{tikzpicture}
		\caption{}
	\end{subfigure}
	\hfill
	\begin{subfigure}[t]{.75\textwidth}
		  \centering
	\begin{tikzpicture}
	\tikzstyle{gauge} = [circle, draw];
	\tikzstyle{flavour} = [regular polygon,regular polygon sides=4, draw];
	\node (g1) [gauge,label=below:{$O_2$}] {};
	\node (g2) [gauge, right of=g1,label=below:{$C_1$}] {};
	\node (g3) [gauge, right of=g2,label=below:{$O_3$}] {};
	\node (g4) [gauge, right of=g3,label=below:{$C_1$}] {};
	\node (g5) [gauge, right of=g4,label=below:{$O_3$}] {};
	\node (gd) [right of=g5] {$\dots$};
	\node (g7) [gauge, right of=gd,label=below:{$O_3$}] {};
	\node (g8) [gauge, right of=g7,label=below:{$C_1$}] {};
	\node (g9) [gauge, right of=g8,label=below:{$O_3$}] {};
	\node (g10) [gauge, right of=g9,label=below:{$C_1$}] {};
	\node (g11) [gauge, right of=g10,label=below:{$O_2$}] {};	\node (f2) [flavour,above of=g2,label=above:{$O_1$}] {};
	\node (f10) [flavour,above of=g10,label=above:{$O_1$}] {};
	\draw (g1)--(g2)--(g3)--(g4)--(g5)--(gd)--(g7)--(g8)--(g9)--(g10)--(g11)
			(g2)--(f2)
			(g10)--(f10)
		;
	\draw [decorate,decoration={brace,mirror,amplitude=5pt}](-.5,-.8) -- (10.5,-.8) node [black,midway,xshift=-0.6cm] { };
		\draw node at (5,-1.3) {\footnotesize $2n-1$};
	\end{tikzpicture}
		\caption{}
	\end{subfigure}
	\hfill
	\begin{subfigure}[t]{.20\textwidth}
		\centering
		\raisebox{.3\height}{
		\begin{tikzpicture}
		\tikzstyle{gauge} = [circle,draw];
			\tikzstyle{flavour} = [regular polygon,regular polygon sides=4,draw];
			\node (g1) [gauge, label=below:{$C_1$}]{};
			\node (f1) [flavour, above of=g1, label=above:{$O_{2n+2}$}]{};
			\draw (g1)--(f1);
		\end{tikzpicture}
		}
		\caption{}
	\end{subfigure}
	\caption{$D_{n+1}$ subregular singularity. (a) Brane subsystem that represents the moduli space generated by a single D3-brane in the leftmost interval of figure \ref{fig:SubregularCn}. (b) After S-duality the moduli space generated by this single D3-brane as a Coulomb branch. (c) Quiver corresponding to brane system (a), its Higgs branch is $\M_H=D_{n+1}$. (d) Quiver corresponding to brane system (b), its Coulomb branch is $\M_C=D_{n+1}$. These quivers are identical to the ones found in figure \ref{fig:Dn}.}
	\label{fig:Dn2}
\end{figure}
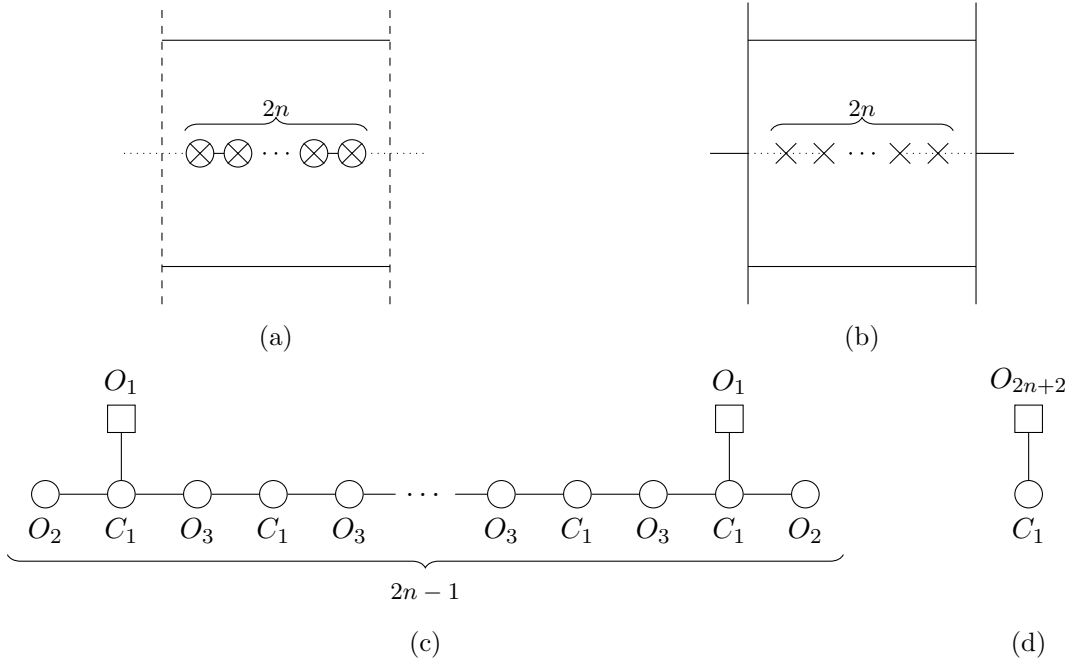

 The fact that the Higgs brane configuration of these models has this local brane subsystem, corresponding to the surface singularity $D_n$ or $D_{n+1}$ constitutes once more a physical realization of the Brieskorn-Slodowy theory \cite{B70,Sl80}.
 
\begin{figure}[t]
		\centering
	\begin{subfigure}[t]{.49\textwidth}
		\centering
		\begin{tikzpicture}\draw[dashed] 	(1.5,-1)--(1.5,3)
				(4.5,-1)--(4.5,3);
		
		\draw (1.5,2.5)--(4.5,2.5);
		\draw (1.5,-.5)--(4.5,-.5);
		
		\draw[dotted] (1,1)--(2,1);
		\draw[dotted] (4,1)--(5,1);

		\draw (2,1) node[circle,fill=white,draw=black] {};
		\draw (2.5,1) node[circle,fill=white,draw=black] {};
		\fill (3,1) circle [radius=.5pt];
		\fill (3.15,1) circle [radius=.5pt];
		\fill (2.85,1) circle [radius=.5pt];
		\draw (3.5,1) node[circle,fill=white,draw=black] {};
		\draw (4,1) node[circle,fill=white,draw=black] {};
				
		\draw (2,1) node[cross] {};
		\draw (2.5,1) node[cross] {};
		\draw (3.5,1) node[cross] {};
		\draw (4,1) node[cross] {};
		
		\draw [decorate,decoration={brace,amplitude=5pt}](1.8,1.3) -- (4.2,1.3);
		\draw node at (3,1.8) {\footnotesize $2n-2$};
		\end{tikzpicture}
		\caption{}
	\end{subfigure}
	\hfill
	\begin{subfigure}[t]{.49\textwidth}
		\centering
		\begin{tikzpicture}	
		\draw 	(1.5,-1)--(1.5,3)
				(4.5,-1)--(4.5,3);
		
		\draw (1.5,2.5)--(4.5,2.5);
		\draw (1.5,-.5)--(4.5,-.5);
		
		\draw[dotted] (1,1)--(1.5,1);
		\draw 			(1.5,1)--(2,1);
		\draw 			(4,1)--(4.5,1);
		\draw 			(2.5,1)--(2.6,1);
		\draw 			(3.4,1)--(3.5,1);
		\draw[dotted] (4.5,1)--(5,1);

		\fill (3,1) circle [radius=.5pt];
		\fill (3.15,1) circle [radius=.5pt];
		\fill (2.85,1) circle [radius=.5pt];
				
		\draw (2,1) node[cross] {};
		\draw (2.5,1) node[cross] {};
		\draw (3.5,1) node[cross] {};
		\draw (4,1) node[cross] {};
		
		\draw [decorate,decoration={brace,amplitude=5pt}](1.8,1.3) -- (4.2,1.3);
		\draw node at (3,1.8) {\footnotesize $2n-2$};
		\end{tikzpicture}
		\caption{}
	\end{subfigure}
	\hfill
	\begin{subfigure}[t]{.75\textwidth}
		  \centering
	\begin{tikzpicture}
	\tikzstyle{gauge} = [circle, draw];
	\tikzstyle{flavour} = [regular polygon,regular polygon sides=4, draw];
	\node (g1) [gauge,label=below:{$O_1$}] {};
	\node (g2) [gauge, right of=g1,label=below:{$C_1$}] {};
	\node (g3) [gauge, right of=g2,label=below:{$O_2$}] {};
	\node (g4) [gauge, right of=g3,label=below:{$C_1$}] {};
	\node (g5) [gauge, right of=g4,label=below:{$O_2$}] {};
	\node (gd) [right of=g5] {$\dots$};
	\node (g7) [gauge, right of=gd,label=below:{$O_2$}] {};
	\node (g8) [gauge, right of=g7,label=below:{$C_1$}] {};
	\node (g9) [gauge, right of=g8,label=below:{$O_2$}] {};
	\node (g10) [gauge, right of=g9,label=below:{$C_1$}] {};
	\node (g11) [gauge, right of=g10,label=below:{$O_1$}] {};	\node (f2) [flavour,above of=g2,label=above:{$O_1$}] {};
	\node (f10) [flavour,above of=g10,label=above:{$O_1$}] {};
	\draw (g1)--(g2)--(g3)--(g4)--(g5)--(gd)--(g7)--(g8)--(g9)--(g10)--(g11)
			(g2)--(f2)
			(g10)--(f10)
		;
	\draw [decorate,decoration={brace,mirror,amplitude=5pt}](-.5,-.8) -- (10.5,-.8) node [black,midway,xshift=-0.6cm] { };
		\draw node at (5,-1.3) {\footnotesize $2n-3$};
	\end{tikzpicture}
		\caption{}
	\end{subfigure}
	\hfill
	\begin{subfigure}[t]{.20\textwidth}
		\centering
		\raisebox{.3\height}{
		\begin{tikzpicture}
		\tikzstyle{gauge} = [circle,draw];
			\tikzstyle{flavour} = [regular polygon,regular polygon sides=4,draw];
			\node (g1) [gauge, label=below:{$O_2$}]{};
			\node (f1) [flavour, above of=g1, label=above:{$C_{n-2}$}]{};
			\draw (g1)--(f1);
		\end{tikzpicture}
		}
		\caption{}
	\end{subfigure}
	\caption{$D_{n}$ subregular singularity. (a) Brane subsystem that represents the moduli space generated by a single D3-brane. (b) After S-duality the moduli space generated by this single D3-brane as a Coulomb branch. (c) Quiver corresponding to brane system (a), its Higgs branch is $\M_H=D_{n}$. (d) Quiver corresponding to brane system (b) (note the choice of $O_2$ as the gauge node), its Coulomb branch is $\M_C=D_{n}$.}
	\label{fig:Dn3}
\end{figure}
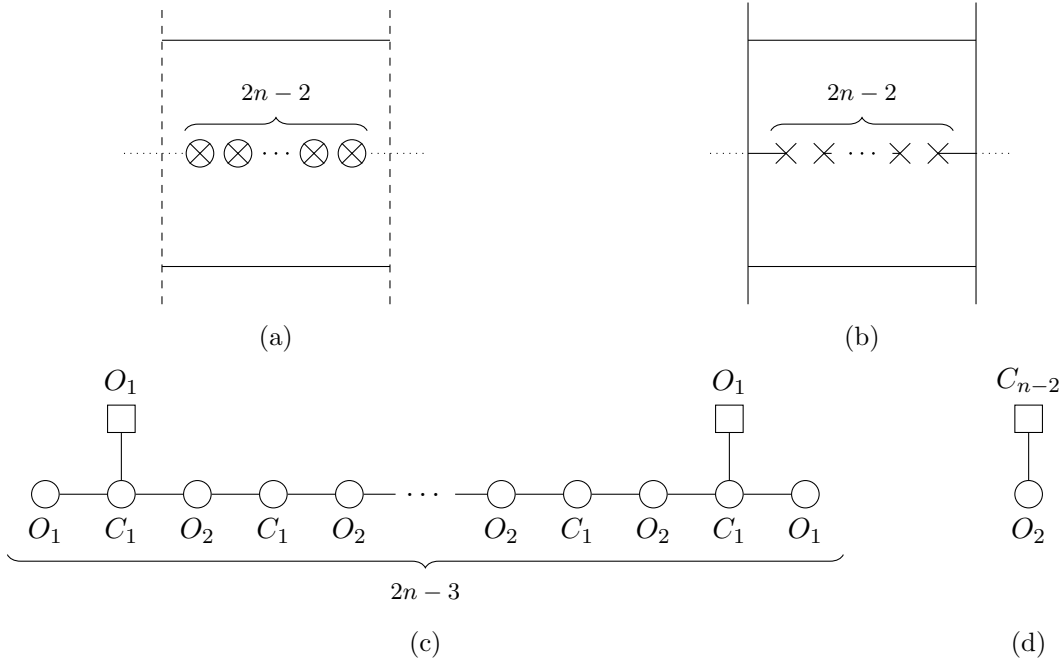

 There exists a third brane subsystem that can appear as a transverse slice $S=D_n$ during the KP transitions, it is depicted in figure \ref{fig:Dn3}(a). The Higgs branch of the quiver in figure \ref{fig:Dn3}(c) can be computed to be $\M_H=D_n$ (see appendix \ref{app:DnHiggs}) and the Coulomb branch of the quiver in figure \ref{fig:Dn3}(d) to be $\M_C=D_n$ (see appendix \ref{sec:monopole}).
 
\subsection{$A_{2n-1}$ transitions}\label{sec:Antransitions}

\begin{figure}[t]
		\centering
		\begin{tikzpicture}
		\draw[dashed] 	(1.5,-1)--(1.5,3)
				(4.5,-1)--(4.5,3);
		
		\draw[dashed] (5.5,-1)--(5.5,3)
				(6.5,-1)--(6.5,3);
				
		\fill (7,1) circle [radius=1pt];
		\fill (7.25,1) circle [radius=1pt];
		\fill (7.5,1) circle [radius=1pt];
		
		\draw[dashed] 
				(8,-1)--(8,3)
				(8.5,-1)--(8.5,3)
				(9,-1)--(9,3)
				(9.5,-1)--(9.5,3);
		
		\draw (1.5,2.8)--(4.5,2.8);
		\draw (1.5,2.6)--(4.5,2.6);
		\draw (1.5,2)--(4.5,2);
		
		\fill (3,2.3) circle [radius=.5pt];
		\fill (3,2.45) circle [radius=.5pt];
		\fill (3,2.15) circle [radius=.5pt];
		
		\draw (1.5,-.8)--(4.5,-.8);
		\draw (1.5,-.6)--(4.5,-.6);
		\draw (1.5,0)--(4.5,0);
		
		\fill (3,-.3) circle [radius=.5pt];
		\fill (3,-.45) circle [radius=.5pt];
		\fill (3,-.15) circle [radius=.5pt];
		
		\draw (4.5,2.6)--(5.5,2.6);
		\draw (4.5,2)--(5.5,2);
		
		\fill (5,2.3) circle [radius=.5pt];
		\fill (5,2.45) circle [radius=.5pt];
		\fill (5,2.15) circle [radius=.5pt];
		
		\draw (4.5,-.6)--(5.5,-.6);
		\draw (4.5,0)--(5.5,0);
		
		\fill (5,-.3) circle [radius=.5pt];
		\fill (5,-.45) circle [radius=.5pt];
		\fill (5,-.15) circle [radius=.5pt];
		
		\draw (5.5,2.6)--(6,2.6);
		\draw (5.5,2)--(6,2);
		
		\fill (6,2.3) circle [radius=.5pt];
		\fill (6,2.45) circle [radius=.5pt];
		\fill (6,2.15) circle [radius=.5pt];
		
		\draw (5.5,-.6)--(6,-.6);
		\draw (5.5,0)--(6,0);
		
		\fill (6,-.3) circle [radius=.5pt];
		\fill (6,-.45) circle [radius=.5pt];
		\fill (6,-.15) circle [radius=.5pt];
		
		\draw (6,2.6)--(6.7,2.6);
		\draw (6,2)--(6.7,2);

		\draw (6,-.6)--(6.7,-.6);
		\draw (6,0)--(6.7,0);

		\draw (7.8,1.9)--(9,1.9);
		\draw (7.8,0.1)--(9,0.1);
		\draw (7.8,2)--(8,2);
		\draw (7.8,0)--(8,0.);
		
		\draw[dotted] (0,1)--(2,1);
		\draw[dotted] (3.5,1)--(4,1);
		
		\draw (4.5,1)--(5.5,1);
		\draw (6.5,1)--(6.7,1);
		\draw (8,1)--(8.5,1)
				(9,1)--(9.5,1);
		
		\draw (1.5,2.7)--(6.5,2.7);
		\draw (1.5,-.7)--(6.5,-.7);
		
		\draw (1.5,1.9)--(6.7,1.9);
		\draw (1.5,.1)--(6.7,.1);
		
		\draw (2,1) node[circle,fill=white,draw=black] {};
		\draw (2.5,1) node[circle,fill=white,draw=black] {};
		\fill (3,1) circle [radius=.5pt];
		\fill (3.15,1) circle [radius=.5pt];
		\fill (2.85,1) circle [radius=.5pt];
		\draw (3.5,1) node[circle,fill=white,draw=black] {};
		\draw (4,1) node[circle,fill=white,draw=black] {};
				
		\draw (2,1) node[cross] {};
		\draw (2.5,1) node[cross] {};
		\draw (3.5,1) node[cross] {};
		\draw (4,1) node[cross] {};
		
		\draw [decorate,decoration={brace,amplitude=5pt}](1.8,1.3) -- (4.2,1.3);
		\draw node at (3,1.6) {\footnotesize $2n+1$};

		\draw [decorate,decoration={brace,amplitude=4pt}](1.45,1.85) -- (1.45,2.85);
		\draw node at (0.7,2.35) {\footnotesize $n$};
		
		\draw [decorate,decoration={brace,mirror,amplitude=4pt}](1.45,.15) -- (1.45,-.85);
		\draw node at (0.7,-.35) {\footnotesize $n$};
		\end{tikzpicture}
	\caption{Higgs branch brane configuration for $\M_H$ the closure of the maximal nilpotent orbit of $\mathfrak{so}(2n+1)$. There are $n$ D3-branes in the leftmost interval between half D5-branes. The number of D3-branes then decreases by one on each next pair of intervals to the right, until it reaches zero.}
	\label{fig:SubregularBn}
\end{figure}
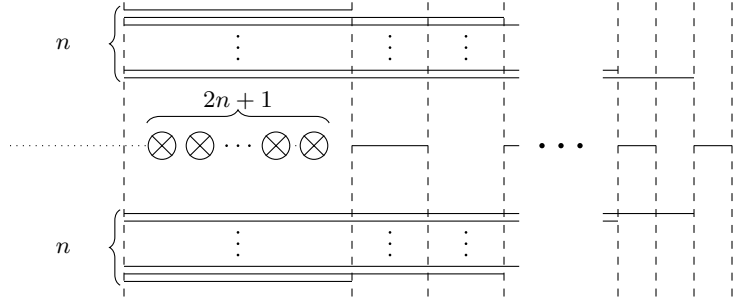

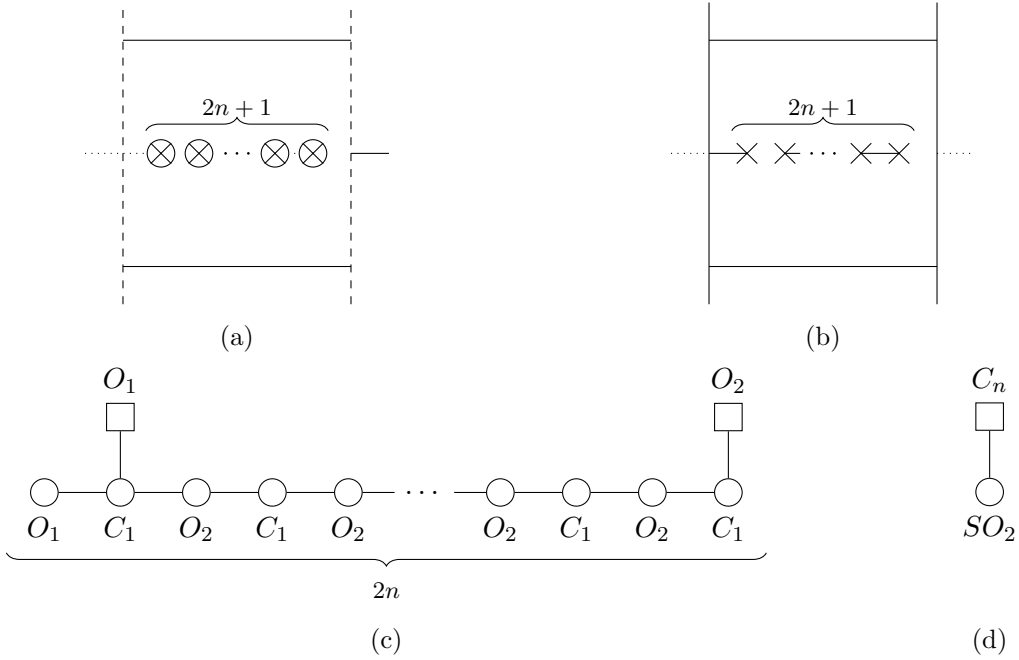
\begin{figure}[t]
		\centering
	\begin{subfigure}[t]{.49\textwidth}
		\centering
		\begin{tikzpicture}
				\draw[dashed] 	(1.5,-1)--(1.5,3)
				(4.5,-1)--(4.5,3);
		
		\draw (1.5,2.5)--(4.5,2.5);
		\draw (1.5,-.5)--(4.5,-.5);
		
		\draw[dotted] (1,1)--(2,1);
		\draw (4.5,1)--(5,1);
		
		\draw[dotted] (3.5,1)--(4,1);
		
		\draw (2,1) node[circle,fill=white,draw=black] {};
		\draw (2.5,1) node[circle,fill=white,draw=black] {};
		\fill (3,1) circle [radius=.5pt];
		\fill (3.15,1) circle [radius=.5pt];
		\fill (2.85,1) circle [radius=.5pt];
		\draw (3.5,1) node[circle,fill=white,draw=black] {};
		\draw (4,1) node[circle,fill=white,draw=black] {};
				
		\draw (2,1) node[cross] {};
		\draw (2.5,1) node[cross] {};
		\draw (3.5,1) node[cross] {};
		\draw (4,1) node[cross] {};
		
		\draw [decorate,decoration={brace,amplitude=5pt}](1.8,1.3) -- (4.2,1.3);
		\draw node at (3,1.6) {\footnotesize $2n+1$};
		
		\end{tikzpicture}
		\caption{}
	\end{subfigure}
	\hfill
	\begin{subfigure}[t]{.49\textwidth}
		\centering
		\begin{tikzpicture}
		\draw 	(1.5,-1)--(1.5,3)
				(4.5,-1)--(4.5,3);
		
		\draw (1.5,2.5)--(4.5,2.5);
		\draw (1.5,-.5)--(4.5,-.5);
		
		\draw[dotted] (1,1)--(1.5,1);
		\draw (1.5,1)--(2,1);
		\draw (2.5,1)--(2.7,1);
		\draw (3.5,1)--(4,1);
		\draw[dotted] (4.5,1)--(5,1);
		
		\fill (3,1) circle [radius=.5pt];
		\fill (3.15,1) circle [radius=.5pt];
		\fill (2.85,1) circle [radius=.5pt];
				
		\draw (2,1) node[cross] {};
		\draw (2.5,1) node[cross] {};
		\draw (3.5,1) node[cross] {};
		\draw (4,1) node[cross] {};
		
		\draw [decorate,decoration={brace,amplitude=5pt}](1.8,1.3) -- (4.2,1.3);
		\draw node at (3,1.6) {\footnotesize $2n+1$};
		\end{tikzpicture}
		\caption{}
	\end{subfigure}
	\hfill
	\begin{subfigure}[t]{.75\textwidth}
		  \centering
	\begin{tikzpicture}
	\tikzstyle{gauge} = [circle, draw];
	\tikzstyle{flavour} = [regular polygon,regular polygon sides=4, draw];
	\node (g1) [gauge,label=below:{$O_1$}] {};
	\node (g2) [gauge, right of=g1,label=below:{$C_1$}] {};
	\node (g3) [gauge, right of=g2,label=below:{$O_2$}] {};
	\node (g4) [gauge, right of=g3,label=below:{$C_1$}] {};
	\node (g5) [gauge, right of=g4,label=below:{$O_2$}] {};
	\node (gd) [right of=g5] {$\dots$};
	\node (g7) [gauge, right of=gd,label=below:{$O_2$}] {};
	\node (g8) [gauge, right of=g7,label=below:{$C_1$}] {};
	\node (g9) [gauge, right of=g8,label=below:{$O_2$}] {};
	\node (g10) [gauge, right of=g9,label=below:{$C_1$}] {};
	\node (f2) [flavour,above of=g2,label=above:{$O_1$}] {};
	\node (f10) [flavour,above of=g10,label=above:{$O_2$}] {};
	\draw (g1)--(g2)--(g3)--(g4)--(g5)--(gd)--(g7)--(g8)--(g9)--(g10)
			(g2)--(f2)
			(g10)--(f10)
		;
	\draw [decorate,decoration={brace,mirror,amplitude=5pt}](-.5,-.8) -- (9.5,-.8) node [black,midway,xshift=-0.6cm] { };
		\draw node at (4.5,-1.3) {\footnotesize $2n$};
	\end{tikzpicture}
		\caption{}
	\end{subfigure}
	\hfill
	\begin{subfigure}[t]{.20\textwidth}
		\centering
		\raisebox{.3\height}{
		\begin{tikzpicture}
		\tikzstyle{gauge} = [circle,draw];
			\tikzstyle{flavour} = [regular polygon,regular polygon sides=4,draw];
			\node (g1) [gauge, label=below:{$SO_2$}]{};
			\node (f1) [flavour, above of=g1, label=above:{$C_{n}$}]{};
			\draw (g1)--(f1);
		\end{tikzpicture}
		}
		\caption{}
	\end{subfigure}
	\caption{$A_{2n-1}$ subregular singularity. (a) Brane subsystem that represents the moduli space generated by a single D3-brane in the leftmost interval of figure \ref{fig:SubregularBn}. In here, the rightmost $\widetilde{O3^-}$ contributes as half a flavor attached to the rightmost $C_1$ gauge node. (b) After S-duality the moduli space generated by this single D3-brane as a Coulomb branch. (c) Quiver corresponding to brane system (a), its Higgs branch is $\M_H=A_{2n-1}$. (d) Quiver corresponding to brane system (b), its Coulomb branch is $\M_C=A_{2n-1}$.}
	\label{fig:Anp1}
\end{figure}

An $A_{2n-1}$ KP transition can be performed in the Higgs branch brane configuration whose $\M_H$ is the closure of the maximal nilpotent orbit of $\mathfrak{so}({2n+1})$, figure \ref{fig:SubregularBn}. A single D3-brane from the leftmost interval between half D5-branes is removed from the system during the KP transition. The moduli space generated by this brane is the transverse slice $S$ that gives name to the transition. It can be seen from the brane picture that this moduli space, figure \ref{fig:Anp1}(a), corresponds to the Higgs branch of the $3d$ $\mathcal N=4$ quiver theory depicted in figure \ref{fig:Anp1}(c). The Hilbert series for the Higgs branch of this quiver can be computed (see appendix \ref{app:AnHiggs} for a detailed computation). The result is that the moduli space $S$ generated by this single D3-brane is indeed a Kleinian singularity of type $A_{2n-1}$. Alternatively, an S-duality transition, \ref{fig:Anp1}(b), shows that this is also the Coulomb branch of the quiver theory in figure \ref{fig:Anp1}(d), with gauge group $G=SO(2)$ and flavor group $F=Sp(n)$. The Hilbert series for the Coulomb branch of this quiver was computed using the \emph{monopole formula} \citep{CHZ13} (the computation is reproduced again in appendix \ref{sec:monopole}). The resulting brane system after the $A_{2n-1}$ KP transition is performed has $\M_H$  equal to the closure of the subregular orbit of $\mathfrak{so}(2n+1)$. This constitutes again a physical realization of the Brieskorn-Slodowy theory \cite{B70,Sl80}.

\subsection{$A_{2n-1}\cup A_{2n-1}$ transitions}\label{sec:AnUAntransitions}

A transition of this type has already been discussed in the set of examples above. This is the transition from the system whose Higgs branch is the closure of the minimal nilpotent orbit of $\mathfrak{so}(4)$ under the adjoint action of the group $O(4)$ to the system whose Higgs branch is the trivial orbit. The transverse slice $S$ in this transition is the whole closure of the minimal orbit $S=A_1\cup A_1$.

Let us study this case in some detail in order to establish the generalization to $A_{2n-1}\cup A_{2n-1}$. The Higgs branch brane configuration with $\M_H=A_1\cup A_1$ is depicted in figure \ref{fig:SO4minKP}(a). In this case the structure of the orientifold planes is ($O3^+$,$O3^-$,$O3^+$). In the middle interval, with orientifold plane $O3^-$, there are two half D5-branes and two half D3-branes.

One can generalize this to construction in figure \ref{fig:Anp1UAnp1}(a), by including $2n$ half NS5-branes in the middle interval between half D5-branes instead of just $2$. The quiver can be read from this brane system, figure \ref{fig:Anp1UAnp1}(c). Hilbert series of the Higgs branches for this family of quivers are computed in appendix \ref{app:AnAnHiggs} and are checked to correspond to the union $\M_H=A_{2n-1}\cup A_{2n-1}$. If an S-Duality is performed, the brane system in figure \ref{fig:Anp1UAnp1}(b) is obtained, with quiver depicted in figure \ref{fig:Anp1UAnp1}(d). Notice how the Coulomb branch does not see the union of both singularities but rather just one of them: $\M_C=A_{2n-1}$. This is a similar phenomenon of what happened in \cite{FH16} (also observed in \cite{GW09}) with the union of two orbits corresponding to the very even partition $\lambda=(2^k)\in \mathcal P_{+1}(2k)$ with $k$ even: the Higgs branch sees the union of the closure of both orbits, while the counterpart Coulomb branch is isomorphic to the closure of only one of the orbits. A generalization of this phenomenon to any very even partition was shown in \cite{HK16} with the computation of the Higgs branch of partition $(4^2)\in \mathcal P_{+1}(8)$ and in \cite{CHZ17} the precise quiver with the counterpart Coulomb branch is given also for any very even partition. As reviewed in section \ref{sec:so4interlude}, for partition $(2^2)$ the union of the closure of the two orbits is also the union of two surface singularities $A_1\cup A_1$. This opens the door to a new generalization of the phenomenon where the Higgs branch is the union of surface singularities $A_{2n-1}\cup A_{2n-1}$ instead of the union of the closure of two nilpotent orbits. This is the generalization that is taking place here.

\begin{figure}[t]
		\centering
	\begin{subfigure}[t]{.49\textwidth}
		\centering
		\begin{tikzpicture}
				\draw[dashed] 
				(1.5,-1)--(1.5,3)
				(4.5,-1)--(4.5,3);
		
		\draw (1.5,2.5)--(4.5,2.5);
		\draw (1.5,-.5)--(4.5,-.5);
		
		\draw (1,1)--(1.5,1);
		\draw (4.5,1)--(5,1);
		\draw[dotted] (2,1)--(2.5,1);
		\draw[dotted] (3.5,1)--(4,1);
		
		\draw (2,1) node[circle,fill=white,draw=black] {};
		\draw (2.5,1) node[circle,fill=white,draw=black] {};
		\fill (3,1) circle [radius=.5pt];
		\fill (3.15,1) circle [radius=.5pt];
		\fill (2.85,1) circle [radius=.5pt];
		\draw (3.5,1) node[circle,fill=white,draw=black] {};
		\draw (4,1) node[circle,fill=white,draw=black] {};
				
		\draw (2,1) node[cross] {};
		\draw (2.5,1) node[cross] {};
		\draw (3.5,1) node[cross] {};
		\draw (4,1) node[cross] {};
		
		\draw [decorate,decoration={brace,amplitude=5pt}](1.8,1.3) -- (4.2,1.3);
		\draw node at (3,1.6) {\footnotesize $2n$};
		
		\end{tikzpicture}
		\caption{}
	\end{subfigure}
	\hfill
	\begin{subfigure}[t]{.49\textwidth}
		\centering
		\begin{tikzpicture}
		\draw 					
				(1.5,-1)--(1.5,3)
				(4.5,-1)--(4.5,3);
		
		\draw (1.5,2.5)--(4.5,2.5);
		\draw (1.5,-.5)--(4.5,-.5);
		
		\draw[dotted] (1,1)--(1.5,1);
		\draw[dotted] (4.5,1)--(5,1);
		\draw (2,1)--(2.5,1);
		\draw (3.5,1)--(4,1);
		
		\fill (3,1) circle [radius=.5pt];
		\fill (3.15,1) circle [radius=.5pt];
		\fill (2.85,1) circle [radius=.5pt];
				
		\draw (2,1) node[cross] {};
		\draw (2.5,1) node[cross] {};
		\draw (3.5,1) node[cross] {};
		\draw (4,1) node[cross] {};
		
		\draw [decorate,decoration={brace,amplitude=5pt}](1.8,1.3) -- (4.2,1.3);
		\draw node at (3,1.6) {\footnotesize $2n$};
		
		\end{tikzpicture}
		\caption{}
	\end{subfigure}
	\hfill
	\begin{subfigure}[t]{.75\textwidth}
		  \centering
	\begin{tikzpicture}
	\tikzstyle{gauge} = [circle, draw];
	\tikzstyle{flavour} = [regular polygon,regular polygon sides=4, draw];
	\node (g2) [gauge,label=below:{$C_1$}] {};
	\node (g3) [gauge, right of=g2,label=below:{$O_2$}] {};
	\node (g4) [gauge, right of=g3,label=below:{$C_1$}] {};
	\node (g5) [gauge, right of=g4,label=below:{$O_2$}] {};
	\node (gd) [right of=g5] {$\dots$};
	\node (g7) [gauge, right of=gd,label=below:{$O_2$}] {};
	\node (g8) [gauge, right of=g7,label=below:{$C_1$}] {};
	\node (g9) [gauge, right of=g8,label=below:{$O_2$}] {};
	\node (g10) [gauge, right of=g9,label=below:{$C_1$}] {};
	\node (f2) [flavour,above of=g2,label=above:{$O_2$}] {};
	\node (f10) [flavour,above of=g10,label=above:{$O_2$}] {};
	\draw (g2)--(g3)--(g4)--(g5)--(gd)--(g7)--(g8)--(g9)--(g10)
			(g2)--(f2)
			(g10)--(f10)
		;
	\draw [decorate,decoration={brace,mirror,amplitude=5pt}](-.5,-.8) -- (8.5,-.8) node [black,midway,xshift=-0.6cm] { };
		\draw node at (4,-1.3) {\footnotesize $2n-1$};
	\end{tikzpicture}
		\caption{}
	\end{subfigure}
	\hfill
	\begin{subfigure}[t]{.20\textwidth}
		\centering
		\raisebox{.3\height}{
		\begin{tikzpicture}
		\tikzstyle{gauge} = [circle,draw];
			\tikzstyle{flavour} = [regular polygon,regular polygon sides=4,draw];
			\node (g1) [gauge, label=below:{$SO_2$}]{};
			\node (f1) [flavour, above of=g1, label=above:{$C_{n}$}]{};
			\draw (g1)--(f1);
		\end{tikzpicture}
		}
		\caption{}
	\end{subfigure}
	\caption{$A_{2n-1}\cup A_{2n-1}$ subregular singularity. (a) Brane subsystem that represents the generalization of the moduli space in figure \ref{fig:SO4minKP}(a). (b) After S-duality the moduli space generated by this single D3-brane as a Coulomb branch. (c) Quiver corresponding to brane system (a), its Higgs branch is $\M_H=A_{2n-1}\cup A_{2n-1}$. (d) Quiver corresponding to brane system (b), its Coulomb branch is $\M_C=A_{2n-1}$.}
	\label{fig:Anp1UAnp1}
\end{figure}

\subsection{$b_n$ transitions}

The $b_n$ KP transitions are not observed in the brane systems since the minimal orbit of $\gso (2n+1)$ is \emph{not special}. The transition that is observed instead is the one from the closure of the next to minimal orbit of $\mathfrak{so}({2n+1})$, corresponding to partition $\lambda=(3,1^{2n-2})$, to the trivial orbit. These are always the composition of an $A_1$ KP transition to the closure of the minimal orbit, corresponding to partition $\lambda=(2^2,1^{2n-3})$ and then a $b_n$ KP transition to the trivial orbit.

From section \ref{sec:311transition} we see that, in the case of transitions $A_1$ followed by $b_2$ (this composition is denoted above as an $\Or_{(3,1^2)}$ transition) the relevant brane configuration is as in figure \ref{fig:SO5subregKP}(a). This allows to generalize to the brane configuration (figure \ref{fig:A1bn}) of an $\Or_{(3,1^{2n-2})}$ KP transition, that we denote as:
\begin{align}
\tilde b_n:=\Or_{(3,1^{2n-2})}
\end{align}

\begin{figure}[t]
		\centering
	\begin{subfigure}[t]{.49\textwidth}
   		\centering
		\begin{tikzpicture}
		\draw[dashed] 	(2,0)--(2,2)
				(3.5,0)--(3.5,2)
				(4,0)--(4,2)
				(4.5,0)--(4.5,2)
				(5.5,0)--(5.5,2)
				(6,0)--(6,2)
				(6.5,0)--(6.5,2)
				(7.5,0)--(7.5,2)
				(8,0)--(8,2)
				;
		\draw [dotted] (1.5,1)--(2.5,1)
				(3,1)--(4.7,1)
				(5.3,1)--(7,1);
		\draw	(7.5,1)--(8,1);
		\draw	(2,1.8)--(3.5,1.8)
				(2,.2)--(3.5,.2)
				(3.5,1.7)--(4,1.7)
				(4,1.8)--(4.5,1.8)
				(3.5,.3)--(4,.3)
				(4,.2)--(4.5,.2)
				(4.5,1.7)--(4.7,1.7)
				(4.5,.3)--(4.7,.3)
				(5.3,1.7)--(5.5,1.7)
				(5.3,.3)--(5.5,.3)
				(5.5,1.8)--(6,1.8)
				(5.5,.2)--(6,.2)
				(6,1.7)--(6.5,1.7)
				(6,.3)--(6.5,.3)
				(6.5,1.8)--(7.5,1.8)
				(6.5,.2)--(7.5,.2)
				;
		\draw 	(2.5,1)node[circ]{}
				(3,1)node[circ]{}
				(7,1)node[circ]{};
		\draw 	(2.5,1)node[cross]{}
				(3,1)node[cross]{}
				(7,1)node[cross]{};
		\fill (5,1) circle [radius=1pt];
		\fill (5.15,1) circle [radius=1pt];
		\fill (4.85,1) circle [radius=1pt];
		\draw [decorate,decoration={brace,amplitude=5pt}](1.8,2.1) -- (7.7,2.1);
		\draw node at (4.75,2.4) {\footnotesize $2n$};
		\end{tikzpicture}
        \caption{}
    \end{subfigure}
	\hfill
	\begin{subfigure}[t]{.49\textwidth}
		\centering
		   		\centering
		\begin{tikzpicture}
		\draw 	(2,0)--(2,2)
				(3.5,0)--(3.5,2)
				(4,0)--(4,2)
				(4.5,0)--(4.5,2)
				(5.5,0)--(5.5,2)
				(6,0)--(6,2)
				(6.5,0)--(6.5,2)
				(7.5,0)--(7.5,2)
				(8,0)--(8,2)
				;
		\draw [dotted] (1.5,1)--(2,1)
				(3,1)--(4.7,1)
				(5.3,1)--(7,1)
				(7.5,1)--(8,1);
		\draw	(2,1)--(2.5,1)
				(3,1)--(3.5,1)
				(4,1)--(4.5,1)
				(5.5,1)--(6,1)
				(6.5,1)--(7,1);
		\draw	(2,1.8)--(3.5,1.8)
				(2,.2)--(3.5,.2)
				(3.5,1.7)--(4,1.7)
				(4,1.8)--(4.5,1.8)
				(3.5,.3)--(4,.3)
				(4,.2)--(4.5,.2)
				(4.5,1.7)--(4.7,1.7)
				(4.5,.3)--(4.7,.3)
				(5.3,1.7)--(5.5,1.7)
				(5.3,.3)--(5.5,.3)
				(5.5,1.8)--(6,1.8)
				(5.5,.2)--(6,.2)
				(6,1.7)--(6.5,1.7)
				(6,.3)--(6.5,.3)
				(6.5,1.8)--(7.5,1.8)
				(6.5,.2)--(7.5,.2)
				;
		\draw 	(2.5,1)node[cross]{}
				(3,1)node[cross]{}
				(7,1)node[cross]{};
		\fill (5,1) circle [radius=1pt];
		\fill (5.15,1) circle [radius=1pt];
		\fill (4.85,1) circle [radius=1pt];
		\draw [decorate,decoration={brace,amplitude=5pt}](1.8,2.1) -- (7.7,2.1);
		\draw node at (4.75,2.4) {\footnotesize $2n$};
		\end{tikzpicture}
		\caption{}
	\end{subfigure}
	\hfill
	\begin{subfigure}[t]{.20\textwidth}
		\centering
		\raisebox{.3\height}{
		\begin{tikzpicture}
		\tikzstyle{gauge} = [circle,draw];
			\tikzstyle{flavour} = [regular polygon,regular polygon sides=4,draw];
			\node (g1) [gauge, label=below:{$C_1$}]{};
			\node (g0) [gauge, left of=g1, label=below:{$O_1$}]{};
			\node (f1) [flavour, above of=g1, label=above:{$O_{2n+1}$}]{};
			\draw (g0)--(g1)--(f1);
		\end{tikzpicture}
		}
		\caption{}
	\end{subfigure}
	\hfill
	\begin{subfigure}[t]{.75\textwidth}
		  \centering
	\begin{tikzpicture}
	\tikzstyle{gauge} = [circle, draw];
	\tikzstyle{flavour} = [regular polygon,regular polygon sides=4, draw];
	\node (g1) [gauge,label=below:{$O_2$}] {};
	\node (g2) [gauge, right of=g1,label=below:{$C_1$}] {};
	\node (g3) [gauge, right of=g2,label=below:{$O_3$}] {};
	\node (g4) [gauge, right of=g3,label=below:{$C_1$}] {};
	\node (g5) [gauge, right of=g4,label=below:{$O_3$}] {};
	\node (gd) [right of=g5] {$\dots$};
	\node (g7) [gauge, right of=gd,label=below:{$O_3$}] {};
	\node (g8) [gauge, right of=g7,label=below:{$C_1$}] {};
	\node (g9) [gauge, right of=g8,label=below:{$O_3$}] {};
	\node (g10) [gauge, right of=g9,label=below:{$C_1$}] {};
	\node (g11) [gauge, right of=g10,label=below:{$SO_2$}] {};	\node (f2) [flavour,above of=g2,label=above:{$O_1$}] {};
	\node (f10) [flavour,above of=g10,label=above:{$O_1$}] {};
	\draw (g1)--(g2)--(g3)--(g4)--(g5)--(gd)--(g7)--(g8)--(g9)--(g10)--(g11)
			(g2)--(f2)
			(g10)--(f10)
		;
	\draw [decorate,decoration={brace,mirror,amplitude=5pt}](-.5,-.8) -- (10.5,-.8) node [black,midway,xshift=-0.6cm] { };
		\draw node at (5,-1.3) {\footnotesize $2n-1$};
	\end{tikzpicture}
		\caption{}
	\end{subfigure}
	\caption{$\tilde{b}_n:=\Or_{(3,1^{2n-2})}$ singularity. (a) Brane subsystem that represents the moduli space generated by $2n-1$ D3-branes. There are a total of $2n+1$ half D5-branes in the system. (b) After S-duality the moduli space generated by these D3-branes as a Coulomb branch. (c) Quiver corresponding to brane system (a), its Higgs branch is $\M_H=\Or_{(3,1^{2n-2})}$ (see \cite{HK16} for example). (d) Quiver corresponding to brane system (b), its Coulomb branch is  $\M_C=\Or_{(3,1^{2n-2})}$ (see \cite{CHZ17}). Note that a node $SO_2$ has been chosen for the leftmost node of the quiver. This is related to the Lusztig's Canonical Quotient of the orbit being $\bar A(\mathcal {O}_{(3,1^{2n-2})})=\mathbb{Z}_2$.}
	\label{fig:A1bn}
\end{figure}

\subsection{$c_n$ transitions}

Similar to the previous case, $c_n$ transitions are not directly observed in the brane configurations. The composition of one $c_{n-1}$ KP transition followed by a $c_n$ KP transition is the one realized by the branes instead, this corresponds to the closure of the next to minimal nilpotent orbit that we denote:
\begin{align}
	\tilde c_n:=\Or_{(2^2,1^{2n-4})}
\end{align}
We see the brane configuration for the case of $c_1$ transition followed by $c_2$ in figure \ref{fig:Sp4subregKP}(a). And this can be generalized to $c_{n-1}$ followed by $c_n$, figure \ref{fig:cnm1cn}.

\begin{figure}[t]
		\centering
	\begin{subfigure}[t]{.49\textwidth}
   		\centering
		\begin{tikzpicture}
		\draw[dashed] 	(2,0)--(2,2)
				(3.5,0)--(3.5,2)
				(4,0)--(4,2)
				(4.5,0)--(4.5,2)
				(5.5,0)--(5.5,2)
				(6,0)--(6,2)
				(6.5,0)--(6.5,2)
				(7.5,0)--(7.5,2)
				(8,0)--(8,2)
				;
		\draw [dotted]
				(2.5,1)--(3,1)
				(7,1)--(8.5,1);
		\draw	(2,1)--(2.5,1)
				(3,1)--(3.5,1)
				(4,1)--(4.5,1)
				(5.5,1)--(6,1)
				(6.5,1)--(7,1);
		\draw	(2,1.8)--(3.5,1.8)
				(2,.2)--(3.5,.2)
				(3.5,1.7)--(4,1.7)
				(4,1.8)--(4.5,1.8)
				(3.5,.3)--(4,.3)
				(4,.2)--(4.5,.2)
				(4.5,1.7)--(4.7,1.7)
				(4.5,.3)--(4.7,.3)
				(5.3,1.7)--(5.5,1.7)
				(5.3,.3)--(5.5,.3)
				(5.5,1.8)--(6,1.8)
				(5.5,.2)--(6,.2)
				(6,1.7)--(6.5,1.7)
				(6,.3)--(6.5,.3)
				(6.5,1.8)--(7.5,1.8)
				(6.5,.2)--(7.5,.2)
				;
		\draw 	(2.5,1)node[circ]{}
				(3,1)node[circ]{}
				(7,1)node[circ]{};
		\draw 	(2.5,1)node[cross]{}
				(3,1)node[cross]{}
				(7,1)node[cross]{};
		\fill (5,1) circle [radius=1pt];
		\fill (5.15,1) circle [radius=1pt];
		\fill (4.85,1) circle [radius=1pt];
		\draw [decorate,decoration={brace,amplitude=5pt}](1.8,2.1) -- (7.7,2.1);
		\draw node at (4.75,2.4) {\footnotesize $2n$};
		\end{tikzpicture}
        \caption{}
    \end{subfigure}
	\hfill
	\begin{subfigure}[t]{.49\textwidth}
		\centering
		   		\centering
		\begin{tikzpicture}
		\draw 	(2,0)--(2,2)
				(3.5,0)--(3.5,2)
				(4,0)--(4,2)
				(4.5,0)--(4.5,2)
				(5.5,0)--(5.5,2)
				(6,0)--(6,2)
				(6.5,0)--(6.5,2)
				(7.5,0)--(7.5,2)
				(8,0)--(8,2)
				;
		\draw [dotted]
				(2.5,1)--(3,1)
				(7,1)--(7.5,1)
				(8,1)--(8.5,1);
		\draw 	(7.5,1)--(8,1);
		\draw[dotted]	(2,1)--(2.5,1)
				(3,1)--(3.5,1)
				(4,1)--(4.5,1)
				(5.5,1)--(6,1)
				(6.5,1)--(7,1);
		\draw	(2,1.8)--(3.5,1.8)
				(2,.2)--(3.5,.2)
				(3.5,1.7)--(4,1.7)
				(4,1.8)--(4.5,1.8)
				(3.5,.3)--(4,.3)
				(4,.2)--(4.5,.2)
				(4.5,1.7)--(4.7,1.7)
				(4.5,.3)--(4.7,.3)
				(5.3,1.7)--(5.5,1.7)
				(5.3,.3)--(5.5,.3)
				(5.5,1.8)--(6,1.8)
				(5.5,.2)--(6,.2)
				(6,1.7)--(6.5,1.7)
				(6,.3)--(6.5,.3)
				(6.5,1.8)--(7.5,1.8)
				(6.5,.2)--(7.5,.2)
				;
		\draw 	(2.5,1)node[cross]{}
				(3,1)node[cross]{}
				(7,1)node[cross]{};
		\fill (5,1) circle [radius=1pt];
		\fill (5.15,1) circle [radius=1pt];
		\fill (4.85,1) circle [radius=1pt];
		\draw [decorate,decoration={brace,amplitude=5pt}](1.8,2.1) -- (7.7,2.1);
		\draw node at (4.75,2.4) {\footnotesize $2n$};
		\end{tikzpicture}
		\caption{}
	\end{subfigure}
	\hfill
	\begin{subfigure}[t]{.20\textwidth}
		\centering
		\raisebox{.3\height}{
		\begin{tikzpicture}
		\tikzstyle{gauge} = [circle,draw];
			\tikzstyle{flavour} = [regular polygon,regular polygon sides=4,draw];
			\node (g1) [gauge, label=below:{$O_2$}]{};
			\node (f1) [flavour, above of=g1, label=above:{$C_{n}$}]{};
			\draw (g1)--(f1);
		\end{tikzpicture}
		}
		\caption{}
	\end{subfigure}
	\hfill
	\begin{subfigure}[t]{.75\textwidth}
		  \centering
	\begin{tikzpicture}
	\tikzstyle{gauge} = [circle, draw];
	\tikzstyle{flavour} = [regular polygon,regular polygon sides=4, draw];
	\node (g2) [gauge,,label=below:{$C_1$}] {};
	\node (g3) [gauge, right of=g2,label=below:{$[SO_2$}] {};
	\node (g4) [gauge, right of=g3,label=below:{$C_1$}] {};
	\node (g5) [gauge, right of=g4,label=below:{$SO_2$}] {};
	\node (gd) [right of=g5] {$\dots$};
	\node (g7) [gauge, right of=gd,label=below:{$SO_2$}] {};
	\node (g8) [gauge, right of=g7,label=below:{$C_1$}] {};
	\node (g9) [gauge, right of=g8,label=below:{$SO_2]$}] {};
	\node (g10) [gauge, right of=g9,label=below:{$C_1$}] {};
	\node (g11) [gauge, right of=g10,label=below:{$O_1$}] {};
	\node (f2) [flavour,above of=g2,label=above:{$O_2$}] {};
	\node (f10) [flavour,above of=g10,label=above:{$O_1$}] {};
	\draw (g2)--(g3)--(g4)--(g5)--(gd)--(g7)--(g8)--(g9)--(g10)--(g11)
			(g2)--(f2)
			(g10)--(f10)
		;
	\draw [decorate,decoration={brace,mirror,amplitude=5pt}](-.5,-.8) -- (9.5,-.8) node [black,midway,xshift=-0.6cm] { };
		\draw node at (4.5,-1.3) {\footnotesize $2n$};
	\end{tikzpicture}
		\caption{}
	\end{subfigure}
	\caption{$\tilde c_n:=\Or_{(2^2,1^{2n-4})}$ singularity. (a) Brane subsystem that represents the moduli space generated by $2n-1$ D3-branes. There are a total of $2n+1$ half D5-branes in the system. (b) After S-duality the moduli space generated by these D3-branes as a Coulomb branch. (c) Quiver corresponding to brane system (a), its Higgs branch is $\M_H=\Or_{(2^2,1^{2n-4})}$ (see for example \cite{HK16}). (d) Quiver corresponding to brane system (b), its Coulomb branch is predicted to be $\M_C=\Or_{(2^2,1^{2n-4})}$. The brackets in the quiver denote a consistent choice of the $O_2$ gauge nodes: a $\mathbb Z_2 $ extension of $SO_2\times SO_2\times \dots \times SO_2  \times SO_2$, according to \emph{Section 6: Diagonal $\mathbb Z_2$ actions on different magnetic lattices} of \cite{CHZ17}.}
	\label{fig:cnm1cn}
\end{figure}

\subsection{$d_n$ transitions}

A $d_n$ KP transition always takes place between the closure of the minimal nilpotent orbit of $\mathfrak{so}({2n})$, corresponding to partition $\lambda=(2^2,1^{2n-4})$, and the trivial orbit. The corresponding brane system is depicted in figure \ref{fig:dnsing}.

\begin{figure}[t]
		\centering
	\begin{subfigure}[t]{.49\textwidth}
   		\centering
		\begin{tikzpicture}
		\draw[dashed] 	
				(2,0)--(2,2)				
				(2.5,0)--(2.5,2)
				(3.5,0)--(3.5,2)
				(4,0)--(4,2)
				(4.5,0)--(4.5,2)
				(5.5,0)--(5.5,2)
				(6,0)--(6,2)
				(6.5,0)--(6.5,2)
				(7.5,0)--(7.5,2)
				(8,0)--(8,2)
				;
		\draw [dotted]
				(3,1)--(4.6,1)
				(5.4,1)--(7,1);
		\draw	(2,1)--(2.5,1)
				(7.5,1)--(8,1);
		\draw	(2.5,1.8)--(3.5,1.8)
				(2.5,.2)--(3.5,.2)
				(3.5,1.7)--(4,1.7)
				(4,1.8)--(4.5,1.8)
				(3.5,.3)--(4,.3)
				(4,.2)--(4.5,.2)
				(4.5,1.7)--(4.7,1.7)
				(4.5,.3)--(4.7,.3)
				(5.3,1.7)--(5.5,1.7)
				(5.3,.3)--(5.5,.3)
				(5.5,1.8)--(6,1.8)
				(5.5,.2)--(6,.2)
				(6,1.7)--(6.5,1.7)
				(6,.3)--(6.5,.3)
				(6.5,1.8)--(7.5,1.8)
				(6.5,.2)--(7.5,.2)
				;
		\draw 	(3,1)node[circ]{}
				(7,1)node[circ]{};
		\draw 	(3,1)node[cross]{}
				(7,1)node[cross]{};
		\fill (5,1) circle [radius=1pt];
		\fill (5.15,1) circle [radius=1pt];
		\fill (4.85,1) circle [radius=1pt];
		\draw [decorate,decoration={brace,amplitude=5pt}](1.8,2.1) -- (8.2,2.1);
		\draw node at (5,2.4) {\footnotesize $2n$};
		\end{tikzpicture}
        \caption{}
    \end{subfigure}
	\hfill
	\begin{subfigure}[t]{.49\textwidth}
		\centering
		   		\centering
		\begin{tikzpicture}
		\draw 	
				(2,0)--(2,2)				
				(2.5,0)--(2.5,2)
				(3.5,0)--(3.5,2)
				(4,0)--(4,2)
				(4.5,0)--(4.5,2)
				(5.5,0)--(5.5,2)
				(6,0)--(6,2)
				(6.5,0)--(6.5,2)
				(7.5,0)--(7.5,2)
				(8,0)--(8,2)
				;
		\draw [dotted]
				(3.5,1)--(4,1)
				(4.5,1)--(4.6,1)
				(5.4,1)--(5.5,1)
				(6,1)--(6.5,1);
		\draw[dotted]	(2,1)--(2.5,1)
				(7.5,1)--(8,1);
		\draw	(3,1)--(3.5,1)
				(4,1)--(4.5,1)
				(5.5,1)--(6,1)
				(6.5,1)--(7,1);
		\draw	(2.5,1.8)--(3.5,1.8)
				(2.5,.2)--(3.5,.2)
				(3.5,1.7)--(4,1.7)
				(4,1.8)--(4.5,1.8)
				(3.5,.3)--(4,.3)
				(4,.2)--(4.5,.2)
				(4.5,1.7)--(4.7,1.7)
				(4.5,.3)--(4.7,.3)
				(5.3,1.7)--(5.5,1.7)
				(5.3,.3)--(5.5,.3)
				(5.5,1.8)--(6,1.8)
				(5.5,.2)--(6,.2)
				(6,1.7)--(6.5,1.7)
				(6,.3)--(6.5,.3)
				(6.5,1.8)--(7.5,1.8)
				(6.5,.2)--(7.5,.2)
				;
		\draw 	(3,1)node[cross]{}
				(7,1)node[cross]{};
		\fill (5,1) circle [radius=1pt];
		\fill (5.15,1) circle [radius=1pt];
		\fill (4.85,1) circle [radius=1pt];
		\draw [decorate,decoration={brace,amplitude=5pt}](1.8,2.1) -- (8.2,2.1);
		\draw node at (5,2.4) {\footnotesize $2n$};
		\end{tikzpicture}
		\caption{}
	\end{subfigure}
	\hfill
	\begin{subfigure}[t]{.20\textwidth}
		\centering
		\raisebox{.3\height}{
		\begin{tikzpicture}
		\tikzstyle{gauge} = [circle,draw];
			\tikzstyle{flavour} = [regular polygon,regular polygon sides=4,draw];
			\node (g1) [gauge, label=below:{$C_1$}]{};
			\node (f1) [flavour, above of=g1, label=above:{$O_{2n}$}]{};
			\draw (g1)--(f1);
		\end{tikzpicture}
		}
		\caption{}
	\end{subfigure}
	\hfill
	\begin{subfigure}[t]{.75\textwidth}
		  \centering
	\begin{tikzpicture}
	\tikzstyle{gauge} = [circle, draw];
	\tikzstyle{flavour} = [regular polygon,regular polygon sides=4, draw];
	\node (g1) [gauge,label=below:{$SO_2$}] {};
	\node (g2) [gauge, right of=g1,label=below:{$C_1$}] {};
	\node (g3) [gauge, right of=g2,label=below:{$O_3$}] {};
	\node (g4) [gauge, right of=g3,label=below:{$C_1$}] {};
	\node (g5) [gauge, right of=g4,label=below:{$O_3$}] {};
	\node (gd) [right of=g5] {$\dots$};
	\node (g7) [gauge, right of=gd,label=below:{$O_3$}] {};
	\node (g8) [gauge, right of=g7,label=below:{$C_1$}] {};
	\node (g9) [gauge, right of=g8,label=below:{$O_3$}] {};
	\node (g10) [gauge, right of=g9,label=below:{$C_1$}] {};
	\node (g11) [gauge, right of=g10,label=below:{$SO_2$}] {};	\node (f2) [flavour,above of=g2,label=above:{$O_1$}] {};
	\node (f10) [flavour,above of=g10,label=above:{$O_1$}] {};
	\draw (g1)--(g2)--(g3)--(g4)--(g5)--(gd)--(g7)--(g8)--(g9)--(g10)--(g11)
			(g2)--(f2)
			(g10)--(f10)
		;
	\draw [decorate,decoration={brace,mirror,amplitude=5pt}](-.5,-.8) -- (10.5,-.8) node [black,midway,xshift=-0.6cm] { };
		\draw node at (5,-1.3) {\footnotesize $2n-3$};
	\end{tikzpicture}
		\caption{}
	\end{subfigure}
	\caption{$d_n$ singularity. (a) Brane subsystem that represents the moduli space generated by $2n-3$ D3-branes. There are $2n$ half D5-branes in the system. (b) After S-duality the moduli space generated by these D3-branes as a Coulomb branch. (c) Quiver corresponding to brane system (a), its Higgs branch is $\M_H=\Or_{(2^2,1^{2n-4})}\subset \mathfrak{so}(2n)$. (d) Quiver corresponding to brane system (b), its Coulomb branch is $\M_C=\Or_{(2^2,1^{2n-4})}\subset \mathfrak{so}(2n)$. Note the choice of $SO_2$ gauge nodes according to \cite{CHZ17}. This is related to the Lusztig's Canonical Quotient of the orbit $\bar A(\mathcal {O}_{(2^2,1^{2n-4})})$ being trivial. }
	\label{fig:dnsing}
\end{figure}

\subsection{Summary}

Tables \ref{tab:subregularKPsingularities} and \ref{tab:minimalKPsingularities} summarize the relevant brane configurations corresponding to each KP transition.

Note that table \ref{tab:subregularKPsingularities} exhausts all possible combinations of \emph{boundary conditions} on a single interval between two half D5-branes. In the first row, the interval is flanked by O3-planes of types $O3^+$ and $O3^+$. In the second row, the types are $\widetilde{O3^+}$ and $\widetilde{O3^+}$. In the third row, the flanking O3-planes are $\widetilde{O3^+}$ and $\widetilde{O3^-}$. In the fourth row, $\widetilde{O3^-}$ and $\widetilde{O3^-}$. Any other combination would involve an $O3^-$ plane, however, after the collapse transition a half NS5-brane would be pushed though the half D5-brane connected to the $O3^-$ plane, changing the nature of the flanking plane to $O3^+$, and the Higgs branch and quiver would be one of the previous four cases contained in the table.

This constitutes an alluring result: all the \emph{surface singularities} employed by Kraft and Procesi are described by the moduli space of a single D3-brane with changing boundary conditions.

\begin{table}[t]
	\centering
	\begin{tabular}{| c | c | c | c |}
	\hline
	$S$ & Brane configuration & $\M_H$ of & $\M_C$ of \\ \hline 
	& & & \\
	$D_n$ & \raisebox{-.5\height}{\begin{tikzpicture}[scale=0.6, transform shape]
				\draw[dashed] 	(1.5,-1)--(1.5,3)
				(4.5,-1)--(4.5,3);
		
		\draw (1.5,2.5)--(4.5,2.5);
		\draw (1.5,-.5)--(4.5,-.5);
		
		\draw[dotted] (1,1)--(2,1);
		\draw[dotted] (4,1)--(5,1);
		
		\draw (2,1)--(2.5,1);
		\draw (3.5,1)--(4,1);
		
		\draw (2,1) node[circle,fill=white,draw=black] {};
		\draw (2.5,1) node[circle,fill=white,draw=black] {};
		\fill (3,1) circle [radius=.5pt];
		\fill (3.15,1) circle [radius=.5pt];
		\fill (2.85,1) circle [radius=.5pt];
		\draw (3.5,1) node[circle,fill=white,draw=black] {};
		\draw (4,1) node[circle,fill=white,draw=black] {};
				
		\draw (2,1) node[cross] {};
		\draw (2.5,1) node[cross] {};
		\draw (3.5,1) node[cross] {};
		\draw (4,1) node[cross] {};
		
		\draw [decorate,decoration={brace,amplitude=5pt}](1.8,1.3) -- (4.2,1.3);
		\draw node at (3,1.8) {\footnotesize $2n-2$};
		
		\end{tikzpicture}} &\raisebox{-.5\height}{\begin{tikzpicture}[scale=0.6,transform shape]
	\tikzstyle{gauge} = [circle, draw];
	\tikzstyle{flavour} = [regular polygon,regular polygon sides=4, draw];
	\node (g1) [gauge,label=below:{$O_2$}] {};
	\node (g2) [gauge, right of=g1,label=below:{$C_1$}] {};
	\node (g3) [gauge, right of=g2,label=below:{$O_3$}] {};
	\node (g4) [gauge, right of=g3,label=below:{$C_1$}] {};
	\node (g5) [gauge, right of=g4,label=below:{$O_3$}] {};
	\node (gd) [right of=g5] {$\dots$};
	\node (g7) [gauge, right of=gd,label=below:{$O_3$}] {};
	\node (g8) [gauge, right of=g7,label=below:{$C_1$}] {};
	\node (g9) [gauge, right of=g8,label=below:{$O_3$}] {};
	\node (g10) [gauge, right of=g9,label=below:{$C_1$}] {};
	\node (g11) [gauge, right of=g10,label=below:{$O_2$}] {};	\node (f2) [flavour,above of=g2,label=above:{$O_1$}] {};
	\node (f10) [flavour,above of=g10,label=above:{$O_1$}] {};
	\draw (g1)--(g2)--(g3)--(g4)--(g5)--(gd)--(g7)--(g8)--(g9)--(g10)--(g11)
			(g2)--(f2)
			(g10)--(f10)
		;
	\draw [decorate,decoration={brace,mirror,amplitude=5pt}](-.5,-.8) -- (10.5,-.8) node [black,midway,xshift=-0.6cm] { };
		\draw node at (5,-1.3) {\footnotesize $2n-3$};
	\end{tikzpicture}} & \raisebox{-.3\height}{\begin{tikzpicture}[scale=0.6,transform shape]
		\tikzstyle{gauge} = [circle,draw];
			\tikzstyle{flavour} = [regular polygon,regular polygon sides=4,draw];
			\node (g1) [gauge, label=below:{$C_1$}]{};
			\node (f1) [flavour, above of=g1, label=above:{$O_{2n}$}]{};
			\draw (g1)--(f1);
		\end{tikzpicture}}\\ 
	& & &\\ \hline
	& & &  \\
	$D_{n}$ &\raisebox{-.4\height}{\begin{tikzpicture}[scale=0.6,transform shape]
				\draw[dashed] 	(1.5,-1)--(1.5,3)
				(4.5,-1)--(4.5,3);
		
		\draw (1.5,2.5)--(4.5,2.5);
		\draw (1.5,-.5)--(4.5,-.5);
		
		\draw[dotted] (1,1)--(2,1);
		\draw[dotted] (4,1)--(5,1);

		\draw (2,1) node[circle,fill=white,draw=black] {};
		\draw (2.5,1) node[circle,fill=white,draw=black] {};
		\fill (3,1) circle [radius=.5pt];
		\fill (3.15,1) circle [radius=.5pt];
		\fill (2.85,1) circle [radius=.5pt];
		\draw (3.5,1) node[circle,fill=white,draw=black] {};
		\draw (4,1) node[circle,fill=white,draw=black] {};
				
		\draw (2,1) node[cross] {};
		\draw (2.5,1) node[cross] {};
		\draw (3.5,1) node[cross] {};
		\draw (4,1) node[cross] {};
		
		\draw [decorate,decoration={brace,amplitude=5pt}](1.8,1.3) -- (4.2,1.3);
		\draw node at (3,1.8) {\footnotesize $2n-2$};
		
		\end{tikzpicture}} &\raisebox{-.5\height}{\begin{tikzpicture}[scale=0.6,transform shape]
	\tikzstyle{gauge} = [circle, draw];
	\tikzstyle{flavour} = [regular polygon,regular polygon sides=4, draw];
	\node (g1) [gauge,label=below:{$O_1$}] {};
	\node (g2) [gauge, right of=g1,label=below:{$C_1$}] {};
	\node (g3) [gauge, right of=g2,label=below:{$O_2$}] {};
	\node (g4) [gauge, right of=g3,label=below:{$C_1$}] {};
	\node (g5) [gauge, right of=g4,label=below:{$O_2$}] {};
	\node (gd) [right of=g5] {$\dots$};
	\node (g7) [gauge, right of=gd,label=below:{$O_2$}] {};
	\node (g8) [gauge, right of=g7,label=below:{$C_1$}] {};
	\node (g9) [gauge, right of=g8,label=below:{$O_2$}] {};
	\node (g10) [gauge, right of=g9,label=below:{$C_1$}] {};
	\node (g11) [gauge, right of=g10,label=below:{$O_1$}] {};
	\node (f2) [flavour,above of=g2,label=above:{$O_1$}] {};
	\node (f10) [flavour,above of=g10,label=above:{$O_1$}] {};
	\draw (g1)--(g2)--(g3)--(g4)--(g5)--(gd)--(g7)--(g8)--(g9)--(g10)--(g11)
			(g2)--(f2)
			(g10)--(f10)
		;
	\draw [decorate,decoration={brace,mirror,amplitude=5pt}](-.5,-.8) -- (10.5,-.8) node [black,midway,xshift=-0.6cm] { };
		\draw node at (5,-1.3) {\footnotesize $2n-3$};
	\end{tikzpicture}} & \raisebox{-.3\height}{\begin{tikzpicture}[scale=0.6,transform shape]
		\tikzstyle{gauge} = [circle,draw];
			\tikzstyle{flavour} = [regular polygon,regular polygon sides=4,draw];
			\node (g1) [gauge, label=below:{$O_2$}]{};
			\node (f1) [flavour, above of=g1, label=above:{$C_{n-2}$}]{};
			\draw (g1)--(f1);
		\end{tikzpicture}}\\ 
	& & & \\ \hline
	& & &  \\
	$A_{2n-1}$ &\raisebox{-.4\height}{\begin{tikzpicture}[scale=0.6,transform shape]
					\draw[dashed] 	(1.5,-1)--(1.5,3)
				(4.5,-1)--(4.5,3);
		
		\draw (1.5,2.5)--(4.5,2.5);
		\draw (1.5,-.5)--(4.5,-.5);
		
		\draw[dotted] (1,1)--(2,1);
		\draw (4.5,1)--(5,1);
		
		\draw[dotted] (3.5,1)--(4,1);
		
		\draw (2,1) node[circle,fill=white,draw=black] {};
		\draw (2.5,1) node[circle,fill=white,draw=black] {};
		\fill (3,1) circle [radius=.5pt];
		\fill (3.15,1) circle [radius=.5pt];
		\fill (2.85,1) circle [radius=.5pt];
		\draw (3.5,1) node[circle,fill=white,draw=black] {};
		\draw (4,1) node[circle,fill=white,draw=black] {};
				
		\draw (2,1) node[cross] {};
		\draw (2.5,1) node[cross] {};
		\draw (3.5,1) node[cross] {};
		\draw (4,1) node[cross] {};
		
		\draw [decorate,decoration={brace,amplitude=5pt}](1.8,1.3) -- (4.2,1.3);
		\draw node at (3,1.8) {\footnotesize $2n+1$};
		
		\end{tikzpicture}
} &\raisebox{-.5\height}{\begin{tikzpicture}[scale=0.6,transform shape]
	\tikzstyle{gauge} = [circle, draw];
	\tikzstyle{flavour} = [regular polygon,regular polygon sides=4, draw];
	\node (g1) [gauge,label=below:{$O_1$}] {};
	\node (g2) [gauge, right of=g1,label=below:{$C_1$}] {};
	\node (g3) [gauge, right of=g2,label=below:{$O_2$}] {};
	\node (g4) [gauge, right of=g3,label=below:{$C_1$}] {};
	\node (g5) [gauge, right of=g4,label=below:{$O_2$}] {};
	\node (gd) [right of=g5] {$\dots$};
	\node (g7) [gauge, right of=gd,label=below:{$O_2$}] {};
	\node (g8) [gauge, right of=g7,label=below:{$C_1$}] {};
	\node (g9) [gauge, right of=g8,label=below:{$O_2$}] {};
	\node (g10) [gauge, right of=g9,label=below:{$C_1$}] {};
	\node (f2) [flavour,above of=g2,label=above:{$O_1$}] {};
	\node (f10) [flavour,above of=g10,label=above:{$O_2$}] {};
	\draw (g1)--(g2)--(g3)--(g4)--(g5)--(gd)--(g7)--(g8)--(g9)--(g10)
			(g2)--(f2)
			(g10)--(f10)
		;
	\draw [decorate,decoration={brace,mirror,amplitude=5pt}](-.5,-.8) -- (9.5,-.8) node [black,midway,xshift=-0.6cm] { };
		\draw node at (4.5,-1.3) {\footnotesize $2n$};
	\end{tikzpicture}} &\raisebox{-.3\height}{\begin{tikzpicture}[scale=0.6,transform shape]
		\tikzstyle{gauge} = [circle,draw];
			\tikzstyle{flavour} = [regular polygon,regular polygon sides=4,draw];
			\node (g1) [gauge, label=below:{$SO_2$}]{};
			\node (f1) [flavour, above of=g1, label=above:{$C_{n}$}]{};
			\draw (g1)--(f1);
		\end{tikzpicture}} \\ 
	& & & \\ \hline
	& & &  \\
	$A_{2n-1}\cup A_{2n-1}$ &\raisebox{-.4\height}{\begin{tikzpicture}[scale=0.6,transform shape]
				\draw[dashed] 				
				(1.5,-1)--(1.5,3)
				(4.5,-1)--(4.5,3);
		
		\draw (1.5,2.5)--(4.5,2.5);
		\draw (1.5,-.5)--(4.5,-.5);
		
		\draw (1,1)--(1.5,1);
		\draw (4.5,1)--(5,1);
		\draw[dotted] (2,1)--(2.5,1);
		\draw[dotted] (3.5,1)--(4,1);
		
		\draw (2,1) node[circle,fill=white,draw=black] {};
		\draw (2.5,1) node[circle,fill=white,draw=black] {};
		\fill (3,1) circle [radius=.5pt];
		\fill (3.15,1) circle [radius=.5pt];
		\fill (2.85,1) circle [radius=.5pt];
		\draw (3.5,1) node[circle,fill=white,draw=black] {};
		\draw (4,1) node[circle,fill=white,draw=black] {};
				
		\draw (2,1) node[cross] {};
		\draw (2.5,1) node[cross] {};
		\draw (3.5,1) node[cross] {};
		\draw (4,1) node[cross] {};
		
		\draw [decorate,decoration={brace,amplitude=5pt}](1.8,1.3) -- (4.2,1.3);
		\draw node at (3,1.8) {\footnotesize $2n$};
		
		\end{tikzpicture}
} &\raisebox{-.5\height}{\begin{tikzpicture}[scale=0.6,transform shape]
	\tikzstyle{gauge} = [circle, draw];
	\tikzstyle{flavour} = [regular polygon,regular polygon sides=4, draw];
	\node (g2) [gauge,label=below:{$C_1$}] {};
	\node (g3) [gauge, right of=g2,label=below:{$O_2$}] {};
	\node (g4) [gauge, right of=g3,label=below:{$C_1$}] {};
	\node (g5) [gauge, right of=g4,label=below:{$O_2$}] {};
	\node (gd) [right of=g5] {$\dots$};
	\node (g7) [gauge, right of=gd,label=below:{$O_2$}] {};
	\node (g8) [gauge, right of=g7,label=below:{$C_1$}] {};
	\node (g9) [gauge, right of=g8,label=below:{$O_2$}] {};
	\node (g10) [gauge, right of=g9,label=below:{$C_1$}] {};
	\node (f2) [flavour,above of=g2,label=above:{$O_2$}] {};
	\node (f10) [flavour,above of=g10,label=above:{$O_2$}] {};
	\draw (g2)--(g3)--(g4)--(g5)--(gd)--(g7)--(g8)--(g9)--(g10)
			(g2)--(f2)
			(g10)--(f10)
		;
	\draw [decorate,decoration={brace,mirror,amplitude=5pt}](-.5,-.8) -- (8.5,-.8) node [black,midway,xshift=-0.6cm] { };
		\draw node at (4,-1.3) {\footnotesize $2n-1$};
	\end{tikzpicture}} & \raisebox{-.3\height}{\begin{tikzpicture}[scale=0.6,transform shape]
		\tikzstyle{gauge} = [circle,draw];
			\tikzstyle{flavour} = [regular polygon,regular polygon sides=4,draw];
			\node (g1) [gauge, label=below:{$SO_2$}]{};
			\node (f1) [flavour, above of=g1, label=above:{$C_{n}$}]{};
			\draw (g1)--(f1);
		\end{tikzpicture}}\\ 
	& & & \\ \hline
	\end{tabular}
	\caption{\emph{Surface} KP singularities $S$ and their brane configurations. The second column shows their brane systems as Higgs branches. The third column shows the quivers obtained from the brane configurations in the second column. The fourth column shows the quivers obtained after performing S-duality on the second column. Note that the quiver in the fourth column of the fourth row does not have $\M_C=A_{2n-1}\cup A_{2n-1}$ but rather $\M_C=A_{2n-1}$, following an effect analogous to \cite{FH16} explained in section \ref{sec:AnUAntransitions}.}
	\label{tab:subregularKPsingularities}
\end{table}

\begin{table}[t]
	\centering
	\begin{tabular}{| c | c | c | c |}
	\hline
	$S$ & Brane configuration & $\M_H$ of & $\M_C$ of \\ \hline 
	& & & \\
	$\tilde b_n$ &\raisebox{-.3\height}{\begin{tikzpicture}[scale=0.6,transform shape]
		\draw[dashed] 	(2,0)--(2,2)
				(3.5,0)--(3.5,2)
				(4,0)--(4,2)
				(4.5,0)--(4.5,2)
				(5.5,0)--(5.5,2)
				(6,0)--(6,2)
				(6.5,0)--(6.5,2)
				(7.5,0)--(7.5,2)
				(8,0)--(8,2)
				;
		\draw [dotted] (1.5,1)--(2.5,1)
				(3,1)--(4.7,1)
				(5.3,1)--(7,1);
		\draw	(7.5,1)--(8,1);
		\draw	(2,1.8)--(3.5,1.8)
				(2,.2)--(3.5,.2)
				(3.5,1.7)--(4,1.7)
				(4,1.8)--(4.5,1.8)
				(3.5,.3)--(4,.3)
				(4,.2)--(4.5,.2)
				(4.5,1.7)--(4.7,1.7)
				(4.5,.3)--(4.7,.3)
				(5.3,1.7)--(5.5,1.7)
				(5.3,.3)--(5.5,.3)
				(5.5,1.8)--(6,1.8)
				(5.5,.2)--(6,.2)
				(6,1.7)--(6.5,1.7)
				(6,.3)--(6.5,.3)
				(6.5,1.8)--(7.5,1.8)
				(6.5,.2)--(7.5,.2)
				;
		\draw 	(2.5,1)node[circ]{}
				(3,1)node[circ]{}
				(7,1)node[circ]{};
		\draw 	(2.5,1)node[cross]{}
				(3,1)node[cross]{}
				(7,1)node[cross]{};
		\fill (5,1) circle [radius=1pt];
		\fill (5.15,1) circle [radius=1pt];
		\fill (4.85,1) circle [radius=1pt];
		\draw [decorate,decoration={brace,amplitude=5pt}](1.8,2.1) -- (7.7,2.1);
		\draw node at (4.75,2.6) {\footnotesize $2n$};
		\end{tikzpicture}}  & \raisebox{-.3\height}{\begin{tikzpicture}[scale=0.6,transform shape]
		\tikzstyle{gauge} = [circle,draw];
			\tikzstyle{flavour} = [regular polygon,regular polygon sides=4,draw];
			\node (g1) [gauge, label=below:{$C_1$}]{};
			\node (g2) [gauge, right of=g1, label=below:{$O_1$}]{};
			\node (f1) [flavour, above of=g1, label=above:{$O_{2n+1}$}]{};
			\draw (g2)--(g1)--(f1);
		\end{tikzpicture}}&\raisebox{-.5\height}{\begin{tikzpicture}[scale=0.6,transform shape]
	\tikzstyle{gauge} = [circle, draw];
	\tikzstyle{flavour} = [regular polygon,regular polygon sides=4, draw];
	\node (g1) [gauge,label=below:{$SO_2$}] {};
	\node (g2) [gauge, right of=g1,label=below:{$C_1$}] {};
	\node (g3) [gauge, right of=g2,label=below:{$O_3$}] {};
	\node (g4) [gauge, right of=g3,label=below:{$C_1$}] {};
	\node (g5) [gauge, right of=g4,label=below:{$O_3$}] {};
	\node (gd) [right of=g5] {$\dots$};
	\node (g7) [gauge, right of=gd,label=below:{$O_3$}] {};
	\node (g8) [gauge, right of=g7,label=below:{$C_1$}] {};
	\node (g9) [gauge, right of=g8,label=below:{$O_3$}] {};
	\node (g10) [gauge, right of=g9,label=below:{$C_1$}] {};
	\node (g11) [gauge, right of=g10,label=below:{$O_2$}] {};	\node (f2) [flavour,above of=g2,label=above:{$O_1$}] {};
	\node (f10) [flavour,above of=g10,label=above:{$O_1$}] {};
	\draw (g1)--(g2)--(g3)--(g4)--(g5)--(gd)--(g7)--(g8)--(g9)--(g10)--(g11)
			(g2)--(f2)
			(g10)--(f10)
		;
	\draw [decorate,decoration={brace,mirror,amplitude=5pt}](-.5,-.8) -- (10.5,-.8) node [black,midway,xshift=-0.6cm] { };
		\draw node at (5,-1.3) {\footnotesize $2n-1$};
	\end{tikzpicture}}\\ 
	& & & \\ \hline
	& & &  \\
	$\tilde c_n$ &\raisebox{-.3\height}{\begin{tikzpicture}[scale=0.6,transform shape]
		\draw[dashed] 	(2,0)--(2,2)
				(3.5,0)--(3.5,2)
				(4,0)--(4,2)
				(4.5,0)--(4.5,2)
				(5.5,0)--(5.5,2)
				(6,0)--(6,2)
				(6.5,0)--(6.5,2)
				(7.5,0)--(7.5,2)
				(8,0)--(8,2)
				;
		\draw [dotted]
				(2.5,1)--(3,1)
				(7,1)--(8.5,1);
		\draw	(2,1)--(2.5,1)
				(3,1)--(3.5,1)
				(4,1)--(4.5,1)
				(5.5,1)--(6,1)
				(6.5,1)--(7,1);
		\draw	(2,1.8)--(3.5,1.8)
				(2,.2)--(3.5,.2)
				(3.5,1.7)--(4,1.7)
				(4,1.8)--(4.5,1.8)
				(3.5,.3)--(4,.3)
				(4,.2)--(4.5,.2)
				(4.5,1.7)--(4.7,1.7)
				(4.5,.3)--(4.7,.3)
				(5.3,1.7)--(5.5,1.7)
				(5.3,.3)--(5.5,.3)
				(5.5,1.8)--(6,1.8)
				(5.5,.2)--(6,.2)
				(6,1.7)--(6.5,1.7)
				(6,.3)--(6.5,.3)
				(6.5,1.8)--(7.5,1.8)
				(6.5,.2)--(7.5,.2)
				;
		\draw 	(2.5,1)node[circ]{}
				(3,1)node[circ]{}
				(7,1)node[circ]{};
		\draw 	(2.5,1)node[cross]{}
				(3,1)node[cross]{}
				(7,1)node[cross]{};
		\fill (5,1) circle [radius=1pt];
		\fill (5.15,1) circle [radius=1pt];
		\fill (4.85,1) circle [radius=1pt];
		\draw [decorate,decoration={brace,amplitude=5pt}](1.8,2.1) -- (7.7,2.1);
		\draw node at (4.75,2.6) {\footnotesize $2n$};
		\end{tikzpicture}
}  &\raisebox{-.3\height}{\begin{tikzpicture}[scale=0.6,transform shape]
		\tikzstyle{gauge} = [circle,draw];
			\tikzstyle{flavour} = [regular polygon,regular polygon sides=4,draw];
			\node (g1) [gauge, label=below:{$O_2$}]{};
			\node (f1) [flavour, above of=g1, label=above:{$C_{n}$}]{};
			\draw (g1)--(f1);
		\end{tikzpicture}} &\raisebox{-.5\height}{\begin{tikzpicture}[scale=0.6,transform shape]
	\tikzstyle{gauge} = [circle, draw];
	\tikzstyle{flavour} = [regular polygon,regular polygon sides=4, draw];
	\node (g1) [gauge,label=below:{$O_1$}] {};
	\node (g2) [gauge, right of=g1,label=below:{$C_1$}] {};
	\node (g3) [gauge, right of=g2,label=below:{$[SO_2$}] {};
	\node (g4) [gauge, right of=g3,label=below:{$C_1$}] {};
	\node (g5) [gauge, right of=g4,label=below:{$SO_2$}] {};
	\node (gd) [right of=g5] {$\dots$};
	\node (g7) [gauge, right of=gd,label=below:{$SO_2$}] {};
	\node (g8) [gauge, right of=g7,label=below:{$C_1$}] {};
	\node (g9) [gauge, right of=g8,label=below:{$SO_2]$}] {};
	\node (g10) [gauge, right of=g9,label=below:{$C_1$}] {};
	\node (f2) [flavour,above of=g2,label=above:{$O_1$}] {};
	\node (f10) [flavour,above of=g10,label=above:{$O_2$}] {};
	\draw (g1)--(g2)--(g3)--(g4)--(g5)--(gd)--(g7)--(g8)--(g9)--(g10)
			(g2)--(f2)
			(g10)--(f10)
		;
	\draw [decorate,decoration={brace,mirror,amplitude=5pt}](-.5,-.8) -- (9.5,-.8) node [black,midway,xshift=-0.6cm] { };
		\draw node at (4.5,-1.3) {\footnotesize $2n$};
	\end{tikzpicture}}\\ 
	& & & \\ \hline
	$d_n$ & \raisebox{-.3\height}{\begin{tikzpicture}[scale=0.6, transform shape]
		\draw[dashed] 	
				(2,0)--(2,2)				
				(2.5,0)--(2.5,2)
				(3.5,0)--(3.5,2)
				(4,0)--(4,2)
				(4.5,0)--(4.5,2)
				(5.5,0)--(5.5,2)
				(6,0)--(6,2)
				(6.5,0)--(6.5,2)
				(7.5,0)--(7.5,2)
				(8,0)--(8,2)
				;
		\draw [dotted]
				(3,1)--(4.6,1)
				(5.4,1)--(7,1);
		\draw	(2,1)--(2.5,1)
				(7.5,1)--(8,1);
		\draw	(2.5,1.8)--(3.5,1.8)
				(2.5,.2)--(3.5,.2)
				(3.5,1.7)--(4,1.7)
				(4,1.8)--(4.5,1.8)
				(3.5,.3)--(4,.3)
				(4,.2)--(4.5,.2)
				(4.5,1.7)--(4.7,1.7)
				(4.5,.3)--(4.7,.3)
				(5.3,1.7)--(5.5,1.7)
				(5.3,.3)--(5.5,.3)
				(5.5,1.8)--(6,1.8)
				(5.5,.2)--(6,.2)
				(6,1.7)--(6.5,1.7)
				(6,.3)--(6.5,.3)
				(6.5,1.8)--(7.5,1.8)
				(6.5,.2)--(7.5,.2)
				;
		\draw 	(3,1)node[circ]{}
				(7,1)node[circ]{};
		\draw 	(3,1)node[cross]{}
				(7,1)node[cross]{};
		\fill (5,1) circle [radius=1pt];
		\fill (5.15,1) circle [radius=1pt];
		\fill (4.85,1) circle [radius=1pt];
		\draw [decorate,decoration={brace,amplitude=5pt}](1.8,2.1) -- (8.2,2.1);
		\draw node at (5,2.6) {\footnotesize $2n$};
		\end{tikzpicture}} & \raisebox{-.3\height}{\begin{tikzpicture}[scale=0.6,transform shape]
		\tikzstyle{gauge} = [circle,draw];
			\tikzstyle{flavour} = [regular polygon,regular polygon sides=4,draw];
			\node (g1) [gauge, label=below:{$C_1$}]{};
			\node (f1) [flavour, above of=g1, label=above:{$O_{2n}$}]{};
			\draw (g1)--(f1);
		\end{tikzpicture}}&\raisebox{-.5\height}{\begin{tikzpicture}[scale=0.6,transform shape]
	\tikzstyle{gauge} = [circle, draw];
	\tikzstyle{flavour} = [regular polygon,regular polygon sides=4, draw];
	\node (g1) [gauge,label=below:{$SO_2$}] {};
	\node (g2) [gauge, right of=g1,label=below:{$C_1$}] {};
	\node (g3) [gauge, right of=g2,label=below:{$O_3$}] {};
	\node (g4) [gauge, right of=g3,label=below:{$C_1$}] {};
	\node (g5) [gauge, right of=g4,label=below:{$O_3$}] {};
	\node (gd) [right of=g5] {$\dots$};
	\node (g7) [gauge, right of=gd,label=below:{$O_3$}] {};
	\node (g8) [gauge, right of=g7,label=below:{$C_1$}] {};
	\node (g9) [gauge, right of=g8,label=below:{$O_3$}] {};
	\node (g10) [gauge, right of=g9,label=below:{$C_1$}] {};
	\node (g11) [gauge, right of=g10,label=below:{$SO_2$}] {};	\node (f2) [flavour,above of=g2,label=above:{$O_1$}] {};
	\node (f10) [flavour,above of=g10,label=above:{$O_1$}] {};
	\draw (g1)--(g2)--(g3)--(g4)--(g5)--(gd)--(g7)--(g8)--(g9)--(g10)--(g11)
			(g2)--(f2)
			(g10)--(f10)
		;
	\draw [decorate,decoration={brace,mirror,amplitude=5pt}](-.5,-.8) -- (10.5,-.8) node [black,midway,xshift=-0.6cm] { };
		\draw node at (5,-1.3) {\footnotesize $2n-3$};
	\end{tikzpicture}} \\ 
	& & &\\ \hline
	\end{tabular}
	\caption{\emph{Minimal} KP singularites $S$ and their brane configurations. $\tilde b_n$ denotes the closure of the next to minimal nilpotent orbit of $\gso (2n+1)$, $\tilde c_n$ denotes the closure of the next to minimal nilpotent orbit of $\gsp (n)$ and $d_n$ is used for the closure of the minimal nilpotent orbit of $\gso (2n)$. The second column shows their brane systems as Higgs branches. The third column shows the quivers obtained from the brane configurations in the second column. The fourth column shows the quivers obtained after performing S-duality on the second column.}
	\label{tab:minimalKPsingularities}
\end{table}

\section{The matrix formalism}\label{sec:12}

\subsection{The formalism}

Once the local brane configurations corresponding to each KP transition have been identified, a formalism that computes KP transitions on generic brane configurations corresponding to orthosymplectic quivers can be implemented, following the same principles behind the \emph{matrix formalism} developed in \cite{CH16}. The formalism encodes brane systems in $2\times N$ matrices with integer elements.

Let the $2\times N$ matrices encode the relevant Higgs branch brane configuration (after the \emph{collapse} transition is performed, i.e. the half D5-branes are pulled away from adjacent $O3^-$ planes without brane creation/annihilation) of each model. Each column of the matrix corresponds to a different interval between half D5-branes. The top row of the matrix counts how many half NS5-branes are in each interval (this encodes the same data as the \emph{interval numbers} $\vec k_s$ of the half NS5-branes). The bottom row counts the number of half D3-branes. An $\widetilde{O3^-}$ plane is also counted as half a D3-brane. Therefore, a column with an odd number in the bottom row corresponds to an interval with an $\widetilde{O3^-}$ plane. Note that such column will have $0$ as the top element, since there cannot be an interval with an $\widetilde{O3^-}$ plane and one or more half NS5-branes remaining in it after the collapse transition.  See figure \ref{fig:matrixExample} for an example of a brane system and its matrix.

\begin{figure}[t]
	\centering
	\begin{subfigure}[]{0.3\textwidth}
	    \centering
	\begin{tikzpicture}[scale=0.6, every node/.style={transform shape}]
			\draw[dashed] 	(2,0)--(2,2)
					(4.5,0)--(4.5,2)
					(5,0)--(5,2)
					(6,0)--(6,2)
					(6.5,0)--(6.5,2)
					(7,0)--(7,2)
					(7.5,0)--(7.5,2);
			\draw 	(1,1) node[cross] {}
					(1.5,1) node[cross] {}
					(2.5,1) node[cross] {}
					(3,1) node[cross] {}
					(3.5,1) node[cross] {}
					(4,1) node[cross] {}
					(5.5,1) node[cross] {};
			\draw 	(1,1) node[circle,draw] {}
					(1.5,1) node[circle,draw] {}
					(2.5,1) node[circle,draw] {}
					(3,1) node[circle,draw] {}
					(3.5,1) node[circle,draw] {}
					(4,1) node[circle,draw] {}
					(5.5,1) node[circle,draw] {};
			\draw [dotted] (0.5,1)--(1,1)
					(1.5,1)--(2.5,1)
					(3,1)--(3.5,1)
					(4,1)--(5.5,1);
			\draw 	(1,1)--(1.5,1)
					(6,1)--(6.5,1)
					(7,1)--(7.5,1);
			\draw 	(2,1.3)--(4.5,1.3)
					(2,1.5)--(4.5,1.5)
					(2,.7)--(4.5,.7)
					(2,.5)--(4.5,.5)
					(4.5,1.4)--(5,1.4)
					(4.5,1.6)--(5,1.6)
					(4.5,.6)--(5,.6)
					(4.5,.4)--(5,.4)
					(5,1.3)--(6,1.3)
					(5,1.5)--(6,1.5)
					(5,.7)--(6,.7)
					(5,.5)--(6,.5)
					(6,1.4)--(6.5,1.4)
					(6,.6)--(6.5,.6)
					(6.5,1.3)--(7,1.3)
					(6.5,.7)--(7,.7);
		\end{tikzpicture}
	\end{subfigure}
	\begin{subfigure}[]{0.1\textwidth}
	    \centering
	\begin{tikzpicture}[scale=0.7, every node/.style={transform shape}]
	\draw (0,0)--(1,0);
	\draw (0,.5)--(1,.5);
	\draw (0.9,-.1)--(1.25,0.25);
	\draw (0.9,.6)--(1.25,0.25);
	\end{tikzpicture}
	\end{subfigure}
	\begin{subfigure}[]{0.30\textwidth}
 $\left(
\begin{array}{cccccccc}
 2 & 4 & 0 & 1 & 0 & 0 & 0 & 0 \\
 0 & 4 & 4 & 4 & 3 & 2 & 1 & 0 \\
\end{array}
\right)$ 
	\end{subfigure}
		\caption{Example of a matrix encoding the data of a Higgs branch brane configuration.}
		\label{fig:matrixExample}
\end{figure}

\subsection{Matrix of the closure of the maximal nilpotent orbit}\label{sec:nc}

\begin{figure}
	\centering
	\begin{subfigure}[t]{1\textwidth}
	\centering
	\begin{tikzpicture}[node distance=40pt]
	\tikzstyle{gauge} = [circle, draw];
	\tikzstyle{flavour} = [regular polygon,regular polygon sides=4, draw];
	\node (g2) [gauge,label=below:{$O_2$}] {};
	\node (g3) [gauge, right of=g2,label=below:{$C_1$}] {};
	\node (g4) [gauge, right of=g3,label=below:{$O_4$}] {};
	\node (g5) [gauge, right of=g4,label=below:{$C_2$}] {};
	\node (gd) [right of=g5] {$\dots$};
	\node (g7) [gauge, right of=gd,label=below:{$O_{2n-4}$}] {};
	\node (g8) [gauge, right of=g7,label=below:{$C_{n-2}$}] {};
	\node (g9) [gauge, right of=g8,label=below:{$O_{2n-2}$}] {};
	\node (g10) [gauge, right of=g9,label=below:{$C_{n-1}$}] {};
	\node (f10) [flavour,above of=g10,label=above:{$O_{2n}$}] {};
	\draw (g2)--(g3)--(g4)--(g5)--(gd)--(g7)--(g8)--(g9)--(g10)
			(g10)--(f10)
		;
	\end{tikzpicture}
		\caption{}
	\end{subfigure}
	\hfill

	\begin{subfigure}[t]{1\textwidth}
	\centering
	\begin{tikzpicture}[node distance=40pt]
	\tikzstyle{gauge} = [circle, draw];
	\tikzstyle{flavour} = [regular polygon,regular polygon sides=4, draw];
	\node (g2) [gauge,label=below:{$O_2$}] {};
	\node (g3) [gauge, right of=g2,label=below:{$C_1$}] {};
	\node (g4) [gauge, right of=g3,label=below:{$O_4$}] {};
	\node (g5) [gauge, right of=g4,label=below:{$C_2$}] {};
	\node (gd) [right of=g5] {$\dots$};
	\node (g7) [gauge, right of=gd,label=below:{$C_{n-2}$}] {};
	\node (g8) [gauge, right of=g7,label=below:{$O_{2n-2}$}] {};
	\node (g9) [gauge, right of=g8,label=below:{$C_{n-1}$}] {};
	\node (g10) [gauge, right of=g9,label=below:{$O_{2n}$}] {};
	\node (f10) [flavour,above of=g10,label=above:{$C_{n}$}] {};
	\draw (g2)--(g3)--(g4)--(g5)--(gd)--(g7)--(g8)--(g9)--(g10)
			(g10)--(f10)
		;
	\end{tikzpicture}
		\caption{}
	\end{subfigure}
	\hfill

	\begin{subfigure}[t]{1\textwidth}
	\centering
	\begin{tikzpicture}[node distance=40pt]
	\tikzstyle{gauge} = [circle, draw];
	\tikzstyle{flavour} = [regular polygon,regular polygon sides=4, draw];
	\node (g2) [gauge,label=below:{$O_1$}] {};
	\node (g3) [gauge, right of=g2,label=below:{$C_1$}] {};
	\node (g4) [gauge, right of=g3,label=below:{$O_3$}] {};
	\node (g5) [gauge, right of=g4,label=below:{$C_2$}] {};
	\node (gd) [right of=g5] {$\dots$};
	\node (g7) [gauge, right of=gd,label=below:{$O_{2n-3}$}] {};
	\node (g8) [gauge, right of=g7,label=below:{$C_{n-2}$}] {};
	\node (g9) [gauge, right of=g8,label=below:{$O_{2n-1}$}] {};
	\node (g10) [gauge, right of=g9,label=below:{$C_{n-1}$}] {};
	\node (f10) [flavour,above of=g10,label=above:{$O_{2n+1}$}] {};
	\draw (g2)--(g3)--(g4)--(g5)--(gd)--(g7)--(g8)--(g9)--(g10)
			(g10)--(f10)
		;
	\end{tikzpicture}
		\caption{}
	\end{subfigure}
	\hfill
	\caption{Quivers whose Higgs branch is the maximal nilpotent orbit of the algebras: (a) $\gso (2n)$, (b) $\gsp (n)$ and (c) $\gso (2n+1)$.}
	\label{fig:nilpotentconequivers}
\end{figure}
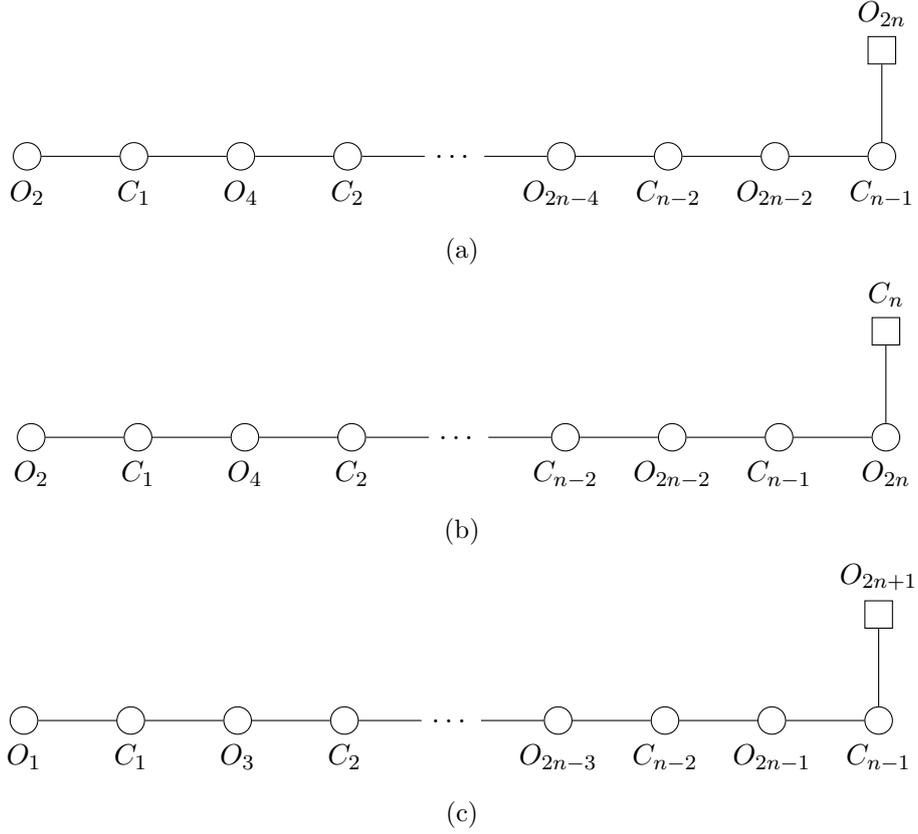

As in \cite{CH16}, the algorithm takes as input the model corresponding to the closure of the maximal nilpotent orbit of each algebra, and then performs all possible KP transitions, obtaining the models corresponding to all the remaining special orbits and the transverse slices corresponding to each transition. The orthosymplectic quiver for the closure of the maximal nilpotent orbit can be taken from\footnote{Note the difference in the choice of $O(n)$ gauge nodes instead of $SO(n)$.} \cite{GW09}, figure \ref{fig:nilpotentconequivers}. These are elements of the set of theories described in section \ref{sec:sigmarho} with maximal partitions $(2n+1)\in \mathcal P_{+1}(2n+1)$, $(2n)\in \mathcal P_{-1}(2n)$ and $(2n-1,1)\in \mathcal P_{+1}(2n)$ . Their Higgs branch brane configurations are depicted in figures \ref{fig:SubregularBn}, \ref{fig:SubregularCn} and \ref{fig:SubregularDn} for $\gso (2n+1)$, $\gsp (n)$ and $\gso (2n)$ respectively.

The corresponding matrices are as follows. For the maximal nilpotent orbit of $\mathfrak{so}({2n+1})$:

\begin{align}
	M_B=\left(\begin{array}{ccccccccc}
		0&2n+1&0&0&\cdots&0&0&0&0\\
		0&2n&2n-1&2n-2&\cdots&3&2&1&0\\
		\end{array}
	\right)
\end{align}

For the maximal nilpotent orbit of $\mathfrak{sp}({n})$:

\begin{align}
	M_C=\left(\begin{array}{cccccccccc}
		1&2n&0&0&\cdots&0&0&0&0&0\\
		0&2n&2n-2&2n-2&\cdots&4&4&2&2&0\\
		\end{array}
	\right)
\end{align}

For the maximal nilpotent orbit of $\mathfrak{so}({2n})$:

\begin{align}
	M_D=\left(\begin{array}{cccccccccccc}
		1&2n-2&1&0&0&\cdots&0&0&0&0&0&0\\
		0&2n-2&2n-2&2n-3&2n-4&\cdots&5&4&3&2&1&0\\
		\end{array}
	\right)
\end{align}

\paragraph{Interval numbers and partitions.}
The partition $\lambda$ that labels each model according to section \ref{sec:sigmarho} becomes manifest in the matrix formalism in the following way:
\emph{The interval numbers of the half NS5-branes $\vec{k}_s$ become the parts of partition $\lambda^t$.}

Given a matrix $M$, an element of the first row and the $j$th column $M_{1j}>0$ determines the presence of $M_{1j}$ half NS5-branes at the $(j-1)^{th}$ interval between half D5-branes (note that $M_{11}$ is reserved for the half NS5-branes to the left of the leftmost half D5-brane, i.e. the $0^{th}$ interval). Therefore it gives rise to $M_{1j}$ parts of value $(j-1)$ in partition $\lambda^t$. Let us examine the examples for the closure of the maximal nilpotent orbits.

For $\gso (2n+1)$: $(M_B)_{12} =2n+1$, this gives $\vec{k}_s=(1,1,\cdots,1,1)$ with $2n+1$ elements with value 1. Therefore $\lambda^t=(1^{2n+1})$. Taking the transpose one obtains $\lambda=(2n+1)$, which is the maximal partition in $\mathcal{P}_{+1}(2n+1)$.

For $\gsp (n)$: $(M_C)_{11}=1$, $(M_C)_{12}=2n$, this gives $\vec{k}_s=(0,1,1,\cdots,1,1)$ with one element with value $0$ and $2n$ elements with value $1$. Therefore $\lambda^t=(1^{2n},0)$. Taking the transpose one obtains $\lambda=(2n)$, which is the maximal partition in $\mathcal{P}_{-1}(2n)$.

For $\gso (2n)$: $(M_D)_{11}=1$, $(M_D)_{12}=2n-2$ and $(M_D)_{13}=1$, this gives $\vec{k}_s=(0,1,1,\cdots,1,1,2)$ with one element with value $0$, $2n-2$ elements with value $1$ and one element with value $2$. Therefore $\lambda^t=(2,1^{2n-2},0)$. Taking the transpose one obtains $\lambda=(2n-1,1)$, which is the maximal partition in $\mathcal{P}_{+1}(2n)$.

\subsection{Kraft-Procesi transitions}

This section defines the operations in the matrices that implement the KP transitions on the brane systems described in section \ref{sec:11}. Given a $2\times N$ matrix $M$ defining a Higgs branch brane configuration after the collapse transition, the following KP transitions can be performed:
\begin{itemize}
	\item \emph{Surface singularities}. If there is an element $M_{1j}$ of the first row of the matrix such that $M_{1j}>1$, and also $M_{2j}> 1$, a single D3-brane in the $(j-1)^{th}$ interval can be \emph{removed} via a KP transition. 
	\item \emph{Minimal singularities}. If there is a set of consecutive elements $\{M_{1j},M_{1(j+1)}, M_{1(j+2)}$,\\ $ \dots , M_{1(j+l)}\}$ such that $M_{1j}=M_{1(j+l)}=1$ and $M_{1(j+1)}=M_{1(j+2)}=\dots =M_{1(j+l-1)}=0$ and also $M_{2(j+k)}\geq 1$ for $k=0,1,2,\dots, l$, a minimal KP transition can be performed.
\end{itemize}

\subsubsection{Surface singularity}\label{sec:surface}
As mentioned before, whenever $M_{1j}>1$ and $M_{2j}>1$ a transverse slice $S\in \M_H$ that is a surface singularity can be \emph{removed} from the Higgs branch via a KP transition. The nature of the singularity is determined by 3 parameters: $M_{1j}$, $M_{2(j-1)}$ and $M_{2(j+1)}$. The possible singularities can be $D_n$, $A_{2n-1}$ and $A_{2n-1}\cup A_{2n-1}$. $M_{1j}$ determines $n$ and the other two numbers determine the nature of the singularity. There are five possibilities:
\begin{enumerate}[(a)]
	\item If $M_{1j}\geq 4$, $M_{2(j-1)}$ is even and $M_{2(j+1)}$ is even: $S=D_n$ where $n=M_{1j}/2+1$.
	\item If $M_{1j}\geq 3$, $M_{2(j-1)}$ is odd and $M_{2(j+1)}$ is even: $S=A_{2n-1}$ where $2n=M_{1j}-1$.
	\item If $M_{1j}\geq 3$, $M_{2(j-1)}$ is even and $M_{2(j+1)}$ is odd: $S=A_{2n-1}$ where $2n=M_{1j}-1$.
	\item If $M_{1j}\geq 2$, $M_{2(j-1)}$ is odd and $M_{2(j+1)}$ is odd: $S=A_{2n-1}\cup A_{2n-1}$ where $2n=M_{1j}$.
	\item If $M_{1j}= 2$, $M_{2(j-1)}$ is even and $M_{2(j+1)}$ is even: The variety is $A_{1}$.
\end{enumerate}

To perform the transition the two half D3-branes that generate $S$ are removed, so the new matrix $M'$ after the KP transition differs from $M$ in:

\begin{align}
	M'_{2j}=M_{2j}-2
\end{align}

There is an exception in case (e) where the orientifold plane changes to $\widetilde {O3^-}$, hence $M'_{2j}=M_{2j}-1$. Furthermore, $M'$ also differs in other elements, in order to encode the transitions that need to be performed to annihilate the fixed D3-brane segments and  the \emph{collapse} transition. These changes are (depending on the previous five possibilities) summarized in table \ref{tab:matrixKP}.

\begin{table}[t]
	\centering
	\begin{tabular}{|c|c|c|}
	\hline
		Type & Initial matrix $M$ & Final matrix $M'$ \\ \hline
		 $D_{k/2+1}$ &$\left(
\begin{array}{ccccc}
 \dots & m_1 & k & m_2 & \dots \\
 \dots & e_1 & l & e_2 & \dots \\
\end{array}
\right) $ & $ \left(
\begin{array}{ccccc}
 \dots & m_1+2 & k-4 & m_2+2 & \dots \\
 \dots & e_1 & l-2 & e_2 & \dots \\
\end{array}
\right)$ \\ \hline
		 $A_{k-2}$ & $\left(
\begin{array}{cccccc}
 \dots & m_1 & 0   & k & m_2 & \dots \\
 \dots & m_3 & o_1 & l & e_1 & \dots \\
\end{array}
\right)$& $\left(
\begin{array}{cccccc}
 \dots & m_1+1 & 0 & k-3 & m_2+2 & \dots \\
 \dots & m_3 & o_1-1 & l-2 & e_1 & \dots \\
\end{array}
\right)$\\ \hline
		$A_{k-2}$& $\left(
\begin{array}{cccccc}
 \dots & m_1 & k & 0   & m_2 & \dots \\
 \dots & e_1 & l & o_1 & m_3 & \dots \\
\end{array}
\right)$& $\left(
\begin{array}{cccccc}
 \dots & m_1+2 & k-3 & 0   & m_2+1 & \dots \\
 \dots & e_1 & l-2 & o_1-1 & m_3 & \dots \\
\end{array}
\right)$\\ \hline
		 $A_{k-1}\cup A_{k-1}$ &$\left(
\begin{array}{ccccccc}
 \dots & m_1 & 0   & k & 0   & m_2 & \dots \\
 \dots & m_3 & o_1 & l & o_2 & m_4 & \dots \\
\end{array}
\right)$ & $\left(
\begin{array}{ccccccc}
 \dots & m_1+1 & 0 & k-2 & 0 & m_2+1 & \dots \\
  \dots & m_3 & o_1-1 & l-2 & o_2-1 & m_4 & \dots \\
\end{array}
\right)$\\\hline
		 $A_{1}$ & $\left(
\begin{array}{ccccc}
  \dots & m_1 & 2 & m_2 & \dots \\
 \dots &  e_1 & l & e_2 & \dots \\
\end{array}
\right)$& $\left(
\begin{array}{ccccc}
 \dots & m_1+1 & 0 & m_2+1 & \dots \\
 \dots & e_1 & l-1 & e_2  & \dots \\
\end{array}
\right)$\\\hline
	\end{tabular}
	\caption{Classification of \emph{surface singularities} type Kraft-Procesi transitions for orthosymplectic brane configurations in the \emph{matrix formalism}. The number $k$ follows the restrictions of the five different cases labelled from (a) to (e) in section \ref{sec:surface}. The number $l$ is any integer greater than $1$. Labels $e_i$ (resp. $o_i$) denote numbers that are even (resp. odd). Labels $m_i$ denote any integer number. $M$ is a matrix before the KP transition is performed. $M'$ depicts the matrix after the KP transition is performed. The KP transition is a local process: the dots represent the parts of the matrices that are not affected by the KP transition.}
	\label{tab:matrixKP}
\end{table}

The matrix configurations for cases (a-d) can be shown to have the claimed singularity $S$ as their Higgs branch and that this can be removed utilizing the Higgs mechanism as in all the examples discussed in the sections above. For (e) there are two possibilities: the even numbers in the second row correspond to $O3^+$ planes in which case the transition can occur, or they correspond to $\widetilde{O3^+}$ in which case there is a supersymmetric obstruction for the transition (this is the same phenomenon as the one described in section \ref{sec:311transition}). If this is the case, and one decides to perform the transition on the matrix nonetheless, a resulting matrix $M'$ is found that cannot be translated into a brane system. This matrix however has the data of some interval numbers that correspond to the partition of a \emph{non-special} orbit.

\subsubsection{Minimal singularities}

In the case of minimal singularities where the initial matrix $M$  has a set of consecutive elements $\{M_{1j},M_{1(j+1)}, M_{1(j+2)}, \dots , M_{1(j+l)}\}$ such that $M_{1j}=M_{1(j+l)}=1$ and $M_{1(j+1)}=M_{1(j+2)}=\dots =M_{1(j+l-1)}=0$ and also $M_{2(j+k)}\geq 1$ for $k=0,1,2,\dots, l$, the singularities can be:
\begin{align}
	\begin{aligned}
		b_n&:=\Or_{(2^2,1^{2n-3})}\subset \gso (2n+1)\\
		c_n&:=\Or_{(2,1^{2n-2})}\subset \gsp (n)\\
		d_n&:=\Or_{(2^2,1^{2n-4})}\subset \gso (2n)\\
	\end{aligned}
\end{align}
The type of singularity $S$ that conforms the transverse slice and the parameter $n$ depend on the matrix elements $M_{2(j-1)}$, $M_{2(j+l+1)}$ and in the number $l$. There are three possibilities:

\begin{enumerate}[(a)]
	\item If $l$ is odd, $M_{2(j-1)}$ is odd and $M_{2(j+l+1)}$ is odd: $S=b_{(l+3)/2}$.
	\item If $l$ is even, $M_{2(j-1)}$ is odd and $M_{2(j+l+1)}$ is odd: $S=d_{(l+4)/2}$.
	\item If $l$ is even, $M_{2(j-1)}$ is even and $M_{2(j+l+1)}$ is even: $S=c_{(l+2)/2}$.
\end{enumerate}

The modification that the original matrix $M$ undergoes after each transition is summarized in table \ref{tab:matrixKPminimal}. Note that for $b_n$ the starting setting is a matrix that cannot be obtained from a brane system. This type of matrix can however be obtained from the $A_1$ transition corresponding to case (e) in the previous section where the supersymmetric configuration is broken. Therefore, the matrix implementation of transition $b_n$ always starts with a matrix corresponding to a \emph{non-special} orbit. A $c_n$ transition can start and can end in matrices that have no brane system counterpart and therefore correspond to \emph{non-special} orbits. The reason why the $A_1$ from previous case (e), the $b_n$ and the $c_n$ transitions are considered in the matrix formalism is that they give extra information that is not problematic: some matrices $M'$ are produced that correspond to non-special orbits and have no physical brane interpretation. However, these matrices can easily be removed from the resulting Hasse diagram and the relevant physical transitions $\tilde b_n$ and $\tilde c_n$ can always be recovered from combinations of the non-physical ones $A_1$, $b_n$ and $c_n$:
\begin{itemize}
	\item $\tilde b_n$ is equal to $A_1$ followed by $b_n$.
	\item $\tilde c_n$ is equal to $c_{n-1}$ followed by $c_n$.
\end{itemize}

\begin{table}[t]
	\centering
	\begin{tabular}{|c|c|c|}
	\hline
		 
		 \multirow{2}{*}{$b_k$ and $d_k$} & $M$ & $\left(
\begin{array}{ccccccccccc}
\dots & m_1 & 0 & 1 & 0 & \dots & 0 & 1 & 0 & m_4 & \dots\\
\dots & m_5 & o_1 & m_6 & m_7 & \dots & m_8 & m_9 & o_2 & m_{10}& \dots \\
\end{array}
\right)$ \\  \hhline{~--}
 & $M'$ & $\left(
\begin{array}{ccccccccccc}
\dots & m_1+1 & 0 & 0 & 0 & \dots & 0 & 0 & 0 & m_4+1 & \dots \\
\dots & m_5 & o_1-1 & m_6-2 & m_7-2 & \dots & m_8-2 & m_9-2 & o_2-1 & m_{10} & \dots\\
\end{array}
\right)$\\\hline
		\multirow{2}{*}{$c_k$}  & $M$ &$\left(
\begin{array}{ccccccccccc}
\dots & m_1 & 1 & 0 & 0 & \dots & 0 & 0 & 1 & m_2 & \dots\\
\dots & e_1 & m_3 & m_4 & m_5 & \dots & m_6 & m_7 & m_8 & e_2 & \dots\\
\end{array}
\right)$ \\ \hhline{~--}
& $M'$ & $\left(
\begin{array}{ccccccccccc}
\dots & m_1+1 & 0 & 0 & 0 & \dots & 0 & 0 & 0 & m_2+1  & \dots\\
\dots & e_1 & m_3-1 & m_4-1 & m_5-1 & \dots & m_6-1 & m_7-1 & m_8-1 & e_2 & \dots\\
\end{array}
\right)$\\\hline
	\end{tabular}
	\caption{Classification of \emph{minimal singularity} type Kraft-Procesi transitions for orthosymplectic brane configurations in the \emph{matrix formalism}. Labels $e_i$ denote numbers that are even while $o_i$ denotes odd numbers. $m_i$ denote any integer number. In all the initial matrices $M$, let $M_{1j}$ and $M_{1(j+l)}$ be the elements with value $1$. In the first two rows if $l$ is odd the singularity is $b_{(l+3)/2}$, if $l$ is even the singularity is $d_{(l+4)/2}$. In the last two rows the singularity is $c_{(l+2)/2}$. $M$ is a matrix before the KP transition is performed. $M'$ depicts the matrix after the KP transition is performed.}
	\label{tab:matrixKPminimal}
\end{table}

\section{Results}\label{sec:13}

In this section we show all the results after applying the matrix formalism to the maximal nilpotent orbit matrices introduced in section \ref{sec:nc}. The partitions in bold denote \emph{special} orbits. These are the orbits for which the \emph{Matrix} encodes the information about a brane system. The partitions that are not in bold correspond to \emph{non-special} nilpotent orbits\footnote{These partitions are also highlighted by marking them in the Hasse diagram with an empty circle, instead of a filled one.}: the \emph{Matrix} does not have a brane system counterpart. 
$dim$ denotes the quaternionic dimension of each orbit. In the matrix formalism this can be obtained with:
\begin{align}
	dim=\sum_{j}\lfloor {M_{2j}}/{2} \rfloor
\end{align}

\begin{table}[t]
	\centering
	\begin{subfigure}[]{0.3\textwidth}
	    \centering

	\end{subfigure}
	\begin{subfigure}[]{0.5\textwidth}
	\begin{tabular}{ c c c }
	\toprule
	\textbf{Matrix} & \textbf{Partition} & $dim$ \\ 
	\midrule \addlinespace[1.5ex]
	$\left(
\begin{array}{cccc}
 0 & 3 & 0 & 0  \\
 0 & 2 & 1 & 0  \\
\end{array}
\right)$ & $\mathbf{3}$ & 1 \\ 
	\addlinespace[1.5ex]
	 $\left(
\begin{array}{cccc}
 2 & 0 & 0 & 1  \\
 0 & 0 & 0 & 0 \\
\end{array}
\right)$ & $\mathbf{1^3}$ & 0 \\ \addlinespace[1.5ex]
	\bottomrule
	\end{tabular}
	\end{subfigure}
		\caption{Results obtained applying the matrix formalism to $\mathfrak{so}(3)$.}
\end{table}

\begin{table}[t]
	\centering
	\begin{subfigure}[]{0.3\textwidth}
	    \centering
	
	\end{subfigure}
	\begin{subfigure}[]{0.5\textwidth}
	\begin{tabular}{ c c c }
	\toprule
	\textbf{Matrix} & \textbf{Partition} & $dim$ \\ 
	\midrule \addlinespace[1.5ex]
	$\left(
\begin{array}{ccccc}
 1 & 2 & 1 & 0 & 0 \\
 0 & 2 & 2 & 1 & 0 \\
\end{array}
\right)$ & $\mathbf{3,1}$ & 2 \\ 
	\addlinespace[1.5ex]
	 $\left(
\begin{array}{ccccc}
 2 & 0 & 2 & 0 & 0 \\
 0 & 1 & 2 & 1 & 0 \\
\end{array}
\right)$ & $\mathbf{2^2}$ & 1 \\ 
	\addlinespace[1.5ex]
 	$\left(
\begin{array}{ccccc}
 3 & 0 & 0 & 0 & 1 \\
 0 & 0 & 0 & 0 & 0 \\
\end{array}
\right)$ & $\mathbf{1^4}$ & 0 \\ \addlinespace[1.5ex]
	\bottomrule
	\end{tabular}
	\end{subfigure}
		\caption{Results obtained applying the matrix formalism to $\mathfrak{so}(4)$.}
\end{table}

\begin{table}[t]
	\centering
	\begin{subfigure}[]{0.3\textwidth}
	    \centering
	
	\end{subfigure}
	\begin{subfigure}[]{0.5\textwidth}
	\begin{tabular}{ c c c }
	\toprule
	\textbf{Matrix} & \textbf{Partition} & $dim$ \\ 
	\midrule \addlinespace[1.5ex]
 $\left(
\begin{array}{cccccc}
 0 & 5 & 0 & 0 & 0 & 0 \\
 0 & 4 & 3 & 2 & 1 & 0 \\
\end{array}
\right)$ & $\mathbf{5}$ & 4 \\
  \addlinespace[1.5ex]
 $\left(
\begin{array}{cccccc}
 2 & 2 & 0 & 1 & 0 & 0 \\
 0 & 2 & 2 & 2 & 1 & 0 \\
\end{array}
\right)$ & $\mathbf{3,1^2}$ & 3 \\
  \addlinespace[1.5ex]
 $\left(
\begin{array}{cccccc}
 3 & 0 & 1 & 1 & 0 & 0 \\
 0 & 1 & 2 & 2 & 1 & 0 \\
\end{array}
\right)$ & ${2^2,1}$ & 2 \\
  \addlinespace[1.5ex]
 $\left(
\begin{array}{cccccc}
 4 & 0 & 0 & 0 & 0 & 1 \\
 0 & 0 & 0 & 0 & 0 & 0 \\
\end{array}
\right)$ & $\mathbf{1^5}$ & 0 \\	
	\bottomrule
	\end{tabular}
	\end{subfigure}
		\caption{Results obtained applying the matrix formalism to $\mathfrak{so}(5)$.}
\end{table}

\begin{table}[t]
	\centering
	\begin{subfigure}[]{0.3\textwidth}
	    \centering
	
	\end{subfigure}
	\begin{subfigure}[]{0.5\textwidth}
	\begin{tabular}{ c c c }
	\toprule
	\textbf{Matrix} & \textbf{Partition} & $dim$ \\ 
	\midrule \addlinespace[1.5ex]
$\left(
\begin{array}{ccccccc}
 1 & 4 & 1 & 0 & 0 & 0 & 0 \\
 0 & 4 & 4 & 3 & 2 & 1 & 0 \\
\end{array}
\right)$ & $\mathbf{5,1}$ & 6 \\
\addlinespace[1.5ex]
 $\left(
\begin{array}{ccccccc}
 3 & 0 & 3 & 0 & 0 & 0 & 0 \\
 0 & 2 & 4 & 3 & 2 & 1 & 0 \\
\end{array}
\right)$ & $\mathbf{3^2}$ & 5 \\
\addlinespace[1.5ex]
 $\left(
\begin{array}{ccccccc}
 3 & 2 & 0 & 0 & 1 & 0 & 0 \\
 0 & 2 & 2 & 2 & 2 & 1 & 0 \\
\end{array}
\right)$ & $\mathbf{3,1^3}$ & 4 \\
\addlinespace[1.5ex]
 $\left(
\begin{array}{ccccccc}
 4 & 0 & 1 & 0 & 1 & 0 & 0 \\
 0 & 1 & 2 & 2 & 2 & 1 & 0 \\
\end{array}
\right)$ & $\mathbf{2^2,1^2 }$& 3 \\
\addlinespace[1.5ex]
 $\left(
\begin{array}{ccccccc}
 5 & 0 & 0 & 0 & 0 & 0 & 1 \\
 0 & 0 & 0 & 0 & 0 & 0 & 0 \\
\end{array}
\right)$ & $\mathbf{1^6}$ & 0 \\ \addlinespace[1.5ex]
	\bottomrule
	\end{tabular}
	\end{subfigure}
	\begin{subfigure}[]{0.5\textwidth}
	\begin{tabular}{ c c c }
	\toprule
	\textbf{Matrix} & \textbf{Partition} & $dim$ \\ 
	\midrule \addlinespace[1.5ex]
 $\left(
\begin{array}{cccccccc}
 0 & 7 & 0 & 0 & 0 & 0 & 0 & 0 \\
 0 & 6 & 5 & 4 & 3 & 2 & 1 & 0 \\
\end{array}
\right)$ & $\mathbf{7}$ & 9 \\
  \addlinespace[1.5ex]
 $\left(
\begin{array}{cccccccc}
 2 & 4 & 0 & 1 & 0 & 0 & 0 & 0 \\
 0 & 4 & 4 & 4 & 3 & 2 & 1 & 0 \\
\end{array}
\right)$ & $\mathbf{5,1^2}$ & 8 \\
  \addlinespace[1.5ex]
 $\left(
\begin{array}{cccccccc}
 4 & 0 & 2 & 1 & 0 & 0 & 0 & 0 \\
 0 & 2 & 4 & 4 & 3 & 2 & 1 & 0 \\
\end{array}
\right)$ & $\mathbf{3^2,1}$ & 7 \\
  \addlinespace[1.5ex]
 $\left(
\begin{array}{cccccccc}
 4 & 1 & 0 & 2 & 0 & 0 & 0 & 0 \\
 0 & 2 & 3 & 4 & 3 & 2 & 1 & 0 \\
\end{array}
\right)$ & $\mathbf{3,2^2}$ & 6 \\
  \addlinespace[1.5ex]
 $\left(
\begin{array}{cccccccc}
 4 & 2 & 0 & 0 & 0 & 1 & 0 & 0 \\
 0 & 2 & 2 & 2 & 2 & 2 & 1 & 0 \\
\end{array}
\right)$ & $\mathbf{3,1^4}$ & 5 \\
  \addlinespace[1.5ex]
 $\left(
\begin{array}{cccccccc}
 5 & 0 & 1 & 0 & 0 & 1 & 0 & 0 \\
 0 & 1 & 2 & 2 & 2 & 2 & 1 & 0 \\
\end{array}
\right)$ & ${2^2,1^3}$ & 4 \\
  \addlinespace[1.5ex]
 $\left(
\begin{array}{cccccccc}
 6 & 0 & 0 & 0 & 0 & 0 & 0 & 1 \\
 0 & 0 & 0 & 0 & 0 & 0 & 0 & 0 \\
\end{array}
\right)$ & $\mathbf{1^7}$ & 0 \\	
	\bottomrule
	\end{tabular}
	\end{subfigure}
	\begin{subfigure}[]{0.5\textwidth}
	\begin{tabular}{ c c c }
	\toprule
	\textbf{Matrix} & \textbf{Partition} & $dim$ \\ 
	\midrule \addlinespace[1.5ex]
$\left(\begin{array}{ccccccccc}
 1 & 6 & 1 & 0 & 0 & 0 & 0 & 0 & 0 \\
 0 & 6 & 6 & 5 & 4 & 3 & 2 & 1 & 0 \\
\end{array}
\right)$ & $\mathbf{7,1}$ & 12 \\
  \addlinespace[1.5ex]
 $\left(
\begin{array}{ccccccccc}
 3 & 2 & 3 & 0 & 0 & 0 & 0 & 0 & 0 \\
 0 & 4 & 6 & 5 & 4 & 3 & 2 & 1 & 0 \\
\end{array}
\right)$ & $\mathbf{5,3}$ & $11$ \\
 \addlinespace[1.5ex]
 $\left(
\begin{array}{ccccccccc}
 4 & 0 & 4 & 0 & 0 & 0 & 0 & 0 & 0 \\
 0 & 3 & 6 & 5 & 4 & 3 & 2 & 1 & 0 \\
\end{array}
\right)$ & $\mathbf{4^2}$ & $10$ \\
  \addlinespace[1.5ex]
 $\left(
\begin{array}{ccccccccc}
 3 & 4 & 0 & 0 & 1 & 0 & 0 & 0 & 0 \\
 0 & 4 & 4 & 4 & 4 & 3 & 2 & 1 & 0 \\
\end{array}
\right)$ & $\mathbf{5,1^3}$ & 10 \\
  \addlinespace[1.5ex]
 $\left(
\begin{array}{ccccccccc}
 5 & 0 & 2 & 0 & 1 & 0 & 0 & 0 & 0 \\
 0 & 2 & 4 & 4 & 4 & 3 & 2 & 1 & 0 \\
\end{array}
\right)$ & $\mathbf{3^2,1^2}$ & 9 \\
  \addlinespace[1.5ex]
 $\left(
\begin{array}{ccccccccc}
 5 & 1 & 0 & 1 & 1 & 0 & 0 & 0 & 0 \\
 0 & 2 & 3 & 4 & 4 & 3 & 2 & 1 & 0 \\
\end{array}
\right)$ & ${3,2^2,1}$ & 8 \\
  \addlinespace[1.5ex]
 $\left(
\begin{array}{ccccccccc}
 5 & 2 & 0 & 0 & 0 & 0 & 1 & 0 & 0 \\
 0 & 2 & 2 & 2 & 2 & 2 & 2 & 1 & 0 \\
\end{array}
\right)$ & $\mathbf{3,1^5}$ & 6 \\
  \addlinespace[1.5ex]
 $\left(
\begin{array}{ccccccccc}
 6 & 0 & 0 & 0 & 2 & 0 & 0 & 0 & 0 \\
 0 & 1 & 2 & 3 & 4 & 3 & 2 & 1 & 0 \\
\end{array}
\right)$ & $\mathbf{2^4}$ & 6 \\
  \addlinespace[1.5ex]
 $\left(
\begin{array}{ccccccccc}
 6 & 0 & 1 & 0 & 0 & 0 & 1 & 0 & 0 \\
 0 & 1 & 2 & 2 & 2 & 2 & 2 & 1 & 0 \\
\end{array}
\right)$ & $\mathbf{2^2,1^4}$ & 5 \\
  \addlinespace[1.5ex]
 $\left(
\begin{array}{ccccccccc}
 7 & 0 & 0 & 0 & 0 & 0 & 0 & 0 & 1 \\
 0 & 0 & 0 & 0 & 0 & 0 & 0 & 0 & 0 \\
\end{array}
\right)$ & $\mathbf{1^8}$ & 0 \\
	\bottomrule
	\end{tabular}
	\end{subfigure}
	\begin{subfigure}[]{0.5\textwidth}
	\begin{tabular}{ c c c }
	\toprule
	\textbf{Matrix} & \textbf{Partition} & $dim$ \\ 
	\midrule \addlinespace[1.5ex]
 $\left(
\begin{array}{cccccccccc}
 0 & 9 & 0 & 0 & 0 & 0 & 0 & 0 & 0 & 0 \\
 0 & 8 & 7 & 6 & 5 & 4 & 3 & 2 & 1 & 0 \\
\end{array}
\right)$ & $\mathbf{9}$ & 16 \\
  \addlinespace[1.5ex]
 $\left(
\begin{array}{cccccccccc}
 2 & 6 & 0 & 1 & 0 & 0 & 0 & 0 & 0 & 0 \\
 0 & 6 & 6 & 6 & 5 & 4 & 3 & 2 & 1 & 0 \\
\end{array}
\right)$ & $\mathbf{7,1^2}$ & 15 \\
  \addlinespace[1.5ex]
 $\left(
\begin{array}{cccccccccc}
 4 & 2 & 2 & 1 & 0 & 0 & 0 & 0 & 0 & 0 \\
 0 & 4 & 6 & 6 & 5 & 4 & 3 & 2 & 1 & 0 \\
\end{array}
\right)$ & $\mathbf{5,3,1}$ & 14 \\
  \addlinespace[1.5ex]
 $\left(
\begin{array}{cccccccccc}
 5 & 0 & 3 & 1 & 0 & 0 & 0 & 0 & 0 & 0 \\
 0 & 3 & 6 & 6 & 5 & 4 & 3 & 2 & 1 & 0 \\
\end{array}
\right)$ & ${4^2,1}$ & 13 \\
  \addlinespace[1.5ex]
 $\left(
\begin{array}{cccccccccc}
 4 & 3 & 0 & 2 & 0 & 0 & 0 & 0 & 0 & 0 \\
 0 & 4 & 5 & 6 & 5 & 4 & 3 & 2 & 1 & 0 \\
\end{array}
\right)$ & $\mathbf{5,2^2}$ & 13 \\
  \addlinespace[1.5ex]
 $\left(
\begin{array}{cccccccccc}
 6 & 0 & 0 & 3 & 0 & 0 & 0 & 0 & 0 & 0 \\
 0 & 2 & 4 & 6 & 5 & 4 & 3 & 2 & 1 & 0 \\
\end{array}
\right)$ & $\mathbf{3^3}$ & 12 \\
  \addlinespace[1.5ex]
 $\left(
\begin{array}{cccccccccc}
 4 & 4 & 0 & 0 & 0 & 1 & 0 & 0 & 0 & 0 \\
 0 & 4 & 4 & 4 & 4 & 4 & 3 & 2 & 1 & 0 \\
\end{array}
\right)$ & $\mathbf{5,1^4}$ & 12 \\
\bottomrule
	\end{tabular}
	\end{subfigure}
	\begin{subfigure}[]{0.5\textwidth}
	\begin{tabular}{ c c c }
	\toprule
	\textbf{Matrix} & \textbf{Partition} & $dim$ \\ 
	\midrule \addlinespace[1.5ex]
 $\left(
\begin{array}{ccccccccccc}
 1 & 8 & 1 & 0 & 0 & 0 & 0 & 0 & 0 & 0 & 0 \\
 0 & 8 & 8 & 7 & 6 & 5 & 4 & 3 & 2 & 1 & 0 \\
\end{array}
\right)$ & $\mathbf{9,1}$ & 20 \\
 \addlinespace[1.5ex]
 $\left(
\begin{array}{ccccccccccc}
 3 & 4 & 3 & 0 & 0 & 0 & 0 & 0 & 0 & 0 & 0 \\
 0 & 6 & 8 & 7 & 6 & 5 & 4 & 3 & 2 & 1 & 0 \\
\end{array}
\right)$ & $\mathbf{7,3}$ & 19 \\
 \addlinespace[1.5ex]
 $\left(
\begin{array}{ccccccccccc}
 3 & 6 & 0 & 0 & 1 & 0 & 0 & 0 & 0 & 0 & 0 \\
 0 & 6 & 6 & 6 & 6 & 5 & 4 & 3 & 2 & 1 & 0 \\
\end{array}
\right)$ & $\mathbf{7,1^3}$ & 18 \\
 \addlinespace[1.5ex]
 $\left(
\begin{array}{ccccccccccc}
 5 & 0 & 5 & 0 & 0 & 0 & 0 & 0 & 0 & 0 & 0 \\
 0 & 4 & 8 & 7 & 6 & 5 & 4 & 3 & 2 & 1 & 0 \\
\end{array}
\right)$ & $\mathbf{5^2}$ & 18 \\
 \addlinespace[1.5ex]
 $\left(
\begin{array}{ccccccccccc}
 5 & 2 & 2 & 0 & 1 & 0 & 0 & 0 & 0 & 0 & 0 \\
 0 & 4 & 6 & 6 & 6 & 5 & 4 & 3 & 2 & 1 & 0 \\
\end{array}
\right)$ & $\mathbf{5,3,1^2}$ & 17 \\
 \addlinespace[1.5ex]
 $\left(
\begin{array}{ccccccccccc}
 6 & 0 & 3 & 0 & 1 & 0 & 0 & 0 & 0 & 0 & 0 \\
 0 & 3 & 6 & 6 & 6 & 5 & 4 & 3 & 2 & 1 & 0 \\
\end{array}
\right)$ & $\mathbf{4^2,1^2}$ & 16 \\
 \addlinespace[1.5ex]
 $\left(
\begin{array}{ccccccccccc}
 5 & 3 & 0 & 1 & 1 & 0 & 0 & 0 & 0 & 0 & 0 \\
 0 & 4 & 5 & 6 & 6 & 5 & 4 & 3 & 2 & 1 & 0 \\
\end{array}
\right)$ & ${5,2^2,1}$ & 16 \\
\bottomrule
	\end{tabular}
	\end{subfigure}
	\begin{subfigure}[]{0.5\textwidth}
	\fontsize{10}{11}\selectfont
	\begin{tabular}{ c c c }
	\toprule
	\textbf{Matrix} & \textbf{Partition} & $dim$ \\ 
	\midrule \addlinespace[1ex]
 $\left(
\begin{array}{cccccccccccc}
 0 & 11 & 0 & 0 & 0 & 0 & 0 & 0 & 0 & 0 & 0 & 0 \\
 0 & 10 & 9 & 8 & 7 & 6 & 5 & 4 & 3 & 2 & 1 & 0 \\
\end{array}
\right)$ & $\mathbf{11}$ & 25 \\
\addlinespace[1ex]
 $\left(
\begin{array}{cccccccccccc}
 2 & 8 & 0 & 1 & 0 & 0 & 0 & 0 & 0 & 0 & 0 & 0 \\
 0 & 8 & 8 & 8 & 7 & 6 & 5 & 4 & 3 & 2 & 1 & 0 \\
\end{array}
\right)$ & $\mathbf{9,1^2}$ & 24 \\
\addlinespace[1ex]
 $\left(
\begin{array}{cccccccccccc}
 4 & 4 & 2 & 1 & 0 & 0 & 0 & 0 & 0 & 0 & 0 & 0 \\
 0 & 6 & 8 & 8 & 7 & 6 & 5 & 4 & 3 & 2 & 1 & 0 \\
\end{array}
\right)$ & $\mathbf{7,3,1}$ & 23 \\
\addlinespace[1ex]
 $\left(
\begin{array}{cccccccccccc}
 6 & 0 & 4 & 1 & 0 & 0 & 0 & 0 & 0 & 0 & 0 & 0 \\
 0 & 4 & 8 & 8 & 7 & 6 & 5 & 4 & 3 & 2 & 1 & 0 \\
\end{array}
\right)$ & $\mathbf{5^2,1}$ & 22 \\
\addlinespace[1ex]
 $\left(
\begin{array}{cccccccccccc}
 4 & 5 & 0 & 2 & 0 & 0 & 0 & 0 & 0 & 0 & 0 & 0 \\
 0 & 6 & 7 & 8 & 7 & 6 & 5 & 4 & 3 & 2 & 1 & 0 \\
\end{array}
\right)$ & $\mathbf{7,2^2}$ & 22 \\
\addlinespace[1ex]
 $\left(
\begin{array}{cccccccccccc}
 6 & 2 & 0 & 3 & 0 & 0 & 0 & 0 & 0 & 0 & 0 & 0 \\
 0 & 4 & 6 & 8 & 7 & 6 & 5 & 4 & 3 & 2 & 1 & 0 \\
\end{array}
\right)$ & $\mathbf{5,3^2}$ & 21 \\
\addlinespace[1ex]
 $\left(
\begin{array}{cccccccccccc}
 4 & 6 & 0 & 0 & 0 & 1 & 0 & 0 & 0 & 0 & 0 & 0 \\
 0 & 6 & 6 & 6 & 6 & 6 & 5 & 4 & 3 & 2 & 1 & 0 \\
\end{array}
\right)$ & $\mathbf{7,1^4}$ & 21 \\
\bottomrule
	\end{tabular}
	\end{subfigure}
	\begin{subfigure}[]{0.5\textwidth}
	\centering
	\fontsize{7.5}{9}\selectfont
	\begin{tabular}{ c l c }
	\toprule
	\textbf{Matrices} & $\lambda$ & $dim$ \\ 
	\midrule	 \addlinespace[1ex]
 $\left(
\begin{array}{ccccccccccccc}
 1 & 10 & 1 & 0 & 0 & 0 & 0 & 0 & 0 & 0 & 0 & 0 & 0 \\
 0 & 10 & 10 & 9 & 8 & 7 & 6 & 5 & 4 & 3 & 2 & 1 & 0 \\
\end{array}
\right)$ & $\mathbf{11,1}$ & 30 \\
  \addlinespace[1ex]
 $\left(
\begin{array}{ccccccccccccc}
 3 & 6 & 3 & 0 & 0 & 0 & 0 & 0 & 0 & 0 & 0 & 0 & 0 \\
 0 & 8 & 10 & 9 & 8 & 7 & 6 & 5 & 4 & 3 & 2 & 1 & 0 \\
\end{array}
\right)$ & $\mathbf{9,3}$ & 29 \\
  \addlinespace[1ex]
 $\left(
\begin{array}{ccccccccccccc}
 3 & 8 & 0 & 0 & 1 & 0 & 0 & 0 & 0 & 0 & 0 & 0 & 0 \\
 0 & 8 & 8 & 8 & 8 & 7 & 6 & 5 & 4 & 3 & 2 & 1 & 0 \\
\end{array}
\right)$ & $\mathbf{9,1^3}$ & 28 \\
  \addlinespace[1ex]
 $\left(
\begin{array}{ccccccccccccc}
 5 & 2 & 5 & 0 & 0 & 0 & 0 & 0 & 0 & 0 & 0 & 0 & 0 \\
 0 & 6 & 10 & 9 & 8 & 7 & 6 & 5 & 4 & 3 & 2 & 1 & 0 \\
\end{array}
\right)$ & $\mathbf{7,5}$ & 28 \\
  \addlinespace[1ex]
 $\left(
\begin{array}{ccccccccccccc}
 5 & 4 & 2 & 0 & 1 & 0 & 0 & 0 & 0 & 0 & 0 & 0 & 0 \\
 0 & 6 & 8 & 8 & 8 & 7 & 6 & 5 & 4 & 3 & 2 & 1 & 0 \\
\end{array}
\right)$ & $\mathbf{7,3,1^2}$ & 27 \\
  \addlinespace[1ex]
 $\left(
\begin{array}{ccccccccccccc}
 6 & 0 & 6 & 0 & 0 & 0 & 0 & 0 & 0 & 0 & 0 & 0 & 0 \\
 0 & 5 & 10 & 9 & 8 & 7 & 6 & 5 & 4 & 3 & 2 & 1 & 0 \\
\end{array}
\right)$ & $\mathbf{6^2}$ & 27 \\
  \addlinespace[1ex]
 $\left(
\begin{array}{ccccccccccccc}
 7 & 0 & 4 & 0 & 1 & 0 & 0 & 0 & 0 & 0 & 0 & 0 & 0 \\
 0 & 4 & 8 & 8 & 8 & 7 & 6 & 5 & 4 & 3 & 2 & 1 & 0 \\
\end{array}
\right)$ & $\mathbf{5^2,1^2}$ & 26 \\
\bottomrule
	\end{tabular}
	\end{subfigure}
		\caption{Results obtained applying the matrix formalism to $\mathfrak{so}(12)$.}
\end{table}

\clearpage

\begin{table}[t]
	\centering
	\begin{subfigure}[]{0.3\textwidth}
	    \centering
	
	\end{subfigure}
	\begin{subfigure}[]{0.5\textwidth}
	\begin{tabular}{ c c c }
	\toprule
	\textbf{Matrix} & \textbf{Partition} & $dim$ \\ 
	\midrule \addlinespace[1.5ex]
\addlinespace[1.5ex]
 $\left(
\begin{array}{ccccc}
 1 & 4 & 0 & 0 & 0 \\
 0 & 4 & 2 & 2 & 0 \\
\end{array}
\right)$ & $\mathbf{4}$ & 4 \\
\addlinespace[1.5ex]
 $\left(
\begin{array}{ccccc}
 3 & 0 & 2 & 0 & 0 \\
 0 & 2 & 2 & 2 & 0 \\
\end{array}
\right)$ & $\mathbf{2^2}$ & 3 \\
\addlinespace[1.5ex]
 $\left(
\begin{array}{ccccc}
 3 & 1 & 0 & 1 & 0 \\
 0 & 2 & 1 & 2 & 0 \\
\end{array}
\right)$ & ${2,1^2}$ & 2 \\
\addlinespace[1.5ex]
 $\left(
\begin{array}{ccccc}
 4 & 0 & 0 & 0 & 1 \\
 0 & 1 & 0 & 1 & 0 \\
\end{array}
\right)$ & $\mathbf{1^4}$ & 0 \\	
	\bottomrule
	\end{tabular}
	\end{subfigure}
		\caption{Results obtained applying the matrix formalism to $\mathfrak{sp}(2)$.}
\end{table}

\begin{table}[t]
	\centering
	\begin{subfigure}[]{0.3\textwidth}
	    \centering
	
	\end{subfigure}
	\begin{subfigure}[]{0.5\textwidth}
	\begin{tabular}{ c c c }
	\toprule
	\textbf{Matrix} & \textbf{Partition} & $dim$ \\ 
	\midrule \addlinespace[1.5ex]
\addlinespace[1.5ex]
 $\left(
\begin{array}{ccccccc}
 1 & 6 & 0 & 0 & 0 & 0 & 0 \\
 0 & 6 & 4 & 4 & 2 & 2 & 0 \\
\end{array}
\right)$ & $\mathbf{6}$ & 9 \\
\addlinespace[1.5ex]
 $\left(
\begin{array}{ccccccc}
 3 & 2 & 2 & 0 & 0 & 0 & 0 \\
 0 & 4 & 4 & 4 & 2 & 2 & 0 \\
\end{array}
\right)$ & $\mathbf{4,2}$ & 8 \\
\addlinespace[1.5ex]
 $\left(
\begin{array}{ccccccc}
 4 & 0 & 3 & 0 & 0 & 0 & 0 \\
 0 & 3 & 4 & 4 & 2 & 2 & 0 \\
\end{array}
\right)$ & $\mathbf{3^2}$ & 7 \\
\addlinespace[1.5ex]
 $\left(
\begin{array}{ccccccc}
 3 & 3 & 0 & 1 & 0 & 0 & 0 \\
 0 & 4 & 3 & 4 & 2 & 2 & 0 \\
\end{array}
\right)$ & ${4,1^2}$ & 7 \\
\addlinespace[1.5ex]
 $\left(
\begin{array}{ccccccc}
 5 & 0 & 0 & 2 & 0 & 0 & 0 \\
 0 & 2 & 2 & 4 & 2 & 2 & 0 \\
\end{array}
\right)$ & $\mathbf{2^3}$ & 6 \\
\addlinespace[1.5ex]
 $\left(
\begin{array}{ccccccc}
 5 & 0 & 1 & 0 & 1 & 0 & 0 \\
 0 & 2 & 2 & 3 & 2 & 2 & 0 \\
\end{array}
\right)$ & $\mathbf{2^2,1^2}$ & 5 \\
\addlinespace[1.5ex]
 $\left(
\begin{array}{ccccccc}
 5 & 1 & 0 & 0 & 0 & 1 & 0 \\
 0 & 2 & 1 & 2 & 1 & 2 & 0 \\
\end{array}
\right)$ & ${2,1^4}$ & 3 \\
\addlinespace[1.5ex]
 $\left(
\begin{array}{ccccccc}
 6 & 0 & 0 & 0 & 0 & 0 & 1 \\
 0 & 1 & 0 & 1 & 0 & 1 & 0 \\
\end{array}
\right)$ & $\mathbf{1^6}$ & 0 \\	
	\bottomrule
	\end{tabular}
	\end{subfigure}
		\caption{Results obtained applying the matrix formalism to $\mathfrak{sp}(3)$.}
\end{table}

\begin{table}[t]
	\centering
	\begin{subfigure}[]{0.3\textwidth}
	    \centering
	
	\end{subfigure}
	\begin{subfigure}[]{0.5\textwidth}
	\begin{tabular}{ c c c }
	\toprule
	\textbf{Matrix} & \textbf{Partition} & $dim$ \\ 
	\midrule \addlinespace[1.5ex]
\addlinespace[1.5ex]
 $\left(
\begin{array}{ccccccccc}
 1 & 8 & 0 & 0 & 0 & 0 & 0 & 0 & 0 \\
 0 & 8 & 6 & 6 & 4 & 4 & 2 & 2 & 0 \\
\end{array}
\right)$ & $\mathbf{8}$ & 16 \\
\addlinespace[1.5ex]
 $\left(
\begin{array}{ccccccccc}
 3 & 4 & 2 & 0 & 0 & 0 & 0 & 0 & 0 \\
 0 & 6 & 6 & 6 & 4 & 4 & 2 & 2 & 0 \\
\end{array}
\right)$ & $\mathbf{6,2}$ & 15 \\
\addlinespace[1.5ex]
 $\left(
\begin{array}{ccccccccc}
 3 & 5 & 0 & 1 & 0 & 0 & 0 & 0 & 0 \\
 0 & 6 & 5 & 6 & 4 & 4 & 2 & 2 & 0 \\
\end{array}
\right)$ & ${6,1^2}$ & 14 \\
\addlinespace[1.5ex]
 $\left(
\begin{array}{ccccccccc}
 5 & 0 & 4 & 0 & 0 & 0 & 0 & 0 & 0 \\
 0 & 4 & 6 & 6 & 4 & 4 & 2 & 2 & 0 \\
\end{array}
\right)$ & $\mathbf{4^2}$ & 14 \\
\addlinespace[1.5ex]
 $\left(
\begin{array}{ccccccccc}
 5 & 2 & 0 & 2 & 0 & 0 & 0 & 0 & 0 \\
 0 & 4 & 4 & 6 & 4 & 4 & 2 & 2 & 0 \\
\end{array}
\right)$ & $\mathbf{4,2^2}$ & 13 \\
\addlinespace[1.5ex]
 $\left(
\begin{array}{ccccccccc}
 5 & 2 & 1 & 0 & 1 & 0 & 0 & 0 & 0 \\
 0 & 4 & 4 & 5 & 4 & 4 & 2 & 2 & 0 \\
\end{array}
\right)$ & $\mathbf{4,2,1^2}$ & 12 \\
\addlinespace[1.5ex]
 $\left(
\begin{array}{ccccccccc}
 6 & 0 & 1 & 2 & 0 & 0 & 0 & 0 & 0 \\
 0 & 3 & 4 & 6 & 4 & 4 & 2 & 2 & 0 \\
\end{array}
\right)$ & $\mathbf{3^2,2}$ & 12 \\
\bottomrule
	\end{tabular}
	\end{subfigure}
		\caption{Results obtained applying the matrix formalism to $\mathfrak{sp}(4)$.}
\end{table}

\begin{table}[t]
	\centering
	\begin{subfigure}[]{0.3\textwidth}
	    \centering
	
	\end{subfigure}
	\begin{subfigure}[]{0.5\textwidth}
	\fontsize{8.5}{10}\selectfont
	\begin{tabular}{ c c c }
	\toprule
	\textbf{Matrix} & \textbf{Partition} & $dim$ \\ 
	\midrule \addlinespace[1ex]
 $\left(
\begin{array}{ccccccccccc}
 1 & 10 & 0 & 0 & 0 & 0 & 0 & 0 & 0 & 0 & 0 \\
 0 & 10 & 8 & 8 & 6 & 6 & 4 & 4 & 2 & 2 & 0 \\
\end{array}
\right)$ & $\mathbf{10}$ & 25 \\
\addlinespace[1ex]
 $\left(
\begin{array}{ccccccccccc}
 3 & 6 & 2 & 0 & 0 & 0 & 0 & 0 & 0 & 0 & 0 \\
 0 & 8 & 8 & 8 & 6 & 6 & 4 & 4 & 2 & 2 & 0 \\
\end{array}
\right)$ & $\mathbf{8,2}$ & 24 \\
\addlinespace[1ex]
 $\left(
\begin{array}{ccccccccccc}
 3 & 7 & 0 & 1 & 0 & 0 & 0 & 0 & 0 & 0 & 0 \\
 0 & 8 & 7 & 8 & 6 & 6 & 4 & 4 & 2 & 2 & 0 \\
\end{array}
\right)$ & ${8,1^2}$ & 23 \\
\addlinespace[1ex]
 $\left(
\begin{array}{ccccccccccc}
 5 & 2 & 4 & 0 & 0 & 0 & 0 & 0 & 0 & 0 & 0 \\
 0 & 6 & 8 & 8 & 6 & 6 & 4 & 4 & 2 & 2 & 0 \\
\end{array}
\right)$ & $\mathbf{6,4}$ & 23 \\
\addlinespace[1ex]
 $\left(
\begin{array}{ccccccccccc}
 5 & 4 & 0 & 2 & 0 & 0 & 0 & 0 & 0 & 0 & 0 \\
 0 & 6 & 6 & 8 & 6 & 6 & 4 & 4 & 2 & 2 & 0 \\
\end{array}
\right)$ & $\mathbf{6,2^2}$ & 22 \\
\addlinespace[1ex]
 $\left(
\begin{array}{ccccccccccc}
 6 & 0 & 5 & 0 & 0 & 0 & 0 & 0 & 0 & 0 & 0 \\
 0 & 5 & 8 & 8 & 6 & 6 & 4 & 4 & 2 & 2 & 0 \\
\end{array}
\right)$ & $\mathbf{5^2}$ & 22 \\
\addlinespace[1ex]
 $\left(
\begin{array}{ccccccccccc}
 7 & 0 & 2 & 2 & 0 & 0 & 0 & 0 & 0 & 0 & 0 \\
 0 & 4 & 6 & 8 & 6 & 6 & 4 & 4 & 2 & 2 & 0 \\
\end{array}
\right)$ & $\mathbf{4^2,2}$ & 21 \\
\bottomrule
	\end{tabular}
	\end{subfigure}
		\caption{Results obtained applying the matrix formalism to $\mathfrak{sp}(5)$.}
\end{table}

\clearpage

\subsection{Recovering the branes}

In order to recover the brane systems and the corresponding quivers from the matrices, the nature of the O3-planes needs to be restored. Since KP transitions don't change the nature of the rightmost orientifold plane, one can set the rightmost orientifold plane of the system to be the same as the rightmost orientifold plane of the initial matrix (corresponding to the closure of the maximal nilpotent orbit). This choice fixes the rest of the O3-planes in the system. Therefore the rightmost orientifold planes for a given matrix in the results above depends in the lie algebra for which the nilpotent orbits are being studied:
\begin{align}
	\begin{aligned}
		\gso(2n+1)&\rightarrow O3^-\\
		\gsp(n)&\rightarrow \widetilde{O3^+}\\
		\gso(2n)&\rightarrow O3^-\\
	\end{aligned}
\end{align}

Table \ref{tab:so7example} illustrates this process, the recovery of the brane systems and corresponding quivers, for the matrices of $\gso (7)$ obtained in the section above.

\section{Conclusions and Outlook}\label{sec:14}

This work shows the physical realization of the mathematical results of Kraft and Procesi \cite{KP82}. This is an interesting link between quantum field theory and the geometry of Lie algebras: the Higgs mechanism can be utilized to reproduce the \emph{slicing}, in the Brieskorn-Slodowy sense \cite{B70,Sl80}, of the moduli space of the $3d\ \mathcal N=4$ theory. This Higgs mechanism has a clear interpretation in terms of the brane dynamics of the superstring embedding of the quantum field theory. 

The current analysis utilizes this Higgs mechanism, that has been given the name of Kraft-Procesi transition, to delve into the geometry of moduli spaces that are closures of nilpotent orbits. However, there is the potential of studying different spaces employing the same technique: The KP transition can now be applied to many other models, in particular, quiver gauge theories whose moduli spaces are extensions of nilpotent orbit closures (the chiral ring has generators with spin under $SU(2)_R$ higher than $s=1$).

There are many challenges ahead: for example, \emph{mirror symmetry} is still not fully realized for the orthosymplectic models described in this paper. The brane systems that we have studied always produce a pair of candidate mirror orthosymplectic quivers, up to the choice of $SO/O(n)$ groups for the orthogonal gauge nodes. The ambiguity rises because until now there weren't many tools that could probe the difference of choice in the Coulomb or in the Higgs branch. On the present work we always present pairs that have a branch in common, the Higgs branch of one is the Coulomb branch of the other. These branches are typically closures of nilpotent orbits or surface Kleinian singularities. A systematic study of this problem that also computes the remaining two branches of each pair needs to be attempted, this can be along the lines of \cite{CHMZ14,CHZ17} for the Coulomb branch and \cite{HK16, FH16, BP17} for the Higgs branch. 

Another very interesting challenge is to find the explicit effect of the KP transitions in the Hilbert series of the moduli spaces. This could be done at the level of the \emph{unrefined} Hilbert series, the \emph{refined} Hilbert series, or the HWG (\emph{Highest Weight Generating function}).

\paragraph{Acknowledgements} S.C. is specially thankful to Rudolph Kalveks for the countless hours he has spent shedding light over the fascinating world of nilpotent orbits and for sharing his invaluable expertise that has made possible all the efficient computations for the Higgs branches. We would also like to thank Bo Feng,  Giulia Ferlito, Claudio Procesi, Marcus Sperling and Zhenghao Zhong for helpful conversations during the development of this project.  S.C. is supported by an EPSRC DTP studentship EP/M507878/1. A.H. is supported by STFC Consolidated Grant ST/J0003533/1, and EPSRC Programme Grant EP/K034456/1.

\begin{table}[t]
	\centering
	\begin{tabular}{ c c}
	  
		\raisebox{-.5\height}{\begin{tikzpicture}[node distance=48pt]
		\node at (0,0)[]{\large{$\mathfrak{so}(7)$}};
		\node at (0,-0.5) [hasse] (1) [] {};
		\node [hasse] (2) [below of=1] {};
		\node [hasse] (3) [below of=2] {};
		\node [hasse] (4) [below of=3] {};
		\node [hasse] (5) [below of=4] {};
		\node [ns] (6) [below of=5] {};
		\node [hasse] (7) [below of=6] {};
		\draw (1) edge [] node[label=left:$A_5$] {} (2)
			(2) edge [] node[label=left:$D_3$] {} (3)
			(3) edge [] node[label=left:$A_1$] {} (4)
			(4) edge [] node[label=left:$A_1\cup A_1$] {} (5)
			(5) edge [] node[label=left:$A_1$] {} (6)
			(6) edge [] node[label=left:$b_3$] {} (7);
	\end{tikzpicture}}
	&
	\fontsize{10}{11}\selectfont
	\begin{tabular}{ c c c  c}
	\toprule
	$\lambda$ & \textbf{Matrix} & \textbf{Branes} & \textbf{Quiver} \\ 
	\midrule \addlinespace[1.5ex]
$\mathbf{7}$& $\left(
\begin{array}{cccccccc}
 0 & 7 & 0 & 0 & 0 & 0 & 0 & 0 \\
 0 & 6 & 5 & 4 & 3 & 2 & 1 & 0 \\
\end{array}
\right)$ & \raisebox{-.4\height}{\begin{tikzpicture}[scale=0.5, every node/.style={transform shape}]
			\draw[dashed] 	(1,0)--(1,2)
					(5,0)--(5,2)
					(5.5,0)--(5.5,2)
					(6,0)--(6,2)
					(6.5,0)--(6.5,2)
					(7,0)--(7,2)
					(7.5,0)--(7.5,2);
			\draw 	(1.5,1) node[cross] {}
					(2,1) node[cross] {}
					(2.5,1) node[cross] {}
					(3,1) node[cross] {}
					(3.5,1) node[cross] {}
					(4,1) node[cross] {}
					(4.5,1) node[cross] {};
			\draw 	(1.5,1) node[circle,draw] {}
					(2,1) node[circle,draw] {}
					(2.5,1) node[circle,draw] {}
					(3,1) node[circle,draw] {}
					(3.5,1) node[circle,draw] {}
					(4,1) node[circle,draw] {}
					(4.5,1) node[circle,draw] {};
			\draw [dotted] (0.5,1)--(1.5,1)
					(2,1)--(2.5,1)
					(3,1)--(3.5,1)
					(4,1)--(4.5,1);
			\draw 	(5,1)--(5.5,1)
					(6,1)--(6.5,1)
					(7,1)--(7.5,1);
			\draw 	(1,1.3)--(5,1.3)
					(1,1.5)--(5,1.5)
					(1,1.7)--(5,1.7)
					(1,.7)--(5,.7)
					(1,.5)--(5,.5)
					(1,.3)--(5,.3)
					(5,1.4)--(5.5,1.4)
					(5,1.6)--(5.5,1.6)
					(5,.6)--(5.5,.6)
					(5,.4)--(5.5,.4)
					(5.5,1.3)--(6,1.3)
					(5.5,1.5)--(6,1.5)
					(5.5,.7)--(6,.7)
					(5.5,.5)--(6,.5)
					(6,1.4)--(6.5,1.4)
					(6,.6)--(6.5,.6)
					(6.5,1.3)--(7,1.3)
					(6.5,.7)--(7,.7);
		\end{tikzpicture}}
& \raisebox{-.4\height}{\begin{tikzpicture}[scale=0.7, every node/.style={transform shape},node distance=20]
			\tikzstyle{gauge} = [circle, draw];
			\tikzstyle{flavour} = [regular polygon,regular polygon sides=4, draw];
			\node (g1) [gauge,label=below:{$O_1$}] {};
			\node (g2) [gauge,right of=g1,label=below:{$C_1$}] {};
			\node (g3) [gauge,right of=g2,label=below:{$O_3$}] {};
			\node (g4) [gauge,right of=g3,label=below:{$C_2$}] {};
			\node (g5) [gauge,right of=g4,label=below:{$O_5$}] {};
			\node (g6) [gauge,right of=g5,label=below:{$C_3$}] {};
			\node (f6) [flavour,above of=g6,label=above:{$O_7$}] {};
			\draw (g1)--(g2)--(g3)--(g4)--(g5)--(g6)--(f6);
		\end{tikzpicture}}
 \\
  \addlinespace[1.5ex]
 $\mathbf{5,1^2}$& $\left(
\begin{array}{cccccccc}
 2 & 4 & 0 & 1 & 0 & 0 & 0 & 0 \\
 0 & 4 & 4 & 4 & 3 & 2 & 1 & 0 \\
\end{array}
\right)$ & 
		\raisebox{-.4\height}{\begin{tikzpicture}[scale=0.5, every node/.style={transform shape}]
			\draw[dashed] 	(2,0)--(2,2)
					(4.5,0)--(4.5,2)
					(5,0)--(5,2)
					(6,0)--(6,2)
					(6.5,0)--(6.5,2)
					(7,0)--(7,2)
					(7.5,0)--(7.5,2);
			\draw 	(1,1) node[cross] {}
					(1.5,1) node[cross] {}
					(2.5,1) node[cross] {}
					(3,1) node[cross] {}
					(3.5,1) node[cross] {}
					(4,1) node[cross] {}
					(5.5,1) node[cross] {};
			\draw 	(1,1) node[circle,draw] {}
					(1.5,1) node[circle,draw] {}
					(2.5,1) node[circle,draw] {}
					(3,1) node[circle,draw] {}
					(3.5,1) node[circle,draw] {}
					(4,1) node[circle,draw] {}
					(5.5,1) node[circle,draw] {};
			\draw [dotted] (0.5,1)--(1,1)
					(1.5,1)--(2.5,1)
					(3,1)--(3.5,1)
					(4,1)--(5.5,1);
			\draw 	(1,1)--(1.5,1)
					(6,1)--(6.5,1)
					(7,1)--(7.5,1);
			\draw 	(2,1.3)--(4.5,1.3)
					(2,1.5)--(4.5,1.5)
					(2,.7)--(4.5,.7)
					(2,.5)--(4.5,.5)
					(4.5,1.4)--(5,1.4)
					(4.5,1.6)--(5,1.6)
					(4.5,.6)--(5,.6)
					(4.5,.4)--(5,.4)
					(5,1.3)--(6,1.3)
					(5,1.5)--(6,1.5)
					(5,.7)--(6,.7)
					(5,.5)--(6,.5)
					(6,1.4)--(6.5,1.4)
					(6,.6)--(6.5,.6)
					(6.5,1.3)--(7,1.3)
					(6.5,.7)--(7,.7);
		\end{tikzpicture}}
 & \raisebox{-.4\height}{\begin{tikzpicture}[scale=0.7, every node/.style={transform shape},node distance=20]
			\tikzstyle{gauge} = [circle, draw];
			\tikzstyle{flavour} = [regular polygon,regular polygon sides=4, draw];
			\node (g3) [gauge,label=below:{$O_1$}] {};
			\node (g4) [gauge,right of=g3,label=below:{$C_1$}] {};
			\node (g5) [gauge,right of=g4,label=below:{$O_3$}] {};
			\node (g6) [gauge,right of=g5,label=below:{$C_2$}] {};
			\node (f6) [flavour,above of=g6,label=above:{$O_7$}] {};
			\draw (g3)--(g4)--(g5)--(g6)--(f6);
		\end{tikzpicture}} \\
  \addlinespace[1.5ex]
 $\mathbf{3^2,1}$ &$\left(
\begin{array}{cccccccc}
 4 & 0 & 2 & 1 & 0 & 0 & 0 & 0 \\
 0 & 2 & 4 & 4 & 3 & 2 & 1 & 0 \\
\end{array}
\right)$ & 
		\raisebox{-.4\height}{\begin{tikzpicture}[scale=0.5, every node/.style={transform shape}]
			\draw[dashed] 	(3,0)--(3,2)
					(3.5,0)--(3.5,2)
					(5,0)--(5,2)
					(6,0)--(6,2)
					(6.5,0)--(6.5,2)
					(7,0)--(7,2)
					(7.5,0)--(7.5,2);
			\draw 	(1,1) node[cross] {}
					(1.5,1) node[cross] {}
					(2,1) node[cross] {}
					(2.5,1) node[cross] {}
					(4.5,1) node[cross] {}
					(4,1) node[cross] {}
					(5.5,1) node[cross] {};
			\draw 	(1,1) node[circle,draw] {}
					(1.5,1) node[circle,draw] {}
					(2,1) node[circle,draw] {}
					(2.5,1) node[circle,draw] {}
					(4.5,1) node[circle,draw] {}
					(4,1) node[circle,draw] {}
					(5.5,1) node[circle,draw] {};
			\draw [dotted] (0.5,1)--(1,1)
					(1.5,1)--(2,1)
					(2.5,1)--(4,1)
					(4.5,1)--(5.5,1);
			\draw 	(1,1)--(1.5,1)
					(2,1)--(2.5,1)
					(4,1)--(4.5,1)
					(6,1)--(6.5,1)
					(7,1)--(7.5,1);
			\draw 	(3,1.5)--(3.5,1.5)
					(3,.5)--(3.5,.5)
					(3.5,1.4)--(5,1.4)
					(3.5,1.6)--(5,1.6)
					(3.5,.6)--(5,.6)
					(3.5,.4)--(5,.4)
					(5,1.3)--(6,1.3)
					(5,1.5)--(6,1.5)
					(5,.7)--(6,.7)
					(5,.5)--(6,.5)
					(6,1.4)--(6.5,1.4)
					(6,.6)--(6.5,.6)
					(6.5,1.3)--(7,1.3)
					(6.5,.7)--(7,.7);
		\end{tikzpicture}}
 & \raisebox{-.4\height}{\begin{tikzpicture}[scale=0.7, every node/.style={transform shape},node distance=20]
			\tikzstyle{gauge} = [circle, draw];
			\tikzstyle{flavour} = [regular polygon,regular polygon sides=4, draw];
			\node (g5) [gauge,right of=g4,label=below:{$O_3$}] {};
			\node (g6) [gauge,right of=g5,label=below:{$C_2$}] {};
			\node (f6) [flavour,above of=g6,label=above:{$O_7$}] {};
			\draw (g5)--(g6)--(f6);
		\end{tikzpicture}} \\
  \addlinespace[1.5ex]
  $\mathbf{3,2^2}$ &$\left(
\begin{array}{cccccccc}
 4 & 1 & 0 & 2 & 0 & 0 & 0 & 0 \\
 0 & 2 & 3 & 4 & 3 & 2 & 1 & 0 \\
\end{array}
\right)$ & 	\raisebox{-.4\height}{\begin{tikzpicture}[scale=0.5, every node/.style={transform shape}]
			\draw[dashed] 	(3,0)--(3,2)
					(4,0)--(4,2)
					(4.5,0)--(4.5,2)
					(6,0)--(6,2)
					(6.5,0)--(6.5,2)
					(7,0)--(7,2)
					(7.5,0)--(7.5,2);
			\draw 	(1,1) node[cross] {}
					(1.5,1) node[cross] {}
					(2,1) node[cross] {}
					(2.5,1) node[cross] {}
					(5,1) node[cross] {}
					(3.5,1) node[cross] {}
					(5.5,1) node[cross] {};
			\draw 	(1,1) node[circle,draw] {}
					(1.5,1) node[circle,draw] {}
					(2,1) node[circle,draw] {}
					(2.5,1) node[circle,draw] {}
					(5,1) node[circle,draw] {}
					(3.5,1) node[circle,draw] {}
					(5.5,1) node[circle,draw] {};
			\draw [dotted] (0.5,1)--(1,1)
					(1.5,1)--(2,1)
					(2.5,1)--(3.5,1)
					(5,1)--(5.5,1);
			\draw 	(1,1)--(1.5,1)
					(2,1)--(2.5,1)
					(4,1)--(4.5,1)
					(6,1)--(6.5,1)
					(7,1)--(7.5,1);
			\draw 	(3,1.5)--(4,1.5)
					(3,.5)--(4,.5)
					(4,1.4)--(4.5,1.4)
					(4,.6)--(4.5,.6)
					(4.5,1.3)--(6,1.3)
					(4.5,1.5)--(6,1.5)
					(4.5,.7)--(6,.7)
					(4.5,.5)--(6,.5)
					(6,1.4)--(6.5,1.4)
					(6,.6)--(6.5,.6)
					(6.5,1.3)--(7,1.3)
					(6.5,.7)--(7,.7);
		\end{tikzpicture}}
		& \raisebox{-.4\height}{\begin{tikzpicture}[scale=0.7, every node/.style={transform shape},node distance=20]
			\tikzstyle{gauge} = [circle, draw];
			\tikzstyle{flavour} = [regular polygon,regular polygon sides=4, draw];			\node (g5) [gauge,right of=g4,label=below:{$O_1$}] {};
			\node (g6) [gauge,right of=g5,label=below:{$C_2$}] {};
			\node (f6) [flavour,above of=g6,label=above:{$O_7$}] {};
			\draw (g5)--(g6)--(f6);
		\end{tikzpicture}} \\
  \addlinespace[1.5ex]
$\mathbf{3,1^5}$ & $\left(
\begin{array}{cccccccc}
 4 & 2 & 0 & 0 & 0 & 1 & 0 & 0 \\
 0 & 2 & 2 & 2 & 2 & 2 & 1 & 0 \\
\end{array}
\right)$ &  	\raisebox{-.4\height}{\begin{tikzpicture}[scale=0.5, every node/.style={transform shape}]
			\draw[dashed] 	(3,0)--(3,2)
					(4.5,0)--(4.5,2)
					(5,0)--(5,2)
					(5.5,0)--(5.5,2)
					(6,0)--(6,2)
					(7,0)--(7,2)
					(7.5,0)--(7.5,2);
			\draw 	(1,1) node[cross] {}
					(1.5,1) node[cross] {}
					(2,1) node[cross] {}
					(2.5,1) node[cross] {}
					(4,1) node[cross] {}
					(3.5,1) node[cross] {}
					(6.5,1) node[cross] {};
			\draw 	(1,1) node[circle,draw] {}
					(1.5,1) node[circle,draw] {}
					(2,1) node[circle,draw] {}
					(2.5,1) node[circle,draw] {}
					(4,1) node[circle,draw] {}
					(3.5,1) node[circle,draw] {}
					(6.5,1) node[circle,draw] {};
			\draw [dotted] (0.5,1)--(1,1)
					(1.5,1)--(2,1)
					(2.5,1)--(3.5,1)
					(4,1)--(6.5,1);
			\draw 	(1,1)--(1.5,1)
					(2,1)--(2.5,1)
					(7,1)--(7.5,1);
			\draw 	(3,1.5)--(4.5,1.5)
					(3,.5)--(4.5,.5)
					(4.5,1.4)--(5,1.4)
					(4.5,.6)--(5,.6)
					(5,1.5)--(5.5,1.5)
					(5,.5)--(5.5,.5)
					(5.5,1.4)--(6,1.4)
					(5.5,.6)--(6,.6)
					(6,1.3)--(7,1.3)
					(6,.7)--(7,.7);
		\end{tikzpicture}} & \raisebox{-.4\height}{\begin{tikzpicture}[scale=0.7, every node/.style={transform shape},node distance=20]
			\tikzstyle{gauge} = [circle, draw];
			\tikzstyle{flavour} = [regular polygon,regular polygon sides=4, draw];			\node (g5) [gauge,right of=g4,label=below:{$O_1$}] {};
			\node (g6) [gauge,right of=g5,label=below:{$C_1$}] {};
			\node (f6) [flavour,above of=g6,label=above:{$O_7$}] {};
			\draw (g5)--(g6)--(f6);
		\end{tikzpicture}} \\
  \addlinespace[4ex]
${2^2,1^3}$& $\left(
\begin{array}{cccccccc}
 5 & 0 & 1 & 0 & 0 & 1 & 0 & 0 \\
 0 & 1 & 2 & 2 & 2 & 2 & 1 & 0 \\
\end{array}
\right)$ &  & \raisebox{-.4\height}{\begin{tikzpicture}[scale=0.7, every node/.style={transform shape},node distance=20]
			\tikzstyle{gauge} = [circle, draw];
			\tikzstyle{flavour} = [regular polygon,regular polygon sides=4, draw];
			\node (g6) [] {};
			\node (f6) [above of=g6] {};
		\end{tikzpicture}} \\
  \addlinespace[2ex]
 $\mathbf{1^7}$ & $\left(
\begin{array}{cccccccc}
 6 & 0 & 0 & 0 & 0 & 0 & 0 & 1 \\
 0 & 0 & 0 & 0 & 0 & 0 & 0 & 0 \\
\end{array}
\right)$ & 	\raisebox{-.4\height}{\begin{tikzpicture}[scale=0.5, every node/.style={transform shape}]
			\draw[dashed] 	(4,0)--(4,2)
					(4.5,0)--(4.5,2)
					(5,0)--(5,2)
					(5.5,0)--(5.5,2)
					(6,0)--(6,2)
					(6.5,0)--(6.5,2)
					(7,0)--(7,2);
			\draw 	(1,1) node[cross] {}
					(1.5,1) node[cross] {}
					(2,1) node[cross] {}
					(2.5,1) node[cross] {}
					(3,1) node[cross] {}
					(3.5,1) node[cross] {}
					(7.5,1) node[cross] {};
			\draw 	(1,1) node[circle,draw] {}
					(1.5,1) node[circle,draw] {}
					(2,1) node[circle,draw] {}
					(2.5,1) node[circle,draw] {}
					(3,1) node[circle,draw] {}
					(3.5,1) node[circle,draw] {}
					(7.5,1) node[circle,draw] {};
			\draw [dotted] (0.5,1)--(1,1)
					(1.5,1)--(2,1)
					(2.5,1)--(3,1)
					(3.5,1)--(7.5,1);
			\draw 	(1,1)--(1.5,1)
					(2,1)--(2.5,1)
					(3,1)--(3.5,1);
		\end{tikzpicture}} & \raisebox{-.2\height}{\begin{tikzpicture}[scale=0.7, every node/.style={transform shape},node distance=20]
			\tikzstyle{gauge} = [circle, draw];
			\tikzstyle{flavour} = [regular polygon,regular polygon sides=4, draw];
			\node (g6) [] {};
			\node (f6) [flavour,above of=g6,label=above:{$O_7$}] {};
		\end{tikzpicture}} \\   \addlinespace[1.5ex]	
	\bottomrule
	\end{tabular}
	\end{tabular}
		\caption{Example of the recovery of the branes and the quiver with $\M_H=\Or_{\lambda}\subset \mathfrak{so} (7)$ from the results obtained applying the \emph{matrix formalism}. The transition from $(\mathbf{3,1^5})$ to $(\mathbf{1^7})$ is $\tilde b_3$, formed by a composition of $A_1$ and $b_3$.}
		\label{tab:so7example}
\end{table}

\appendix

\section{The Higgs branch computation and the Hilbert series for $D_n$}\label{app:DnHiggs}

The Kleinian singularity
\begin{align}
	D_n:=\mathbb{C}^2/Dic_{n-2}
\end{align}
where $Dic_{k}$ is the \emph{dicyclic group} of order $|Dic_k|=4k$, has the following Hilbert series:

\begin{align}
\begin{aligned}
	H&=\frac{(1-t^{4n-4})}{(1-t^4)(1-t^{2n-2})(1-t^{2n-4})}\\
\end{aligned}
\end{align}

During this work two different $3d\  \mathcal N=4$ orthosymplectic quivers were found whose Higgs branch is $\M_H=D_n$ (table \ref{tab:subregularKPsingularities}): when the Higgs branch brane configuration is an interval between two half D5-branes flanked by $O3^+$ planes and with $2n-2$ half NS5-branes inside; and when the brane system is the same interval but this time flanked by two $\widetilde{O3^+}$ planes instead. Let us examine both possibilities:

\subsection{$O3^+$ planes flanking the interval}

\subsubsection{$D_3$}

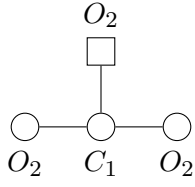
\begin{figure}[t]
	\centering
	\begin{tikzpicture}
		\tikzstyle{gauge} = [circle, draw];
	\tikzstyle{flavour} = [regular polygon,regular polygon sides=4, draw];
	\node (g2) [gauge,label=below:{$O_2$}] {};
	\node (g3) [gauge, right of=g2,label=below:{$C_1$}] {};
	\node (g4) [gauge, right of=g3,label=below:{$O_2$}] {};
	\node (f3) [flavour,above of=g3,label=above:{$O_{2}$}] {};
	\draw (g2)--(g3)--(g4)
			(g3)--(f3)
		;
	\end{tikzpicture}
	\caption{Quiver obtained from brane systems  in the first row of table \ref{tab:subregularKPsingularities}. Its Higgs branch is $\M_H=D_3$.}
	\label{fig:D3slo}
\end{figure}

For $D_3$ the brane configuration has a quiver depicted in figure \ref{fig:D3slo}. The Hilbert series for the Higgs branch of this quiver is computed following the descriptions\footnote{See also \cite{BP17} for a recent discussion on the computation of Higgs branches of orthosymplectic quivers with $O_n$ nodes.} in \cite{HK16}:

\begin{align}
\begin{aligned}
	H(t)=&\oint \frac{dx}{x}\frac{dy}{y}\frac{dz}{z}(1-z^2)
\frac{1}{4}\sum_{\substack{(a_1,a_2)=\{(1,-1),(x,1/x)\}\\(b_1,b_2)=\{(1,-1),(y,1/y)\}}}\left(\frac{PE[2(z+1/z)t]}{PE[(z^2+1+1/z^2)t^2]}\times\right.\\
	&\left.\frac{PE[a_1(z+1/z)t]PE[a_2(z+1/z)t]}{PE[a_1 a_2 t^2]}\times\frac{PE[b_1(z+1/z)t]PE[b_2(z+1/z)t]}{PE[b_1 b_2 t^2]}\right)\\
	=&\frac{1-t^8}{(1-t^4)(1-t^4)(1-t^2)}
\end{aligned}
\end{align}

where $PE[\sum_{i,j} c_{ij}z^it^j]:=\prod_i (1-z^it^j)^{-c_{ij}}$ is the \emph{plethystic exponential}, where $c_{ij}$ are integer numbers and $z$ and $t$ are \emph{fugacities} (see \cite{FHH07}). Note that it is important to perform the $PE[]$ operation \emph{before} evaluating the discrete sum. The fugacities $x$, $z$ and $y$ correspond to the gauge group factors $O_2$, $C_1$ and $O_2$ respectively.

The resulting Hilbert series corresponds to the surface singularity $D_3$.

\subsubsection{$D_4$}

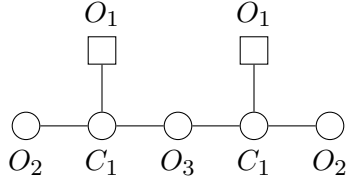
\begin{figure}[t]
	\centering
		\begin{tikzpicture}
	\tikzstyle{gauge} = [circle, draw];
	\tikzstyle{flavour} = [regular polygon,regular polygon sides=4, draw];
	\node (g2) [gauge,label=below:{$O_2$}] {};
	\node (g3) [gauge, right of=g2,label=below:{$C_1$}] {};
	\node (g4) [gauge, right of=g3,label=below:{$O_3$}] {};
	\node (g5) [gauge, right of=g4,label=below:{$C_1$}] {};
	\node (g6) [gauge, right of=g5,label=below:{$O_2$}] {};
	\node (f3) [flavour,above of=g3,label=above:{$O_{1}$}] {};
	\node (f5) [flavour,above of=g5,label=above:{$O_{1}$}] {};
	\draw (g2)--(g3)--(g4)--(g5)--(g6)
			(g3)--(f3)
			(g5)--(f5)
		;
	\end{tikzpicture}
	\caption{Quiver obtained from brane systems  in the first row of table \ref{tab:subregularKPsingularities}. Its Higgs branch is $\M_H=D_4$.}
	\label{fig:D4slo}
\end{figure}

For $n=4$, the Hilbert series of the Higgs branch of the quiver in figure \ref{fig:D4slo} (computed according to \cite{HK16}) gives:

\begin{align}
\begin{aligned}
	H(t)=&\oint \frac{dx}{x}\frac{dy}{y}\frac{dz}{z}\frac{dw}{w}\frac{dp}{p}(1-z^2)(1-w^2)(1-p^2)
\frac{1}{8}\times\\
	&\sum_{\substack{(a_1,a_2)=\{(1,-1),(x,1/x)\}\\(b_1,b_2)=\{(1,-1),(y,1/y)\}\\c=\{1,-1\}}}\left( PE[(a_1+a_2 + 1)(z+z^{-1})t]\times\right.\\
	&PE[c(p^2+1+p^{-2})(z+z^{-1}+w+w^{-1})t+(b_1+b_2 + 1)(w+w^{-1})t]\times\\
	&\left.PE[(a_1 a_2+z^2+1+z^{-2}+p^2+1+p^{-2}+w^2+1+w^{-2}+b_1 b_2)t^2]^{-1}\right)\\
	=&\frac{1-t^{12}}{(1-t^4)(1-t^6)(1-t^4)}
\end{aligned}
\end{align}

Where the fugacities for the gauge group $O_2\times C_1\times O_3\times C_1\times O_2$ are $x,z,p,w,y$ corresponding to the different factors in the same order. This is the Hilbert series of the surface singularity $D_4$.

\subsubsection{$D_5$}

\begin{figure}[t]
	\centering
		\begin{tikzpicture}
	\tikzstyle{gauge} = [circle, draw];
	\tikzstyle{flavour} = [regular polygon,regular polygon sides=4, draw];
	\node (g2) [gauge,label=below:{$O_2$}] {};
	\node (g3) [gauge, right of=g2,label=below:{$C_1$}] {};
	\node (g4) [gauge, right of=g3,label=below:{$O_3$}] {};
	\node (g5) [gauge, right of=g4,label=below:{$C_1$}] {};
	\node (g6) [gauge, right of=g5,label=below:{$O_3$}] {};
	\node (g7) [gauge, right of=g6,label=below:{$C_1$}] {};
	\node (g8) [gauge, right of=g7,label=below:{$O_2$}] {};
	\node (f3) [flavour,above of=g3,label=above:{$O_{1}$}] {};
	\node (f7) [flavour,above of=g7,label=above:{$O_{1}$}] {};
	\draw (g2)--(g3)--(g4)--(g5)--(g6)--(g7)--(g8)
			(g3)--(f3)
			(g7)--(f7)
		;
	\end{tikzpicture}
	\caption{Quiver obtained from brane systems in the first row of table \ref{tab:subregularKPsingularities}. Its Higgs branch is $\M_H=D_5$.}
	\label{fig:D5slo}
\end{figure}
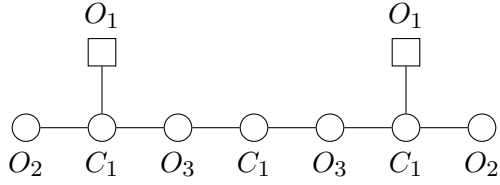

Let the quiver for $D_5$ be depicted in figure \ref{fig:D5slo}. The Hilbert series of the Higgs branch can be computed in a manner analogous to the previous two cases, the result is:
\begin{align}
\begin{aligned}
	H(t)=&\frac{1-t^{16}}{(1-t^4)(1-t^8)(1-t^6)}
\end{aligned}
\end{align}

This time we do not show the integral over the gauge group explicitly, but it is analogous to those in the cases above. This is the Hilbert series for the surface singularity $D_5$.

\subsection{$\widetilde{O3^+}$ planes flanking the interval}

\subsubsection{$D_3$}

\begin{figure}[t]
	\centering
	\begin{tikzpicture}
		\tikzstyle{gauge} = [circle, draw];
	\tikzstyle{flavour} = [regular polygon,regular polygon sides=4, draw];
	\node (g2) [gauge,label=below:{$O_1$}] {};
	\node (g3) [gauge, right of=g2,label=below:{$C_1$}] {};
	\node (g4) [gauge, right of=g3,label=below:{$O_1$}] {};
	\node (f3) [flavour,above of=g3,label=above:{$O_{2}$}] {};
	\draw (g2)--(g3)--(g4)
			(g3)--(f3)
		;
	\end{tikzpicture}
	\caption{Quiver obtained from brane systems  in the second row of table \ref{tab:subregularKPsingularities}. Its Higgs branch is $\M_H=D_3$.}
	\label{fig:D3slo2}
\end{figure}
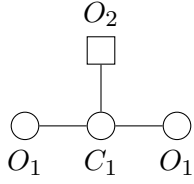

Let us now study the case illustrated in the second row of table \ref{tab:subregularKPsingularities}. For $D_3$ one computes the Hilbert series of the Higgs branch of quiver in figure \ref{fig:D3slo2}, following the descriptions in \cite{HK16}:

\begin{align}
\begin{aligned}
	H(t)=&\oint \frac{dz}{z}(1-z^2)
\frac{1}{4}\sum_{\substack{a=\{1,-1\}\\b=\{1,-1\}}}\left(\frac{PE[(2+a+b)(z+1/z)t]}{PE[(z^2+1+1/z^2)t^2]}\right)\\
	=&\frac{1-t^8}{(1-t^4)(1-t^4)(1-t^2)}
\end{aligned}
\end{align}

The fugacities $a$, $z$ and $b$ correspond to the gauge group factors $O_1$, $C_1$ and $O_1$ respectively. This is the Hilbert series of the surface singularity $D_3$.

\subsubsection{$D_4$}

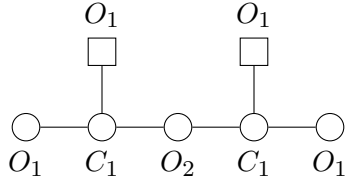
\begin{figure}[t]
	\centering
		\begin{tikzpicture}
	\tikzstyle{gauge} = [circle, draw];
	\tikzstyle{flavour} = [regular polygon,regular polygon sides=4, draw];
	\node (g2) [gauge,label=below:{$O_1$}] {};
	\node (g3) [gauge, right of=g2,label=below:{$C_1$}] {};
	\node (g4) [gauge, right of=g3,label=below:{$O_2$}] {};
	\node (g5) [gauge, right of=g4,label=below:{$C_1$}] {};
	\node (g6) [gauge, right of=g5,label=below:{$O_1$}] {};
	\node (f3) [flavour,above of=g3,label=above:{$O_{1}$}] {};
	\node (f5) [flavour,above of=g5,label=above:{$O_{1}$}] {};
	\draw (g2)--(g3)--(g4)--(g5)--(g6)
			(g3)--(f3)
			(g5)--(f5)
		;
	\end{tikzpicture}
	\caption{Quiver obtained from brane systems  in the second row of table \ref{tab:subregularKPsingularities}. Its Higgs branch is $\M_H=D_4$.}
	\label{fig:D4slo2}
\end{figure}

For $n=4$, the Hilbert series of the Higgs branch of the quiver in figure \ref{fig:D4slo2} gives:

\begin{align}
\begin{aligned}
	H(t)=&\oint \frac{dz}{z}\frac{dw}{w}\frac{dp}{p}(1-z^2)(1-w^2)
\frac{1}{8}\times\\
	&\sum_{\substack{a=\{1,-1\}\\b=\{1,-1\}\\(c_1,c_2)=\{(1,-1),(p,1/p)\}}}\left( PE[(a + 1)(z+z^{-1})t]\times\right.\\
	&PE[(c_1+c_2)(z+z^{-1}+w+w^{-1})t+(b+1)(w+w^{-1})t]\times\\
	&\left.PE[(z^2+1+z^{-2}+w^2+1+w^{-2}+c_1c_2)t^2]^{-1}\right)\\
	=&\frac{1-t^{12}}{(1-t^4)(1-t^6)(1-t^4)}
\end{aligned}
\end{align}

Where the fugacities for the gauge group $O_1 \times C_1\times O_2\times C_1\times O_1$ are $a,z,p,w,b$ corresponding to the different factors in the same order. This is the Hilbert series of the surface singularity $D_4$.

\subsubsection{$D_5$}

\begin{figure}[t]
	\centering
		\begin{tikzpicture}
	\tikzstyle{gauge} = [circle, draw];
	\tikzstyle{flavour} = [regular polygon,regular polygon sides=4, draw];
	\node (g2) [gauge,label=below:{$O_1$}] {};
	\node (g3) [gauge, right of=g2,label=below:{$C_1$}] {};
	\node (g4) [gauge, right of=g3,label=below:{$O_2$}] {};
	\node (g5) [gauge, right of=g4,label=below:{$C_1$}] {};
	\node (g6) [gauge, right of=g5,label=below:{$O_2$}] {};
	\node (g7) [gauge, right of=g6,label=below:{$C_1$}] {};
	\node (g8) [gauge, right of=g7,label=below:{$O_1$}] {};
	\node (f3) [flavour,above of=g3,label=above:{$O_{1}$}] {};
	\node (f7) [flavour,above of=g7,label=above:{$O_{1}$}] {};
	\draw (g2)--(g3)--(g4)--(g5)--(g6)--(g7)--(g8)
			(g3)--(f3)
			(g7)--(f7)
		;
	\end{tikzpicture}
	\caption{Quiver obtained from brane systems in the second row of table \ref{tab:subregularKPsingularities}. Its Higgs branch is $\M_H=D_5$.}
	\label{fig:D5slo2}
\end{figure}
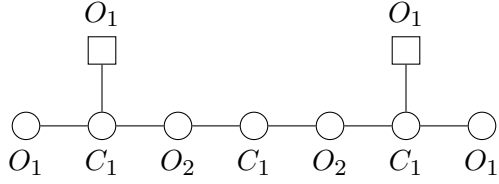

Let the quiver for $D_5$ be depicted in figure \ref{fig:D5slo2}. The Hilbert series of the Higgs branch can be computed in a manner analogous to the previous two cases, the result is:
\begin{align}
\begin{aligned}
	H(t)=&\frac{1-t^{16}}{(1-t^4)(1-t^8)(1-t^6)}
\end{aligned}
\end{align}

This time the integral over the gauge group is not shown explicitly. This is the Hilbert series for the surface singularity $D_5$.

\section{The Higgs branch computation and the Hilbert series for $A_{2n-1}$ singularities}\label{app:AnHiggs}

The surface singularity
\begin{align}
A_{2n-1}:=\mathbb{C}^2/\mathbb{Z}_{2n}
\end{align}
has a Hilbert series of the form:
\begin{align}
\begin{aligned}
	H&=\frac{1-t^{4n}}{(1-t^2)(1-t^{2n})(1-t^{2n})} \\
\end{aligned}
\end{align}

\subsection{$A_3$}

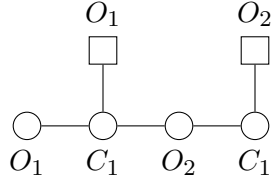
\begin{figure}[t]
	\centering
		\begin{tikzpicture}
	\tikzstyle{gauge} = [circle, draw];
	\tikzstyle{flavour} = [regular polygon,regular polygon sides=4, draw];
	\node (g2) [gauge,label=below:{$O_1$}] {};
	\node (g3) [gauge, right of=g2,label=below:{$C_1$}] {};
	\node (g4) [gauge, right of=g3,label=below:{$O_2$}] {};
	\node (g5) [gauge, right of=g4,label=below:{$C_1$}] {};	
	\node (f3) [flavour,above of=g3,label=above:{$O_{1}$}] {};
	\node (f5) [flavour,above of=g5,label=above:{$O_{2}$}] {};
	\draw (g2)--(g3)--(g4)--(g5)
			(g3)--(f3)
			(g5)--(f5)
		;
	\end{tikzpicture}
	\caption{Quiver obtained from brane system in the third row of table \ref{tab:subregularKPsingularities}. Its Higgs branch is $\M_H=A_3$.}
	\label{fig:A3slo}
\end{figure}

For $A_3$ one computes the Hilbert series of the Higgs branch of quiver in figure \ref{fig:A3slo}, following the descriptions in \cite{HK16}, the result is:

\begin{align}
\begin{aligned}
	H(t)=&\oint \frac{dz}{z}\frac{dw}{w}\frac{dp}{p}(1-z^2)(1-w^2)
\frac{1}{4} \sum_{\substack{a=\{1,-1\}\\(b_1,b_2)=\{(1,-1),(p,1/p)\}}}\left( PE[(a+1)(z+z^{-1})t]\times\right.\\
	&PE[(b_1+b_2)(z+z^{-1}+w+w^{-1})t+2(w+w^{-1})t]\times\\
	&\left.PE[(z^2+1+z^{-2}+b_1 b_2+w^2+1+w^{-2})t^2]^{-1}\right)\\
	=&\frac{1-t^{8}}{(1-t^2)(1-t^4)(1-t^4)}
\end{aligned}
\end{align}

The fugacities $z$, $p$ and $w$ correspond to the gauge group factors $C_1\times O_2 \times C_1$ in the same order. This is the Hilbert series of the surface singularity $A_3$.

\subsection{$A_5$}

\begin{figure}[t]
	\centering
		\begin{tikzpicture}
	\tikzstyle{gauge} = [circle, draw];
	\tikzstyle{flavour} = [regular polygon,regular polygon sides=4, draw];
	\node (g2) [gauge,label=below:{$O_1$}] {};
	\node (g3) [gauge, right of=g2,label=below:{$C_1$}] {};
	\node (g4) [gauge, right of=g3,label=below:{$O_2$}] {};
	\node (g5) [gauge, right of=g4,label=below:{$C_1$}] {};	
	\node (g6) [gauge, right of=g5,label=below:{$O_2$}] {};
	\node (g7) [gauge, right of=g6,label=below:{$C_1$}] {};	
	\node (f3) [flavour,above of=g3,label=above:{$O_{1}$}] {};
	\node (f7) [flavour,above of=g7,label=above:{$O_{2}$}] {};
	\draw (g2)--(g3)--(g4)--(g5)--(g6)--(g7)
			(g3)--(f3)
			(g7)--(f7)
		;
	\end{tikzpicture}
	\caption{Quiver obtained from brane system in the third row of table \ref{tab:subregularKPsingularities}. Its Higgs branch is $\M_H=A_5$.}
	\label{fig:A5slo}
\end{figure}
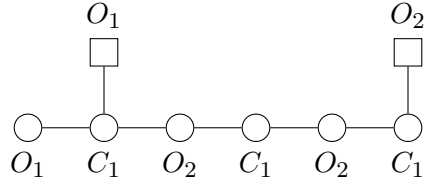

The Hilbert series of the Higgs branch of the quiver in figure \ref{fig:A5slo} gives:

\begin{align}
\begin{aligned}
	H(t)=&\oint \frac{dz}{z}\frac{dw}{w}\frac{dy}{y}\frac{dp}{p}\frac{dq}{q}(1-z^2)(1-w^2)(1-y^2)
\frac{1}{8} \sum_{\substack{a=\{1,-1\}\\(b_1,b_2)=\{(1,-1),(p,1/p)\}\\(c_1,c_2)=\{(1,-1),(q,1/q)\}}}\left( PE[(a+1)(z+z^{-1})t]\times\right.\\
	&PE[(b_1+b_2)(z+z^{-1}+w+w^{-1})t+(c_1+c_2)(w+w^{-1}+y+y^{-1})t+2(y+y^{-1})t]\times\\
	&\left.PE[(z^2+1+z^{-2}+b_1 b_2+w^2+1+w^{-2}+c_1 c_2+y^2+1+y^{-2})t^2]^{-1}\right)\\
	=&\frac{1-t^{12}}{(1-t^2)(1-t^6)(1-t^6)}
\end{aligned}
\end{align}

Where the fugacities for the gauge group $ C_1\times O_2\times C_1\times O_2\times C_1$ are $z,p,w,q,y$ corresponding to the different factors in the same order. This is the Hilbert series of the surface singularity $A_5$.

\section{The Higgs branch computation and the Hilbert series for $A_{2n-1}\cup A_{2n-1}$ singularities}\label{app:AnAnHiggs}

The union of two surface singularities of the form
\begin{align}
A_{2n-1} \cup A_{2n-1}
\end{align}
where the two cones intersect at the origin, has a Hilbert series:
\begin{align}
\begin{aligned}
	H&=\frac{1-t^{4n}}{(1-t^2)(1-t^{2n})(1-t^{2n})} +\frac{1-t^{4n}}{(1-t^2)(1-t^{2n})(1-t^{2n})} -1\\
\end{aligned}
\end{align}

\subsection{$A_3\cup A_3$}

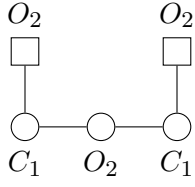
\begin{figure}[t]
	\centering
		\begin{tikzpicture}
	\tikzstyle{gauge} = [circle, draw];
	\tikzstyle{flavour} = [regular polygon,regular polygon sides=4, draw];
	\node (g3) [gauge, right of=g2,label=below:{$C_1$}] {};
	\node (g4) [gauge, right of=g3,label=below:{$O_2$}] {};
	\node (g5) [gauge, right of=g4,label=below:{$C_1$}] {};	
	\node (f3) [flavour,above of=g3,label=above:{$O_{2}$}] {};
	\node (f5) [flavour,above of=g5,label=above:{$O_{2}$}] {};
	\draw (g3)--(g4)--(g5)
			(g3)--(f3)
			(g5)--(f5)
		;
	\end{tikzpicture}
	\caption{Quiver obtained from brane system in the fourth row of table \ref{tab:subregularKPsingularities}. Its Higgs branch is $\M_H=A_3\cup A_3$.}
	\label{fig:A3A3slo}
\end{figure}

For $A_3\cup A_3$ one computes the Hilbert series of the Higgs branch of quiver in figure \ref{fig:A3A3slo}, following the descriptions in \cite{HK16}, the result is:

\begin{align}
\begin{aligned}
	H(t)=&\oint \frac{dz}{z}\frac{dw}{w}\frac{dp}{p}(1-z^2)(1-w^2)
\frac{1}{2} \sum_{\substack{(b_1,b_2)=\{(1,-1),(p,1/p)\}}}\left( PE[2(z+z^{-1})t]\times\right.\\
	&PE[(b_1+b_2)(z+z^{-1}+w+w^{-1})t+2(w+w^{-1})t]\times\\
	&\left.PE[(z^2+1+z^{-2}+b_1b_2+w^2+1+w^{-2})t^2]^{-1}\right)\\
	=&\frac{1-t^{8}}{(1-t^2)(1-t^4)(1-t^4)}+\frac{1-t^{8}}{(1-t^2)(1-t^4)(1-t^4)}-1
\end{aligned}
\end{align}

The fugacities $z$, $p$ and $w$ correspond to the factors of the gauge $C_1\times O_2\times C_1$ in the same order. This is the Hilbert series of surface singularity $A_3\cup A_3$.

\subsection{$A_5\cup A_5$}

\begin{figure}[t]
	\centering
		\begin{tikzpicture}
	\tikzstyle{gauge} = [circle, draw];
	\tikzstyle{flavour} = [regular polygon,regular polygon sides=4, draw];
	\node (g3) [gauge, right of=g2,label=below:{$C_1$}] {};
	\node (g4) [gauge, right of=g3,label=below:{$O_2$}] {};
	\node (g5) [gauge, right of=g4,label=below:{$C_1$}] {};	
	\node (g6) [gauge, right of=g5,label=below:{$O_2$}] {};
	\node (g7) [gauge, right of=g6,label=below:{$C_1$}] {};	
	\node (f3) [flavour,above of=g3,label=above:{$O_{2}$}] {};
	\node (f7) [flavour,above of=g7,label=above:{$O_{2}$}] {};
	\draw (g3)--(g4)--(g5)--(g6)--(g7)
			(g3)--(f3)
			(g7)--(f7)
		;
	\end{tikzpicture}
	\caption{Quiver obtained from brane system in the fourth row of table \ref{tab:subregularKPsingularities}. Its Higgs branch is $\M_H=A_5\cup A_5$.}
	\label{fig:A5A5slo}
\end{figure}
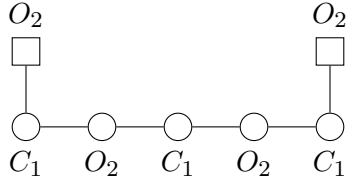

The Hilbert series of the Higgs branch of the quiver in figure \ref{fig:A5A5slo} gives:

\begin{align}
\begin{aligned}
	H(t)=&\oint \frac{dz}{z}\frac{dw}{w}\frac{dy}{y}\frac{dp}{p}\frac{dq}{q}(1-z^2)(1-w^2)(1-y^2)
\frac{1}{4} \sum_{\substack{(b_1,b_2)=\{(1,-1),(p,1/p)\}\\(c_1,c_2)=\{(1,-1),(q,1/q)\}}}\left( PE[2(z+z^{-1})t]\times\right.\\
	&PE[(b_1+b_2)(z+z^{-1}+w+w^{-1})t+(c_1+c_2)(w+w^{-1}+y+y^{-1})t+2(y+y^{-1})t]\times\\
	&\left.PE[(z^2+1+z^{-2}+b_1 b_2+w^2+1+w^{-2}+c_1 c_2+y^2+1+y^{-2})t^2]^{-1}\right)\\
	=&\frac{1-t^{12}}{(1-t^2)(1-t^6)(1-t^6)}+\frac{1-t^{12}}{(1-t^2)(1-t^6)(1-t^6)}-1
\end{aligned}
\end{align}

Where the fugacities for the gauge group $ C_1\times O_2\times  C_1\times O_2\times C_1$ are $z,p,w,q,y$ corresponding to the different factors in the same order. This is the Hilbert series of the surface singularity $A_5\cup A_5$.

\section{The monopole formula and the Hilbert series for $D_n$ and $A_{2n-1}$ singularities}\label{sec:monopole}

\subsection{$D_n$:  $Sp(1)$ with $n$ flavors}\label{sec:CoulombDn}

Let us review the computation of the Hilbert series of the Coulomb branch of the quiver in figure \ref{fig:CDn} utilizing the \emph{monopole formula}\footnote{The reader is directed to \cite{HS16,HananySperling16} for a geometrical interpretation of the monopole formula and an analysis of its algebraic properties.}, \cite{CHZ13}:

\begin{figure}[t]
		\centering
		\begin{tikzpicture}
		\tikzstyle{gauge} = [circle,draw];
			\tikzstyle{flavour} = [regular polygon,regular polygon sides=4,draw];
			\node (g1) [gauge, label=below:{$C_1$}]{};
			\node (f1) [flavour, above of=g1, label=above:{$O_{2n}$}]{};
			\draw (g1)--(f1);
		\end{tikzpicture}
	\caption{Quiver whose Coulomb branch is a $D_{n}$ subregular singularity.}
	\label{fig:CDn}
\end{figure}
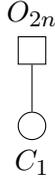

\begin{align}
	H(t)=\sum_{m\in\mathbb{N}} P(t^2,m) t^{2\Delta(m)}
\end{align}
with
\begin{align}
	\Delta(m) = n|m|-2|m|
\end{align}
and $P(t^2,m)$ is the \emph{dressing factor}. For $m=0$ the $Sp(1)$ is unbroken and $P(t^2,0)=PE[t^4]$, for $m\neq 0$ a $U(1)$ group is unbroken and $P(t^2,m\neq 0)=PE[t^2]$. The Weyl group $\mathbb{Z}_2$ is taken into account by only summing over half the integers lattice. 

Hence:
\begin{align}
\begin{aligned}
	H&=\frac{1}{1-t^4}+\frac{1}{1-t^2}\sum_{m=1}^{\infty}  t^{2(n-2)m}\\
	&=\frac{1}{1-t^4}+\frac{1}{(1-t^2)(1-t^{2(n-2)})}-\frac{1}{1-t^2}\\
	&=\frac{1-t^2+t^{2n-2}-t^{2n}}{(1-t^4)(1-t^2)(1-t^{2n-4})}\\
	&=\frac{(1-t^{4n-4})}{(1-t^4)(1-t^{2n-2})(1-t^{2n-4})}\\
\end{aligned}
\end{align}
This is the Hilbert series of the Kleinian singularity:

\begin{align}
	D_n:=\mathbb{C}^2/Dic_{n-2}
\end{align}
Where $Dic_{k}$ is the \emph{dicyclic group} of order $|Dic_k|=4k$.

\subsection{$D_n$:  $O(2)$ with $n-2$ flavors}

\begin{figure}[t]
		\centering
		\begin{tikzpicture}
		\tikzstyle{gauge} = [circle,draw];
			\tikzstyle{flavour} = [regular polygon,regular polygon sides=4,draw];
			\node (g1) [gauge, label=below:{$O_2$}]{};
			\node (f1) [flavour, above of=g1, label=above:{$C_{n-2}$}]{};
			\draw (g1)--(f1);
		\end{tikzpicture}
	\caption{Quiver whose Coulomb branch is a $D_{n}$ subregular singularity.}
	\label{fig:CDn2}
\end{figure}
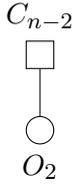

Let us now use the monopole formula \cite{CHZ13,CHMZ14} to compute the Hilbert series of the Coulomb branch for the quiver in figure \ref{fig:CDn2}:
\begin{align}
	H(t)=\sum_{m\in\mathbb{N}} P(t^2,m) t^{2\Delta(m)}
\end{align}
with
\begin{align}
	\Delta(m) = (n-2)|m|
\end{align}
Then:
\begin{align}
\begin{aligned}
	H&=\frac{1}{1-t^4}+\frac{1}{1-t^2}\sum_{m=1}^{\infty}  t^{2(n-2)m}\\
	&=\frac{(1-t^{4n-4})}{(1-t^4)(1-t^{2n-2})(1-t^{2n-4})}\\
\end{aligned}
\end{align}

This is again the Hilbert series for the variety $D_n$.

\subsection{$A_{2n-1}$: $SO(2)$ with $n$ flavors}\label{sec:CoulombAn}

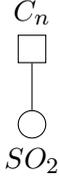
\begin{figure}[t]
		\centering
		\begin{tikzpicture}
		\tikzstyle{gauge} = [circle,draw];
			\tikzstyle{flavour} = [regular polygon,regular polygon sides=4,draw];
			\node (g1) [gauge, label=below:{$SO_2$}]{};
			\node (f1) [flavour, above of=g1, label=above:{$C_n$}]{};
			\draw (g1)--(f1);
		\end{tikzpicture}
	\caption{Quiver whose Coulomb branch is an $A_{2n-1}$ subregular singularity.}
	\label{fig:CAn}
\end{figure}

Let us once again use the monopole formula \cite{CHZ13} to compute the Hilbert series of the Coulomb branch for the quiver in figure \ref{fig:CAn}:
\begin{align}
	H=\sum_{m\in\mathbb{Z}} P(t^2,m) t^{2\Delta(m)}
\end{align}
with
\begin{align}
	\Delta(m) = n|m|
\end{align}
Then:
\begin{align}
\begin{aligned}
	H&=\frac{1}{1-t^2} \left(2 \sum_{m=0}^{\infty}t^{2nm}-1\right)\\
	&=\frac{1}{(1-t^2)} \frac{2}{(1-t^{2n})}-\frac{1}{(1-t^2)} \\
	&=\frac{1-t^{4n}}{(1-t^2)(1-t^{2n})(1-t^{2n})} \\
\end{aligned}
\end{align}

This is the Hilbert series for the variety:
\begin{align}
A_{2n-1}:=\mathbb{C}^2/\mathbb{Z}_{2n}
\end{align}

\bibliography{main}
\bibliographystyle{JHEP}

\end{document}